\documentclass[a4paper,11pt]{article}
\usepackage[ansinew]{inputenc}
\usepackage[T1]{tipa}
\usepackage[german,french]{babel}
\usepackage{lmodern}

\usepackage{epsfig}
\usepackage{amsmath}
\usepackage{amssymb}
\usepackage{latexsym}
\usepackage{ifthen}
\usepackage{graphics}
\usepackage{fancyhdr}
\usepackage{psfrag}

\usepackage[font=small,labelfont=bf]{caption}
\usepackage{float}
\usepackage{subfig}
\newsavebox{\kanaU}
\newcommand{\IAP}{Institut für Angewandte Physik}
\newcommand{\JWG}{Johann Wolfgang Goethe\hspace{.06em}-\hspace{-.06em}Universität}
\newcommand{\mat}[1]{{\bf #1}}
\newcommand{\trans}[1]{{\rm \uppercase{#1}}}
\renewcommand{\vec}[1]{{\boldsymbol{\bf\lowercase{#1}}}}
\renewcommand{\triangle}{\vartriangle\negthickspace}
\newcommand{\arxiv}[1]{}
\title{\underline{Untersuchungen}\\[1ex]
\underline{zur Sprechtraktakustik}\\[4ex] 
       {\Large Dissertation\\
        zur Erlangung des Doktorgrades\\
		der Naturwissenschaften\\
		vorgelegt beim Fachbereich Physik\\
		der \JWG\\
		in Frankfurt am Main\\
		\vspace{2cm}
		von\\
		Frank Ranostaj\\
        aus Frankfurt am Main\\
		\vspace{2cm}
        Frankfurt am Main, 2012 
       }
      }
\date{}

\begin{document}
\selectlanguage{german}
\pagestyle{empty}%
\maketitle%
\arxiv{
\vspace*{6cm}\noindent%
vom Fachbereich Physik der\\%
\vspace*{4cm}\noindent
Johann Wolfgang Goethe-Universität als Dissertation angenommen.\\%
\vspace*{1.5cm}
Dekan: Prof.~Dr.~M.~Huth\\
\vspace*{1.5cm}\noindent
Gutachter: Prof.~Dr.~A.~Lacroix, Prof.~Dr.~H.~Reininger\\
\vspace*{1.5cm}\noindent
Datum der Disputation: \\
\pagebreak
}
\clearpage{\pagestyle{empty}\cleardoublepage}
\pagestyle{fancyplain}
\pagenumbering{roman}
\newpage
\setcounter{page}{1}
\tableofcontents
\newpage
\setcounter{page}{-1}
\pagenumbering{arabic}
\floatstyle{plain}
\newfloat{excerpt}{h}{}[section]
\newfloat{afigure}{H}{}[section]
\newfloat{atable}{h}{}[section]
\floatname{excerpt}{Ausschnitt}
\floatname{afigure}{Bild}
\floatname{atable}{Tabelle}

    \renewcommand{\topfraction}{0.9}	
    \renewcommand{\bottomfraction}{0.8}	
    \setcounter{topnumber}{2}
    \setcounter{bottomnumber}{2}
    \setcounter{totalnumber}{4}     
    \setcounter{dbltopnumber}{2}    
    \renewcommand{\dbltopfraction}{0.9}	
    \renewcommand{\textfraction}{0.07}	
    \renewcommand{\floatpagefraction}{0.7}	
    \renewcommand{\dblfloatpagefraction}{0.7}	


\renewcommand{\textfraction}{0.0}
\newpage
\newcommand{\ws}[0]{\Phi}
\newcounter{Dalpha}
\newcounter{subafigure}
\newcounter{subafigure@save}
\renewcommand{\thesubafigure}{\alph{subafigure}}
\newcounter{linenumbercounter}
\setcounter{linenumbercounter}{0}
\newcommand{\lnr}[0]{%
\addtocounter{linenumbercounter}{1}%
{\tiny\thelinenumbercounter~}}
\newcommand{\name}[1]{{\it#1}}
\newcommand{\Tu}[3]
{
	\expo{#1}{h_x}
	\expo{#2}{h_y}
	\expo{#3}{h_z}
	\Du{#1}{#2}{#3}
}
\newcommand{\expo}[2]
{
	\ifthenelse{ #1 > 0} 
	{ 
		\ifthenelse{ #1 = 1} 
		{#2} {#2^#1}
	}{}
}
\newcommand{\Du}[3]
{
\setcounter{Dalpha}{0}
\addtocounter{Dalpha}{#1}
\addtocounter{Dalpha}{#2}
\addtocounter{Dalpha}{#3}
	\frac
	{ 
		\expo{ \theDalpha }{\partial}\ws
	}
	{
		\expo{#1}{\partial x}
		\expo{#2}{\partial y}
		\expo{#3}{\partial z}
	}
}

\newcommand{\TuT}[3]
{
	\Tu{#1}{#2}{#3} +
	\Tu{#3}{#1}{#2} +
	\Tu{#2}{#3}{#1}
}

\hyphenation{Sprech-traktes}
\hyphenation{Sprech-trakts}
\hyphenation{Sprech-trakt}
\hyphenation{Sprech-trakt-akustik}
\hyphenation{Mehr-tor-adaptoren}
\hyphenation{zeit-unab-hängige}
\hyphenation{akus-tischen}
\hyphenation{Ein-be-ziehung}
\hyphenation{Nach-bildung}
\hyphenation{Resonanz-eigen-schaften}
\hyphenation{Benutzer-schnitt-stelle}
\hyphenation{Schall-erzeugung}
\hyphenation{Mess-systeme}
\hyphenation{An-derer-seits}

\setlength{\unitlength}{1cm}
\newpage
\section{Einleitung}\label{Einleitung}
Das Sprechen ist ein vielschichtiger Vorgang.\footnote{ 
\cite{Le99} hebt beispielweise folgende Schichten hervor: \it 
\begin{itemize} 
\item conceptual preparation, 
\item lexical selection, 
\item phonological encoding, 
\item phonetic encoding,
\item articulation 
\end{itemize}\vspace*{-2.5ex}}
Diese Arbeit betrachtet hierin die Akustik der Sprachentstehung. Physikalisch kann man Sprache als die Schallabstrahlung während des Sprechens, Phonation als die Schallerzeugung, und Artikulation als die zeitliche Variation der sprachformenden Sprechtraktgeometrie sehen. Diese Sicht führt zu einer über zweihundert Jahre alten Zielsetzung der Sprachforschung \cite{Kra1781}:
\begin{quote} {{\it Hae undae sonorae ex larynge in tubam adfixam incidentes inde vario modo et sub variis directionibus reflectuntur, et instar vocis hominum per tubam stentoream propagantur.\\}
{\footnotesize ... Diese Schallwellen aus der Larynx [sollen\footnote{Der Autor räumt an anderer Stelle der Studie ein, dass ihm das Vorhaben noch nicht ganz gelungen sei.}] in dem davorliegenden Rohr in verschiedener Art und unter verschiedenen Richtungen reflektiert werden, so wie sie bei der menschlichen Stimme durch den Sprechtrakt geleitet werden.
}} \end{quote}
Die Schwierigkeiten, dieses Vorgehen umzusetzen und durch Modelle die Sprechtraktakustik nachzuvollziehen, sind in wesentlichen Bereichen die gleichen geblieben.  
Die Artikulation ist wie die Sprache variantenreich und kann sehr dynamisch sein. 
Der Ausgangspunkt der Betrachtung, die Kenntnis der Sprechtraktgeometrie, ist unter anderem deshalb nicht unmittelbar zu erlangen. 
Trotz vielfältiger Fortschritte in der Untersuchungsmethodik ist es noch immer nicht möglich, die Geometrie in drei Raumdimensionen und deren zeitlichen Verlauf vollständig zu erfassen.

Erst durch eine Untergliederung der Sprache in charakteristische Effekte ergeben sich Lautgruppen, für die es gelingt, angepasste Verfahren zur treffenden Bestimmung der Sprechtraktgeometrie zu finden. Zwei Beispiele verdeutlichen dies: Für Vokale und vokalähnliche Laute findet sich eine gute Übereinstimmung der Sprechtraktquerschnittsflächeninhalte und des Sprachsignals, indem man die Ausbreitung ebener Schallwellen längs des Sprechtraktes betrachtet; es gelingt zudem, aus einem Sprachsignal auf die Artikulation und deren zeitlichen Verlauf zurückzuschließen \cite{Sc09}. Andererseits schließt das Modell ebener Schallwellen seiner Definition nach die Betrachtung von Quermoden aus, beispielsweise bei Nasallauten in den Nasengängen. Auch Dämpfungen der Schallwellen augrund der lateralen Querschnittskontur ergeben sich nicht aus diesem Modell. In die entgegengesetzte Richtung zielen Untersuchungen von Raummoden im Schallfeld mittels Finite-Elemente-Approximationen der Sprechtraktgeometrie, die sich häufig auf Tomographien von Kernspinresonanzen stützen. 
Diese Tomographien sind wenig zum Erfassen von dynamischen Vorgängen geeignet, da selbst nur bereichsweise quantitative Messungen stundenlange Messzeiten erfordern.   
Zudem ist der Aufwand für die Umsetzung der tomographischen Daten in eine Diskretisierung durch Finite-Elemente erheblich: Er wird in \cite{Mo02}{} als {\it major obstacle} bezeichnet.

Diese Arbeit liefert Beiträge zur Modellierung des Artikulationsprozesses, die die in den Beispielen aufgezeigten Unzulänglichkeiten an wichtigen Stellen überwinden. Zunächst wird ein Überblick über verschiedene Diskretisierungsmöglichkeiten zur akustischen Untersuchung dreidimensionaler Strukturen gegeben. Ein Verfahren, Finite-Differenzen, wird anschließend genauer betrachtet: Es zeigt sich, dass es die Anforderungen zur Analyse der Artikulation in fast idealer Weise erfüllt.
Für eine möglichst geeignete Datenbasis werden drei verschiedene Tomographie-Methoden, namentlich Kryosektion, Computer- und Kernspinresonanz-Tomo\-graphie, für den Nasalbereich miteinander verglichen -- wobei sich erhebliche Unterschiede in der Qualität der Datensätze zeigen. Gemeinsam ist den Datensätzen eine Untergliederung in sogenannte {\it Voxel}, quaderförmige Raumbereiche, an die die Finiten-Differenzen mittels dem hier entwickelten und als {\it partielle Volumen}{} bezeichneten Verfahren angepasst werden. Mit diesem Vorgehen gelingt die direkte Übernahme der tomographischen Daten ohne Informationsverlust. 

Zur Bestimmung der akustischen Eigenschaften erfolgt die Integration der Wellengleichung im Zeitbereich. Hierbei zeigt sich, dass Erweiterungen wie Wandreibung und Wärmeleitung als lineare Dämpfungsmechanismen der Schallausbreitung im Sprechtrakt in diese Modelle einfach zu integrieren sind und der Nasaltrakt mit hoher Detailtreue modelliert wird. Eine effiziente Implementierung des Finite-Differenzen-Algorithmus hält die Rechenzeit hierfür in Grenzen.

Vokaltraktkonfigurationen in dynamischen Artikulationsphasen werden mit {\sc Speak} ermittelt. {\sc Speak} ist ein im Rahmen dieser Arbeit entwickeltes Programm, das über umfangreiche Analyse-, Synthese- und Visualisierungsmöglichkeiten für typische Prozesse der Sprechakustik verfügt. Der damit bestimmte Verlauf des  Querschnittsflächeninhalts wird mit einer Kontur aus Magnetresonanz-Tomographien versehen, um zu einem wirklichkeitsnahen dreidimensionalen Modell zu gelangen. Für einen Laterallaut wird das Verfahren beispielhaft angewendet. Anhand dieser Ergebnisse werden für weitere Lautgruppen die Vorteile der Herangehensweise aufgezeigt und diskutiert. \\[1ex] 
\noindent
Die in dieser Arbeit verwendeten Begriffe orientieren sich an der Empfehlung \cite{ITG94}.

\pagebreak
\part{Grundlagen}
In den folgenden Abschnitten wird zunächst die gebräuchliche phonetische Segmentierung und eine darauf basierende Klassifizierung von gesprochener Sprache vorgestellt, anhand derer sowohl die Akustik der Lautentstehung als auch die Anatomie des Sprechtraktes erörtert wird. Darauf folgend werden Modelle des Sprechtraktes betrachtet, mit denen zunehmend genauer die Akustik und die relevante Anatomie nachgebildet werden. Anhand einiger Beispiele werden die Erkenntnisse zur Lautentstehung aufgezeigt, die man aus diesen Modellen gewinnt. Das Rohrmodell, welches  vereinfachend die Ausbreitung ebener Schallwellen entlang des Vokaltraktes beschreibt und wesentliche Lauteigenschaften erklärt, wird dann eingehender betrachtet. Abschließend werden ein Überblick über die Morphologie des Nasaltraktes gegeben und verschiedene Methoden diskutiert, die Schallausbreitung hierfür dreidimensional zu berechnen.

\section{Laute}\label{Sprachelemente}
Für eine akustisch motivierte Modellierung des Sprechens ist es naheliegend und hilfreich, zunächst die während des Sprechens ablaufenden artikulatorischen Vorgänge und die sich daraus ergebenden akustischen Effekte zu betrachten und qualitativ zu verstehen. Dazu wird eine Unterteilung der kontinuierlichen Sprachäußerung in Elemente vorgenommen, den Lauten oder Phonen. Die Elemente werden dabei in Lautklassen zusammengefasst, die in den akustischen Effekten differieren, hervorgerufen durch unterschiedliche schallanregende Mechanismen und artikulierende Organe. Ein Querschnitt durch den Sprechtrakt in  Bild \ref{Kopf} zeigt die Lage der  beteiligten Organe. 
\begin{afigure}[t]
\begin{center}
\psfrag{Kop1}[r][r]{\scriptsize Nasaltrakt}
\psfrag{Kop2}[r][r]{\scriptsize Gaumen ({\it Palatum})}
\psfrag{Kop3}[r][r]{\scriptsize Gaumensegel ({\it Velum})}
\psfrag{Kop4}[r][r]{\scriptsize Gaumenzäpfchen ({\it Uvula})}
\psfrag{Kop5}[r][r]{\scriptsize Zunge ({\it Lingua})}
\psfrag{Kop6}[r][r]{\scriptsize Rachen ({\it Pharynx})}
\psfrag{Kop7}[r][r]{\scriptsize Kehldeckel ({\it Epiglottis})}
\psfrag{Kop8}[r][r]{\scriptsize Stimmritze ({\it Glottis})}
\psfrag{Kop9}[r][r]{\scriptsize Luftröhre ({\it Trachea})}
\psfrag{Kop0}[r][r]{\scriptsize Speiseröhre}
\psfrag{Kopp}[r][r]{\scriptsize Kehlkopf ({\it Larynx})}
\psfrag{Kopb}{\scriptsize Lippe ({\it Labium})}
\psfrag{Kopc}{\scriptsize Zähne ({\it Dentes})}
\psfrag{Kopd}{\scriptsize Alveolen}
\epsfig{file=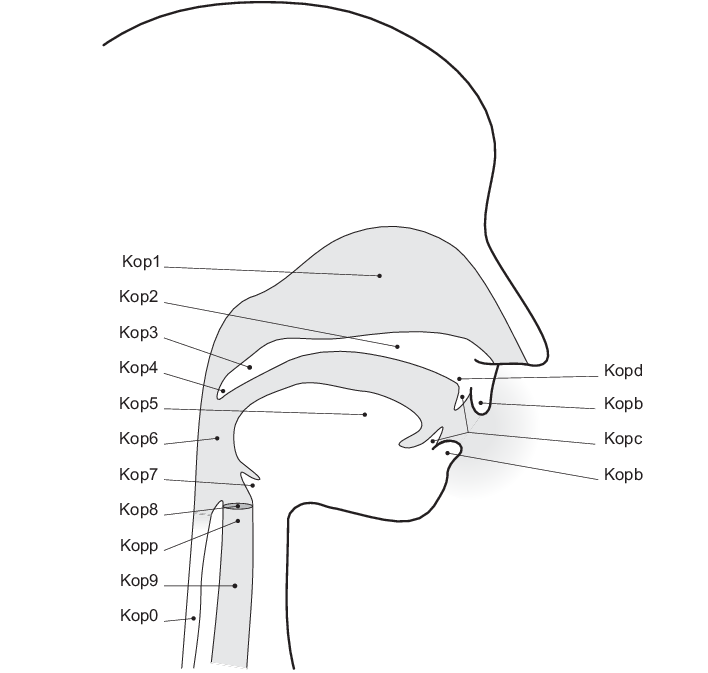, width=.8\linewidth}
\caption{Anatomie des Sprechtrakts}\label{Kopf}
\end{center}
\end{afigure}

Phonetischen Konventionen folgend kann man die Laute in Vokale und Konsonanten unterteilen, wie in dem nachfolgenden Diagramm, Bild~\ref{Vokaltrapez}, und Tabelle \ref{Konsonanten} in der überwiegend verwendeten Notation  des \name{International Phonetic Alphabet} nach \cite{IPA} dargestellt.  Beispiele für die Aussprache der Phone werden in Tabelle \ref{Laute} für die deutsche Sprache gezeigt.  In Tabelle \ref{Konsonanten} erkennt man, wie sich die Konsonanten hinsichtlich Artikulationsort und Artikulationsart unterscheiden.  

Die Unterteilung in Vokale und Konsonanten ist beispielsweise von phonotaktischer und perzeptiver Bedeutung: Vokale tendieren zum Silbenzentrum und weisen eine hohe  Sonorität auf, wie \cite{Si1881, Ze95} verdeutlichen.
Im Hinblick auf ein akustisches Modell des Sprechtrakts ist jedoch eine Gruppierung naheliegender, die sich zunächst an den physikalischen Prozessen der Schallentstehung und im weiteren an der Schallausbreitung orientiert. Die Artikulationsstelle ist dabei nachrangig: sie verschiebt letztlich nur bestimmte Effekte innerhalb einer Gruppe, ohne sie grundlegend zu ändern. Anhand dieser Gruppierung wird im Folgenden ein Überblick über die verschiedenen Laute gegeben.

\subsection{Vokale, Diphthonge und Approximanten}\label{Vok}
\newcommand{\ppair}[2]{\makebox(0,0){#1\hspace{1em}#2}\circle*{.1} }
\stepcounter{footnote}
\begin{afigure}[th]
\begin{center}
\begin{picture}(6,5)\put(0,-.7){
{
	\put(1.5,4){\ppair{i}{y}}
	\put(4,4){\ppair{\textbari}{\textbaru}}
	\put(6.5,4){\ppair{\textturnm}{u}}
	\put(2,3){\ppair{e}{\o}}
	\put(4,3){\ppair{\textreve}{\textbaro}}
	\put(6,3){\ppair{\textramshorns}{o}}
	\put(2.5,2){\ppair{\textepsilon}{\oe}}
	\put(4,2){\ppair{\textrevepsilon}{\textcloserevepsilon}}
	\put(5.5,2){\ppair{\textturnv}{\textopeno}}
	\put(3,1){\ppair{a}{\textscoelig}}
	\put(5,1){\ppair{\textscripta}{\textturnscripta}}

	\put(4,1.5){\makebox(0,0){\textturna}}
	\put(4,2.5){\makebox(0,0){\textschwa}}
	\put(2.5,3.5){\ppair{\textsci}{\textscy}}
	\put(5.5,3.5){\makebox(0,0){\textupsilon}}
	\put(2.75,1.5){\makebox(0,0){\ae\hspace{1.6em}}}

	\put(5,1){\line(1,2){1.5}}
	\put(3,1){\line(-1,2){1.5}}
	\put(4,4){\line(0,-1){1.3}}
	\put(4,1.7){\line(0,1){.6}}
	\put(4,1){\line(0,1){.3}}
	\put(3.5,1){\line(1,0){1}}
	\put(3,2){\line(1,0){.5}}
	\put(4.5,2){\line(1,0){.5}}
	\put(2.5,3){\line(1,0){1}}
	\put(4.5,3){\line(1,0){1}}
	\put(2,4){\line(1,0){1.5}}
	\put(4.5,4){\line(1,0){1.5}}

	\put(1.5,4.6){\makebox(0,0){\scriptsize vorne}}
	\put(4,4.6){\makebox(0,0){\scriptsize mitte}}
	\put(6.5,4.6){\makebox(0,0){\scriptsize hinten}}
	\put(-1.2,1){\makebox(0,0)[l]{\scriptsize tief}}
	\put(-1.2,2){\makebox(0,0)[l]{\scriptsize mitteltief}}
	\put(-1.2,3){\makebox(0,0)[l]{\scriptsize mittelhoch}}
	\put(-1.2,4){\makebox(0,0)[l]{\scriptsize hoch}}
	\put(-1.2,5.08){\makebox(0,0)[l]{\scriptsize Zungenhöhe}}
	\put(4,5.08){\makebox(0,0){\scriptsize Zungenrückenposition}}
}}
\end{picture}
\end{center}
\caption{Vokale. Die Achsen stellen die übliche artikulatorische Vokalmetrik dar: die horizontale Achse repräsentiert die Zungenrückenposition und die vertikale Achse die Zungenhöhe bzw.{} Kieferstellung. Bei paarweise dargestellten Vokalen sind die Lippen bei der Artikulation des linken gespreizt und des rechten gerundet. Trotz gleicher Notation variiert je nach Sprache die Lage der Vokale in dem gezeichneten Vokaltrapez. Die Anzahl der Vokale in einer Sprache kann von 2-3, je nach Klassifizierung, bis über 20 reichen, vgl.{} \cite{GoA06, Tr95}. Entsprechend genau muss der Klang der Vokale unterschieden werden.\\
Diese Darstellung gibt auch Hinweise auf den Klang des Vokals selbst. Der Zusammenhang ergibt sich aus prominenten Frequenzbereichen, die den Klang des Vokals prägen und als Formanten bezeichnet und fortlaufend nummeriert werden. Eine hohe Zungenhöhe bewirkt einen tiefen Formanten F1 und eine tiefe Zungenhöhe einen hohen Formanten F1. Die horizontale Achse repräsentiert den Formanten F2. Hier bewirkt eine vordere Zungenrückenposition einen höheren Formanten F2 und eine hintere Zungenrückenposition einen tieferen Formanten F2.}\label{Vokaltrapez} 
\end{afigure}
In dieser Gruppe erfolgt die Anregung von Schallwellen durch die Schwingung der Stimmbänder, die nahezu periodisch durch einen in der Lunge erzeugten Luftdruck geöffnet und durch den dann entstehenden Luftstrom aufgrund des Bernoulli-Effekts wieder geschlossen werden. Die Schallabstrahlung erfolgt durch den geöffneten Mund.

Artikulatorisch unterscheiden sich die Vokale durch Mundöffnung und \mbox{-rundung} sowie durch die Stellung der Zunge. Die Zungenstellung beeinflusst den Querschnittsverlauf des Rachenraumes, damit die Reflexion der Schallwellen in diesem Bereich und so letztlich den Klang. Die Artikulatoren sind im zeitlichen Zentrum der Äußerung eines Vokals nahezu unbewegt. Die zweidimensionale Darstellung in Bild \ref{Vokaltrapez} verdeutlicht den Einfluss der Zungenstellung auf die Artikulation von Vokalen.\footnote{
Die Entwicklung des Vokaltrapezes zeigt \cite{Ru28}.} 
Die Lautdauer von Vokalen kann kontextabhängig variieren, aber auch bedeutungstragend sein. Ein Beispiel hierfür ist nach \cite{Du06} \glqq Lamm\grqq~[lam] und \glqq lahm\grqq~[la\textlengthmark m], wenngleich in \cite{KrKE64} eine Verschiebung bei gelängter Artikulation zu [l\textscripta\textlengthmark m] erkannt wird. 

Diphthonge sind Vokalübergänge. Wie bei den Vokalen ist das artikulatorische Organ die Zunge, die durch eine gleitende Bewegung den Laut bildet. In der deutschen Sprache gibt es eine Reihe von Diphthonge, die auf [\textturna] enden, beispielsweise in \glqq Ohr\grqq{} [\texttoptiebar{o\textturna}] oder \glqq hart\grqq{} [h\texttoptiebar{a\textturna}t], und drei periphere Diphthonge, [\texttoptiebar{a\textsci}], [\texttoptiebar{a\textupsilon}] und [\texttoptiebar{\textopeno\textscy}], beispielsweise nach \cite{Ko99} in \glqq Eis\grqq, \glqq Haus\grqq~und \glqq Kreuz\grqq. Wie aus den Beispielen zu erkennen ist, bestehen Diphthonge aus zwei Vokalen, die aufgrund ihrer starken Koartikulation zusammengezogen werden: Der erste Vokal liefert die Anfangsstellung und der zweite Vokal die Endstellung der Zunge; der Laut wird durch einen kontinuierlichen Übergang artikuliert. 
 
Approximanten, in der deutschen Sprache nach bspw.~\cite{KrKE64, Ko99} nur durch den Laut [j]  vertreten, unterschieden sich von Vokalen durch eine starke Verengung des Vokaltrakts durch die Zunge. 

\begin{atable}[t]
\begin{center}
\begin{tabular}{c|ccc|ccc|ccc|c}              
IPA&Beispiel& &IPA&Beispiel& &IPA&Beispiel& &IPA&Beispiel\\ \cline{1-2}\cline{4-5} \cline{7-8}  \cline{10-11} 
& & & & & & & & & & \\[-2ex] 
a & hat &&                             \o & \underline{Ö}konom &&          m & Mast &&                                   f & Fass                        \\                                   
\textepsilon & h\underline{ä}tte &&    \oe & g\underline{ö}ttlich &&       n & Naht &&                                   v & \underline{w}as             \\   
\textturna &Ob\underline{er}  &&       u& kulant &&                        \ng & la\underline{ng} &&                     s & Hast                        \\                                   
\textschwa & halt\underline{e} &&      \textupsilon & P\underline{u}lt &&  p & Pakt  &&                                  z & Ha\underline{s}e            \\                                   
e & Methan &&                          \textscy & f\underline{ü}llen &&    b & Ball  &&                                  \textesh & \underline{sch}al    \\ 
i & vital &&                           y& Physik &&                        t & Tal &&                                    \textyogh & \underline{G}enie   \\
\textsci & B\underline{i}rke &&        j & ja &&                           d & dann &&                                   \c{c} & i\underline{ch}         \\          
o & Moral &&                           l & Last &&                         k & kalt &&                                   x & Ba\underline{ch}*  \\    
\textopeno & P\underline{o}st &&         &  &&                                \begin{IPA}\symbol{'147}\end{IPA} & Gunst &&  h & hat                         \\              
\end{tabular}\end{center}
\caption{Beispiele der Realisierung von Lauten in der deutschen Sprache, vgl.{} \cite{Du06}. Die ersten beiden Spalten zeigen Vokale und Approximanten, die dritte Spalte Nasale und Plosive und die letzte Spalte zeigt Frikative. 
Nicht in der Tabelle enthalten ist der glottale Plosiv [\textglotstop]  wie in \glqq Ve\underline{r-}ein\grqq{} und das /r/, welches in verschiedenen Varianten ausgesprochen werden kann. \\
* Hier differieren \cite{Du06} und \cite{Ko99} zumindest in der Notation, letzterer verwendet [\textchi]. 
}\label{Laute}
\end{atable}

\subsection{Nasalvokale, Nasale und Laterallaute}
\begin{atable}[t]
\begin{center}
\tabcolsep4pt
\begin{tabular}{@{\scriptsize}l|c|c|ccc|c|c|c|c|c}
&\scriptsize Bilabial &\scriptsize\shortstack{ Labio-\\dental} &\multicolumn{3}{|c|}{\scriptsize \shortstack{ Dental\\  Alveolar \\  Postalveolar}} 
&\scriptsize\shortstack{Retro-\\flex} &\scriptsize Palatal &\scriptsize Velar &\scriptsize Uvular &\scriptsize Glottal\\ \hline
& & & & & & & & & & \\[-1.5ex] 
Plosiv &p b& &\multicolumn{3}{|c|}{t d}&\textrtailt~\textrtaild  & c \textbardotlessj & k \begin{IPA}\symbol{'147}\end{IPA} & q \textscg &\textglotstop \\
Nasal & m& \textltailm &\multicolumn{3}{|c|}{n} & \textrtailn &\textltailn & \ng & \textscn & ---\\
Vibrant & \textscb & &\multicolumn{3}{|c|}{r}& & & --- & \textscr& ---\\
Tap/Flap& & &\multicolumn{3}{|c|}{\textfishhookr}& \textrtailr & & --- & & ---\\
Frikativ& \textphi~\textbeta & f v & \texttheta~\dh{} & s z &\textesh~\textyogh& \textrtails~\textrtailz & \c{c} \textctj& x~\textgamma& \textchi~\textinvscr &h \texthth \\
\shortstack[l]{Lateral-\\Frikativ} & --- & --- & \multicolumn{3}{|c|}{\textbeltl~\textlyoghlig} & & & & & ---\\
Approximant& & \textscriptv &\multicolumn{3}{|c|}{ \textturnr }&\textturnrrtail & j&\textturnmrleg& & --- \\
\shortstack[l]{Lateral-\\Approximant}& --- & --- &\multicolumn{3}{|c|}{l} &\textrtaill & \textturny & \textscl & & --- \\ \hline
& & & & & & & & & & \\[-1.8ex] 
Klicklaut& \textbullseye & &\textpipe& \textdoublepipe& ! & & \textdoublebarpipe & & &\\
Implosiv& & \texthtb & \multicolumn{2}{|c}{\texthtd} & & &\texthtbardotlessj & \texthtg & \texthtscg & \\ 
\end{tabular}
\end{center}
\caption{Konsonanten. Die Tabelle  gibt eine Übersicht nach Lauterzeugungsart, vertikal, und Lauterzeugungsstelle, horizontal.
Im oberen Teil der Tabelle sind die aus dem Luftstrom der Lunge erzeugten Laute, \name{Pulmonale}, aufgeführt.  Bei gleicher Erzeugungsart und -stelle unterscheiden sie sich als stimmhafte (rechts) und stimmlose (links)  Konsonanten. Ist nur eine Realisierung möglich, so ist diese mit Ausnahme des glottalen Plosivs \textglotstop~stimmhaft. Nicht realisierbare Kombinationen aus Anregungsstelle und -art sind durch einen Querstrich gekennzeichnet.}\label{Konsonanten}
\end{atable}
Auch bei dieser Gruppe von Lauten erfolgt die Schallerzeugung durch die Glottisschwingung. Bei allen drei Lautklassen spaltet sich die Schallausbreitung jedoch im Vokaltrakt auf, was zu Interferenzen führt. Diese Besonderheiten und die daraus resultierenden charakteristischen Eigenschaften werden an verschiedenen Stellen dieser Arbeit erörtert. 

Wie der Name schon andeutet, ist bei Nasalvokalen die Nase involviert: Durch Absenken  des Velums wird für den  Schall eine Passage zu den Nasengängen hin geöffnet, so dass der Schall über den Mund und über die Nase abgestrahlt wird. Sie kommen überwiegend in Wörtern vor, die dem Französischen entlehnt sind. 

Bei Nasalen ist im Unterschied der Mundraum an einer Stelle geschlossen; der Schall wird komplett über die Nase abgestrahlt.  In der deutschen Sprache existieren drei Nasale, [m], [n] und [\ng]. Diese unterscheiden sich durch die Verschlussstelle des Mundraumes, so dass sich jeweils andere Hohlräume  ergeben.  Deren Resonanzen beeinflussen das Spektrum des nasal abgestrahlten Schalls.

Die Lautklasse der Lateral-Approximanten umfasst in der deutschen Sprache lediglich den Laut [l]. Bei seiner Artikulation berührt die Zungenspitze die Alveolen; seitlich an ihr führen zwei Passagen vorbei, die unterhalb und oberhalb der Zunge wieder zusammenlaufen.

\subsection{Plosive, Vibranten, Taps und Flaps}\label{Plosiv}
Diese Laute zeichnen sich durch eine schnelle Zungenbewegung aus, die einen Verschluss des Vokaltrakts bewirkt und diesen unmittelbar darauf wieder freigibt. Dadurch hebt sich diese Lautgruppe von den anderen durch eine charakteristische, stark ausgeprägte Modulation der Schallamplitude ab. Diese ist gut im zeitlichen Verlauf einer Schallaufzeichnung beobachtbar.

Bei einem Plosiv öffnet sich eine verschlossene Stelle im Stimmtrakt durch ein Zusammenspiel von nachlassender Andruckkraft und des durch die Lunge erzeugten Luftdrucks. Durch die so hervorgerufene, schnelle Verschlusslösung erzeugt die vorher angestaute Luft dabei einen explosionsartigen Knall. Plosive können stimmlos oder stimmhaft artikuliert werden, bei letzteren setzt kurz vor oder unmittelbar nach der Verschlusslösung die Glottisschwingung ein.\footnote{Hier differieren die Darstellungen aus  \cite{IPA} und \ref{Kempelen-Plosiv}, S.~\pageref{Kempelen-Plosiv}. In \cite{LiA64} wird gezeigt, dass der Zeitpunkt des Stimmeinsatzes sprecher- und sprachabhängig ist; dies wird auch durch neuere Studien gestützt, vgl.~\cite{BaO98}.} 

Vibranten sind gleichsam periodisch wiederholte Plosive. Dabei wird die Zunge an der Artikulationsstelle angedrückt und mehrfach durch die sich dabei wieder aufstauende Luft gelöst. Im deutschen Sprachraum wird nach \cite{KrKE64} das /r/ auf unterschiedliche Weise realisiert, neben dem standardsprachlichen Frikativ (\name{Engelaut}) als alveolarer oder uvularer Vibrant. Die beiden Vibranten werden umgangssprachlich auch als \glqq gerolltes r\grqq~bezeichnet.

Taps und Flaps werden durch eine einmalige Zungenbewegung gebildet und ähneln insofern den Plosiven. Im Unterschied zu diesen wird jedoch auch die Freigabe des Verschlusses im Wesentlichen durch Muskelkraft bewirkt und der Verschluss muss nicht vollständig sein. Auch perzeptiv tritt die durch den Verschluss bewirkte kurzzeitige Unterbrechung oder Dämpfung des Schalls in den Vordergrund, vgl.~\cite{Ze07}. Beispiele finden sich in europäischen Sprachen mit dem spanischen \glqq pero\grqq~und dem dänischen \glqq rat\grqq, die mit dem Laut [\textfishhookr] gebildet werden, wie in \cite{IPA} ausgeführt wird.  

\subsection{Implosive und Klicklaute}\label{Implosiv}
Auch wenn beide Lautklassen in dieser Arbeit keine weitere Bedeutung haben und in europäischen Sprachen nach \cite{Ma08a} nicht als Phone vorkommen, seien sie übersichtshalber kurz erwähnt. Beide Laute werden nicht von der Lunge aspiriert, ähneln aber in den übrigen diskutierten Eigenschaften den im vorangegangenen Abschnitt beschriebenen Plosiven. 

Die Erzeugung der Implosive erfolgt über einen der Erzeugung der Plosive entgegengesetzten Prozess. Nach der Verschlussbildung wird der Kehlkopf abgesenkt und dadurch ein Unterdruck erzeugt. 

Bei Klicklauten, ihre Artikulation beschreibend auch als Schnalzlaute bezeichnet, wird der Schall durch die Zunge erzeugt. Die Zunge bildet an der Artikulationsstelle durch eine Lösebewegung einen expandierenden Hohlraum, in dem ein Unterdruck entsteht, welcher beim Öffnen das typische Geräusch verursacht. In der deutschen Sprache werden Klicklaute lediglich zum Ausdruck von Sprachgestiken genutzt, wie beispielsweise dem erstaunten missbilligenden \glqq tz tz tz\grqq, aus dem Laut [\textpipe] gebildet. Als Phon kommen sie in afrikanischen Sprachen vor, wie \cite{Tr95, Tr03} ausführen, in einigen zur Unterscheidung von über vierzig Phonemen.

\subsection{Frikative}\label{Frik}
Die Lautklasse der Frikative zeichnet sich  durch eine Verengung im Vokaltrakt aus, die dort zu einer schnelleren Luftströmung und in Folge zu einem Wechsel von einem laminaren  in einen turbulenten Zustand führt. Durch die dabei aperiodisch entstehenden Wirbel wird eine rauschartige Schallanregung erzeugt. Je nach Ort der Verengung werden unterschiedliche Frikative artikuliert. Die in der deutschen Sprache vorkommenden Frikative sind in der Tabelle \ref{Laute} gezeigt. Von diesen weisen die stimmhaften Frikative [v], [z] und [\textyogh] eine zusätzliche periodische Phonation auf, die dem Rauschen überlagert ist. 

\subsection{Akustische Eigenschaften}\label{aktEig}
Durch die Artikulation ändern sich die akustischen Eigenschaften der Laute. Um diesen Effekt zu illustrieren, wird  eine Auswahl von sechs stimmhaften Lauten betrachtet, die in den Bildern \ref{aiu} und \ref{mnl} gezeigt sind. Die stimmhafte Phonation entsteht durch die nahezu periodische Schwingung der Stimmbänder, die die Grundfrequenz bildet und aufgrund der abrupten Wechsel von geöffneter zu geschlossener Phase sehr obertonreich ist. Diese Periodizität lässt sich gut in dem Zeitverlauf der mittels Mikrofon erfassten Schallsignale der verschiedenen Laute erkennen, sie liegt in den hier gezeigten Beispielen zwischen 8 und 12 Millisekunden, was einer Grundfrequenz zwischen 83 und 125 Hertz entspricht. Ebenfalls gut zu erkennen sind die je nach Laut unterschiedlichen Signalverläufe innerhalb einer Periode.

In dem jeweils rechten Diagramm ist der Betrag der diskreten Fouriertransformation einer Periode gezeigt. In diesen Darstellungen im Frequenzbereich ist eine Reihe von Charakteristika sichtbar. Allen Bildern gemeinsam sind die der Periodizität entsprechenden  Kammstrukturen und der Abfall der Betragsgänge zu höheren Frequenzen hin, welcher aus Eigenschaften der Phonation und aus der Schallabstrahlung\footnote{in Abschnitt \ref{Präemphase} wird darauf näher eingegangen} des Vokaltrakts resultiert. Die Betragsgänge der Spektren unterscheiden sich neben einem lautabhängig unterschiedlich stark ausprägten Abfall zu höheren Frequenzen in lauttypischen lokalen Maxima, den Formanten.\footnote{Die Formanten weisen gute Übereinstimmung mit Werten aus \cite{PB52} und \cite{Ma08c} auf, lediglich der zweite Formant des [u] fällt etwas zu hoch im Vergleich zu der Literatur (870 Hz bzw.{} 600 Hz) aus.}   Bei den Nasalen ist ein gleichmäßigerer Verlauf zu erkennen, der aus einer stärkeren Dämpfung der Resonanzen herrührt, wie in dieser Arbeit gezeigt wird.

\begin{afigure}[thbp]
\psfrag{0}{}
\psfrag{1000}[tc][tc]{\scriptsize 1}
\psfrag{2000}[tc][tc]{\scriptsize 2}
\psfrag{3000}[tc][tc]{\scriptsize 3}
\psfrag{4000}[tc][tc]{\scriptsize 4}
\psfrag{5000}[tc][tc]{\scriptsize 5}
\psfrag{6000}[tc][tc]{\scriptsize 6}
\psfrag{7000}[tc][tc]{\scriptsize {\it f}/[kHz]}
\psfrag{8000}[tc][tc]{\scriptsize 8}
\psfrag{0.05}{}
\psfrag{0.1}{}
\psfrag{0.15}{}
\psfrag{0.2}{}
\psfrag{0.3}{}
\psfrag{0.4}{}
\psfrag{0.5}{}
\psfrag{-0.05}{}
\psfrag{-0.1}{}
\psfrag{-0.15}{}
\psfrag{-0.2}{}
\psfrag{-0.3}{}
\begin{center}
\psfrag{5}{}
\psfrag{10}[tc][tc]{\scriptsize 10}
\psfrag{15}{}
\psfrag{20}[tc][tc]{\scriptsize 20}
\psfrag{25}{}
\psfrag{30}[tc][tc]{\scriptsize 30}
\psfrag{35}{}
\psfrag{40}[tc][tc]{\scriptsize \scriptsize {\it t}/[ms]}
\psfrag{45}{}
\psfrag{50}[tc][tc]{\scriptsize 50}
\epsfig{file=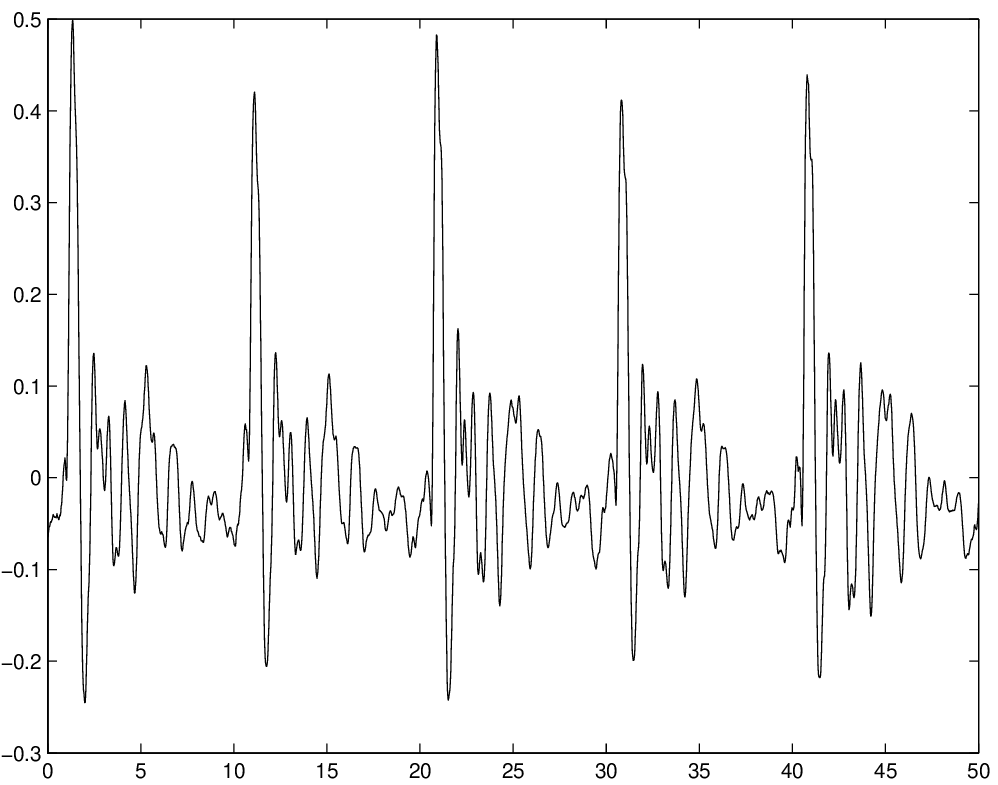, width=0.46\linewidth}
\begin{picture}(0,0)
\put(-5.7,-.06){\scriptsize 0}
\end{picture}
\hspace*{1em}
\psfrag{1}[tc][tc]{\scriptsize 1}
\psfrag{2}[tc][tc]{\scriptsize 2}
\psfrag{3}[tc][tc]{\scriptsize 3}
\psfrag{4}[tc][tc]{\scriptsize 4}
\psfrag{5}[tc][tc]{\scriptsize 5}
\psfrag{6}[tc][tc]{\scriptsize 6}
\psfrag{7}[tc][tc]{\scriptsize {\it f}/[kHz]}
\psfrag{8}[tc][tc]{\scriptsize 8}
\psfrag{10}[r][r]{\scriptsize 10}
\psfrag{20}[r][r]{\scriptsize 20}
\psfrag{30}[r][r]{\scriptsize 30}
\psfrag{40}[r][r]{\scriptsize 40}
\psfrag{50}[r][r]{\scriptsize 50}
\psfrag{60}[r][r]{\scriptsize 60}
\epsfig{file=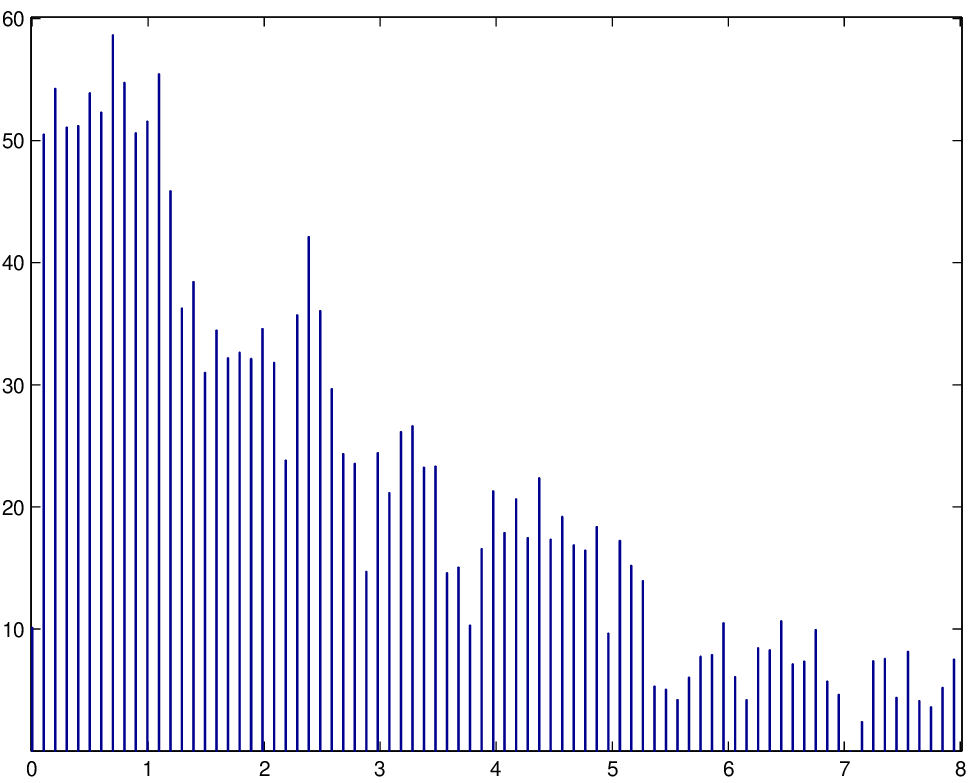, width=0.45\linewidth}
\begin{picture}(0,0)
\put(-5.7,-.06){\scriptsize 0}
\put(-5.83, .1){\scriptsize 0}
\put(-6.2, 3.96){\scriptsize [dB]}
\end{picture}
\vspace*{4ex}
 
\psfrag{5}{}
\psfrag{10}[tc][tc]{\scriptsize 10}
\psfrag{15}{}
\psfrag{20}[tc][tc]{\scriptsize 20}
\psfrag{25}{}
\psfrag{30}[tc][tc]{\scriptsize 30}
\psfrag{35}{}
\psfrag{40}[tc][tc]{\scriptsize \scriptsize {\it t}/[ms]}
\psfrag{45}{}
\psfrag{50}[tc][tc]{\scriptsize 50}
\epsfig{file=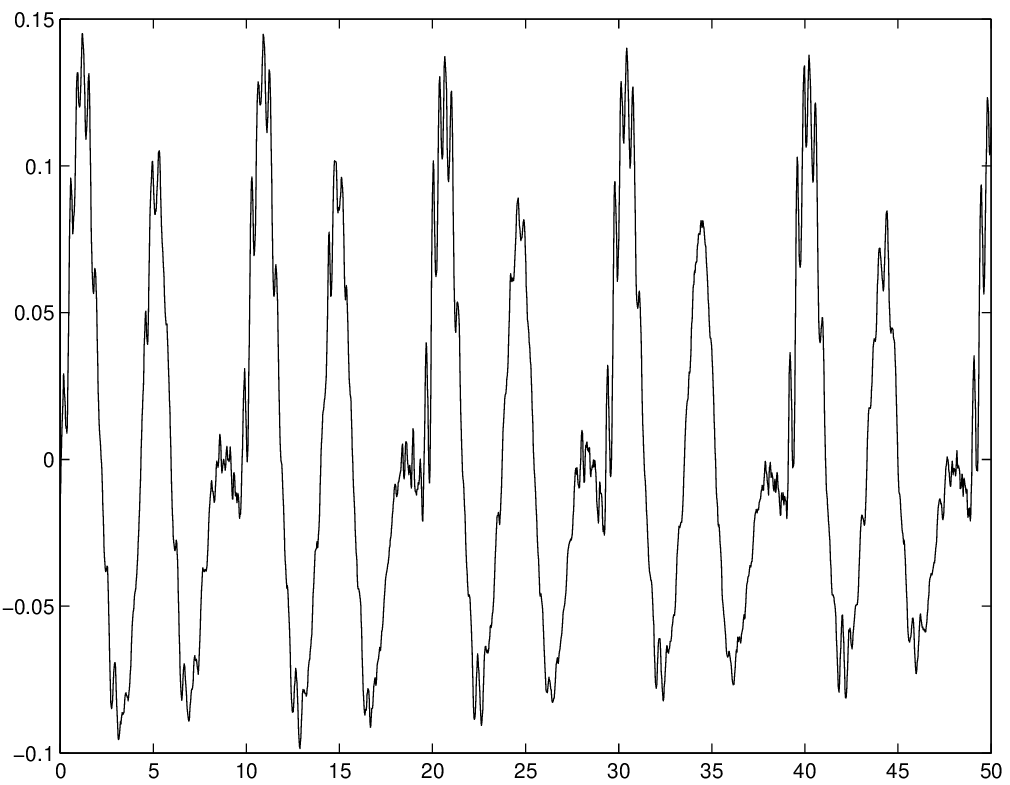, width=0.46\linewidth}
\begin{picture}(0,0)
\put(-5.7,-.06){\scriptsize 0}
\end{picture}
\hspace*{1em}
\psfrag{10}[r][r]{\scriptsize 10}
\psfrag{20}[r][r]{\scriptsize 20}
\psfrag{30}[r][r]{\scriptsize 30}
\psfrag{40}[r][r]{\scriptsize 40}
\psfrag{50}[r][r]{\scriptsize 50}
\psfrag{60}[r][r]{\scriptsize 60}
\epsfig{file=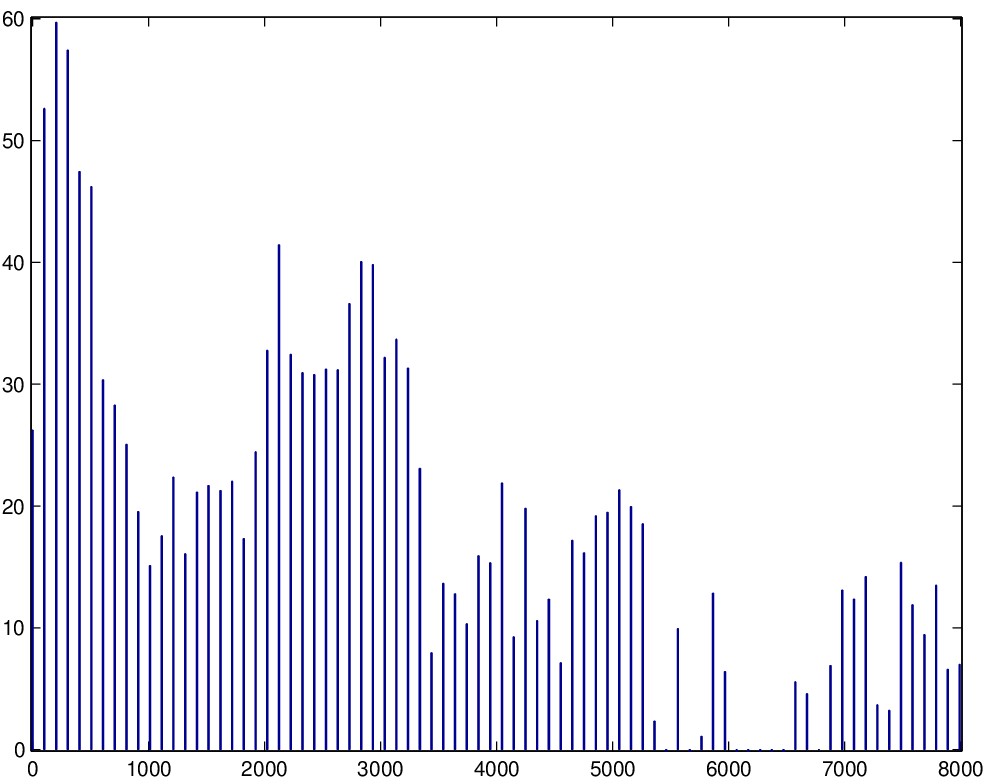, width=0.45\linewidth}
\begin{picture}(0,0)
\put(-5.7,-.06){\scriptsize 0}
\put(-5.83, .1){\scriptsize 0}
\put(-6.2, 3.94){\scriptsize [dB]}
\end{picture}
\vspace*{4ex}

\psfrag{5}{}
\psfrag{10}[tc][tc]{\scriptsize 10}
\psfrag{15}{}
\psfrag{20}[tc][tc]{\scriptsize 20}
\psfrag{25}{}
\psfrag{30}[tc][tc]{\scriptsize 30}
\psfrag{35}{}
\psfrag{40}[tc][tc]{\scriptsize \scriptsize {\it t}/[ms]}
\psfrag{45}{}
\psfrag{50}[tc][tc]{\scriptsize 50}
\epsfig{file=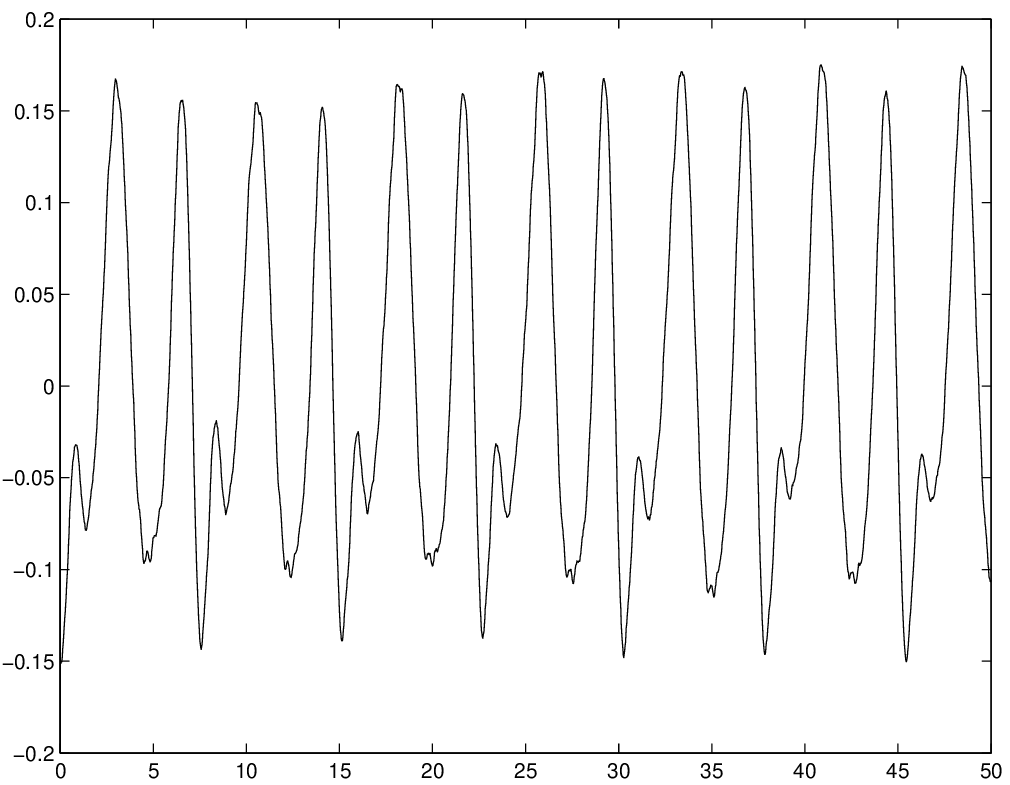, width=0.46\linewidth}
\begin{picture}(0,0)
\put(-5.7,-.06){\scriptsize 0}
\end{picture}
\hspace*{1em}
\psfrag{10}[r][r]{\scriptsize 10}
\psfrag{20}[r][r]{\scriptsize 20}
\psfrag{30}[r][r]{\scriptsize 30}
\psfrag{40}[r][r]{\scriptsize 40}
\psfrag{50}[r][r]{\scriptsize 50}
\psfrag{60}[r][r]{\scriptsize 60}
\epsfig{file=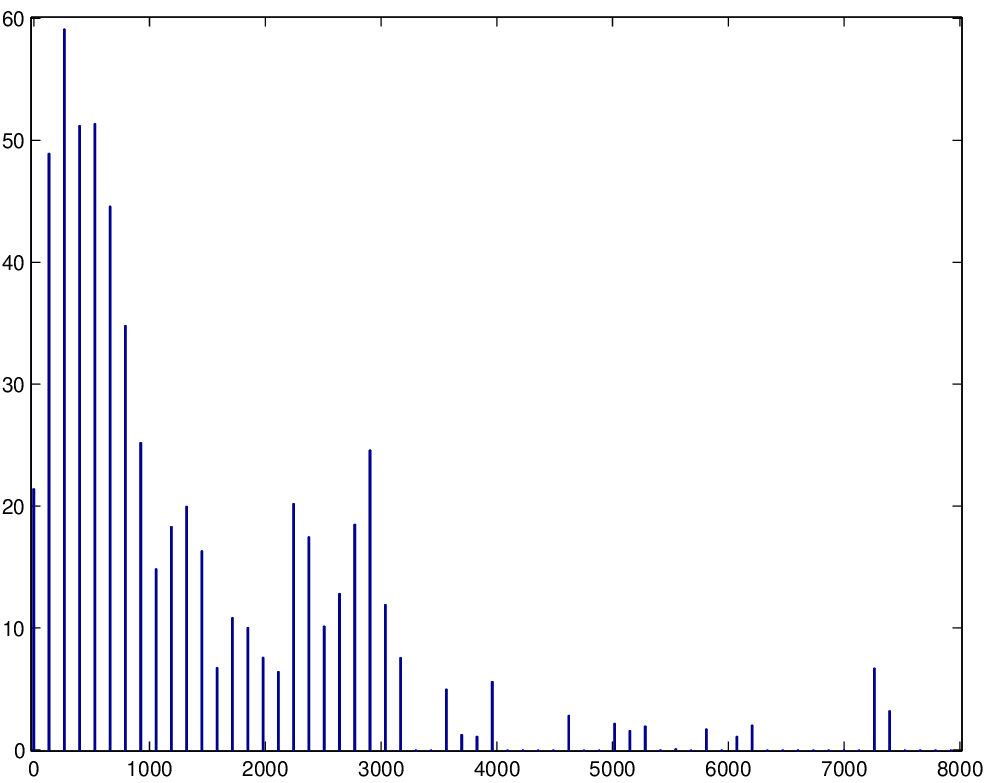, width=0.45\linewidth}
\begin{picture}(0,0)
\put(-5.7,-.06){\scriptsize 0}
\put(-5.83, .1){\scriptsize 0}
\put(-6.2, 3.94){\scriptsize [dB]}
\end{picture}
\vspace*{4ex}
\caption{Zeitverlauf und Spektrum der Vokale [a], [i] und [u], von oben nach unten. Links sind 50~ms des zeitlichen Verlauf des mittels Mikrofon erfassten Signals zu sehen. Rechts daneben sind die logarithmierten Betragsgänge einer Periode daraus gezeigt, 20~dB Unterschied entsprechen einer 10fachen Amplitude. Gut zu erkennen sind die Formanten von [a] bei 700~Hz und 1,1~kHz und 2,4~kHz, von [i] bei 200~Hz, 2,1~kHz und 2,9~kHz sowie von [u] bei 270~Hz, 1,3~kHz und 2,3~kHz.}\label{aiu}
\end{center}
\end{afigure}

\begin{afigure}[thbp]
\psfrag{0}{}
\psfrag{1000}[tc][tc]{\scriptsize 1}
\psfrag{2000}[tc][tc]{\scriptsize 2}
\psfrag{3000}[tc][tc]{\scriptsize 3}
\psfrag{4000}[tc][tc]{\scriptsize 4}
\psfrag{5000}[tc][tc]{\scriptsize 5}
\psfrag{6000}[tc][tc]{\scriptsize 6}
\psfrag{7000}[tc][tc]{\scriptsize {\it f}/[kHz]}
\psfrag{8000}[tc][tc]{\scriptsize 8}
\psfrag{0.05}{}
\psfrag{0.1}{}
\psfrag{0.15}{}
\psfrag{0.2}{}
\psfrag{0.3}{}
\psfrag{0.4}{}
\psfrag{0.5}{}
\psfrag{-0.05}{}
\psfrag{-0.1}{}
\psfrag{-0.15}{}
\psfrag{-0.2}{}
\psfrag{-0.3}{}
\begin{center}
\psfrag{5}{}
\psfrag{10}[tc][tc]{\scriptsize 10}
\psfrag{15}{}
\psfrag{20}[tc][tc]{\scriptsize 20}
\psfrag{25}{}
\psfrag{30}[tc][tc]{\scriptsize 30}
\psfrag{35}{}
\psfrag{40}[tc][tc]{\scriptsize \scriptsize {\it t}/[ms]}
\psfrag{45}{}
\psfrag{50}[tc][tc]{\scriptsize 50}
\epsfig{file=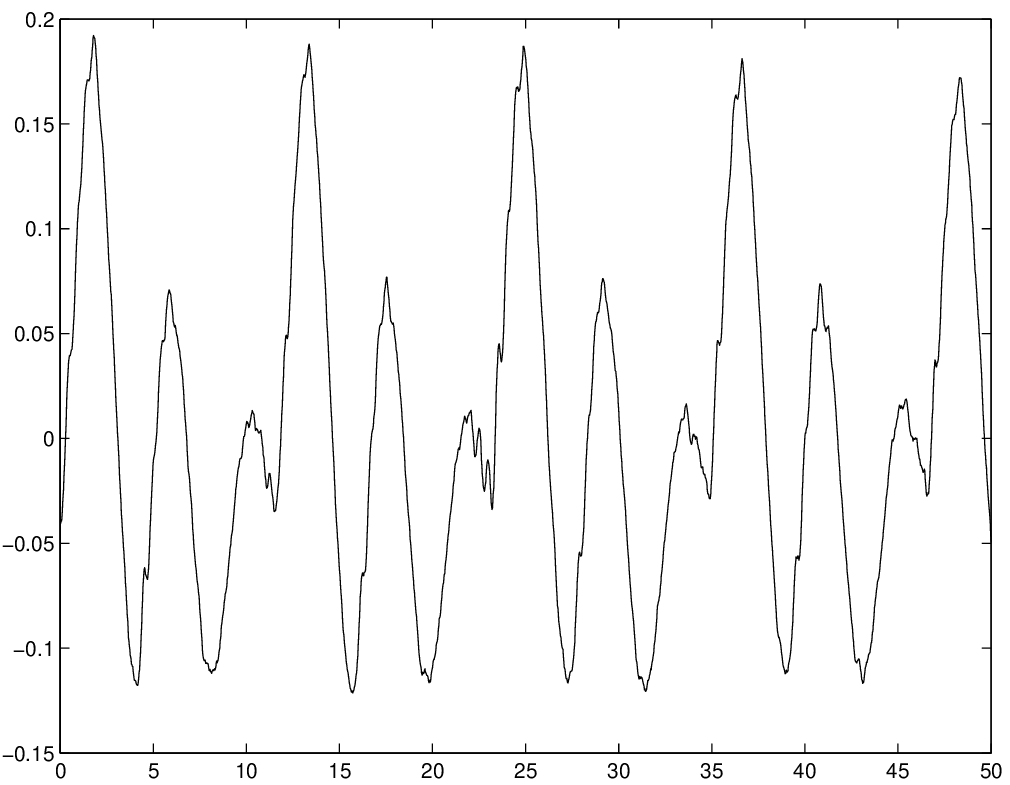, width=0.46\linewidth}
\begin{picture}(0,0)
\put(-5.7,-.06){\scriptsize 0}
\end{picture}
\hspace*{1em}
\psfrag{10}[r][r]{\scriptsize 10}
\psfrag{20}[r][r]{\scriptsize 20}
\psfrag{30}[r][r]{\scriptsize 30}
\psfrag{40}[r][r]{\scriptsize 40}
\psfrag{50}[r][r]{\scriptsize 50}
\psfrag{60}[r][r]{\scriptsize 60}
\epsfig{file=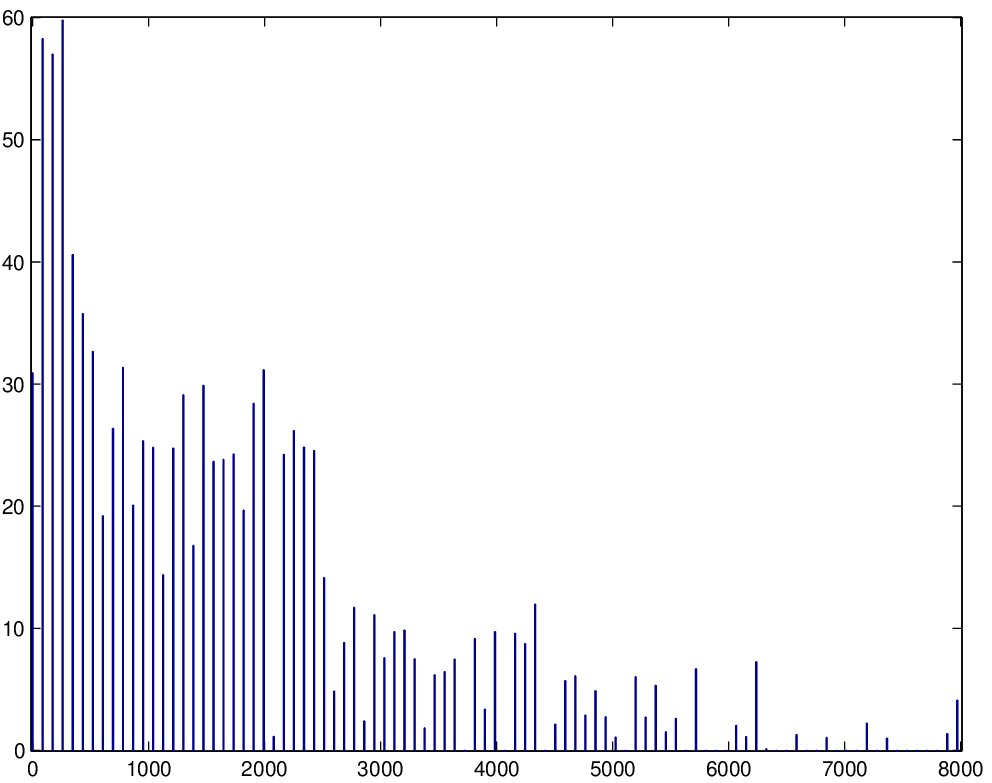, width=0.45\linewidth}
\begin{picture}(0,0)
\put(-5.7,-.06){\scriptsize 0}
\put(-5.83, .1){\scriptsize 0}
\put(-6.2, 3.94){\scriptsize [dB]}
\end{picture}
\vspace*{4ex}
 
\psfrag{5}{}
\psfrag{10}[tc][tc]{\scriptsize 10}
\psfrag{15}{}
\psfrag{20}[tc][tc]{\scriptsize 20}
\psfrag{25}{}
\psfrag{30}[tc][tc]{\scriptsize 30}
\psfrag{35}{}
\psfrag{40}[tc][tc]{\scriptsize {\it t}/[ms]}
\psfrag{45}{}
\psfrag{50}[tc][tc]{\scriptsize 50}
\epsfig{file=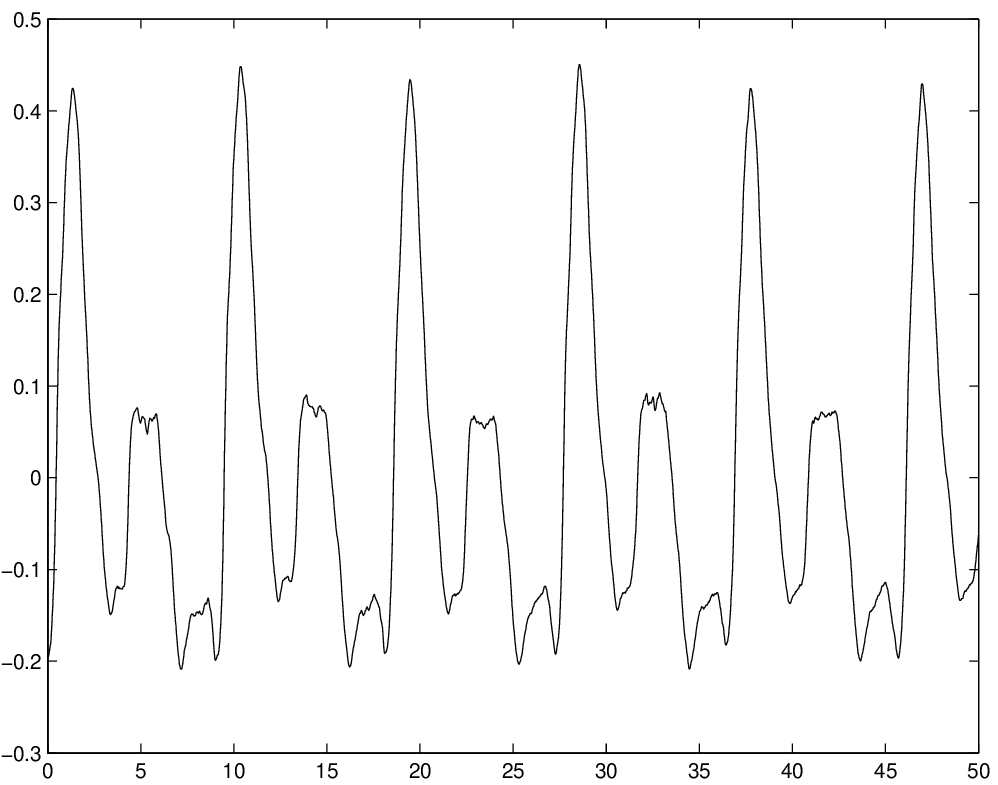, width=0.46\linewidth}
\begin{picture}(0,0)
\put(-5.7,-.06){\scriptsize 0}
\end{picture}
\hspace*{1em}
\psfrag{10}[r][r]{\scriptsize 10}
\psfrag{20}[r][r]{\scriptsize 20}
\psfrag{30}[r][r]{\scriptsize 30}
\psfrag{40}[r][r]{\scriptsize 40}
\psfrag{50}[r][r]{\scriptsize 50}
\psfrag{60}[r][r]{\scriptsize 60}
\epsfig{file=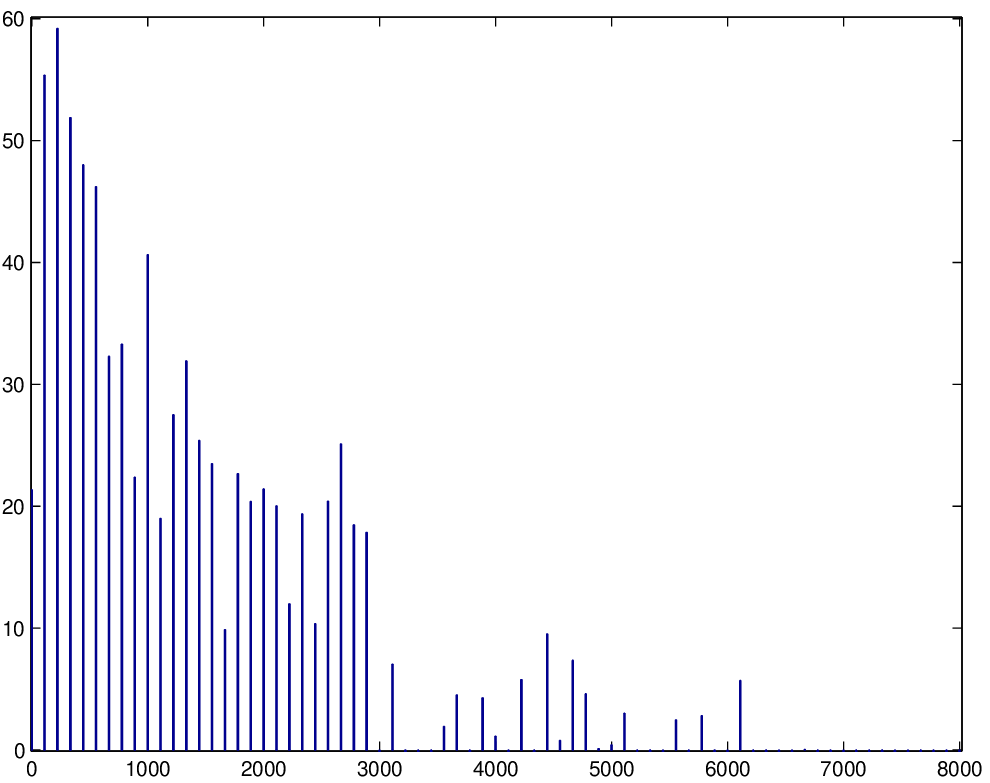, width=0.45\linewidth}
\begin{picture}(0,0)
\put(-5.7,-.06){\scriptsize 0}
\put(-5.83, .1){\scriptsize 0}
\put(-6.2, 3.94){\scriptsize [dB]}
\end{picture}
\vspace*{4ex}

\psfrag{5}{}
\psfrag{10}[tc][tc]{\scriptsize 10}
\psfrag{15}{}
\psfrag{20}[tc][tc]{\scriptsize 20}
\psfrag{25}{}
\psfrag{30}[tc][tc]{\scriptsize 30}
\psfrag{35}{}
\psfrag{40}[tc][tc]{\scriptsize \scriptsize {\it t}/[ms]}
\psfrag{45}{}
\psfrag{50}[tc][tc]{\scriptsize 50}
\epsfig{file=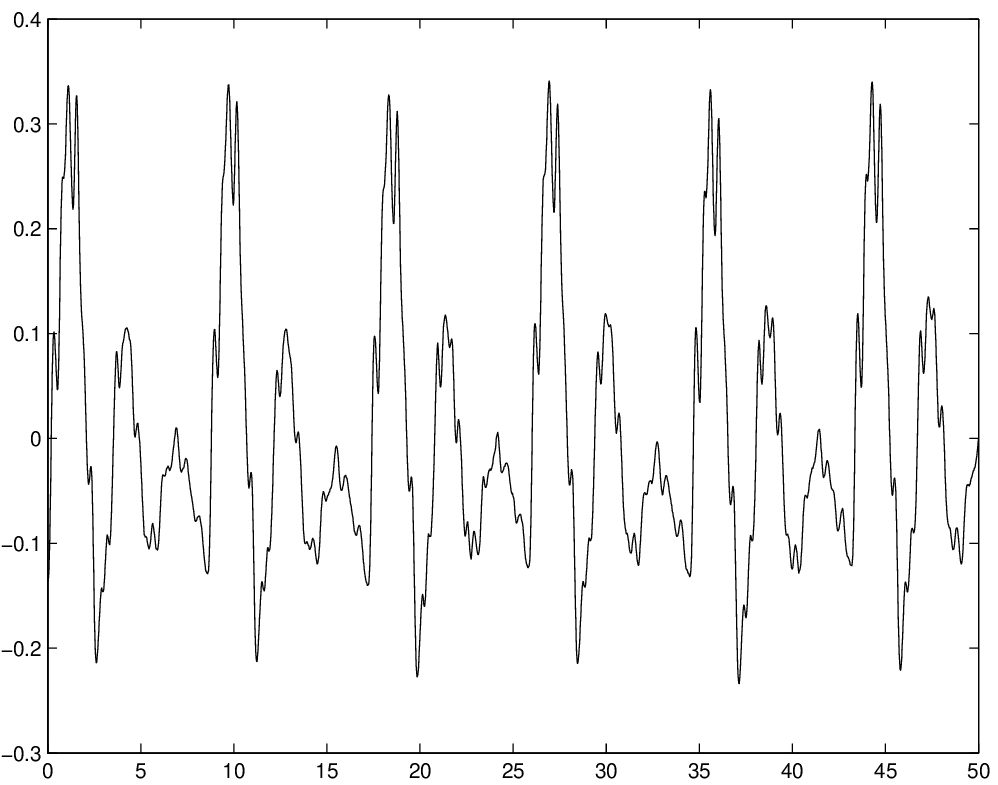, width=0.46\linewidth}
\begin{picture}(0,0)
\put(-5.7,-.06){\scriptsize 0}
\end{picture}
\hspace*{1em}
\psfrag{10}[r][r]{\scriptsize 10}
\psfrag{20}[r][r]{\scriptsize 20}
\psfrag{30}[r][r]{\scriptsize 30}
\psfrag{40}[r][r]{\scriptsize 40}
\psfrag{50}[r][r]{\scriptsize 50}
\psfrag{60}[r][r]{\scriptsize 60}
\epsfig{file=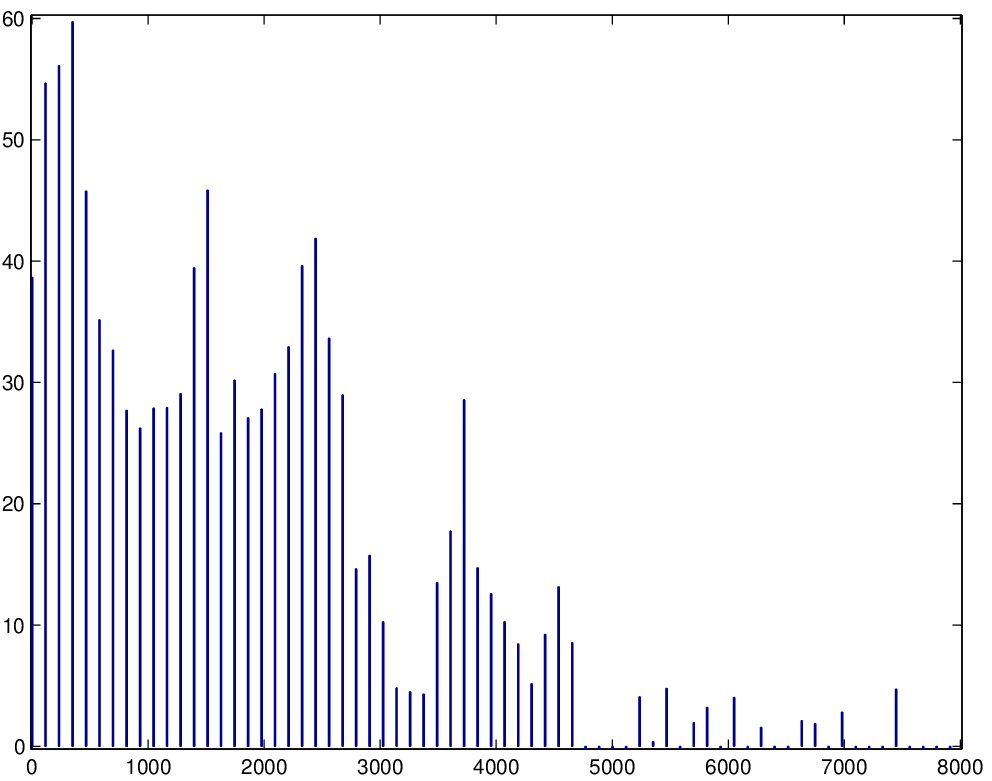, width=0.45\linewidth}
\begin{picture}(0,0)
\put(-5.7,-.06){\scriptsize 0}
\put(-5.83, .1){\scriptsize 0}
\put(-6.2, 3.94){\scriptsize [dB]}
\end{picture}
\vspace*{4ex}
\caption{Zeitverlauf und Spektrum der Konsonanten [m], [\ng] und [l], von oben nach unten. Die Darstellung entspricht Bild \ref{aiu}. Bei [m] und [\ng] sind Formanten bei 250 Hz bzw.{} 220 Hz gut zu erkennen, Frequenzen über 2,5~kHz bzw.{} 2,9~kHz sind stark bedämpft. Der Laut [l] weist neben den Formanten bei 330~Hz, 1,5~kHz und 2,4~kHz eine charakteristische Vertiefung im Spektrum bei ungefähr 3,2~kHz und 4,3 kHz auf, deren Ursache in Abschnitt \ref{Querschnittsform} betrachtet wird.}\label{mnl}
\end{center}
\end{afigure}

\pagebreak
\section{Modelle des Sprechapparats}\label{HistSynth}
In diesem Abschnitt erfolgt ein Rückblick auf die Entwicklung von Apparaturen zur künstlichen Spracherzeugung. Dabei wird deutlich, wie mit der technischen Entwicklung auch Fortschritte im Verständnis der physikalischen und akustischen Vorgänge des Sprechens erzielt wurden und wie eine detaillierte Betrachtung mit einer Verbesserung der Modelle und deren Spracherzeugung einhergeht.\footnote{
Wenn es auch einfache, frühere Beispiele gibt wie die Zeitansage der British Telecom 1936 oder die Kursansage der New York Stock Exchange Ende der 1960er Jahre \cite{Ho80}, so spalten sich ab etwa 1990 diese Wege, als man mit dem sogenannten PSOLA-Verfahren (Akronym von \underline{P}itch \underline{s}ynchronous \underline{O}ver\underline{l}ap-\underline{A}dd) in der Lage war, aufgezeichnete Sprache in der Tonhöhe zu verschieben \cite{VMT91}. Darauf aufbauend entstanden Sprachsynthesen, die auf immer größere Inventare aufgezeichneter Sprache zurückgriffen und deren Elemente mit möglichst geringer Beeinflussung aneinander setzten \cite{TDut94, wwwMBROLA}. Durch diesen phänomenologischen Ansatz zur Sprachsynthese wurden implizit viele Effekte beim Sprechen erfasst.}

Wie dieser Abschnitt zeigt, besteht bereits Ende des 18.~Jahrhunderts ein Grundverständnis der Sprachproduktion.\footnote{
Einige der phonetischen Erkenntnisse aus dem vorangegangenen Kapitel reichen deutlich weiter zurück. Ein gutes Beispiel ist der Bericht über die Ursachen von Lauten (\raisebox{-.8ex}[0pt][0pt]{\epsfig{file=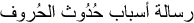}}) von Ibn Sina aus der Zeit der ersten Jahrtausendwende (Übersetzt in \cite{Sa09}). So werden in dem Bericht die unterschiedlichen Artikulations- bzw.{} Konstriktionsstellen (\raisebox{-.75ex}[0pt][0pt]{\epsfig{file=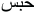}}) für eine Reihe von Konsonanten genannt. Ibn Sina erkennt drüber hinaus, dass bei einem [a] der Vokaltrakt relativ frei (\raisebox{-1.1ex}[0pt][0pt]{\epsfig{file=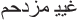}}) bleibt, während für ein [u] die Verengung an den Lippen wesentlich ist. \\
Dieser Bericht ist hier insbesondere erwähnenswert, da neben der Abhandlung der Artikulation und der Anatomie des Sprechapparats auch ein Vergleich zwischen Sprechlauten und anderen Geräuschen gezeigt wird. Aus heutiger Sicht sind einige der Analogien zwar einfache aber qualitativ treffende Modelle der Lautentstehung. So vergleicht Ibn Sina neben weiteren Beispielen den Klang des [d\textsuperscript{\textrevglotstop}] mit dem einer platzenden Blase und den Klang des [h] mit dem Geräusch eines starken Luftstroms.    
} Im 19.~und Anfang des 20.~Jahrhunderts werden für eine Reihe dieser Prozesse physikalische Modelle entwickelt und damit einhergehend gelingt eine zunehmend bessere quantitative Beschreibung von bestimmten Lauten, insbesondere derer aus Abschnitt \ref{Vok}. Etwa Mitte des 20.~Jahrhunderts gelingt es hier durch numerische Methoden, direkt aus dem Sprachschall auf den Artikulationsvorgang zu schließen.  Diese Verfahren werden seitdem weiter verfeinert und durch zahlreiche Untersuchungsmethoden ergänzt, um ein quantitatives Verständnis über alle Lautklassen und Artikulationseffekte hinweg zu erreichen.

\subsection{Mechanische Apparate}\label{mech}
Wenngleich es vor und im 18.~Jahrhundert einige Berichte über \glqq sprechende\grqq~Apparate gab, so sind deren Mechanismen nur selten beschrieben oder tragen nicht zum Verständnis der Sprachproduktion bei, wie \cite{Ru28, FlR73, Ge94, HaS95} darlegen.\footnote{
Auch später wird in \cite{ChG28} beispielsweise notiert: \selectlanguage{french} \glqq  
{\it On sait encore que {\sc Friedrich von Knauss} à Vienna, avait construit avant 1770, trois têtes parlantes, et que cette année-là il en fit une quatrième plus somptueusement présentée, car c'était un cadeau destiné par le couple impérial d'Autriche au grand duc de Toscane ; celui-ci la pla\c{c}a dans sa galarie à Florence. Mais on ne possede aucun dètail sur la technique de ces travaux.}\selectlanguage{german}\grqq{} ohne Referenzen. Auch in \cite{Kn1780, Fi1868, Ku30} finden sich keine Belege für die \glqq sprechenden Köpfe\grqq.
}
Vier bedeutende Ausnahmen finden sich, auf die im Folgenden eingegangen wird.

Ein Experiment von Robert Hooke um 1680, das sprachähnliche Laute hervorbringt, wird in \cite{Wa1705}  kurz beschrieben:
\begin{quote}
\it By the striking of the Teeth of several Brass Wheels, proportionally cut as to their numbers, and turned very fast round, in which it was observable, that the equal or proportional stroaks of the Teeth {\rm[an einem schallabstrahlenden Gegenstand]}, that is, 2 to 1, 4 to 3, \&c.{} made the Musical Notes, but the unequal stroaks of the Teeth more answer'd the sound of the Voice in speaking. 
\end{quote}
Man kann hieraus folgern, dass die vergleichbaren Stimmlaute zu den harmonischen Grundtönen ein Spektrum ganzzahliger Obertöne besitzen, ihnen somit ein einziger periodischer Phonationsprozess zugrunde liegt --- worauf im Abschnitt \ref{aktEig} bereits vorgegriffen worden ist.

Eines der ersten dokumentierten Experimente zur Klärung der physiologischen Unterschiede von Vokalen unternahm Kratzenstein um 1770, indem er sechs unterschiedlich geformte Resonatoren (\name{Tubae}) konstruiert, die auf einer Zungenpfeife aufgesetzt werden, um die verschiedenen Vokale zu reproduzieren, vgl.~Bild \ref{Tuba}. 
Neben einer detaillierten Beschreibung der an der Sprachproduktion beteiligten Organe in \cite{Kra1781} erkennt er den Zusammenhang zwischen Vokal, Zungenposition und gebildetem Hohlraum; er gibt diese quantitativ für die untersuchten Vokale auf S.~15 wieder. Auf S.~35 beschreibt Kratzenstein die Reflektionen der Schallwellen im Sprechtrakt und sein Ziel, diese nachzubilden\label{KratzZitat}: \begin{quote} {\it Hae undae sonorae ex larynge in tubam adfixam incidentes inde vario modo et sub variis directionibus reflectuntur, et instar vocis hominum per tubam stentoream propagantur.} \end{quote}
\begin{afigure}[hbt]%
\begin{center}%
\epsfig{height=2.1cm,file=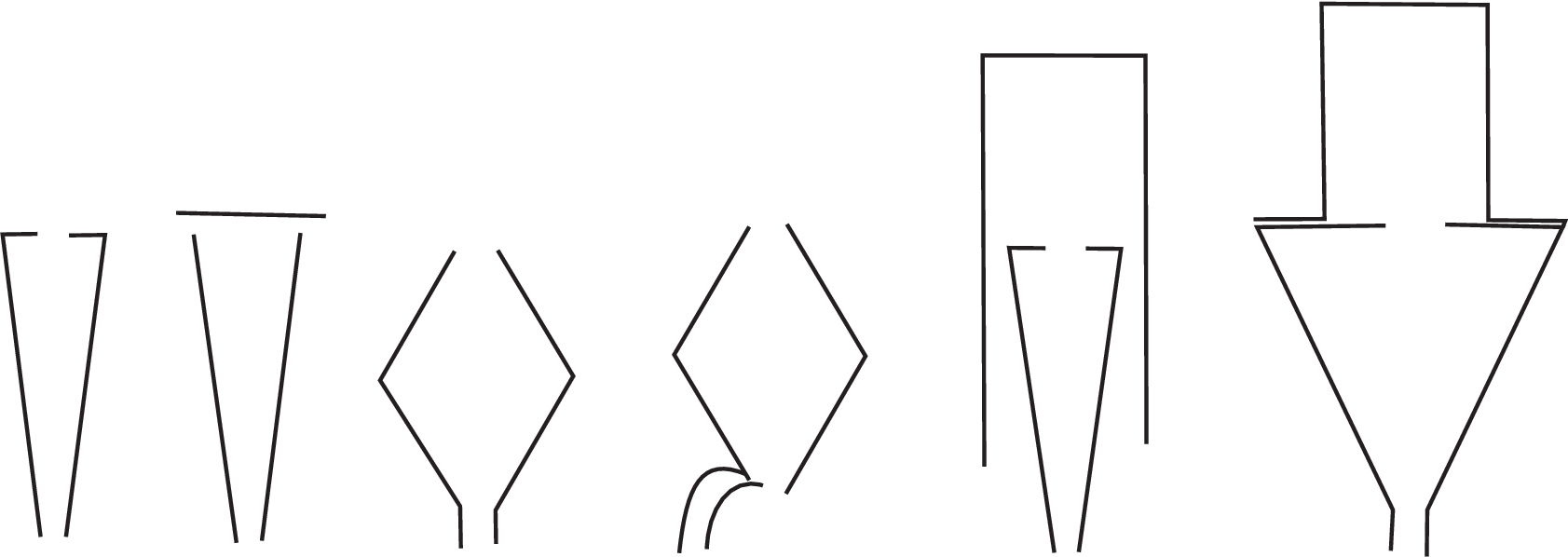}
\end{center}%
\vspace{-1ex}
\caption{Resonatoren nach \cite{Kra1781}, zur Synthese der Vokale a (in zwei Varianten), e, i, o und u, von links nach rechts. In dieser Arbeit diskutiert Kratzenstein auch die Unterschiede zur Anatomie des menschlichen Sprechtraktes.}\label{Tuba}%
\end{afigure}%

\label{mical}Kurze Zeit später verfasste ein namhaftes Komitee der \name{Académie royale des sciences} einen Bericht \cite{Vi1783, Lü10}\footnote{Der zumindest in der Kopie schwer zu entziffernde handschriftliche Bericht ist im Anhang, Abschnitt \ref{Tetes} transkribiert beigefügt.} über die    
\name{Têtes Parlantes} des \name{Abbé Micals}. Ein mehrgliedriger Mechanismus in den \glqq sprechenden Köpfen\grqq ~erzeugt die beiden Sätze\footnote{Nach \cite{ChG28} vier Sätze, gezeigt anhand einer Referenz und der Illustration des Aufbaus.}: 
\begin {quotation}{\noindent \it Le Roi a donné la paix à L'Europe.\\ La paix fait le bonheur des peuples.}\end{quotation}
Der Mechanismus ähnelt teilweise dem einer Orgel: Die Luft wird über einen Blasebalg zugeführt und durch mehrere Ventile in unterschiedliche, flaschenähnliche Kavitäten (\name{boîte}) geleitet. Diese formen den Klang. Die Steuerung erfolgt durch einen --- vermutlich mit Nocken versehenen, rotierenden --- Zylinder, der über Hebel die Ventile betätigt. Für die Erzeugung von Vokalen werden verschiedene Kavitäten benutzt, die sich in ihrer Gestalt, Größe und/oder Öffnung unterscheiden (Z.~56 ff.):
\begin{quote}
\it 
1.~Das {\rm a} prononciert sich in einer der großen Flaschen {\rm[...]}.
Der Klang des Buchstaben {\rm a} in der natürlichen Prononciation resultiert aus einer analogen Disposition, während der die Zunge fixiert
im Innern des \glqq Mundes\grqq{} {\rm[ist]}, ihr Rücken erhebt sich ein bisschen, die zwei Backen sind insgesamt so geöffnet, dass man den gleichen
Klang hört.\\ 
2.~Der Buchstabe {\rm o} verändert sich in einer Flasche der gleichen Größe und der gleichen Form wie der Buchstabe {\rm a}, mit dem
Unterschied, dass die obere Hälfte nicht immobil ist, sondern nur durch eine runde Öffnung durchbohrt. Im Effekt, wenn man den 
Buchstaben {\rm a} prononciert, und man die Öffnung des Mundes ändert, ohne die Situation der Zunge zu ändern, macht sich der Klang {\rm o}
anstatt des ersten hörbar. \\
3. Die Öffnung des Mundes, wenn man den Buchstaben {\rm e} prononciert, hält die Mitte von denen, die für
den Buchstaben {\rm a} und für den Buchstaben {\rm o} eingenommenen werden; auch die Vase, in der (I) der Buchstabe {\rm e} sich hörbar macht, hat eine größere
Öffnung als die erstgenannte, und eine kleinere als die letztgenannte, aber unterscheidet sich noch dadurch, dass sie keine detaillierte und mobile obere 
Hälfte hat, und dadurch, dass sie insgesamt kürzer ist als die beiden ersten. Die Proportion ihrer Öffnungen sind übereinstimmend mit
denjenigen, die Hr.~Kratzenstein \mbox{beobachtet} und bestimmt hat, der den Preis der Akademie von Petersburg im Jahr 1781 auf einem
ähnlichen Gebiet {\rm[...]} gewann.  
\end{quote}
In analoger Weise wird auch der Laterallaut [l] erzeugt. Die Anregung erfolgt bei diesen Lauten durch eine  Zunge[npfeife] mit einem Metallplättchen, das verschiebbar ist und die Tonhöhe bestimmt: Im Bericht wird sie mit den Stimmlippen verglichen. Frikative werden durch das Zischen der Luft in Engstellen und Plosive durch Verschlusslösungen gebildet. Wenngleich die erzeugten Sätze nicht in allen Teilen deutlich ausgesprochen seien, fand der Apparat den Beifall des Komitees und es wurde die {\it approbation de l'Academie} zuerkannt, derer er aufgrund der geistreichen ({\it ingénieuse}) Konstruktion sehr würdig sei.

Aus der gleichen Zeit stammt die \name{sprechende Maschine} von Wolfgang Ritter von Kempelen. Sie besteht aus einem Blasebalg, einem Lederrohr und drei schallerzeugenden \name{Instrumenten}. Der Blasebalg treibt je nach Laut einen der Schallerzeuger an. Für Vokale wird eine Rohrblattpfeife genutzt, was zu obertonreichem Schall führt. Dieser wird durch das nachfolgende, variabel verdeckbare Lederrohr derart verändert, dass ihm die den Vokalen entsprechenden Formantenstrukturen aufgeprägt werden. Dabei stellt das Lederrohr  einen Resonator dar, der demjenigen des Mundraums ähnelt. Die zwei verbleibenden \name{Instrumente} erzeugen die Frikative [s] und [\textesh].  Die Bedienung der Maschine stellt gewisse Anforderungen an die Geschicklichkeit des Experimentators. Mit der linken Hand muss die Öffnung des Lederrohrs entsprechend den Lauten verdeckt werden, sie dient gleichsam als Lippen und Zunge. Mit dem rechten Unterarm wird der Blasebalg betrieben, und die rechte Hand muss zudem die Ventile für die Frikative oder einen Mechanismus für das /r/ bedienen und zur Simulation unnasalierter Laute die dafür vorgesehenen Öffnungen zuhalten. Auch mit dieser Maschine ist es möglich, nicht nur einzelne Laute einer Sprache, sondern auch Wörter und kürzere Sätze zu erzeugen. Von Kempelen schreibt, man könne
\begin{quote}
{\it\glqq in einer Zeit von drei Wochen eine bewundernswerte Fertigkeit im Spielen erlangen, besonders wenn man sich auf die lateinische, französische oder italienische Sprache verlegt ... .\grqq} 
\end{quote}
bemerkt jedoch an anderer Stelle:
\begin{quote}
{\it\glqq Vor allem muß ich gestehen, daß ich vier {\rm [der Konsonanten]} nämlich {\rm D G K T} noch nicht bestimmt in meiner Maschine habe, sondern daß ich hierzu immer das {\rm P} brauche. ... Wenn es aber auch ein feines Gehör bemerkt, so kömmt der Maschine doch immer ihre kindliche Stimme zu statten, {\rm [der man Artikulationsfehler nachsieht].} \grqq} 
\end{quote}
Von Kempelen entwickelte die Maschine während seiner Studien zur Sprache und beschreibt die Ergebnisse in \cite{Ke1791}. Dieses Buch widmet sich in den ersten drei Teilen einer Definition von Sprache, etymologischen und philosophischen Betrachtungen der Sprachentstehung und der morphologisch-physiologischen Betrachtung der Artikulatoren. Der folgende phonetische Teil behandelt nach einer sprachenübergreifenden Lautsystematik die Entstehung der meisten in Tabelle \ref{Laute} gezeigten Laute. Von Kempelen erkennt zutreffend, dass der Unterschied zwischen \glqq weichen\grqq~und \glqq harten\grqq~Plosiven in dem Zeitpunkt des Stimmeinsatzes --- vor oder nach der Verschlusslösung --- liegt.\label{Kempelen-Plosiv} Er zeigt, wie die Phonation stimmhafter Plosive durch einen Luftstrom aufgrund der Druckunterschiede von Lunge und Mundhöhle entsteht. Gestützt auf einen diese Artikulation nachbildenden Mechanismus führt er aus, dass sich [b], [d] und [g]  durch verschiedene Verschlussstellen und damit durch verschiedene Klänge der jeweils unterschiedlich großen Hohlräume vor und hinter der Verschlusslösungsstelle unterscheiden. Bei den Frikativen [f], [s] und [\textesh] (F, S, SCH) folgert er anhand von Experimenten, dass die Form der phonierenden Stelle wesentlich für deren charakteristischen Klang ist und erläutert, wie sie in der Sprechmaschine nachgebildet ist. Desweiteren beobachtet er, dass bestimmte Laute kontextabhängig gewählt werden. So unterschiedet sich das \glqq ch\grqq, welches einem [e] oder [i] folgt, deutlich von demjenigen, dass sich einem [a], [o] oder [u] anschließt: Im ersten Fall wird es als [\c{c}] artikuliert, im zweiten Fall als [x]\footnote{Von Kempelen sieht hier die gleiche Lage der Konstriktion wie bei einem [k], \cite{Ko99} erkennt eine etwas weiter hinten liegende Konstriktionsstelle: [\textchi].}. Ein weiteres gezeigtes Beispiel ist das \glqq ng\grqq, das den Laut [\ng] bildet. Von Kempelen weist auf Koartikulation hin, wie der Nasalierung von Vokalen, denen ein [n] folgt, und begründet das mit einer kinetischen Vereinfachung. Ebenso erklärt er die kontextabhängige Lautwahl. Besonders eingehend betrachtet er die Bildung der Vokale. Hier erkennt er zwei wesentliche Merkmale, die die Vokale unterscheiden: Die Öffnung des Mundes und die Öffnung des Zungenkanals\footnote{Kempelen gibt keine präzise Definition des Zungenkanals. Vergleicht man die von ihm angegebene Lautfolge U O A E I für dessen zunehmende Verengung mit dem Diagramm \ref{Vokaltrapez}, so entspricht das einem Ablaufen im Uhrzeigersinn.}. Im letzten Teil des Buches beschreibt er detailliert die bereits erörterte Maschine und die Lauterzeugung damit.

Die Vokalformanten selbst wurden in den 1820er Jahren durch Willis untersucht [Wi1828], indem er an eine Rohrblattpfeife ein auf der anderen Seite offenes Rohr mit verstellbarer Länge anschloss. Er erkannte, dass sich je nach Rohrlänge unterschiedliche Vokale ergaben, und führte dies auf die Eigenresonanz des Rohres zurück, die er tabellarisch angab. Eine spätere, vergleichbare Untersuchung von Jones konnte nach \cite{Pa30}, S.~17, diese jedoch nur teilweise bestätigen. Beide Untersuchungen sind in Tabelle \ref{Willis_Jones} zusammengefasst wiedergegeben. Paget zeigt später in \cite{Pa30}, dass Vokale durch zumindest zwei Resonanzen charakterisiert sind. Er konstruiert mit diesem Wissen eine Reihe verbesserter Resonatoren für Vokale und einige Konsonanten.
\begin{atable}[t]
\begin{center}
\begin{tabular}{c|cr|c|c}
Rohrlänge (inch)&Ton& (Hz)& Vokal (Willis) & Vokal (Jones)\\ \hline      
& & & &  \\[-2ex] 
6,5 & & $\sim$378&              & \textturna \\
4,7 & $c''$ &    523   & u & \\
3,8 &$\flat e''$ & 659 & o & \ae \\
3,1 &$g''$ &     784   &\textopeno & \\
2,2 &$\flat d''[']$&1109   & \textscripta & \textscripta-\textturnv \\
1,8 &$f'''$ &   1396   & u & \\
1,0 &$d^{IV}$ & 2349   &  \ae & \\
0,6 &$c^{V}$ &  4186   & h\underline{ay} & e \\
0,4 &$g^{V}$ &  6272   & i & i
\end{tabular}
\end{center}\caption{Nach \cite{Wi1828}, S.~243, ergänzt durch die Untersuchungen von Jones und durch Frequenzangabe der Töne sowie einer Darstellung im IPA. Die Frequenzen entsprechen etwa $\lambda/4$-Resonanzen der Rohrlänge.}\label{Willis_Jones}
\end{atable} 

\begin{afigure}[t]
\begin{center}
\psfrag{a}[l][l]{\scriptsize Nasengang, -löcher}
\psfrag{b}[l][l]{\scriptsize Ledertubus  (Vokaltrakt)}
\psfrag{e}[r][r]{\scriptsize Druckkammer}
\psfrag{f}[r][r]{\scriptsize Blattpfeife (Glottis)}
\epsfig{height=3.3cm,file=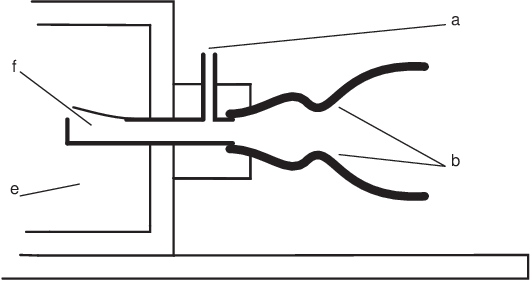}
\end{center}
\caption{Wheatstones Nachbau von Kempelens Sprechapparat nach \cite{Fl65}. Links neben dem gezeigten Ausschnitt ist ein Blasebalg zum Antrieb an die Druckkammer angeschlossen. In der Druckkammer befinden sich zudem weitere Pfeifen und Steuerhebel zur Erzeugung von Frikativen.}\label{Wheat}
\end{afigure}
  
In dieser Zeit baute Wheatstone die Maschine von Kempelens nach. Bemerkenswert ist dabei der Ansatz, anstelle eines starren ein verformbares Lederrohr einzusetzen, vgl.{} Bild \ref{Wheat}. Dadurch kam er den akustischen Eigenschaften des menschlichen Rachentrakts erheblich näher.  Er bestätigte damit den Zusammenhang zwischen der Form des Rachenraumes und den verschiedenen Hauptresonanzfrequenzen, den Formanten\footnote{Die Resonanztheorien des Sprechtrakts von Helmholtz, Hermann, Rayleigh, Scripture, Trendelenburg, Wheatstone und Willis sind in \cite{Ru28, CK41} zusammengefasst.}. Im  19.~und noch Anfang des 20.~Jahrhunderts wurden aus der Konzeption von Kempelens weiterentwickelte Geräte gebaut. Beispiele sind die \name{Euphonia} von \name{Joseph Faber}, bei der der Vokaltrakt aus Gummi hergestellt ist und durch Tasten über Drähte   der natürlichen Artikulation entsprechend geformt wird\footnote{In \cite {Sc1842} wird desweiteren berichtet, dass die {\it Sprechmaschine {\rm [...]} vollständiger als die bisher dazu gemachten Versuche die menschliche Stimme {\rm [...]} nachahmt {\rm und} ziemlich deutlich spricht.} } und die Apparatur von Riesz, skizziert in Bild \ref{Riesz} nach \cite{Fl65}. Diese zeichnen sich durch die Verwendung einer wesentlich naturgetreueren Form des Ansatzrohres aus und haben eine gewisse Ähnlichkeit mit einem in dieser Arbeit verwendeten Sprechtraktmodell.
Insbesondere ist der Vokaltraktbereich zwischen den Stimmbändern und den Lippen in mehrere Abschnitte unterteilt, deren Querschnittsflächeninhalt man den Lauten entsprechend einstellen kann.
\begin{afigure}[t]
\begin{center}
\psfrag{Rachen}[t][t]{{\scriptsize Rachen }}
\psfrag{Gaumen}[t][t]{\scriptsize\shortstack{Gaumen-\\segel}}
\psfrag{L}[Bl][Bl]{{\scriptsize Lippen}} 
\psfrag{Zahne}{{\scriptsize Zähne}} 
\psfrag{Luf}[b][bc]{{\scriptsize Luftzufuhr}} 
\psfrag{Nasaltrakt}[r][r]{{\scriptsize Nasaltrakt}} 
\psfrag{S}[r][r]{{\scriptsize Stimmbänder}}
\epsfig{width=7cm, 
height=4.0cm,file=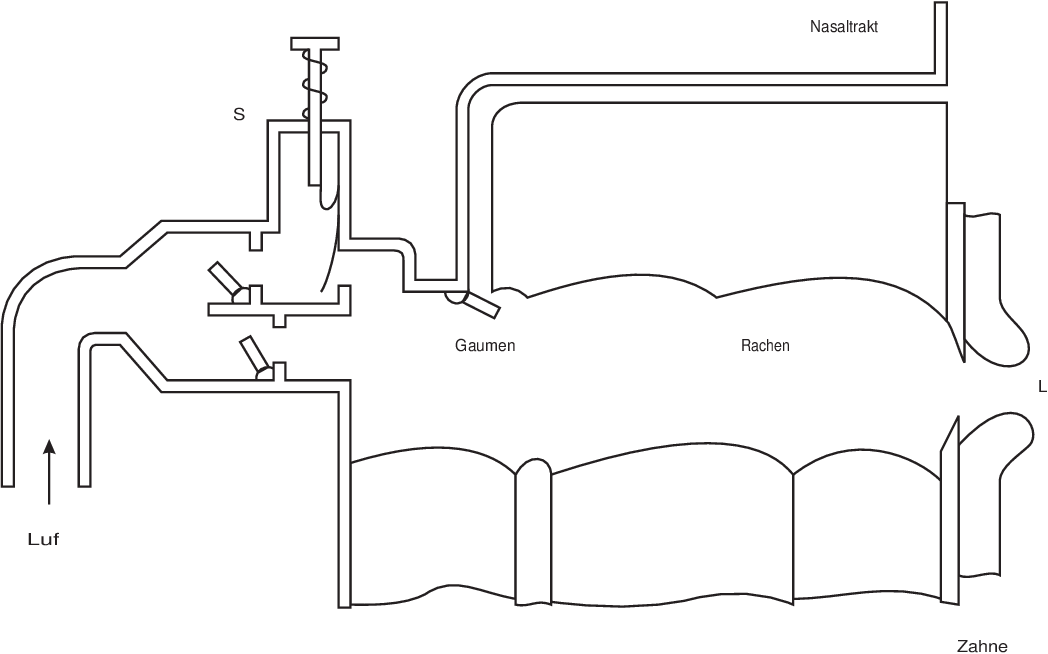}
\end{center}
\caption{Riesz'scher Sprechapparat, 1937. Die sechs verschiebbaren Segmente der Unterseite können in einer anderen Ausführung des Apparats mit Tasten, ähnlich einer Trompete, gesteuert werden.}\label{Riesz}
\end{afigure}
Über den Apparat von Riesz wird a.~a.~O.~berichtet:
\begin{quote}
{\it When operated by a skilled person, the machine could be made to simulate connected speech. One of the particulary good utterances was reported to be ``cigarette''.

\hfill \small --- Personal communication, R.~R.~Riesz.}
\end{quote}

\arxiv{
\savebox{\kanaU}{\raisebox{-.2ex}[0pt][0pt]{\scalebox{0.76}{\psfig{file=Kana-U-corel}}}}%
\begin{afigure}[t]%
\begin{center}%
\epsfig{width=.8\linewidth, file=Chiba_U.eps}%
\end{center}%
\caption{Bildtafel aus \cite{CK41} -- Bestimmung der Vokaltraktkonfiguration für den Laut [\textturnm], der Aussprache des japanischen Schriftzeichens~\usebox{\kanaU}. 
Der Querschnitt durch den Vokaltrakt ist oben links gezeigt; er ist mit der Mundöffnung nach  links und der Glottis nach unten orientiert. In den Querschnitt eingezeichnet sind Lage und Winkel der senkrecht dazu bestimmten Vokaltraktkonturen. Die Konturen selbst sind in der Bildtafel rechts und unten wiedergegeben und durch Flächeninhaltsangaben ergänzt.  Das mittig abgebildete Spektrum der während der Untersuchung aufgezeichneten Lautäußerung umfasst einen Frequenzbereich bis 3 kHz.
 }%
\label{Chiba1}%
\end{afigure}
}
Die quantitative Verwendung der Querschnittsflächeninhalte zur Bestimmung der Vokaltraktkonfiguration gelingt in der Arbeit von Chiba und Kajiyama \cite{CK41}. Hierin werden für die Vokale [i], [e], [a], [o] und [\textturnm] die Vokaltraktkontur anhand von Röntgenaufnahmen bestimmt, wobei auf den Artikulatoren angebrachte dünne Golddrähte, Stanniolbändchen oder aufgebrachtes Bariumsulfatpulver zur Hervorhebung von Konturen genutzt werden. Die während der Aufnahme aufgezeichnete Lautäußerung wird mit einem akustisch vermessenen Modell und einem berechneten vereinfachenden zwei-Resonator-Modell verglichen. Das aus Gips geformte akustische Modell spiegelt  den Verlauf der Querschnittsflächeninhalte wider. Ein Beispiel, bei dem die Spektren zwischen Lautäußerung, den Modellen und mit den in Abschnitt~\ref{aktEig} für das [u] ermittelten qualitativ übereinstimmt\arxiv{, ist in Bild~\ref{Chiba1} gezeigt}.

\subsection{Elektronische Systeme}
In den vierziger Jahren wurden mit dem Beginn der Entwicklung der magnetischen Signalaufzeichnung erste Zeitbereichsverfahren untersucht. Man zeichnete Sprache auf Tonbändern auf, segmentierte diese, indem man das Tonband in Abschnitte unterteilte, und fügte sie entsprechend der zu synthetisierenden Äußerung wieder zusammen. Es zeigte sich, dass Phonem-Segmente zu einer gänzlich unverständlichen Sprachwiedergabe führten. Eine Segmentierung in Diphone war zwar erfolgreicher, aber auf Grund der Größe des sich ergebenden Inventars\footnote{Mit dem Inventar wird die Menge alle Phoneme bzw.{} Diphone, die durch die Synthese realisiert werden sollen, bezeichnet. Ein Phoneminventar umfasst ca.~50 Elemente, ein Diphoninventars folglich ungefähr $50 \cdot  50=2500$ Elemente. Weiterführendes findet sich bspw.{} in \cite{FEng97}.} anfangs nur exemplarisch handhabbar, wie in \cite{KW56, Cr64} diskutiert. Weitere, ähnliche Untersuchungen wie \cite{Gr76} zeigen die Bedeutung der Lautlängen.

Hauptsächlich in den 1920er bis 1960er Jahren wurden, durch Fortschritte in der  Analog\-elektronik ermöglicht, Formantensynthesizer entwickelt und untersucht. Einen der ersten Synthesizer\footnote{Einen elektromechanischen Versuch sehr ähnlicher Konzeption zeigt Helmholtz bereits in \cite{HeTonempfindungen}, Abschnitt \glqq Künstliche Vokale\grqq.} realisierte Stewart und beschreibt ihn 1922 in \cite{St22}: Mittels zweier über einen \name{Buzzer} angeregte Resonanzkreise, deren Resonanzfrequenzen an die beiden unteren Formanten angeglichen werden kann, lassen sich Vokale und Diphthonge reproduzieren. \cite{Cr25, St35, Le36} zeigen kurze Zeit darauf mit weiterentwickelten elektronischen Analysesystemen, dass Sprache weitere Formanten enthält.
Spätere Synthesizer verfügen meist über zwei Signalgeneratoren, zur Erzeugung von periodischen Signalen und von Rauschen, zwischen denen je nach Phon umgeschaltet werden kann. Diesen folgt ein Filtersystem, wobei sich die Synthesizer hier in drei Typen unterscheiden lassen. Der 1939 von Dudley entwickelter Synthesizer \name{Voder} verwendet eine Filterbank, die aus Bandpässen mit festen, aneinanderfolgenden Frequenzbändern besteht. Über in einer Tastatur angeordnete Potentiometer können diese Frequenzbänder in ihrem Pegel verändert werden, wodurch die Formantstruktur bereichsweise gemittelt nachgebildet werden kann.  Wenngleich auf der Weltausstellung 1939 und 1940 gezeigt, wird beispielsweise über den Voder in \cite{Ma99} festgestellt:\begin{quote}
{\it\glqq ... the synthetic speech, to judge from the recordings that still survive, was not highly intelligble.\grqq}
\end{quote}
\noindent%
\begin{afigure}[t]%
\begin{center}%
\epsfig{width=.97\linewidth,file=Dudley-Vod\arxiv{er}.eps}%
\vspace*{-8mm}%
\end{center}%
\caption{Filterbanksynthese nach \cite{Du38}}\label{Voder}%
\end{afigure}%
Bereits einige Zeit zuvor wurden ähnliche Filterbänke zur Sprachübertragung genutzt: Das von Schmidt in \cite{Sc32} beschriebene Verfahren verwendet zwei gleiche Filterbänke, eine zur Analyse oder Kodierung von Sprachsignalen und eine zweite Filterbank, die das Sprachsignal resynthetisiert bzw.{} dekodiert. Die Ausgangssignale der ersten Filterbank werden mit verringerter Bandbreite übertragen und steuern die zweite Filterbank an. Die Steuerung des Synthesizers durch ein natürliches Sprachsignal vermied vermutlich eine Reihe von Abweichungen, die durch die beschränkten Möglichkeiten einer Tastatur des zuvorgenannten Synthesizers unvermeidlich waren, und dürfte zu einem natürlicheren Zeitverlauf der Formantenstruktur und damit zu einer verständlicheren synthetisierten Sprache geführt haben.\label{Formant} Eine in der Frequenzauflösung verfeinerte Variante war der von Cooper rund zehn Jahre später entwickelte \name{Pattern Playback}-Synthesizer, welcher optoelektronisch über einen Film gesteuert wurde; auf dem Film sind die Intensitäten einzelner Frequenzbänder durch die Transparenz paraleller Streifen kodiert. Durch Abfahren des Films waren damit reproduzierbare Synthesen möglich, Beispiele finden sich unter \cite{PPwww}. Die für die damalige Zeit gute Verständlichkeit (\name{higly intelligible} \cite{Co53}) beruht ebenfalls auf der inhärenten Resynthese -- die Filme wurden durch Analyse von Sprache gewonnen. 

Auf einer anderen Filterstruktur basieren die in den 1950ern entwickelten \name{Orator Verbis Electris (OVE)} Synthesizer von Fant, der in \cite{Fa62} beschrieben ist. Diese bestehen aus verstimmbaren Schwingkreisen, deren Mittenfrequenz und Pegel in unterschiedlichen Varianten manuell oder durch ein Steuerwerk vorgegeben werden können; zudem enthalten sie ein Filter, das bestimmte Frequenzen unterdrückt, wie es für bestimmte Laute typisch ist. Wenngleich hiermit sicherlich wichtige Formanten stationärer Laute genau wiedergegeben werden können, liegt die Schwierigkeit dieses Verfahrens in der treffenden Bestimmung und Nachbildung der Formantenbewegung, der zeitlichen Änderung von Güte und Mittenfrequenz. In ähnlicher Weise arbeitet auch der \name{Parametric Artificial Talker (PAT)} von Lawrence aus der gleichen Zeit. Synthesen und Bilder finden sich unter \cite{wwwKe}.

Der dritte Filtertyp basiert auf einer Kette von LC-Gliedern, über die sich die elektrischen Signale ähnlich ausbreiten wie eine ebene Schallwelle entlang dem Vokaltrakt.  Dadurch werden nicht mehr einzelne Formanten betrachtet, sondern die Formantenstruktur bzw.{} die Hüllkurve des Spektrums als Ganzes werden mit dieser Filterstruktur reproduziert. Dunn nutzt 1950 diese Analogie für ein Sprechtraktmodell und stellt in  \cite{Du50} die Vorteile dieses Ansatzes fest:
\begin{quote} {\it\glqq A line with distributed constants is approximated through the use of 25 lumped sections, each representing a cylinder 0.5 cm long and 6 cm**2 in cross section. The whole is then divided into two 'cavities' by the use of a lumped, but variable, inductance which can be inserted between any two sections of the line. This represents the 'tongue hump' constriction. Another variable inductance at the end of the line represents the constriction at the lips. {\rm [...]} The whole series of English vowels can be produced by this apparatus -- not perfectly, but distinctly better than we were able to make with three independent tuned circuits.
\grqq}
\end{quote}

\subsection{Digitale Signalverarbeitung}\label{DigitalSynthese}
Die digitale, zeitdiskrete Signalverarbeitung erlaubt eine einfachere und präzisere Modellierung und Steuerung. Kelly und Lochbaum übertrugen 1962 in  \cite{KeL62} das Modell der Wellenausbreitung längs des im Querschnittsflächeninhalt variierenden Vokaltrakts auf passende digitale Filter, die Kreuzgliedketten\footnote{die Filterstruktur wird in~\ref{TheoLinRohr} beschrieben};  sie greifen damit in vereinfachter Form (ohne Berücksichtigung des Nasaltraktes) den Ansatz von Chiba, Kajiyama und Dunn erneut auf, indem sie die akustischen Vorgänge zeitdiskret und digital beschreiben.  Im Unterschied zu den elektronischen Systemen von Dunn gelingt zudem die Modellierung eines wesentlich natürlicheren Querschnittsverlaufs, da jedem Glied ein Querschnitt zugewiesen werden kann. Der Querschnittsverlauf wurde aus Röntgenaufnahmen ermittelt, die Fant angefertigt hatte. Durch die rechnergesteuerte Synthese gelingt auch die Erzeugung von Lautübergängen mit diesem Modell, wobei die breitbandige Anregung wieder wahlweise durch Rauschen oder periodische Signale  erfolgt.  

Einen wichtigen Fortschritt bringt die Ende der 1960er Jahre gewonnene Erkenntnis, wie man anhand von Sprachsignalen die Filterkoeffizienten beziehungsweise Querschnittsverläufe ermitteln kann. Grundlegende Arbeiten über die Eigenschaften von Sprachsignalen und deren Bezug zur Sprechtraktgeometrie stammen von Mermelstein und Schroeder, die in \cite{MeS65} zunächst nur auf Formanten betrachten, von Saito und Itakura, die in \cite{SI66, SI68, SI69} einen statistischen Ansatz basierend auf \name{Maximum Likelihood}  verfolgen und in der letztgenannten Arbeit die \name{partielle Korrelation -- PARCOR} vorstellen, und von Atal und Schröder, die in \cite{AS67, AS70} die \name{Linear Prediction} zur komprimierten Sprachübertragung einsetzen. Nach \cite{Ma72} gehen diese Verfahren, \cite{SI69, AS70}, auf \cite{Pr1795} zurück. Eine alternative Betrachtungsweise wird von Burg aufgezeigt, die \name{Entropie-Maximierung} in \cite{Bu67}, anhand derer er in \cite{Bu68} ein insbesondere für kurze Signalabschnitte geeignetes Verfahren entwirft, mit dem man gut aus Sprachabschnitten diese Koeffizienten schätzen kann, wie \cite{GM78} darlegt und Bild~\ref{Rohrquerschnitte} illustriert. 
 Letztlich ist aber die Übereinstimmung durch das zugrundeliegenden Modell begrenzt, wie \cite{La05} in einem  Überblick unter Einbeziehung von Teilen dieser Arbeit aufzeigt: 
 \begin{quote}{\it  Yet, if one looks at the vowel spectra in more detail it turns out that appearently even for vowels the all-pole model has its deficiencies.} \end{quote}

\begin{afigure}[t]%
\begin{center}%
\epsfig{width=.31\linewidth,file=a-tub\arxiv{e}.eps}\hspace*{.02\linewidth}%
\epsfig{width=.31\linewidth,file=i-tub\arxiv{e}.eps}\hspace*{.02\linewidth}%
\epsfig{width=.328\linewidth,file=u-tub\arxiv{e}.eps}%
\end{center}%
\vspace{-.5ex}%
\caption{Querschnitte durch Kreuzgliedkettenfiltern entsprechende Rohrmodelle des Vokaltrakts für die in Bild \ref{aiu} gezeigten Laute [a], [i] und [u], von links nach rechts. Diese sind mit der in Teil \ref{SPEAK} beschriebenen Software unter Verwendung der Burg-Methode, einer doppelten Preemphase und einer Abtastrate von 44,1~kHz bestimmt worden, die Glottis ist jeweils links, der Mund rechts. Skaliert auf die Vokaltraktlänge weisen sie große Ähnlichkeit mit MRI-Untersuchungen dieser Laute auf, vgl.~bspw.~\cite{St2008}. Abszisse und Ordinate sind zur Hervorhebung der Kontur nicht maßstäblich.}\label{Rohrquerschnitte}%
\end{afigure}%

\subsection{Detaillierte anatomische Modelle}\label{DetailMod} 
Um das Verhalten bestimmter Artikulatoren genauer zu betrachten und zu verstehen, wurde in der jüngeren Vergangenheit für diese Artikulatoren eine Reihe detaillierter Modelle entwickelt. Ein Beispiel hierfür ist ein dreidimensionales Zungenmodell, das von einem wenige Parameter umfassenden System in \cite{Me73} weiter verfeinert wurde, um die inhärente Kinematik zu berücksichtigen; eine Übersicht gibt bspw.~\cite{BiJK06}. Jedoch ergibt die indirekte Kontrolle über die Zungenbewegung letztlich noch keine befriedigende Artikulation, wie \cite{GWPP2003} zeigt, oder erfordert lautweises Nachjustieren von Parametern, wie in \cite{BiJK06} ausgeführt. Ergänzt werden diese Modelle durch eine Reihe spezieller Untersuchungen der Zungenbewegung, beispielweise durch akustische Impedanzmessung nach \cite{Sc67, KoNR02}, \arxiv{Bild \ref{Lip-Imped},} durch Ultraschall, wie in \cite{ZhHH08, WrS08} beschrieben, oder mittels elektromagnetischer Artikulatographie, wie \cite{Sc83} zeigt. Letztere erfasst mittels kleiner aufgeklebter Spulen auch die Lippenformation und Velum- und Kiefernstellung mit Hilfe eines um den Probanden erzeugten ortsabhängigen magnetischen Wechselfeldes. 
\arxiv{
\begin{afigure}[b]%
\begin{center}%
\epsfig{width=.75\linewidth,file=F03-Lip_Impedance_Measurement.eps}%
\end{center}%
\caption{Akustische Lippenimpedanz-Messung. Links: Topfförmige Schallquelle, darüber Rohr mit konstanter, bekannter akustischer Impedanz, auf dem sich zwei Messmikrofone zur Bestimmung von Schalldruck und Schallschnelle befinden. Bildquelle \cite{Sc67}.}\label{Lip-Imped}%
\end{afigure}%
}

Zur Untersuchung der Anatomie der Artikulatoren kommen auch weitere etablierte Methoden aus der medizinischen Diagnostik zum Einsatz. Insbesondere radiologische Untersuchungsverfahren, wie Röntgendurchleuchtung und -kinematographie, wurden zeitweise verwendet. So wurde bereits 1897, weniger als zwei Jahre nach Entdeckung der Röntgenstrahlung, eine der ersten Untersuchungen in \cite{Sc1897} publiziert:
\begin{quote}
{\it ... Durchleuchtet man den Kopf seitlich, so sieht man auf dem Schirmbilde den Nasenrachenraum und den Pharynx als hellen Schatten hervortreten {\rm [...]} . Lässt man nun die zu untersuchende Person einen Vocal phoniren, so sieht man, wie das Gaumensegel sich hebt, und zwar ganz verschieden in den Nasenrachenraum sich hinlegt je nach dem Vocal, den man aussprechen lässt. ...}
\end{quote} 
Die Anwendung von Filmaufnahmen sind jedoch mit der Kenntnis über Risiken der Röntgenstrahlung sehr eingeschränkt; einige dieser Aufnahmen sind unter \cite{MuVBT95} bereitgestellt. Kürzlich gelang jedoch die Aufzeichnung einschichtiger Magnetresonanztomographien des Sprechtrakts in Intervallen von 20~ms, wie \cite{UeZVKMF10} zeigt.

Bereits Helmholtz argumentiert in \cite{HeTonempfindungen}, dass die Glottisschwingung unabhängig von der akustischen Konfiguration des Sprechtrakts ist und zeigt am Beispiel von Zungenpfeifen, welches sich auf die Glottisfunktion stimmhafter Laute übertragen lässt, dass die hohe Schallintensität aus einer zyklischen Unterbrechung der Luftströmung herrührt. Tondorff in \cite{To25} und in Folge van den Berg et.~al.~in \cite{BeZD57, Be58} erkennen den Bernoulli-Effekt als wesentlichen Beitrag zur Glottis-Schwingung. Darauf aufbauende 1- und 2-Massen-Modelle der Stimm\-lippenvibration werden in \cite{FlL68} bzw.{} in \cite{IsM72, IsF72} gezeigt.
Seitdem ist eine Vielzahl von Modellen der Glottis beschrieben worden, die deren Schwingverhalten durch Finite-Elemente genauer nachbilden, in \cite{Vr03} die akustischen und aerodynamischen Effekte durch numerische Lösung der Navier-Stokes-Gleichung behandeln, oder wie in \cite{LOli93} einfach den zeitlichen Schalldruckverlauf genauer beschreiben. Letzteres wird auch in dieser Arbeit verwendet, vgl.~Kapitel \ref{Speak-Quelle}. Gestützt werden diese Modelle durch Untersuchungen der Glottisschwingung, insbesondere mittels Hochgeschwindigkeitskameras und Elektroglottograph: Verfahren die erstmals in \cite{TrW35,Be37} bzw.{} in \cite{Fa57} beschrieben werden. Eine Übersicht gibt \cite{BaLMG83}.

Auch für die Akustik der Nasenhohlräume wurden bestimmte Modelle entwickelt. So haben Lindqvist und Sundberg in \cite{LiS72} das akustische Verhalten des Nasaltrakts untersucht, indem Schall oberhalb des Velum mittels eines dünnen Rohres eingeleitet und der an den Nasenlöchern austretende Schall erfasst wurde. Unter Berücksichtigung des Einflusses der Schallquelle erhält man so das Übertragungsverhalten. Bei dieser Methode ist es jedoch
schwierig, die natürlichen Verhältnisse für die Abschlussimpedanz am Velum zu schaffen: Ist es abgesenkt, entspricht seine Artikulationsstellung derjenigen der nasalierten Laute, aber der Vokaltrakt beeinflusst die Messung -- genau umgekehrt wäre es bei einem angehobenen Velum. Weitere Untersuchungen betreffen in \cite{Ma82, DaHS94} die Modellierung von Nebenhöhlen, in \cite{DaH95} die Bestimmung ihrer Resonanzfrequenzen, in \cite{SuNS95} die Relevanz von Quermoden in den Nasengängen  und in \cite{SuNS96} die Auswirkung von Asymmetrien zwischen den Nasengängen. Bei letzteren werden Finite Elemente zur Bestimmung der Schallausbreitung  eingesetzt, mit den bereits in Abschnitt \ref{Einleitung} erörterten Nachteilen. 

Finite-Differenzen zur Berechnung der Sprechtraktakustik sind vor den in dieser Arbeit durchgeführten Untersuchungen vereinzelt und nur für die Mundhöhle angewendet worden. In \cite{Mc87} werden sie in 1-dimensionaler Form  als Alternative zu dem von Kelly und Lochbaum vorgeschlagenen Verfahren zur Berechnung der Ausbreitung ebener Wellen diskutiert. Eine zweidimensionale Betrachtung findet sich in \cite{Ri95, CuMC95, AlS95}; in letzterer werden rotationssymmetrische Lösungen untersucht. Mit Wellenleitern wurde zwischenzeitlich eine zwei- und dreidimensionale Modellierung der Akustik der Mundhöhle versucht, wie \cite{CoMHT06, Sp08, Fr09} zeigen, wobei jedoch eine prinzipbedingte Anisotropie der Wellenausbreitung verbleibt. Diesen Fehler vermeiden die in \cite{Mo02} gezeigten Wellenleiter höherer Ordnung, die eine akkurate Beschreibung des Mundbereichs erlauben. Zur vereinfachten Anwendung von Finiten Elementen  ist in \cite{SaMM03} ein automatischer Mesh-Generator für einen elliptisch konturierten Vokaltrakt gezeigt. Kürzlich ist in \cite{TaMK} die Akustik von Kunststoffmodellen, deren Geometrie sich an den Mund-Rachenbereich des Vokaltrakts anlehnt, vermessen worden und zeigte im Vergleich zu einer  einfachen Finite-Differenzen-Simulation derselben eine gute Übereinstimmung. 


\pagebreak

\section{Ausbreitung ebener Wellen:\\ Das Rohrmodell des Sprechtrakts}\label{Rohrmodell}
\newsavebox{\EinAusgang}
\savebox{\EinAusgang}(0,0)[lb]
{
	\put(2.9,1.9){\makebox(0,0)[b]{$a_1$}}
	\put(2.9,0.74){\makebox(0,0)[b]{$b_1$}}
	\put(9.2,1.9){\makebox(0,0)[b]{$b_2$}}
	\put(9.2,0.74){\makebox(0,0)[b]{$a_2$}}
}
Wie im vorangegangenen Teil deutlich wurde, ist der Vokaltrakt aufgrund seiner Variabilität das wesentliche Element der natürlichen Spracherzeugung. In diesem Teil werden die Grundlagen für seine Nachbildung mit zeitdiskreten digitalen Filtern vorgestellt. Die Schallwellen werden dazu vereinfacht mit einer ebenen Wellenfront angenommen und deren Ausbreitung entlang des Vokaltrakts betrachtet.

\begin{afigure}
\begin{center} \epsfig{file=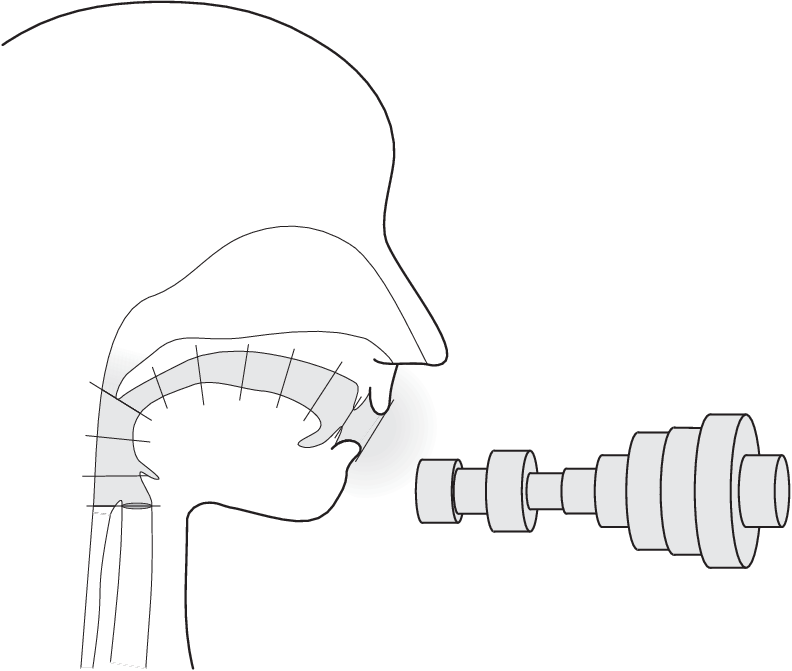,width=.8\linewidth}\hspace{.1\linewidth}\end{center}
\caption{Diskretisierung des Vokaltraktes --- schematisch}\label{disV}
\end{afigure}

Für die als Rohrmodell bezeichnete Abstraktion unterteilt man zunächst den Vokaltrakt in gleichlange Abschnitte, wie links in Bild \ref{disV}~angedeutet.
Dabei idealisiert man die Abschnitte in homogene Bereiche, die Krümmung des Vokaltrakts bleibt unberücksichtigt, und abrupte Querschnittssprünge. Das resultierende Modell ist rechts in Bild \ref{disV} dargestellt.  Für dieses Modell  lassen sich handhabbare Filter finden, die in den nachfolgenden Abschnitten beschrieben werden. Anhand dieser Filter lässt sich das Übertragungsverhalten bestimmen. Umgekehrt lassen sich  auch die Filterkoeffizienten aus dem Betragsspektrum von Sprachproben schätzen, wie in den weiteren Abschnitten gezeigt wird.

\subsection{Matrixdarstellung von Wellenleiterelemente}\label{WelleEle}
Schallharte Rohrsysteme kann man, solange ihr Querschnitt klein gegenüber den Wellenlängen der betrachteten 
Schwingungen ist\footnote{Die niederfrequenteste Radialmode eines Zylinders, die Besselmode $j(1,0)$, ergibt bei einem Durchmesser von 3,6~cm eine Resonanzfrequenz von 5,8~kHz, die Eigenresonanz einer Kugel gleichen Durchmessers liegt bei 6,3~kHz. }, als eindimensionale Wellenleiter auffassen; es wird nur die Ausbreitung ebener Wellen berücksichtigt. 

Ein geeignetes Mittel zur Beschreibung eindimensionaler Wellenleiter sind zum einen Adaptoren, die Querschnitts- beziehungsweise Impedanzsprünge und Verzweigungen darstellen können. Zum anderen werden homogene Abschnitte des Wellenleiters durch Leitungs- bzw.{} Laufzeitelemente erfasst. Ihnen gemeinsam ist ihr lineares Übertragungsverhalten, welches sich in Form von Matrizen beschreiben lässt, vgl.{} \cite{ALac96}. Die dabei zugrunde liegende Idee ist die Separation der Wellenausbreitung in eine hin- und eine zurücklaufende Welle, da diese Lösungen der Differentialgleichung sind, die den homogenen Wellenleiter beschreibt. Adaptoren verknüpfen dann die hin- und zurücklaufende Welle.

Im Folgenden werden zwei wichtige Typen von Matrizen eingeführt. Für ein lineares System mit zwei Eingängen $a_1$~und $a_2$, die in dem Vektor $\vec{A}= \begin{pmatrix} a_1 \\ a_2 \end{pmatrix}$ zusammengefasst werden, und zwei Ausgängen, $\vec{B}=\begin{pmatrix} b_1 \\ b_2 \end{pmatrix}$, kann man deren Beziehung durch eine Streumatrix $\mat{S}$ angeben:
\[  \vec{B}=
\begin{pmatrix} b_1 \\ b_2 \end{pmatrix} 
 =  \begin{pmatrix}
s_{11} & s_{12}  \\
s_{21} & s_{22}  \end{pmatrix} 
\begin{pmatrix} a_1 \\ a_2 \end{pmatrix} 
=\mat{S}\vec{A}\:. \]

Eine andere Darstellungsform ist die Betriebskettenmatrix $\mat{T}$, sie erlaubt das Aufmultiplizieren verketteter Adaptoren.  Die Definition von $\mat{T}$ ist:
\[ 
\begin{pmatrix} b_1 \\ a_1 \end{pmatrix} = \mat{T}
\begin{pmatrix} a_2 \\ b_2 \end{pmatrix}. \]
Hieraus ergibt sich folgender Zusammenhang  zwischen  $\mat{T}$ und $\mat{S}$:
\[ \mat{T} =
 \frac{1}{s_{21}} \begin{pmatrix}
	-\det\mat{S} & s_{11}  \\
	-s_{22} & 1  \end{pmatrix}\:,\qquad
\mat{S} =
 \frac{1}{t_{22}} \begin{pmatrix}
	t_{12} & \det\mat{T}   \\
	1 & - t_{21} \end{pmatrix}\:
.\]
	
Als Ein- und Ausgangssignal sind physikalischen Größen geeignet, die sich durch eine linearen Funktion oder Differentialoperator aus dem akustischen Potential $\Phi$ bilden lassen, wie  der Schalldruck $p = -\rho\, \partial /\partial t\, \Phi$, die Schallschnelle $v= \nabla \Phi$, der Schallfluss \mbox{$u=F\nabla \Phi$} und die nach \cite{Ei96} vorteilhafte Wurzelleistung $\sqrt{pu} = -\rho\sqrt{\frac{F}{Z_0}}\frac{\partial}{\partial t} \Phi$ $= \sqrt{Z_0 F}\,\nabla \Phi$. Mit $t$ ist dabei die Zeit, mit $F$ die Rohrquerschnittsfläche, mit $\rho$ die mittlere Luftdichte und mit $Z_0$ der Wellenwiderstand bezeichnet.

\pagebreak
\subsection{Leitungselement}\label{Leitungselement}
\begin{afigure}
\begin{center}
\begin{picture}(12,2.5)
\put(0,0){\usebox{\EinAusgang}}
\put(2,0){\epsfig{file=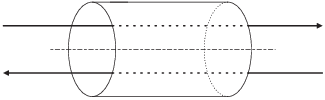, width=8cm}}
\end{picture}
\end{center}
\caption{Leitungselement}\label{Rohr}
\end{afigure}
Bild \ref{Rohr} zeigt ein Leitungselement der Länge $l$ mit konstanter Querschnittsfläche, dass in zwei Richtungen von Schall durchlaufen wird. 
Bedingt durch die endliche Signalausbreitungsgeschwindigkeit, die Schallgeschwindigkeit $c$, tritt eine zeitliche Verschiebung, beschrieben durch die Laufzeit $\triangle t=\frac{l}{c}$, zwischen den Eingängen $\vec{a}$ und den Ausgängen $\vec{b}$ auf:
\begin{align*}
b_1(t)&=a_2(t-\!\triangle t)\:,\\
b_2(t)&=a_1(t-\!\triangle t)\:.
\end{align*}

Um dieses Verhalten mittels zeitdiskreter Filter zu beschreiben, wird die Abtastperiode der Filter -- zunächst\footnote{Es wird sich in Abschnitt \ref{Kreuz} zeigen, dass man die Periodenlänge vorteilhaft verdoppeln kann.} -- so gewählt, dass sie der Laufzeit der Leitungslänge entspricht.
Mit den Abbildungen $b_{n,k}=b_n(k\negthickspace\triangle t)$ und $a_{n,k}=a_n(k\negthickspace\triangle t)$  unter Berücksichtigung des Abtasttheorems, d.~h.~die Bandbreite des zeitkontinuierlichen Signals sei kleiner der halben Abtastfrequenz, gewinnt man eine zeitdiskrete Darstellung:
\begin{align*}
b_{1,k} &= a_{2,k-1}\:,\\
b_{2,k}&=a_{1,k-1}\:.
\end{align*}

Transformiert man diese Gleichung in den $\cal{Z}$-Bereich, in Abschnitt \ref{ZT} wird auf die Zusammenhänge eingegangen, so folgt aus dem Verschiebungssatz: 
\begin{align*}
\trans{B_1}(z)&= z^{-1}\trans{A_2}(z)\:,\\
\trans{B_2}(z)&= z^{-1}\trans{A_1}(z)\:.
\end{align*}
Dies führt zu den Streu- und Betriebskettenmatrizen:
\[
\mat{S}=z^{-1}\begin{pmatrix} 0 & 1 \\ 1 & 0 \end{pmatrix}\:, 
\qquad
\mat{T}=\begin{pmatrix} z^{-1} & 0 \\ 0 & z \end{pmatrix}\:. 
\]

Die Matrizen des Leitungselements können erweitert werden, um eine beispielsweise durch Reibung, thermische Austauscheffekte oder Wandvibration hervorgerufene Dämpfung  zu berücksichtigen. In der einfachsten Form kann man hierfür eine linearen Dämpfungsterm mit dem Parameter $\alpha$ verwenden, wodurch sich die erweiterten Streumatrix und entsprechende Betriebskettenmatrix ergeben:
\[\mat{S}=e^{-\alpha}z^{-1}\begin{pmatrix}0&1 \\ 1&0 \end{pmatrix}, 
\qquad
\mat{T}=\begin{pmatrix}e^{-\alpha}z^{-1} &0 \\ 0&e^\alpha z \end{pmatrix}\:. 
\]
Der Dämpfungsfaktor ist von der Querschnittsfläche abhängig, und kann zur Verbesserung der phänomenologischen Approximation der Dampfungsursachen auch frequenzabhängig formuliert werden. Auf die Frequenzabhängigkeit der Dämpfung wird in Abschnitt \ref{frequenzDaempf} weiter eingegangen.

Einem homogenen Leitungselement lässt sich eine akustische Impedanz $Z^{ak}$ zuordnen, die durch das Verhältnis von Schalldruck $p$ zu Schallfluss $u$, dem Produkt aus Schallschnelle $v$ und Rohrquerschnittsfläche $F$, definiert ist. Die akustische Impedanz steht damit zur Feldimpedanz $Z^0$, dem Verhältnis von Schalldruck und Schallschnelle, in folgender Beziehung:
\[ 	
Z^{ak} = \frac{p}{u} =	\frac{p}{Fv}= \frac{1}{F}Z^0 \:,
\] 
die jeweils für hin- und rücklaufende Welle, also jeweils für die untere und obere Gleichung der drei anfangs erörterten Gleichungsdubletten gültig ist.

\subsection{Querschnittssprung}\label{Theo2Adapt}
\begin{afigure}[ht]
\begin{center}
\begin{picture}(12,5.0)
\put(0,1.155){\usebox{\EinAusgang}}
\put(2,0){\epsfig{file=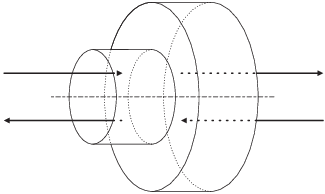, width=8cm}}
\put(4.29,.5){\makebox(0,0)[b]{$F_1$}}
\put(7.19,.5){\makebox(0,0)[b]{$F_2$}}
\end{picture}
\end{center}
\caption{Querschnittssprung}\label{Quersprung}
\end{afigure}

Bei dem in Bild~\ref{Quersprung} dargestellten Querschnittssprung\footnote{Der Querschnittssprung soll keine axiale Ausdehnung besitzen, das betrachtete Volumen ist folglich gleich Null. } von der Fläche $F_1$ auf die Fläche $F_2$ wird ein Teil der einlaufenden Welle reflektiert, der andere Teil transmittiert. Dies wird am Beispiel von Wellen in Druckdarstellung genauer betrachtet.

Das in den Querschnittssprung ein\-strömende Volumen muss gleich dem ausströmenden sein;\footnote{Dieses folgt aus dem Gaußschen Integralsatz $\int_V\negthickspace\nabla v\;dV=\oint_{\partial V}\negthickspace v\;d\vec{a}$, da nach Voraussetzung $V=0$ sein soll. Somit ist auch $\oint_{\partial V}\negthickspace v\;da = 0$.} die Flüsse $u$ sind das Produkt aus Querschnittsfläche und Geschwindigkeit (Schnelle) $v$: 
\[u_1 = -u_2\quad\Longleftrightarrow\quad F_1v_1 = -F_2v_2.\]
Die Zerlegung von $v$ in die Teilschnellen $v_1 = v_{a1}-v_{b1}$ und $v_2 = v_{a2}-v_{b2}$, orientiert an den jeweiligen Pfeilen in Bild \ref{Quersprung},  und die Verknüpfung $v = \frac{1}{\rho c}p$ mit der Schallgeschwindigkeit $c$ führen zu
\[F_1(p_{a1}-p_{b1}) =- F_2(p_{a2}-p_{b2}). \]
Da zudem der Druck $p$ als intensive Größe eindeutig ist, muss die Summe der linksseitigen Teildrücke gleich der der rechtsseitigen sein, also
\[ p_{a1}+p_{b1} = p_{a2}+p_{b2}. \]
Das sich aus den letzten beiden Gleichungen ergebende Gleichungssystem, nach $\vec{p_b}$ aufgelöst, ergibt:
\begin{equation*}\begin{array}{rrr}
p_{b1}\;=&\negthickspace\frac{F_1-F_2}{F_1+F_2}\,p_{a1}\;+&\negthickspace\frac{2F_2}{F_1+F_2}\,p_{a2}\\[2ex]
p_{b2}\;=&\negthickspace\frac{2F_1}{F_1+F_2}\,p_{a1}\;+&\negthickspace\frac{-(F_1-F_2)}{F_1+F_2}\,p_{a2}\makebox[0pt][l]{.}
\end{array}
\end{equation*}

Daraus ergibt sich die Streumatrix der Tabelle~\ref{2TorTab} in Druckdarstellung\footnote{Die anderen Tabelleneinträge ergeben sich durch Auflösen nach $\vec{u_b}$, $\vec{l_b}$ und so fort. Sie stellen verschiedene Sichtweisen des gleichen physikalischen Vorgangs dar.} 
mit dem Reflektionsfaktor
\[ r= \frac{F_1-F_2}{F_1+F_2}\:. \]
Der Reflektionsfaktor kann Werte aus $ \left[-1,1\right] $ annehmen. Besondere Beachtung verdienen der Randwert $+1$, hier ist die zweite Fläche gleich 
Null, man spricht von einem schallharten Abschluss, und der Randwert $-1$, die zweite Fläche ist infinit groß, ein schallweicher Abschluss. In beiden Fällen wird 
die Welle vollständig reflektiert, im zweiten mit umgekehrter Phasenlage.

Der Reflektionsfaktor kann auch durch die akustischen Impedanzen des rechts- und linksseitigen Rohrs, $Z^{ak}_1$ bzw.{} $Z^{ak}_2$, ausgedrückt werden.
\[ r= \frac{F_1-F_2}{F_1+F_2} = \frac{Z^0/Z^{ak}_1-Z^0/Z^{ak}_2}{Z^0/Z^{ak}_1+Z^0/Z^{ak}_2} =-\frac{Z^{ak}_1-Z^{ak}_2}{Z^{ak}_1+Z^{ak}_2} \:. \]
Ein Zusammenhang, der über Rohrmodell hinausgehend weiter von Nutzen ist.

\begin{atable}
\begin{center}\begin{tabular}{l|c|c}
Darstellung & $\mat{S}$ & $\mat{T}$ \\ \hline & & \\[-1.5ex] 
Druck & $\begin{pmatrix}r&1-r\\1+r&-r\end{pmatrix}$  & $\frac{1}{1+r}\begin{pmatrix}1&r\\r&1\end{pmatrix}$ \\[2ex]
Fluss & $\begin{pmatrix}r&1+r\\1-r&-r\end{pmatrix}$  & $\frac{1}{1-r}\begin{pmatrix}1&r\\r&1\end{pmatrix}$ \\[2ex]
Wurzelleistung & $\begin{pmatrix}r&\sqrt{1-r^2}\\ \sqrt{1-r^2} & -r\end{pmatrix}$  & $\frac{1}{\sqrt{1-r^2}}\begin{pmatrix}1&r\\r&1\end{pmatrix}$\\[2ex]
--- & $\begin{pmatrix}r&1-r^2\\1&-r\end{pmatrix}$  & $\begin{pmatrix}1&r\\r&1\end{pmatrix}$ 
\end{tabular}\end{center}
\caption{Streu- und Betriebskettenmatrixdarstellung des 2-Tor Adaptors}\label{2TorTab}
\end{atable}

\subsection{Mehrtor-Adaptor}\label{Mehrtor}
Analog dem 2-Tor-Adaptor zur Beschreibung des Querschnittssprungs gilt für Mehrtor-Adaptoren mit $n$ Ein- und Ausgängen:
\[ p_{ak}+p_{bk} = p_{al}+p_{bl} \quad \forall \: k,l \in \{1,2 ... n \}\:, \] 
\[ \sum_{i=1}^n u_{ai}-u_{bi} =  0\:. \]
Es ergibt sich somit die Streumatrix: 
\[  \vec{B}=
\begin{pmatrix} b_1 \\ b_2 \\ \vdots\\ b_n \end{pmatrix} 
 =  \begin{pmatrix}
s_{11} & s_{12} & \cdots & s_{1n} \\
s_{21} & s_{22} & \cdots & s_{2n} \\
\vdots & \vdots & \ddots & \vdots \\
s_{n1} & s_{n2} & \cdots & s_{nn} \\ 
\end{pmatrix} 
\begin{pmatrix}a_1 \\ a_2 \\ \vdots\\ a_n \end{pmatrix} 
=\mat{S}\,\vec{A}\:. \]
In $\mat{S}$ ist $s_{kk} =\frac{2F_k}{F_\Sigma}-1$
und für $ l \not= k$ in Druckdarstellung
 $s_{lk}=\frac{2F_k}{F_\Sigma}$, mit $F_\Sigma=\sum\limits_{i=1}^n F_{i}$. Die Flussdarstellung ergibt sich analog: $ s_{lk}=\frac{2F_l}{F_\Sigma}$.

Mit dem Mehrtor-Adaptor lassen sich Verzweigungen des Wellenleiters beschreiben. Diese, für eine Reihe von Fragestellungen zur Sprechtraktakustik relevante Erweiterung des Sprechtraktmodells wird in Abschnitt \ref{TheoBaumRohr} eingehend betrachtet. Alternativ ist die Ankopplung von  nicht unmittelbar akustisch motivierten Filtern über den Mehrtoradaptor möglich, wie in Abschnitt~\ref{FIR-Filter} gezeigt ist. Desweiteren kann der Mehrtoradaptor verwendet werden, um weitere Anregungen, wie das Rauschen der Frikative, an bestimmten Stellen in das Modell des  Vokaltrakts einzuspeisen, um damit die Schallquelle an der Verengungsstelle von Frikativen phänomenologisch und die Schallausbreitung wirklichkeitsnah nachzubilden.

\subsection{Schallabstrahlung}\label{Abstrahlung}
Bei der Schallabstrahlung am Mund und in gleicher Weise von den Nasen\-löchern findet ein Übergang von der akustischen Impedanz des Rohrquerschnitts zu der des Freifeldes statt. Um diesen Übergang von einem Rohr mit endlicher Querschnittsfläche in einen sehr viel größeren Halbraum zu beschreiben, kann in erster Näherung ein Querschnitts\-sprung-Adaptor mit dem Reflektionsfaktor $r=-1$ gewählt werden, wie in Abschnitt~\ref{Theo2Adapt}~beschrieben.

Befindet sich der Querschnitt der Austrittsöffnung in der Größenordnung der betrachteten Wellenlängen, so gibt diese Näherung die physikalischen Vorgänge allerdings nur unvollkommen wieder, da an dieser Stelle ein Übergang der ebenen Wellen des Rohrmodells in die kugelförmigen Wellen des Halbraums stattfindet. Dies wirkt sich besonders auf Moden höherer Frequenz aus, die durch den daraus resultierenden Strahlungswiderstand stärker gedämpft werden. Laine hat in \cite{ULai82} für die Impedanz das folgende Modell vorgeschlagen:
\[ z_{pz}=\frac{a(1-z^{-1})}{1+bz^{-1}}\:,\]
wobei die Parameter $a$ und $b$ von dem Verhältnis des Öffnungsradius zu dem  Produkt aus Abtastperiode und Schallgeschwindigkeit abhängen.  

Ein einfacher Hochpass bildet zwischen den beiden zuvor genannten Beschreibungen einen Kompromiss bezüglich physikalisch treffender Modellbildung und Filterkomplexität. Ein wichtiger Vorteil des einfachen Hochpasse liegt darin, dass sowohl dessen Koeffizient als auch die Sprechtraktkonfiguration in einfacher Weise aus einem Sprachsignal geschätzt werden können, wie in Abschnitt \ref{Parameter} gezeigt wird.

\subsection{Kreuzgliedketten}\label{Kreuz}\label{TheoLinRohr}
In diesem Abschnitt werden die aus Adaptoren und Leitungselementen kombinierbaren Filter betrachtet, die das akustische Übertragungsverhalten des Sprechtrakts nachbilden. 
\begin{afigure}[t]
\begin{center}
\begin{picture}(11.2,3.9)
\newsavebox{\kreuz}
\savebox{\kreuz}(0,0)[bl]{
\setlength{\unitlength}{.3cm}
\newcommand{\addcircle}[1]{ \thinlines\put(#1,0){\circle{#1}\makebox(0,0){$+$}}}
\thicklines
 \put(0,7){\vector(1,0){1.05}}  

 \put(3.3,7){\vector(1,0){1.75}}  

 \put(6.95,7){\line(1,0){4.05}}
 \put(10,7){\vector(-2,-3){3.45}}  
 \put(4,1){\addcircle{2}}
 \put(4,7){\addcircle{2}}
 \put(10,1){\circle*{.3}}
 \put(10,7){\circle*{.3}}
 \put(11,1){\vector(-1,0){4.05}}  
 \put(5.05,1){\vector(-1,0){1.7}}
 \put(10,1){\vector(-2,3){3.45}}  
 \put(1,0){\thinlines\dashbox{0.2}(2.3,2){$\,z^{-1}$}}
 \put(6.5,3){\makebox(0,0)[r]{$r$}}
 \put(6.5,5){\makebox(0,0)[r]{$-r$}}
 \put(1,1){\line(-1,0){1}}

 \put(1,6){\thinlines\framebox(2.3,2){$\,z^{-1}$}}

 \thinlines
\put(0.4,-4){\makebox(3.0,13){}}
\put(3.9,-4){\makebox(7.0,13){}}
	
}

\newsavebox{\altkreuz}
\savebox{\altkreuz}(0,0)[bl]{
\setlength{\unitlength}{.3cm}
\newcommand{\addcircle}[1]{ \thinlines\put(#1,0){\circle{#1}\makebox(0,0){$+$}}}
\thicklines
 \put(0,7){\vector(1,0){1.05}}  

 \put(3.3,7){\vector(1,0){1.75}}  

 \put(6.95,7){\line(1,0){4.05}}
 \put(10,7){\vector(-2,-3){3.45}}  
 \put(4,1){\addcircle{2}}
 \put(4,7){\addcircle{2}}
 \put(10,1){\circle*{.3}}
 \put(10,7){\circle*{.3}}
 \put(11,1){\vector(-1,0){4.05}}  
 \put(5.05,1){\vector(-1,0){1.7}}
 \put(10,1){\vector(-2,3){3.45}}  
 \put(1,0){\thinlines\framebox(2.3,2){$\,z^{-1}$}}
 \put(6.5,3){\makebox(0,0)[r]{$r$}}
 \put(6.5,5){\makebox(0,0)[r]{$-r$}}
 \put(1,1){\line(-1,0){1}}

 \put(1,6){\thinlines\dashbox{0.2}(2.3,2){$\,z^{-1}$}}

 \thinlines
\put(0.4,-4){\makebox(3.0,13){}}
\put(3.9,-4){\makebox(7.0,13){}}
	
}

\put(0,0){\usebox{\kreuz}}
\put(3.3,0){\usebox{\altkreuz}}
\put(7.5,0){\usebox{\altkreuz}}
\put(6.7,2.1){\LARGE$\cdots$}
\put(.6,.5){\makebox(0,0){$\bf T_1$}}
\put(2.25,.5){\makebox(0,0){$\bf T_2$}}
\put(3.9,.5){\makebox(0,0){$\bf T_3$}}
\put(5.55,.5){\makebox(0,0){$\bf T_4$}}
\put(8.1,.5){\makebox(0,0){$\bf T_{n-1}$}}%
\put(9.75,.5){\makebox(0,0){$\bf T_{n}$}}%

\put(-.11,0)
{
	\thicklines
	\put(0.41,1.5){\vector(-1,0){.37}}
	\put(11.33,1.5){\line(-1,0){.5}}
	\put(10.72,3,3){\vector(1,0){.5}}
	\thinlines
	\put(0,3.3){\circle{.225}}
	\put(11.33,3.3){\circle{.225}}
	\put(0,1.5){\circle*{.09}}
	\put(11.33,1.5){\circle*{.09}}
	\put(0,3.5){\makebox(0,0)[b]{$\trans{a_1}$}}
	\put(0,1.7){\makebox(0,0)[b]{$\trans{b_1}$}}
	\put(11.33,3.5){\makebox(0,0)[b]{$\trans{b_2}$}}
	\put(11.33,1.7){\makebox(0,0)[b]{$\trans{a_2}$}}
}
\end{picture}
\end{center}
\caption{Signalflussgraph eines unverzweigten Rohrsystems. Der Signalflussgraph basiert auf Tabelle \ref{2TorTab}, letzte Zeile, und zeigt die typischen namensgebenden kreuzförmigen Elemente, die kettenartig angeordnet sind.}\label{unR}
\end{afigure}
Diese Filter bestehen aus einer alternierenden Folge von 2-Tor Adap\-toren und Leitungselementen. Dies veranschaulicht der Signalflussgraph in Bild \ref{unR}, bei dem 
der Eingang $a_1$  und der Ausgang $b_2$ ist. 
Die Übertragungsfunktion 
\[{\rm H(}z{\rm )}=\frac{\trans{B_2}}{\trans{A_1}}= \frac{1}{\trans{T_{22}}}\] 
erhält man durch Multiplizieren der Betriebskettenmatrizen:
\[  \mat{T}=\begin{pmatrix}\trans{T_{11}}& \trans{T_{12}}\\ \trans{T_{21}}&\trans{T_{22}}\end{pmatrix}={\bf T_1 T_2 T_3 T_4} \cdots {\bf T_{n-1} T_n}\:. \] 
Die Übertragungsfunktion besitzt nur Pole. Aufgrund der Beschränkung von $r$ liegen die Pole innerhalb des Einheitskreises, somit ist das System stabil.\footnote{Man schließt die beiden in diesem Fall physikalisch nicht sinnvollen Extremalwerte $r= +1$ und $r= -1$ aus.}

In Abschnitt \ref{Äquivalenz} wird ein Beispiel für die Berechnung der Übertragungsfunktion gegeben. An diesem Beispiel fallen die nur gradzahlige Potenzen von $z$  in der Übertragungsfunktion auf. Die daraus folgende symmetrische Übertragungsfunktion deckt sich nur in der unteren Hälfte mit Messungen, vgl.{} Bild~\ref{aiu} und Bild~\ref{mnl}. Die Grenzen des Modells sind in der oberen Hälfte der Übertragungsfunktion überschritten, da der Vokaltrakt eben nicht stückweise homogen ist. Dies kann durch eine Beschränkung der Betrachtung auf die untere Hälfte der Übertragungsfunktion oder durch einen anderen Ansatz zur Beschreibung der Laufzeitglieder behoben werden. Für letzteren halbiert man die Laufzeit für hin- rücklaufende Welle von $z^{-1}$ auf $z^{-1/2}$.  Realisierbare und den Vokaltrakt treffend beschreibende  Filter erhält man daraus, in dem die Laufzeitglieder der hin- und rücklaufenden Welle alternierend in dem Signalflussgraphen wieder zu $z^{-1}$ zusammenfasst werden, also im Ergebnis die in Bild \ref{unR} gestrichelt gezeichneten Laufzeitglieder entfernt werden. Vertiefendes zeigen \cite{ALac96} und die dort genannten Referenzen. In den Darstellungen dieser Arbeit wird überwiegend die erste Alternative genutzt, um einen direkten Vergleich mit den hauptsächlich betrachteten Finiten Differenzen zu ermöglichen, die Implementierungen verwenden stetz die mit geringerem Rechenaufwand behafteten halbierten Laufzeiten. 


\subsection{Parameterschätzung}\label{Parameter}\label{Präemphase}
Der Querschnittsverlauf des Sprechtrakts beziehungsweise die Parameter des äquivalenten Kreuzglied-Kettenfilters können anhand von Sprachsignalen bestimmt werden. Diese Methode wird im Folgenden genauer beschrieben. Um zutreffende Querschnittsverläufe zu erhalten, muss hierfür der Einflüsse des Sprechtrakts von anderen Einflüssen getrennt werden, die eine spektrale Färbung des Sprachsignals hervorrufen. 

Die spektrale Färbung (Spektrum) des Sprachsignals $\trans{S}(z)$ wird durch die Glottis $\trans{G}(z)$, die Abstrahlung $\trans{R}(z)$ und den Sprechtrakt $\trans{H}(z)$ hervorgerufen:
\[
\trans{S}(z) =\trans{R}(z)\trans{H}(z)\trans{G}(z).
\] 
Für die Schätzung der Parameter eines Filters, welches den Vokaltrakt beschreibt, ist folglich von dem Spektrum des Sprachsignals die spektrale Färbung durch Glottis und Abstrahlung durch eine Vorfilterung zu entfernen. Der spektrale Effekt von Anregung und Abstrahlung wird in guter Näherung durch ein Produkt von Filtern erster Ordnung  beschrieben, die eine Hoch- oder Tiefpasscharakteristik aufweisen. Diese Charakteristik unterscheidet sich deutlich von den Formanten des Sprechtrakts, die durch Resonanzstrukturen gebildet werden, also Produkte  Filter zweiter Ordnung sind. Entsprechend kann man einfach in guter Näherung durch eine Einschränkung auf einen Koeffizienten der im Folgenden beschriebenen \name{Linear Prediction} die Anregungs- und Abstrahlcharakteristik bestimmen, indem man diese ein- oder mehrfach anwendet. 
Bild~\ref{L-BURG-PRE} zeigt das Resultat einer Schätzung des Betragsgangs des Vokaltrakts für eine doppelte, adaptive Preemphase\footnote{Auf eine Übersetzung des gebräuchlichen engl.{} Ausdrucks wird verzichtet.}.  

\begin{afigure}[t]
\begin{center}

\psfrag{-40}[r][r]{\scriptsize 0}
\psfrag{-30}[r][r]{\scriptsize 10}
\psfrag{-20}[r][r]{\scriptsize 20}
\psfrag{-10}[r][r]{\scriptsize 30}
\psfrag{0}[r][r]{\scriptsize 40}
\psfrag{10}[r][r]{\scriptsize 50}
\psfrag{20}[r][r]{\scriptsize 60}
\psfrag{30}[r][r]{\scriptsize [dB]}
\psfrag{a}[t][t]{\scriptsize 0}
\psfrag{b}[t][t]{\scriptsize 4}
\psfrag{d}[t][t]{\put(-3, -0.2){\scriptsize $f/\mathrm{[kHz]}$}\scriptsize 8}
\epsfig{file=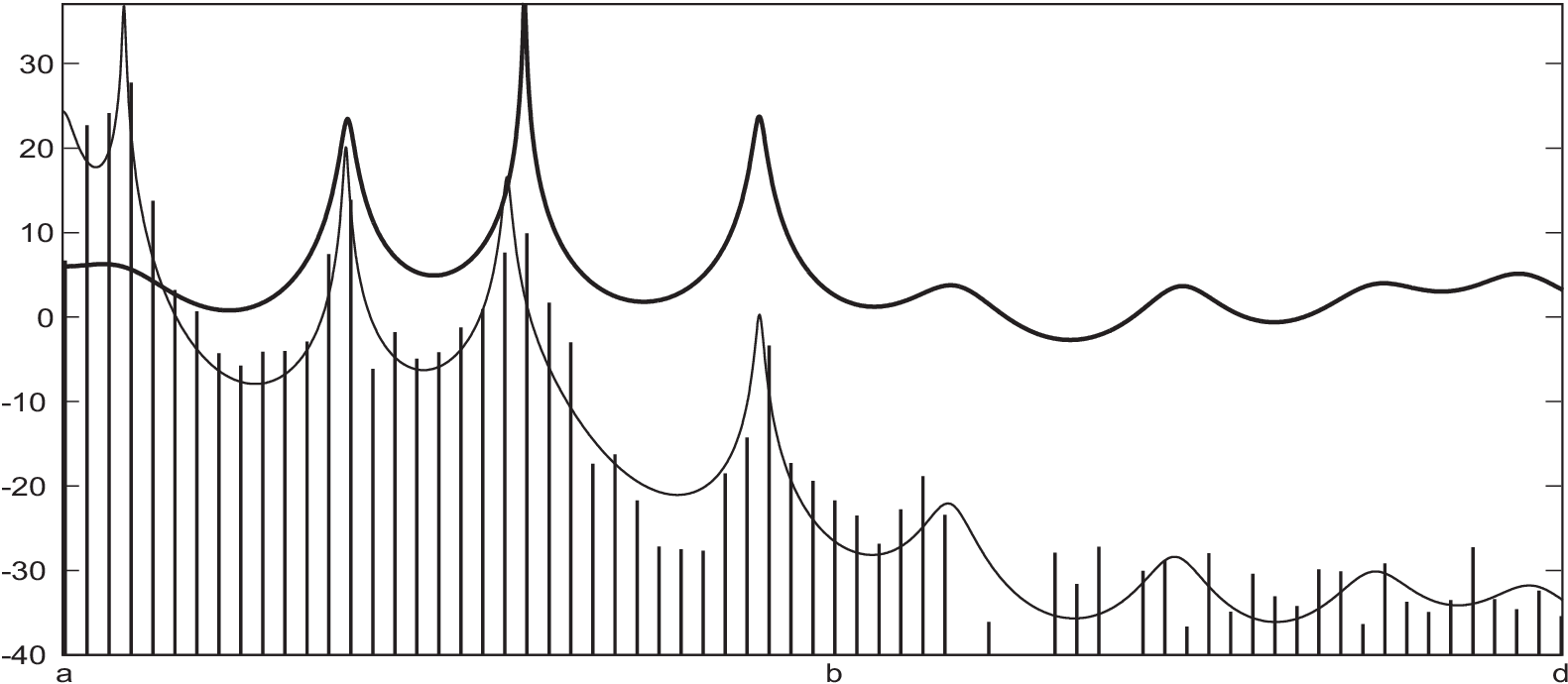, width=.7\linewidth}\vspace{-.5ex}
\end{center}
\caption{Parameterschätzung mittels Burg-Methode aus einem Lautspektrum. Gezeigt ist das Betragsspektrum des Lauts~[l] in Balkendarstellung, die Analyse mittels Burg-Methode als dünne Linie und selbige nach Anwendung einer zweifachen Preemphase, dicke Linie. Es ist gut zu erkennen, dass die Preemphase den spektralen Abfall eliminiert, der nicht aus dem Querschnittsverlauf des Sprechtrakts selbst resultiert, aber dessen Resonanzeigenschaften erhält. }\label{L-BURG-PRE}
\end{afigure}

\label{TheoLPC}
Die Idee hinter der Linear Prediction ist die Vorhersage des weiteren Signals anhand eines Abschnitts zurückliegender Signalwerte durch ein lineares System, welches die Signalwerte linear kombiniert: je genauer diese gelingt, um so besser bildet das System die betrachtete Signalquelle nach. Entsprechend gut kann man von den ermittelten Systemeigenschaften auf die ursprüngliche Signalquelle schließen. In Abschnitt \ref{DigitalSynthese} sind bereits einige dieser Verfahren genannt. Sie unterscheiden sich hinsichtlich des Maßes, mit dem der Abstand zwischen vorhergesagtem und wirklichem Signal gemessen wird und in der algorithmischen Herangehensweise. \cite{La05} betrachtet die unterschiedlichen Verfahren und kommt zu dem Schluss, dass sie sich im Ergebnis für die hier betrachteten Laute trotz verschiedener Ansätze kaum unterscheiden. Exemplarisch wird deshalb nur auf zwei Verfahren eingegangen, die auch in Teil \ref{SPEAK} verwendet werden.

Der Burg-Algorithmus und das Verfahren nach Itakura und Saito nehmen als Signalquelle ein autoregressives System an, wie das Rohrmodell des Sprechtrakts, dessen Übertragungsfunktion nur Pole aufweist. Das Systemverhalten lässt sich durch ein rekursionsfreies Filter invertieren, der durch Umkehren der oberen Signalflussrichtung in Bild \ref{unR} gebildet wird; oben liegende Laufzeitelemente werden überlicherweise nach unten verschoben, die Ordnung $n/2$ des Filters entspricht der Anzahl der Laufzeit- bzw.{} Kreuzglieder des Rohrmodells. Für dieses rekursionsfreie Filter lassen sich aus Signalabschnitten der Länge $l$ die Filterkoeffizienten $r_i$ schrittweise bestimmen. Nach Burg ist
\[r_i = \frac{-2 \sum\limits_{k=1}^l f_{i,k}b_{i,k}} {\sum\limits_{k=1}^l f_{i,k}^2 + \sum\limits_{k=1}^l b_{i,k}^2}
\]
und mit PARCOR nach Itakura und Saito ist 
\[ r_i= \frac{-\sum\limits_{k=1}^l f_{i,k}b_{i,k}} {\sqrt{ \sum\limits_{k=1}^l f_{i,k}^2 \cdot \sum\limits_{k=1}^l b_{i,k}^2}}\:,
\]
womit $f_{i-1,k} = f_{i,k} +r_ib_{i,k-1}$ und $b_{i-1,k} = b_{i,k-1} +r_if_{i,k}$, $b_{i-1,1} =0$ für den nächsten Schritt gebildet wird. Dabei sind $r_i$, $f_i$ und $b_i$ dem Adaptor $\bf T_{2i}$ zugeordnet, letztere sind die rechtsseitigen Signale im oberen bzw.{} unteren Signalpfad. Zu Beginn wird $r_{n/2} = -1$, $b_{n/2,k}=0$ gesetzt und $f_{n/2,k}$ mit den Signalwerten des betrachteten Abschnitts belegt. Die Algorithmen enden mit der Berechnung von $r_1$. Beide Verfahren sind  mit der Einbeziehung der Signalenergie im oberen und im unteren Pfad numerisch stabil, sie unterscheiden sich lediglich in deren Mittelung, die bei Burg arithmetisch und nach Itakura und Saito geometrisch erfolgt. Man erkennt an den Gleichungen zur Reflexionsfaktorbestimmung die Arbeitsweise beider Verfahren: Je größer der  Korrelationskoeffizient zwischen den Signalen $f_i$, $b_i$  ist, oder allgemeiner die Kreuzenergie im Verhältnis zum Mittel, umso größer ist der inverse Reflexionsfaktor, wodurch diese Korrelation in dem folgenden Mischschritt beseitigt wird. Da beide Signale anfangs gleich sind, werden sie so Schritt um Schritt spektral weißer. 

\subsection{Verzweigte Rohrsysteme}\label{TheoBaumRohr}
Mit verzweigten Rohrsystemen lassen sich weitere akustische Prinzipien des Sprechtrakts während der Artikulation bestimmter Laute untersuchen. Dies wird im Folgenden exemplarisch gezeigt und dabei die zugrundeliegende Methodik betrachtet.

So lassen sich die beiden um den vorderen Zungenbereich herumführenden Passagen bei dem später noch eingehender betrachteten Laterallaut [l] durch ein in diesen Bereich aufgespaltenes Rohrsystem beschreiben, wie bspw.{} \cite {ZhEWT03} ausführt: Durch die unterschiedlichen Schall- bzw.{} Signallaufzeiten in beiden Passagen aufgrund von natürlichen Asymmetrien ergeben sich bei bestimmten Frequenzen Interferenzen, die eine Schallabstrahlung verhindert oder reduziert und als Nullstellen im Spektrum hervortreten. 

Nach \cite{MüM04} lassen sich mit einer Abzweigung Quermoden in einem rotationssymmetrischen Rohrsystem vereinfacht betrachten, solange sie bestimmten Proportionen genügen.
An diesem einfachen Beispiel lassen auf kurzem Weg die Auswirkungen von Quermoden bzw.{} Abzweigungen zeigen. Als Abzweigung wird ein einseitig geschlossenes, homogenes Rohr mit der Gesamtlaufzeit $z^{-n}$ verwendet; diese Abzweigung setzt an einem Rohr mit gleichem Querschnitt an. Für den dreifach querschnittsgleichen Dreitoradaptor in Druckdarstellung gilt nach Abschnitt \ref{Mehrtor}
\[  
\begin{pmatrix} b_1 \\ b_2 \\ b_3 \end{pmatrix} 
 =  \frac{1}{3}\begin{pmatrix}
-1 & 2 & 2 \\
2 & -1 & 2 \\
2 & 2 & -1 \\ 
\end{pmatrix} 
\begin{pmatrix}a_1 \\ a_2 \\ a_3 \end{pmatrix} 
\]
und führt im $\cal{Z}$-Bereich mit der am dritten Tor angesetzten Gesamtlaufzeit  $A_3 = z^{-n}B_3$ zu 
\[  
\begin{pmatrix} B_1 \\ B_2  \end{pmatrix} 
 =  \frac{1+z^{-n}}{3+z^{-n}}\begin{pmatrix}
-1 & 2  \\
2 & -1  \\ 
\end{pmatrix} 
\begin{pmatrix}A_1 \\ A_2  \end{pmatrix} 
\]
und schließlich zur Übertragungsfunktion $H(z)= B_2/A_1=\frac{2(1+z^{-n})}{3+z^{-n}} $. Sie unterscheidet sich durch das Auftreten von Nullstellen und in dem Gruppenlaufzeit, die in Bild \ref{Bessel-Rohr} gezeigt ist, von einer durch Querschnittssprung-Adaptoren beschriebenen Ausbuchtung. Diese zusätzliche Gruppenlaufzeit der Bessel-Mode bewirkt folglich eine akustische Verlängerung des einfachen Rohrs und damit eine Frequenzverschiebung von Resonanzen aus den Querschnittsverlauf zu tieferen Frequenzen hin.
\newsavebox{\dreisymb}
\newsavebox{\kreuzsymb}
\savebox{\dreisymb}[1em]{\includegraphics[width=1em]{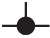}}
\savebox{\kreuzsymb}[1em]{\raisebox{0.15ex}{\includegraphics[width=1em]{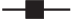}}}
\begin{afigure}[b]
	\begin{minipage}[b]{.4\linewidth}
		\psfrag{R}{\small$Y$}
		\psfrag{E}{\small$X$}
		\includegraphics[width=\linewidth]{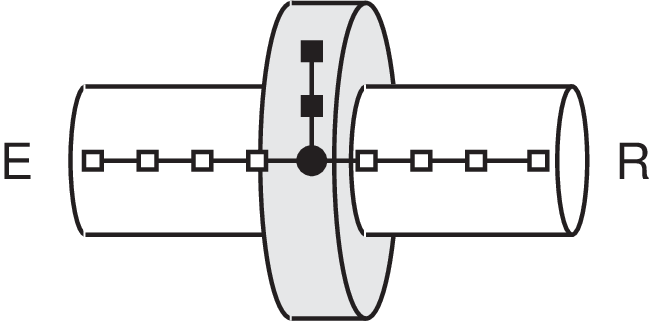}
	\end{minipage}
	\hfill
	\begin{minipage}[b]{.55\linewidth}
	\psfrag{5}{}
\psfrag{A}[tc][tc]{\scriptsize 0}
\psfrag{0.2}[tc][tc]{\scriptsize 0,2}
\psfrag{0.4}[tc][tc]{\scriptsize 0,4}
\psfrag{0.6}[tc][tc]{\scriptsize 0,6}
\psfrag{0.8}[tc][tc]{\scriptsize 0,8}
\psfrag{B}[tc][tc]{\scriptsize 1,0}
\psfrag{0}[r][r]{\scriptsize 0}
\psfrag{1}[r][r]{\scriptsize 1}
\psfrag{2}[r][r]{\scriptsize 2}
\psfrag{3}[r][r]{\scriptsize 3}
	\centering \includegraphics[width=.9\linewidth, height=.38\linewidth]{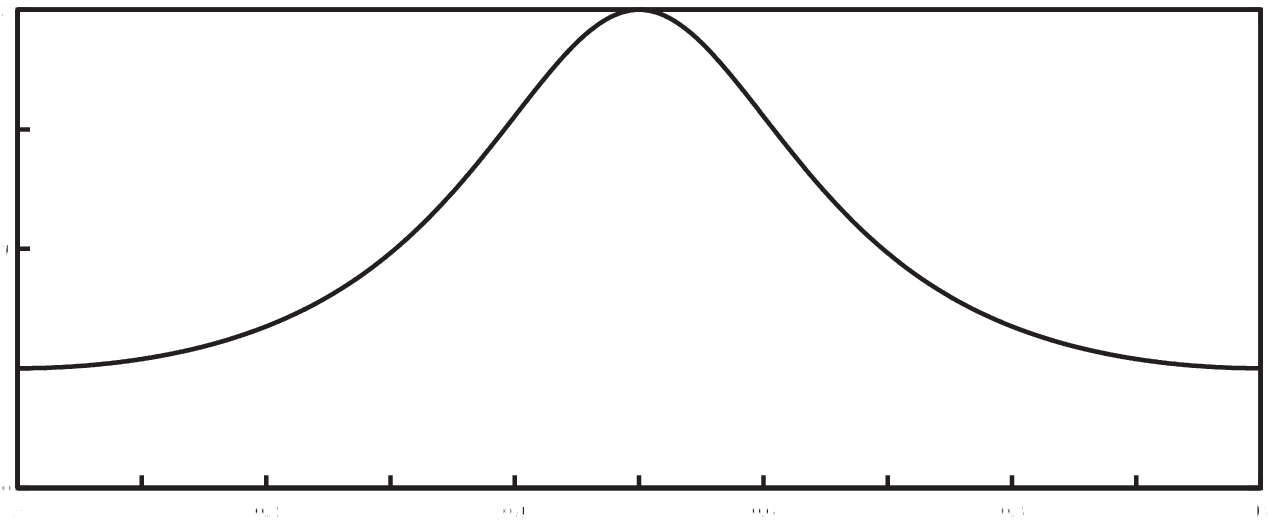}\put(-.85,-.2){\scriptsize $\omega/\pi$}
	\end{minipage}\
		\caption{Links: Rotationssymmetrische Rohrerweiterung, deren Besselmode mit der überlagerten Filterstruktur berücksichtigt wird. Das Symbol \usebox{ 
\dreisymb}  kenzeichnet den Dreitoradaptor, \usebox{\kreuzsymb}  ein Zweitor-Kreuzgliedelement und dazwischenliegende Linien Laufzeiten. Betrachtet werden schattierte/gefüllte Elemente; $X$ und $Y$ bezeichnen Ein- und Ausgang. Rechts: Zusätzliche Gruppenlaufzeit für $n=2$. }\label{Bessel-Rohr}
\end{afigure}

Wie in Bild \ref{disV} erkennbar ist und beispielsweise in \cite{MLiu98} eingehend betrachtet wird, kann der Nasaltrakt, als Rohr modelliert, über ein Dreitoradapter zur Nachbildung der Öffnung des Gaumensegels mit dem pharyngal-oralen Bereich des Sprechtrakts gekoppelt werden. Diese Idee weiterführend ist zu Beginn dieser Arbeit in \cite{RaSL99} untersucht, ob man durch ein verzweigtes Rohrsystem, dessen Topologie an der des Nasaltrakts orientiert, die Akustik des Nasaltrakts erfassen kann.  Dies motiviert die in Bild \ref{NasalFilter} gezeigte Filterstruktur. Die Nachbildung des Übertragungsverhaltens  gelingt mit dieser Struktur bis etwa 3 kHz befriedigend, für höhere Frequenzen zeigen sich erhebliche Abweichungen. Auf eine weitere Betrachtung, ob die dem Filter zugrundeliegenden Rohrquerschnitte die Geometrie des Nasaltrakts widerspiegeln, wird deshalb verzichtet.
\begin{afigure}[b]
	\begin{minipage}[b]{.58\linewidth}
		\includegraphics[width=\linewidth]{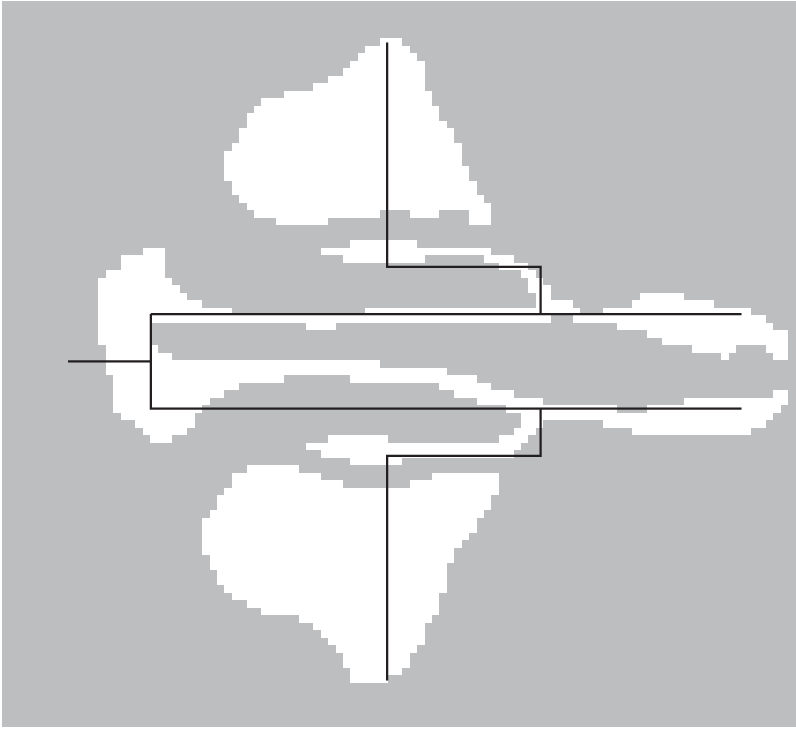} 
		
	\end{minipage}
	\hfill
	\begin{minipage}[b]{.36\linewidth}
		\psfrag{R}{\small$Y^r$}
		\psfrag{L}{\small$Y^l$}
		\psfrag{E}{\small$X$}
		\centering \includegraphics[width=.87\linewidth]{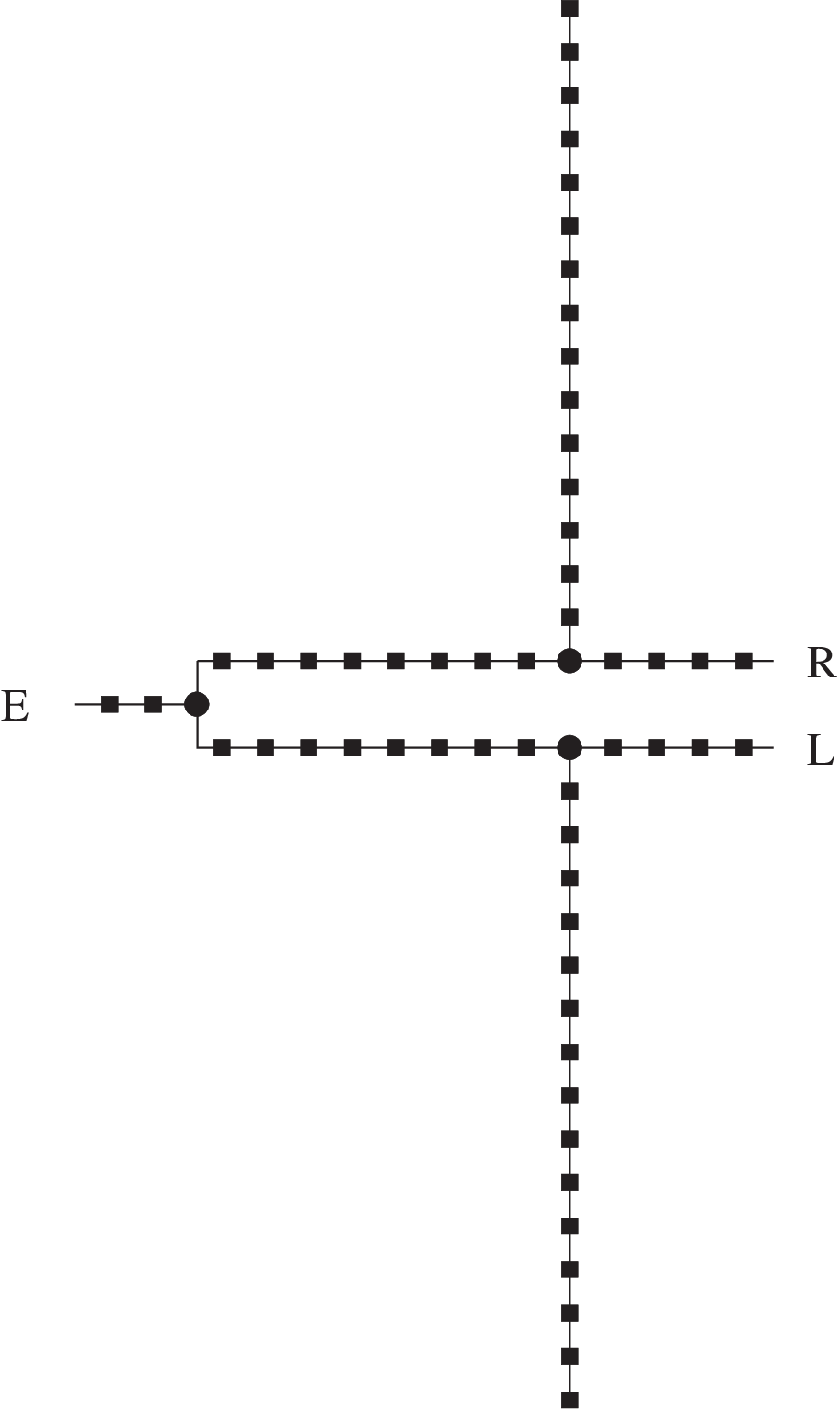} 
	\end{minipage}
\caption{Links: Horizontaler Schnitt durch den Nasaltrakt, überlagert mit einer stilisierten Filterstruktur. Diese beginnt links am Velum, spaltet sich an der Nasenscheidewand auf, erfasst über Abzweigungen die größten Nebenhöhlen ({\it Sinus maxillaris}, oben und unten) und führt bis zu den beiden Nasenlöchern. 
Rechts: Filterstruktur, verdeutlicht  mit den Symbolen \usebox{\kreuzsymb}, \usebox{\dreisymb}  für Zwei- und Dreitoradaptor.}
\label{NasalFilter}
\end{afigure}

Die Berechnung der Übertragungsfunktion erfolgt bei einfach verzweigten Rohrsystemen, indem die Betriebskettenmatrixen der Verzeigung aufmultipliziert werden und 
das Produkt zur Elimination des dritten Tors des Dreitoradapters genutzt wird, wie  in dem ausgeführten Beispiel. Der reduzierte Dreitoradaptor kann dann in eine Betriebskettenmatrix umgeformt werden und das Übertragungsverhalten nach Abschnitt \ref{TheoLinRohr} bestimmt werden. Bei mehrfach verzweigten Systemen, wie in \cite{RaSL99}, wird das Verfahren iterativ angewendet; für zyklische Systeme wird eine Matrix mit den Bezugsgrößen der Knotenpunkte gebildet und gelöst --- analog der bekannten Knoten-Maschen-Analyse linearer elektrischer Schaltungen. Eine Übersicht gibt \cite{Ra99}; die Parameterbestimmung gelingt mit Gradientenverfahren.

\newpage
\section[Mehrdimensionale Integration der Wellengleichung]{Mehrdimensionale Integration der Wellen-\\gleichung}
Wenngleich das Rohrmodell die Ausbreitung ebener Wellen exakt beschreibt, hat sich bereits im vorangegangenen Abschnitt angedeutet, dass der Zerlegung der Sprechtraktakustik in Bereiche ebener Wellen Grenzen gesetzt sind. Diese werden in den nächsten Abschnitten näher betrachtet und der umgekehrte Weg untersucht, bei dem die Wellenausbreitung nur approximativ, dafür aber die dreidimensionale Geometrie exakt erfasst wird.
\subsection{Motivation}\label{nasaltrakt}
In vielen Sprachen tritt die Lautklasse der Nasale (im Deutschen [m], [n], [\ng]) häufig auf. Um deren Lautbildung zu verstehen und mittels eines akustisch motivierten Modells zu reproduzieren, ist eine genaue Kenntnis der Schallausbreitung im Nasaltrakt notwendig. Da die räumliche Konfiguration des Nasaltrakts im Gegensatz zu der des Vokaltrakts zeitlich konstant ist, kann sie mit vergleichsweise langwierigen medizinischen Untersuchungsmethoden ermittelt werden. Die räumliche Konfiguration ist damit dreidimensional abbildenden  Verfahren, wie Kernspin-Resonanz-Tomographie, Computer-Tomographien oder Kryo-Sektionen zugänglich, und es lassen sich die akustischen Eigenschaften mittels numerischer Verfahren daraus bestimmen. Schematisch ist der Nasaltrakt in Bild \ref{nasaltrakt_bild} dargestellt, er bildet die Verbindung des Rachens mit den Nasenlöchern. 
\begin{afigure}[b]
\begin{center}
\psfrag{KopA}[r][r]{\scriptsize Stirnhöhle ({\it Sinus frontalis})}
\psfrag{KopC}[r][r]{\scriptsize Kieferhöhle ({\it Sinus maxillaris})}
\psfrag{KopB}[r][r]{\scriptsize Keilbeinhöhle ({\it Sinus sphenoideus})}
\epsfig{file=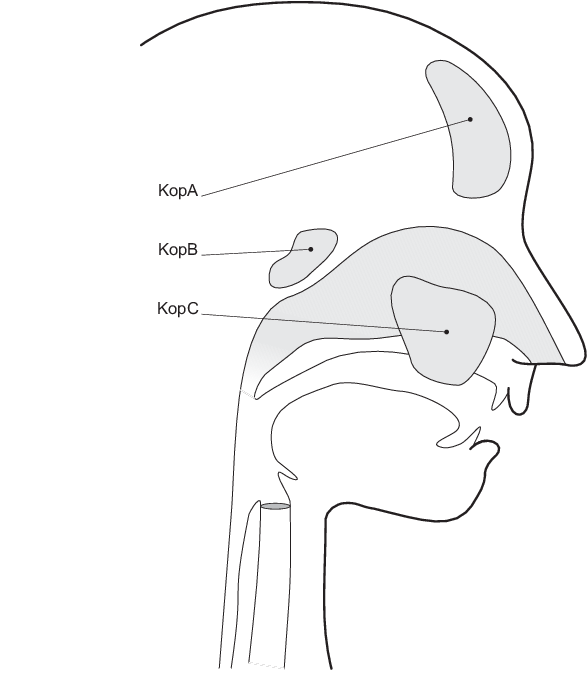,width=0.5\linewidth}\vspace{-1ex}
\end{center} 
\caption{Nasaltrakt und Nebenhöhlen}\label{nasaltrakt_bild}
\end{afigure}%
Der durch die durch die Nasenscheidewand ({\it Septum}) längsgeteilte Verbindungsgang ({\it Meatus nasi communis}) wird durch jeweils drei muschelförmige Knorpel- bzw.~Knocheneinbuchtungen ({\it Concha inferior}, {\it C.~media} und {\it C.~superior}) verengt.  Die Akustik des Nasaltrakts wird, wie bereits erwähnt durch mehrere Nebenhöhlen beeinflusst, die mit ihm über dünne Kanäle verbunden sind.

Desweiteren ergeben sich auch in der Mundhöhle bei Frequenzen ab etwa 4 kHz teils deutliche Abweichungen vom Modell der ebenen Welle, wie es ausführlich in \cite{Mo02} erörtert wird\arxiv{. Das daraus hier wiedergegebene Bild \ref{vokaltrak_FEM} illustriert den Effekt}: Man erkennt die starke Schrägstellung der Intensitätsbereichskontouren im Bereich der vorderen Mundhöhle, insbesondere deren Verwerfung bei den Lippen; auch die Ausbildung der Kontourverläufe am Gaumen lässt sich nicht mit dem Modell ebener Wellen beschreiben. Die gezeigte Simulation wird auch durch in \cite{Mo02} zitierte Messungen belegt und deren Effekt auf die Resonanzeigenschaften quantifiziert. 

\arxiv{\begin{afigure}[t]
\begin{center}
\epsfig{file=ExampleComputed.eps, bb=-160 -400 1293 1210, width=.5\linewidth}
\end{center} 
\caption{Mittels Finiter Elemente berechnete Schallintensität bei 5,3 kHz während der Artikulation des Lautes [a]. Jede Graustufe entspricht einem Intensitätsbereich von 6~dB; dunklere Bereiche weisen eine höhere Intensität auf. Dargestellt ist ein Querschnitt des Rachens ({\it Pharynx}), links unten, und der Mundhöhle ({\it Cavum oris}), darüber mittig, an die sich rechts ein halbkreisscheibenförmig dargestellter halbkugelförmiger Abschnitt zur Erfassung der Lippen-Schallabstrahlung anschließt. }\label{vokaltrak_FEM}
\end{afigure}}

\subsection{Die Wellengleichung}\label{Wellengleichung}
Für die folgende grundlegende Evaluierung wird die Schallausbreitung in ihrer einfachsten Form betrachtet, der akustischen Wellengleichung. Wärmeleitung wird zunächst außer Acht gelassen, so dass eine adiabatische Zustandsänderung erfolgt; ebenso bleiben Reibungen unberücksichtigt und die Betrachtung beschränkt sich auf Terme erste Ordnung.

Ausgehend von der 1.~Akustischen Grundgleichung 
\[\nabla p=-\rho \frac{\partial \vec{u}}{\partial t}\:,\]
welche beinhaltet, dass ein Druckgradient ein Medium beschleunigt, und der 2.~Akustischen Grundgleichung, der Kombination aus Kontinuitäts- und linearisierter Adiabatengleichung,
\[\nabla \vec{u}=-\frac{1}{\kappa \overline{p}}\frac{\partial p}{\partial t}\:.\]
ergibt sich die akustische Wellengleichung
\[\Delta p = \frac{\rho}{\kappa\overline{p}}\frac{\partial^2 p}{\partial t^2}\:,\]
Dabei ist $\kappa$ der Adiabatenkoeffizient, $\rho$ die mittlere Dichte und $\overline{p}$ der mittlere Druck sowie $\frac{\kappa\overline{p}}{\rho}=c^2$ das Quadrat der Schallgeschwindigkeit in Luft. $\vec{u}$, $p$ und $t$ symbolisieren wie in den letzten Abschnitten den Fluss, den Druck und die Zeit. 

Die analytische Lösung der Wellengleichung ist nur für bestimmte einfache Randbedingungen möglich, wie für quader-, kugel- oder zylinderförmige Hohlräume. Für die komplizierteren Geometrien des Sprechtrakts werden deshalb numerische Verfahren eingesetzt.

Die Betrachtungen der numerischen Verfahren in den folgenden Abschnitten \ref{raum_diskret}-\ref{TheoRand} vereinfachen sich, wenn man die Bezugsgröße $p$ durch die Abbildung $c^2\longrightarrow 1$ in eine dimensionslose Form $\ws$ bringt. Auf die Einführung neuer Symbole für die dimensionslose, normierte Zeit und den dimensionslosen, normierten Raum wird dabei verzichtet, da die Symbolik im Kontext eindeutig ist. Die Wellengleichung hat nun die Form:
\[\Delta \ws - \frac{\partial^2 \ws}{\partial t^2}=0\:.\]\\[-1.5ex]	


\subsection{Räumliche Diskretisierung}\label{raum_diskret}
Abhängig von den Randbedingungen ist die numerische Lösung der Wellengleichung mit vielen Verfahren möglich, \cite{HSch93, HOer95, Me08, Wolf} geben einen Überblick. Diese Verfahren führen eine Diskretisierung des Raums oder seiner Oberflächen ein, sowie eine Diskretisierung der Zeit- oder Frequenzkoordinate. Einige dieser Verfahren sind bereits in Abschnitt \ref{DetailMod} genannt und in einigen Aspekten diskutiert worden. Das Ziel der folgenden Betrachtung ist es, ein möglichst gut handhabbares Verfahren zur Untersuchung der Sprechtraktakustik zu ermitteln. Wenngleich kein strenges Maß, setzt sich  die Handhabbarkeit hierbei  aus dem Aufwand für Implementierung, dem Laufzeitverhalten, einer evtl.{} erforderliche Aufbereitung von Untersuchungsdaten und aus der zu erwartenden Genauigkeit zusammen.

\subsubsection*{Waveguide-Mesh}
Der naheliegende Weg, das erfolgreiche Rohrmodell zur Beschreibung der Ausbreitung ebener Wellen auf drei Dimensionen zu erweitern, also beispielsweise ein kubisches Gitter aus uniformen Rohrelementen zu bilden, die an den Knotenstellen mit 6-Tor-Adaptoren verbunden sind, führt zu dem Waveguide-Mesh. Bild \ref{netz} (c) zeigt ein zweidimensionales Abbild dieses Netzes unter Berücksichtigung einer einfachen Randstruktur. 

Die bereits genannte unzutreffende, inhärente Anisotropie der Wellenausbreitung schließt dieses Verfahren für eine quantitative Untersuchung der Akustik aus.
Um diese Anisotropie zu vermeiden, könnte man unter Beibehaltung der Einheitskantenlänge nun versuchen, das Netz feiner und makroskopisch isotrop zu gestalten. Der hierfür erforderliche Netzgenerator scheint aber nicht wesentlich einfacher als ein Netzgenerator zur Untergliederung in Finite Elemente zu sein, das Laufzeitverhalten durch das feinere Netz aber deutlich schlechter.\vspace{-1ex}

\subsubsection*{Finite-Elemente-Methode}
Der Raum wird bei dieser Vorgehensweise in endliche viele Elemente untergliedert, bspw.{} Tetraeder. Deren Abmessungen sind im Allgemeinen unterschiedlich und der Randbedingung angepasst. Auf dem Volumen jedes Tetraeders wird die Zustandsgröße durch eine möglichst einfache Formfunktion angenähert. Zwischen angrenzenden Tetraedern wird dann eine Stetigkeit der Formfunktion durch Gleichheit der Eckwerte und je nach verwendeter Formfunktion weiteren Werten gefordert. 
Ein an Randbedingungen angepasstes, unstrukturiertes Netz der Finiten-Elemente ist in Bild \ref{netz}~(a) gezeigt. 
\begin{afigure}[t]
\begin{center}
\subfloat[\mbox{Unstrukturiertes Netz}]{\epsfig{file=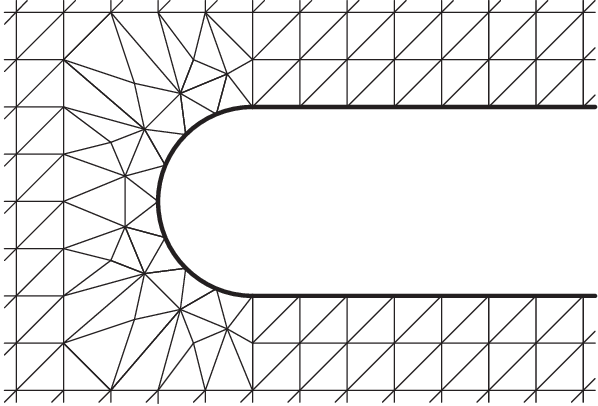, height=3.8cm, angle=270}}\;\;\;
\subfloat[Strukturiertes Netz]{\epsfig{file=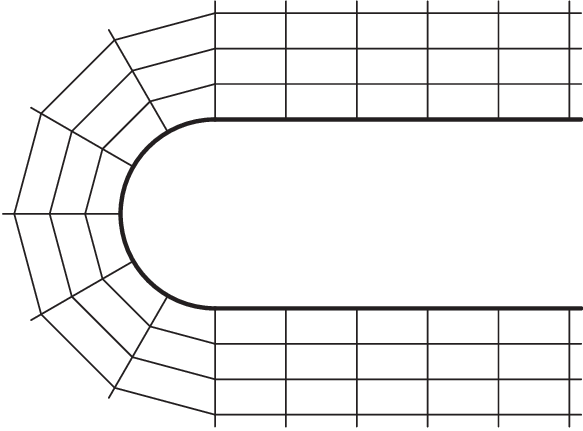, height=3.8cm, angle=270}}\;\;\;
\subfloat[Kartesisches Netz]{\epsfig{file=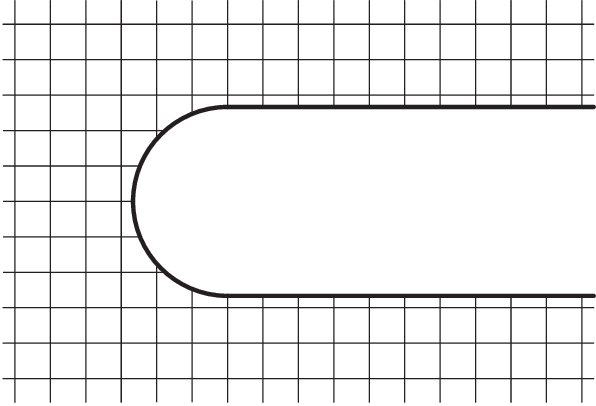, height=3.8cm, angle=270}}
\caption{Netze verschiedener Diskretisierungsmethoden}\label{netz}
\end{center}
\end{afigure}

Der wesentliche Vorteil der Elemente Methode ist, dass das Netz für filigrane Randstrukturen verfeinert werden kann, während es bspw.{} im Zentrum von Hohlräumen grob, mit wenig Elemente ausgeführt wird. Auf diese Weise kann die Anzahl der Elemente bei bestimmten Aufgabenstellungen deutlich reduziert werden, und die algorithmische höhere Komplexität pro Element rentiert sich.
Jedoch sollten die Tetraeder gewissen Kriterien genügen, bspw.{} dem von Delauny in \cite{De34}, um den Raum hinreichend homogen abzudecken, was Algorithmen für die automatische Generierung der Tetraedernetze aufwendig macht: Eine automatische Erzeugung ist bisher nur für einen Vokaltrakt mit einer stark vereinfachten elliptischen Kontur gezeigt worden. Zudem ist bei den hier zu betrachtenden Untersuchungsdaten der Rand nicht unmittelbar in den Datensätzen enthalten, sondern müsste aus räumlich variierenden Volumendichten abgeleitet werden. Letztlich bleibt auch fraglich, ob der genannte Vorteil der Finiten Elemente greift, da gerade der Nasaltrakt eine Vielzahl filigraner Strukturen aufweist --- insbesondere, wenn man eine physikalische treffende, dünne Randschicht mit Dämpfung vorsehen möchte, die mit groben Elementen nicht erfasst werden können.

\subsubsection*{Finite-Differenzen-Methode}
Bei der Finite-Differenzen-Methode legt man in das Volumen ein kubisches Gitternetz der Weite $h$. 
Man approximiert die partiellen Differentialgleichungen durch Differenzengleichungen, welche auf den an den Gitterpunkten definierten Zustandsgrößen basieren. 
Im einfachsten Fall wird aus
\begin{align*}
\frac{\partial}{\partial \vec{\xi}}\ws(\vec{x})&\approx\frac{\triangle}{\triangle  \vec{\xi}}\ws(\vec{x})=\frac{\ws(\vec{x}+h\vec{\xi})-\ws(\vec{x}-h\vec{\xi})}{2h}\\
\intertext{und aus}
\frac{\partial^2}{\partial \vec{\xi}^2}\ws(\vec{x})&\approx\frac{\triangle ^2}{\triangle \vec{\xi}^2}\ws(\vec{x})=\frac{\ws(\vec{x}+h\vec{\xi})-2\ws(\vec{x})+\ws(\vec{x}-h\vec{\xi})}{h^2},
\end{align*}
für zweite Ableitungen, wie sie bei der Wellen-Differentialgleichung vorkommen. Dabei ist $\vec{\xi}$ ein normierter Basisvektor des betrachteten Raums. 
Auf das Verfahren wird unter~\ref{AnFin} genauer eingegangen. 

Man erhält ein kartesisches Netz, Bild \ref{netz}~(c), 
dessen Gitterweite sich an quaderförmige, vorzugsweise kubische Volumenelemente der tomographischen Datensätze anpassen lässt. Die Datensätze können damit direkt übernommen werden. Um wesentliche anatomische Details zu erfassen, weisen diese Datensätze  eine räumliche Auflösung von 1~mm und darunter auf. Diese Auflösung ist deutlich kleiner als die Schallwellenlänge von 4~cm bei der höchsten betrachteten Frequenz 8~kHz. Es ist somit naheliegend, dass die Differenzenapproximation der partiellen Differentialgleichung einen geringen Fehler aufweist --- und es bestätigt sich bei der eingehenden Betrachtung in den folgenden Abschnitten: Anisotropie, Dispersion und eine Skalierungsabhängigkeit können vernachlässigt werden. Darüber hinaus erlaubt die vergleichsweise feine Diskretisierung eine Reihe von Anpassungen, die treffend akustisch relevante Effekte erfassen, wie sich im Weiteren zeigt. Vorteilhaft sind weiterhin die einfache Implementierung und der geringe Berechnungsaufwand einer einzelnen Differenzen-Approximation.

\subsubsection*{Weitere Methoden}
Es gibt eine Vielzahl weiterer Verfahren. Diese weisen jedoch meist andere Zielrichtungen auf, wie bspw.{} die \name{Boundary Element Methode} und die \name{Source-Simulation-Technique}, sind aufwendig zu implementieren, wie die \name{Spektral-Element-Methode} und die \name{Pseudospektral-Methode}, oder haben keinen offensichtlichen Vorteil für die hier untersuchte Aufgabenstellung, wie die \name{Finite-Volumen Methode} mit der Netzstruktur (b) in Bild~\ref{netz}, weshalb diese Verfahren nicht weiter betrachtet werden. 



\subsection{Betrachtung im Frequenzbereich}\label{Frequenzbereich}
Für zeitlich unveränderliche, stationäre Randbedingungen ist eine zeitliche und räumliche Separation der Differentialgleichung möglich. Diese Bedingungen sind im Nasaltrakt und genähert in der Mundhöhle bei bestimmten Lauten erfüllt. Hier kann dann eine direkte Berechnung im Frequenzbereich mit dem harmonischen Ansatz 
\[
	p=p_{x,y,z}\,e^{-i\omega t}
\]
erfolgen. Setzt man diese in die akustische Wellengleichung ein, so erhält man die Helmholtzgleichung
\[
	\left(\Delta+\frac{\omega^2}{c^2}\right) p_{x,y,z} = 0 \;.
\]
Das damit und durch die räumliche Diskretisierung entstehende Gleichungssystem kann dann entweder direkt oder durch iterative Algorithmen, wie Relaxation, gelöst werden.

Alternativ besteht auch die Möglichkeit, die im vorhergehenden Abschnitt gefundenen Differenzenoperatoren als zeitdiskretes Filter zu betrachten, ähnlich den Rohrsegmenten aus Abschnitt \ref{Rohrmodell}. Das Übertragungsverhalten ergibt sich dann durch die Kopplung dieser Filter entsprechend der räumlichen Diskretisierung. 
Der Unterschied zwischen diesen, von den Finiten-Differenzen abgeleiteten Filtern und den Kreuzgliedkettenfiltern liegt letztlich nur in den betrachteten Größen. Erstere betrachten Wellengrößen, welche aus der D’Alembert’schen Lösung der Wellengleichung resultieren, letztere erfassen die Wellengleichung direkt. Unterschiede und Gemeinsamkeiten werden anhand eines Beispiels in Abschnitt \ref{Äquivalenz} nochmals verdeutlicht und dort zur Verifizierung genutzt. 

Damit das Simulationssystem nicht auf statische Vokaltraktkonfigurationen beschränkt bleibt, wird auf die zeitliche Separation verzichtet. Die Rechenzeit der direkten Simulation ist, insbesondere nach den in Abschnitt \ref{opt} vorgestellten und vorgenommenen Optimierungen, ausreichend kurz. Die unterschiedlichen Sichtweisen helfen jedoch beispielsweise, die in den  Abschnitten \ref{partVol} und \ref{ReflexionsfreierAbschluss} betrachteten Erweiterungen zu entwickeln. 
 
\subsection{Integration in Zeitrichtung} \label{TheoZeit}
Ist diese Separation des zeitabhängigen Teils der Lösung nicht möglich oder nicht praktikabel, gibt es nach \cite{HOer95} eine Reihe numerischer Methoden um die Lösung zeitschrittweise zu bestimmen, die folgend  kurz zusammengefasst sind.

Die Idee des \name{Euler-Verfahrens} ist, aus der Tangentensteigung den nächsten Funktionswert zu ermitteln: 
\[\frac{\partial \ws}{\partial t}\approx\frac{\triangle \ws}{\triangle t}\quad\Longrightarrow\quad 
\begin{cases}\ws_{t+1} =\ws_t+\triangle t f(\ws_t)&\text{explizit,}\\ \ws_{t+1} =\ws_t+\triangle t f(\ws_{t+1})&\text{implizit.}\end{cases}\] 
Bei der impliziten Methode lässt sich $\ws_{t+1}$ nur durch Lösen eines Gleichungssystems bestimmen. Dies bedeutet einen erheblichen Mehraufwand. Jedoch führt das Verfahren in jedem Fall zu einem stabilen System.
Die Genauigkeit des Euler-Verfahrens kann erhöht werden, indem zwischen den Stützstellen die Ableitung bestimmt wird:
\[ \ws_{t+1}= \ws_t+\frac{\triangle t}{2}\left(f(\ws_t)+f(\ws_{t+1})\right).\]
Dieses nach \name{Crank-Nicolson} benannte Verfahren ist ebenfalls implizit, folglich numerisch stabil und rechenaufwendig.
Der numerische Aufwand der impliziten Verfahren kann mit der \name{Prädiktor-Korrektor-Methode} gemindert werden, indem $\ws'_{t+1}$ durch das explizite Euler-Verfahren in einem Prädik\-tionsschritt vorausgesagt und danach in einem Korrekturschritt ähnlich dem Crank-Nicolson-Verfahren genauer bestimmt wird:  
\[ \ws_{t+1}= \ws_t+\frac{\triangle t}{2}\left(f(\ws_t)+f(\ws'_{t+1})\right).\]

Es zeigt sich jedoch, dass aufgrund der feinen Diskretisierung der Raumkoordinaten für die Integration in Zeitrichtung bereits das einfache Euler-Verfahren hinreichend genau ist. Insbesondere verdeutlichen die am Ende des Abschnitts~\ref{AnFin} dargestellte Dispersionsrelationen, dass der höhere Aufwand zur Berechnung durch die anderen genannten Verfahren für die Untersuchungen in dieser Arbeit nicht gerechtfertigt ist.

\subsection{Finite-Differenzen-Methode im Zeitbereich}\label{AnFin}
In diesem Abschnitt werden die Eigenschaften der Finite-Differenzen-Methode im Zeitbereich eingehend betrachtet. Hierfür werden zunächst Dif\-ferenzen-Operatoren verschiedener Ordnungen entwickelt und hinsichtlich ihrer Stabilität für ein Euler-Verfahren zur zeitlichen Integration analysiert. Abschließend wird für diese Operatoren der Finiten Differenzen-Methode die Abweichung zur Lösung der partiellen Differentialgleichung untersucht. Für eine kompakte Darstellung des Weges wird dabei ein kubisches Diskretisierungsgitter zugrunde gelegt, bei dem aufgrund der hohen Symmetrie nur wenige Fälle betrachtet werden müssen.  

Um den Laplaceoperator $\Delta \ws =\frac{\partial^2 \ws}{\partial x^2}+\frac{\partial^2 \ws}{\partial y^2}+\frac{\partial^2 \ws}{\partial y^2}$ der Wellengleichung zu approximieren, führt man eine Taylorentwicklung der Funktion $\ws$ aus. Im Folgenden ist sie dargestellt bis zur 4.~Ordnung:
\begin{align*}
\ws(\vec{h})\approx \ws	& +\TuT{1}{0}{0}\\
		& +\frac{1}{2}\left(\TuT{2}{0}{0}\right)\\
		& \qquad +\TuT{1}{1}{0}\\
		& +\frac{1}{6}\left(\TuT{3}{0}{0}\right)+\Tu{1}{1}{1}\\
		& \qquad+\frac{1}{2}\left(\TuT{2}{1}{0}\right)\\
		& \qquad+\frac{1}{2}\left(\TuT{1}{2}{0}\right)\\
		& +\frac{1}{24}\left(\TuT{4}{0}{0}\right)\\
		& \qquad+\frac{1}{6}\left(\TuT{3}{1}{0}\right)\\
		& \qquad+\frac{1}{6}\left(\TuT{1}{3}{0}\right)\\
		& \qquad+\frac{1}{4}\left(\TuT{2}{2}{0}\right)\\
		& \qquad+\frac{1}{2}\left(\TuT{2}{1}{1}\right).
\end{align*}
Man wählt ein Gitter mit der Weite $h=1$,  und bestimmt die benachbarten Werte, indem ihre Koordinaten in die Taylorreihe eingesetzt werden. Dabei ist es zweckmäßig,  symmetrisch angeordnete Punkte zusammenzufassen. 
\noindent
\unitlength4mm
\newcommand{\cir}[2]
{
	\put(#1,#2){\circle*{.3}}
}
\begin{afigure}
\begin{center}
\begin{picture}(16,16)
\cir{0}{8}
\cir{8}{0}
\cir{6}{6}
\cir{10}{10}
\cir{8}{16}
\cir{16}{8}

\multiput(3,3)(2,2){2}
{\multiput(0,0)(8,0){2}
{\multiput(0,0)(0,8){2}
{\cir{0}{0} }}}

\cir{4}{8}
\cir{8}{4}
\cir{12}{8}
\cir{8}{12}
\cir{7}{7}
\cir{9}{9}

\cir{3}{7}
\cir{7}{3}
\cir{4}{4}
\cir{12}{12}
\cir{12}{4}
\cir{4}{12}
\cir{7}{11}
\cir{11}{7}
\cir{9}{13}
\cir{13}{9}
\cir{5}{9}
\cir{9}{5}

\put(7.8,7.8){\rule{2mm}{2mm}}

\thicklines
\multiput(3,3)(4,0){3}{\line(0,1){8}}
\multiput(3,3)(0,4){3}{\line(1,0){8}}
\multiput(3,11)(1,1){3}{\line(1,0){8}}
\multiput(11,3)(1,1){3}{\line(0,1){8}}
\multiput(11,3)(0,4){3}{\line(1,1){2}}
\multiput(3,11)(4,0){2}{\line(1,1){2}}
\thinlines
\put(4,4){\line(0,1){8}\line(1,0){8}}
\put(5,5){\line(0,1){8}\line(1,0){8}}
\put(5,9){\line(1,0){8}}
\put(9,5){\line(0,1){8}}
\put(3,3){\line(1,1){2}}
\put(3,7){\line(1,1){2}}
\put(7,3){\line(1,1){2}}

\put(0,8){\line(1,0){16}}
\put(8,0){\line(0,1){16}}
\put(6,6){\line(1,1){4}}
\end{picture}
\end{center}
\caption{Gitterausschnitt}\label{Gitterausschnitt}
\end{afigure}
Betrachtet man den in Bild~\ref{Gitterausschnitt} abgebildeten Gitterausschnitt, 
mit dem zentralen Punkt (quadratisch markiert)  
\begin{alignat*}{2}
&\ws_{Zentrum}&=\;&\ws\:,\\ 
\intertext{so ergibt sich für die Eckpunkte}
&\ws_{Ecke}&=\;&\frac{1}{3}D^4\ws+2D^{22}\ws+4D^2\ws+8\ws\:,\\
\intertext{für die Kantenpunkte}
&\ws_{Kante}&=\;&\frac{1}{3}D^4\ws+D^{22}\ws+4D^2\ws+12\ws\:,\\
\intertext{für die Flächenpunkte}
&\ws_{Fl"ache}&=\;&\frac{1}{12}D^4\ws+D^2\ws+6\ws\:,\\
\intertext{und für die entfernteren flächenzentrierten Punkte}
&\ws_{Fl"ache2}&\;=\;&\frac{4}{3}D^4\ws+4D^2\ws+6\ws\:.
\end{alignat*}
Dabei haben die Differentialoperatoren die Form:
\begin{eqnarray*}
D^i&=&\frac{\partial^i}{\partial x^i}+\frac{\partial^i}{\partial y^i}+\frac{\partial^i}{\partial z^i}\:,\\
D^{ii}&=&\frac{\partial^{2i}}{\partial y^i\partial z^i}+\frac{\partial^{2i}}{\partial x^i\partial z^i}+\frac{\partial^{2i}}{\partial x^i\partial y^i}\:.
\end{eqnarray*}

Die einfachste  Möglichkeit, den Laplaceoperator, der in dieser Darstellung die Form $D^2$ hat, aus Linearkombinationen der Gitterpunkte zu bilden, ist:
\[ D^2 \approx \ws_{Fl"ache} - 6 \ws_{Zentrum}. \]
Der verbleibende Fehler $\frac{1}{12}D^4$ ist vierter Ordnung. Er kann durch Hinzufügen von Eck- und Kantenpunkten nicht eliminiert werden, da diese von den Flächen- und Zentrumspunkten linear abhängig sind. Eine bessere Approximation erhält man durch Berücksichtigung der Flächen zweiter Ordnung:
\[ D^2 \approx -\ws_{Fl"ache2}+16\ws_{Fl"ache}-90\ws_{Zentrum} \] womit Fehler sechster Ordnung bleiben.
Es verbleiben somit die zwei im letzten Abschnitt gefundenen Operatoren der Differenzen-Methode, dargestellt als Iterationsgleichung:
\begin{align*}
\ws_{t+1,x,y,z}=K(&\ws_{t,x+1,y,z}+\ws_{t,x,y+1,z}+\ws_{t,x,y,z+1}\\+&\ws_{t,x-1,y,z}+\ws_{t,x,y-1,z}+\ws_{t,x,y,z-1})\\
&-\ws_{t-1,x,y,z}-(6K\negmedspace-2)\ws_{t,x,y,z},\\
\intertext{der 9-Punkt Operator, und}
\ws_{t+2,x,y,z}=K(&\ws_{t,x+2,y,z}+\ws_{t,x,y+2,z}+\ws_{t,x,y,z+2}\\+&\ws_{t,x-2,y,z}+\ws_{t,x,y-2,z}+\ws_{t,x,y,z-2} )\\
-16K(&\ws_{t,x+1,y,z}+\ws_{t,x,y+1,z}+\ws_{t,x,y,z+1}\\+&\ws_{t,x-1,y,z}+\ws_{t,x,y-1,z}+\ws_{t,x,y,z-1})\\
+16(&\ws_{t-1,x,y,z}+\ws_{t+1,x,y,z})-\ws_{t-2,x,y,z}\\&+(90K\negmedspace-30)\ws_{t,x,y,z},
\end{align*}
der 17-Punkt Operator.\footnote{Die Operatoren lassen sich mit $K=\frac{1}{3}$ noch weiter vereinfachen, da dann die Koeffizienten $6K\negmedspace-2$ bzw.~$90K\negmedspace-30$ gleich null sind, der Term $\ws_{t,x,y,z}$ entfällt. Die Berechnung kann dann im Fall des 9-Punkt-Operators zeitlich alternierend auf jeden zweiten Gitterpunkt beschränkt werden, also beispielsweise für $(t+x+y+z)\negthickspace\mod 2 = 0$.}   Letzterer benötigt den doppelten Rechenaufwand und hat, da die Zeitebenen $t+2$ und $t-2$ berücksichtigt werden müssen, den doppelten Speicherbedarf. 

Für die Zeitdiskretisierung wird das explizite Euler-Verfahren eingesetzt. Hierfür wurden bereits in \cite{CoFL28} Stabilitätskriterien gefunden, unter anderem durch eine Kausalitätsbetrachtung: Zur Berechnung des zentralen Elements muss der Zeitschritt  so klein gewählt werden, dass es vom Schall eines azentralen Elementes des Operators bei der Schallgeschwindigkeit $c$ innerhalb von $\Delta t$ erreicht werden kann. Eine kompaktere Darstellung findet sich in \cite{LiSB98} mit dem Kriterium:
\[c\frac{\Delta t}{h} \leqslant \sqrt{\frac{a_1}{a_2}}\quad,\]
wobei $a_1$ die Summe aller Koeffizientenbeträge der Differenzenapproximation in Zeitrichtung und $a_2$ in Raumrichtung ist. Diese sind für den 9-Punkt-Operator 4 bzw.{} $12K$ und somit folgt unter Berücksichtigung der Normierung $c=h=\Delta t = 1$:
\[ K\leqslant \frac{1}{3}\quad.\]  
Das gleiche Ergebnis erzielt man mit dem 17-Punkt-Operator aufgrund von  $a_1=64$ und $a_2=192K$.

Die Qualität der erhaltenen Lösungen misst man anhand von Dispersion, Isotropie und Skalierungsunabhängigkeit.
Da die erhaltenen Differenzengleichungen wie die partiellen Differentialgleichungen linear sind, genügt es, hierfür die Dispersionsrelation $\varphi(k)$ zu betrachten.
Die Anisotropie kann hierbei durch den Vergleich verschieden gerichteter Wellenvektoren ermittelt werden. Sinnvollerweise wählt man dazu die Extremalwerte, die Wellenausbreitung entlang einer Gitterachse und die Wellenausbreitung entlang einer Raumdiagonale. 

Zunächst wird die Wellenausbreitung entlang einer Gitterachse betrachtet. Hier führt der Ansatz einer ebenen Welle mit der Wellenzahl $k=2\pi\lambda^{-1}$,
\[\ws = \sin(\varphi_At+k x),\]
in den 9-Punkt Operator ($\varphi_a$) und in den 17-Punkt Operator ($\varphi_A$) eingesetzt, zu:
\begin{align*}
\cos \varphi_a &= 1+ K\left[\cos(k)-1\right]\\
\cos \varphi_A &= 4-\sqrt{9+K\left[\cos^2(k)-8\cos(\pi k)+7\right]}.
\end{align*}

Eine in Richtung einer Raumdiagonalen fortschreitenden ebenen Welle, die durch den Ansatz
\[\ws = \sin\left(\varphi_Rt+k \frac{x+y+z}{\sqrt{3}}\right)\]
beschrieben wird, ergibt in die Operatoren eingesetzt
\begin{align*}
\cos \varphi_r &= 1+3K\left[\cos\left(\frac{1}{\sqrt{3}}k\right)-1\right]\\
\intertext{und}
\cos \varphi_R &=4-\sqrt{9+3K\left[\cos^2\left(\frac{1}{\sqrt{3}}k\right)-8\cos\left(\frac{1}{\sqrt{3}}k\right)+7\right]}.
\end{align*}
Es ergibt sich ein interessanter Spezialfall, falls $K=\frac{1}{3}$ gewählt wird. Die beiden Gleichungen vereinfachen sich dann zu 
\[ 
\varphi_r = \varphi_R =\frac{k}{\sqrt{3}},
\]
einer proportionalen Beziehung zwischen Wellenzahl und Phasengeschwindigkeit. Es tritt keine Dispersion auf.

\begin{afigure}[t]
	\psfrag{0}[][]{$0$}
	\psfrag{Pi14}[][]{$\frac{1}{4}\pi$}
	\psfrag{Pi12}[][]{$\frac{1}{2}\pi$}
	\psfrag{Pi34}[][]{$\frac{3}{4}\pi$}
	\psfrag{Pi1}[][]{$\pi$}
	\psfrag{PiS3}[][]{$\sqrt{3}\pi$}
	\psfrag{Lk}{$k$}
	\psfrag{FLine}[r][r]{$\varphi_r;\varphi_R$}
	\psfrag{FSlash}[r][r]{$\varphi_A$}
	\psfrag{FDot}[r][r]{$\varphi_a$}
	\epsfig{file=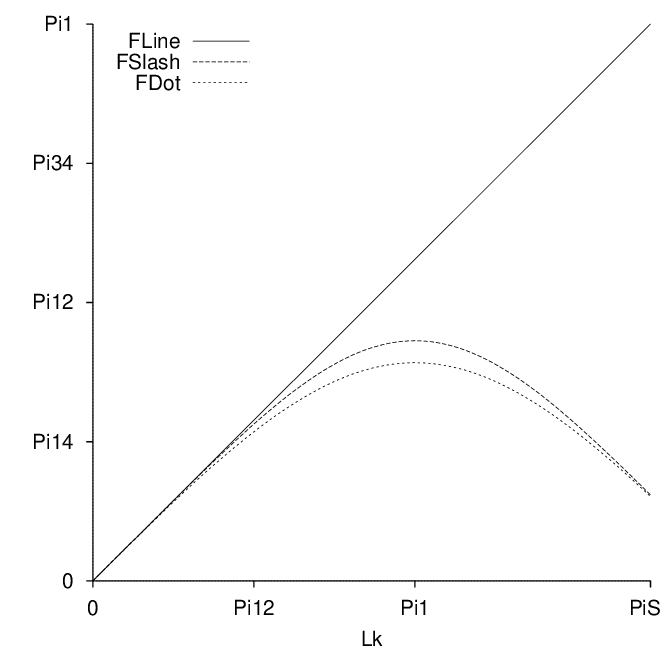, scale=.98}
\caption{Dispersionsrelation des 9- und 17-Punkt Operators mit $K=\frac{1}{3}$}\label{Dispersion} 
\end{afigure}

Man erkennt in Bild~\ref{Dispersion} im Bereich niedriger Frequenzen, $k < \frac{\pi}{2}$ sowohl für den 17-Punkt Operator, $\varphi_R$ und $\varphi_A$, als auch für den 9-Punkt Operator, $\varphi_r$ und $\varphi_a$ einen proportionalen Verlauf der Dispersionsrelation: die auftretende Dispersion ist gering. Eine Anisotropie, in dem Diagramm als \glqq Aufspaltung\grqq~der Funktionenschar zu sehen, tritt ebenfalls erst bei höheren Frequenzen auf.

Für den hörbaren Frequenzbereich $k <\frac{\pi}{10}$ ergibt sich, dass die numerisch hervorgerufenen Fehler sehr gering sind.
Insbesondere zeigt sich, dass der 9-Punkt Operator hinreichend genau ist. Der Mehraufwand für den 17-Punkt Operator ist nicht gerechtfertigt.

Ein weiterer wichtiger Punkt ist die Vermeidung von Aliasing-Artefakten. Diese können durch die unterschiedlichen räumlichen Auflösungen hervorgerufen werden. Eine Möglichkeit besteht in der Bandbreitenbeschränkung der Anregung auf Wellenzahl kleiner $\frac{\pi}{2}$, eine andere in der Bandbreitenbeschränkung der Ergebnisse auf Wellenzahlen kleiner $(2-\sqrt{3})\pi$; beide werden genutzt.

\newpage
\subsection{Randbedingungen}\label{TheoRand}
Während im letzten Abschnitt die Differentialgleichung im allgemeinen betrachtet wurde, sind für eine konkrete Problemstellung die Randbedingungen wichtig, unter denen sie zu lösen ist.

Diese Randbedingungen bilden den Ausgangspunkt der Beschreibung des Vokaltrakts und werden im Laufe der Arbeit weiter an Erfordernisse des Sprechtrakts bzw.{} dessen Datensätze angepasst.

Bei einem schallweichen Abschluss, der in guter Näherung die Schallabstrahlung von der Querschnittsfläche des Mundes oder der Nasenlöcher beschreibt, ist der Schalldruck gleich Null, vgl.{} Abschnitt \ref{Abstrahlung}. Dieser Abschluss kann mit $\varphi =0$ durch  \name{Dirichlet-Randbedingung} realisiert werden, die einen Funktionswert $\varphi$ auf dem Rand $R$ festlegen:
\[\ws_R=\varphi_R\:.\]
Um diese Randbedingung im Kontext der Differenzenmethode zu formulieren, wird für den Differenzen-Operator exemplarisch ein Rand in positiver x-Richtung betrachtet, in der Mitte zwischen der beliebigen Position $x$ und der und eins verschobenen Position $x+1$. Eine zweidimensionale Darstellung genügt: In Bild~\ref{RandOpr} soll das rechte Teilgebiet dem Rand zugehören und an dem ausgezeichneten Punkt die Randbedingung erfüllen. Durch eine Linearisierung des Funktionsverlauf von dem zentralen Punkt des Operators $\ws_{x}$ über den Rand hinaus zu der nächsten Operatorstützstelle $\ws_{x+1}$ erhält man:
\[ \frac{\ws_x+\ws_{x+1}}{2} =\varphi\qquad\Longleftrightarrow\qquad \ws_{x+1}=2\varphi-\ws_{x}\:.\]
Mit $\varphi =0$ wird zur Festlegung eines schallweichen Randes die rechte Beziehung, $\ws_{x+1}= -\ws_{x}$, in die 9-Punkt-Operatorgleichung aus Abschnitt \ref{AnFin} eingesetzt.

Senkrecht zu den schallharten Wänden des Vokaltrakts findet kein Schallfluss statt und die Ableitung des Schalldrucks ist in diese Richtung an den Wänden folglich gleich null. Dies wird durch \name{Neumann-Randbedingung} beschrieben, die am Rand den Wert $\gamma$ der Ableitung  entlang der Normale $\vec{n}_R$ des Randes festlegen: 
\[\frac{\partial \ws_R}{\partial\vec{n}_R} = \gamma_R\:.\] 
\begin{afigure}[t]
\begin{center}


{\setlength{\unitlength}{5.8mm}
\vspace{.5\unitlength}
\begin{picture}(8,8)
\put(6,0){\epsfig{file=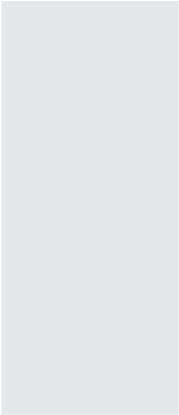, height=8\unitlength}}

\put(4,0){\circle{.3}}
\put(4,8){\circle{.3}}
\put(8,4){\circle{.3}}
\put(0,4){\circle{.3}}
\put(4,4){\circle{.3}}
\thicklines
\put(7.85,4){\vector(-1,0){3.7}}
\put(4,4.15){\line(0,1){3.7}}
\put(0.15,4){\line(1,0){3.7}}
\put(4,0.15){\line(0,1){3.7}}

\thinlines
\put(2,0){\line(0,1){8}}
\put(6,0){\line(0,1){8}}
\put(0,2){\line(1,0){8}}
\put(0,6){\line(1,0){8}}
\put(6,4){\circle*{.4}}

\put(4.3,4.4){\scriptsize $\ws_{x,y}$}
\put(-.7,4.4){\scriptsize $\ws_{x-1,y}$}
\put(3.3,-0.4){\scriptsize $\ws_{x,y-1}$}
\put(7.3,4.4){\scriptsize $\ws_{x+1,y}$} 
\put(3.3,8.4){\scriptsize $\ws_{x,y+1}$}
\end{picture}
}
\end{center}
\caption{Zweidimensionales Schema des Finite-Differenzen-Operators im Randbereich. 
Dicke Linien verdeutlichen die Struktur des Operators, dünne Linien die Ränder der Diskretisierung eines Ausschnitts des zugrunde liegenden Datensatzes. Kreise heben die zentrale Lage der Bezugsgröße des Operators in der Diskretisierung des Datensatzes hervor. Der jenseits des Randes liegende Bereich ist grau dargestellt; der ausgefüllt hervorgehobene Kreis zeigt die Lage des Randwertes $\varphi_R$ und der Pfeil die Flächennormale  $\vec{n}_R$ des Randes.}
\label{RandOpr}
\end{afigure}
Der Funktionsverlauf wird wiederum linearisiert und die Ableitung senkrecht zum Rand durch eine Differenz approximiert. Ein  Vergleich mit Bild~\ref{RandOpr} zeigt, dass sich der Rand im Zentrum der Differenzenapproximation befindet. Folglich erhält man ohne weitere Schritte die Beziehung:
\[ \ws_x-\ws_{x+1} = \gamma \qquad\Longleftrightarrow \qquad \ws_{x+1} = -\gamma+\ws_x\:, \]
die zur Modellierung einer schallharten Fläche mit $\gamma=0$ in den 9-Punkt-Operator eingesetzt wird.  


 Anhand von Korrelationsanalysen, wie sie beispielsweise Bild \ref{Rohrquerschnitte} zugrunde liegen, erkennt man, dass an der Glottisposition nur eine kleine Einschnürung des Vokaltrakts auftritt. Der dahinterliegende glottisseitige Abschluss des Vokaltrakts wird deshalb  meistens reflexionsfrei modelliert, da die Schallwellen auch im weiteren subglottalen Bereich wenige Rückreflexionen erfahren. Ein reflexionsfreier Abschluss ist auch hilfreich zur separaten akustischen Analyse von Teilbereichen, wie dem Nasaltrakt: Hierfür wird die Öffnungsfläche des Velums mit diesen nicht reflektierenden Randbedingung versehen.  
\cite{KSch99} schlägt zur Beschreibung eines reflexionsfreien Abschlusses die Beziehung
\[
\Phi_{x+1,t} = \Phi_{x,t-1}
\]
vor. Diese Beziehung unterdrückt Reflexionen von Schallwellen, indem sie nur eine Teilmenge der D'Alambertschen Lösungen der Wellengleichung zulässt, nämlich die Wellenausbreitung in positive $x$-Richtung, $\Phi_{x-t}$. Rückreflektierte Wellen mit umgekehrter Ausbreitungsrichtung, $\Phi_{x+t}$, sind durch diese Randbedingung nicht mehr möglich. Diese Beziehung liefert für einen eindimensionale Formulierung der Finiten Differenzen eine exakte Randbedingung, ist jedoch aufgrund der Verwendung des D'Alambertschen Integration nicht (oder nur approximativ) auf eine mehrdimensionale Formulierung übertragbar. Mit einem allgemeineren Ansatz wird eine für eine dreidimensionale Formulierung besser geeignete Beschreibung in Abschnitt \ref{ReflexionsfreierAbschluss} entwickelt.

\addtocontents{toc}{\protect\newpage}
\pagebreak
\part{Akustik des Nasaltrakts}
In diesem Teil der Arbeit wird die Extraktion der  Akustik des Nasaltrakts aus seiner räumlichen Gestalt gezeigt. Hierfür werden zunächst verschiedene tomographische Verfahren zur Bestimmung der räumlichen Gestalt des Nasaltrakts evaluiert. Diese Verfahren liefern eine Abfolge zweidimensionaler Schichtbilder, in denen eine Dichte kodiert ist. Es zeigt sich, dass  diese Dichteverteilung eines Verfahrens präzisen Aufschluss über die Gestalt der Hohlräume des Nasaltrakts gibt. Jedoch ist auch bei  diesem Verfahren eine räumliche Auflösung bestimmter, wesentlicher Details nicht möglich. Um trotzdem quantitativ die Akustik berechnen zu können, wird die Formulierung der Finiten Differenzen erweitert, so dass die bestimmte mittlere Dichte der einzelnen Raumbereiche einbezogen wird. Diese Formulierung wird in einem weiteren Schritt zur Berücksichtigung der im Nasaltrakt stark ausgeprägten Schalldämpfung durch Wechselwirkung mit den Hohlraumwänden ergänzt.
Mit der erweiterten Formulierung der Finten Differenzen ist eine direkte Übernahme der tomographischen Daten möglich; die Bestimmung der Akustik erfolgt durch Lösung der Wellengleichung in einem optimierten Zeitschrittverfahren. Der Simulation wird ein speziell für diese Akustik entwickelte Messung gegenübergestellt. 

\section{Tomographische Daten}
Die räumliche Gestalt des Nasaltrakts kann mit verschiedenen Verfahren  bestimmt werden. Der Rückgriff auf eine publizierte, bestehenden Untersuchung --- auf diese wird in den folgendem Abschnitt eingegangen --- zeigt, dass ein nicht an den Erfordernissen ausgerichtetes Verfahren zu erheblichen Unsicherheiten führen kann. Zu den wichtigsten Erfordernissen zählen 
\begin{itemize}
\item{hohe räumliche Auflösung in allen drei Raumrichtungen,}
\item{hoher Kontrast zwischen Hohlraum und Gewebe,}
\item{geringe Artefakte und}
\item{geringe Belastung für die untersuchte Person.}
\end{itemize}
Die erste Eigenschaft ist zwingend erforderlich, um wesentliche Details des Nasaltrakts zu erfassen. Beispielsweise sind die Nasengänge durch Einbuchtungen --- wie später gezeigt wird --- bereichsweise Weiten nur im Millimeterbereich, in dem gleichen Bereich liegt der Durchmesser der Verbindunggänge zu den Nasenhöhlen. Ein hoher Kontrast erleichtert eine algorithmische Aufbereitung der Untersuchung, insbesondere die Verwendung von Schwellwerten zur Klassifikation von Hohlräumen und eine geringe Belastung für die untersuchte Person erleichtert die Handhabung des Verfahrens.

Neben der ursprünglich verwendeten Kryosektion werden zwei weitere, auf gänzlich unterschiedlich physikalischen Prinzipien beruhende tomographische Verfahren aus der medizinischen Diagnostik betrachtet. Diese beiden Verfahren, die Kernspinresonanz-Tomographie und die Röntgenabsorption-Tomographie, werden hinsichtlich der genannten Erfordernisse verglichen.

\subsection{Kryosektion}\label{Kryo}
Die in \cite{Ra99} verwendeten Datensätze des \name{Visual Human}-Projektes, die in \cite{VHum} verfügbar sind, stammen von aus optisch, mittels Scanner abgetasteten  Kryosektionen. In der Ebene des Scanners wird eine Auflösung von 0,33 mm erreicht.  Die räumliche Auflösung der Kryosektionen ist entlang deren Schichtung mit 1 mm deutlich gröber.\footnote{
Ein zweiter Datensatz des Visual-Human-Projects hat eine gleiche feine Auflösung in allen drei Raumrichtungen. Aber auch bei diesem sind erhebliche, präparationsbedingte Artefakte vorhanden.} 
Durch eine Unterabtastung in der Scanner-Ebene wird eine einheitliche Auflösung von 1 mm in alle Raumrichtung hergestellt. Jedoch führt der Kryo-Prozess  mit der nachfolgenden Präparation, zu einer Reihe von Artefakten: die Nasengänge treten beispielsweise nicht hervor, Hohlräume sind teils mit einer blauen Substanz gefüllt, teilweise schwarz. Deshalb erfolgt die Separation von Gewebe und Hohlräumen anhand der Bilddaten halbautomatische gefolgt von einer Berichtigung durch einen Facharzt für HNO-Medizin anhand seiner Erfahrungen, vgl.{} S.~\pageref{Dank}.  Bild \ref{Img-Stack} illustriert diesen Prozess. 
\begin{afigure}[ht]
\begin{center}
\epsfig{width=\linewidth,file=OrgSta\arxiv{ck}.eps}
\caption{Übergang von Kryosektionen, links, zum Volumendatensatz, rechts: Durch eine Unterabtastung werden die Daten auf ein kubisches Gitter der Kantenlänge 1 mm gebracht, wie an der gröberen Randstruktur zu erkennen ist. Die weiß dargestellten Hohlräume sind halbautomatisch klassifiziert.}\label{Img-Stack}
\end{center}
\end{afigure}

\subsection{Magnetresonanztomographie}
Die Magnetresonanz- oder Kernspintomographie ist ein Verfahren, in dem Atomkerne, hier Wasserstoffkerne räumlich aufgelöst in ihrer Konzentration dargestellt werden. Die Wasserstoffkerne werden hierfür in einem starken äußeren Magnetfeld mittels einer elektromagnetischen Welle in einer zu der Magnetfeldstärke passenden Frequenz ausgerichtet und die durch ihre Relaxation entstehenden Radiowellen erfasst. Mittels eines dem Magnetfeld überlagerten Gradienten wird das Verfahren ortsauflösend; die Tomographien entstehen durch wiederholte Messung bei verschobenen Gradienten, vgl.{} bspw.{} \cite{BrS10}. Während das die Hohlräume umgebende Gewebe zu einem Großteil aus Wasser besteht und ein entsprechend starkes Signal liefert, emittiert die Luft in den Hohlräume praktisch keine Signal.
\begin{afigure}[b]
\begin{center}
\epsfig{width=0.6\linewidth,file=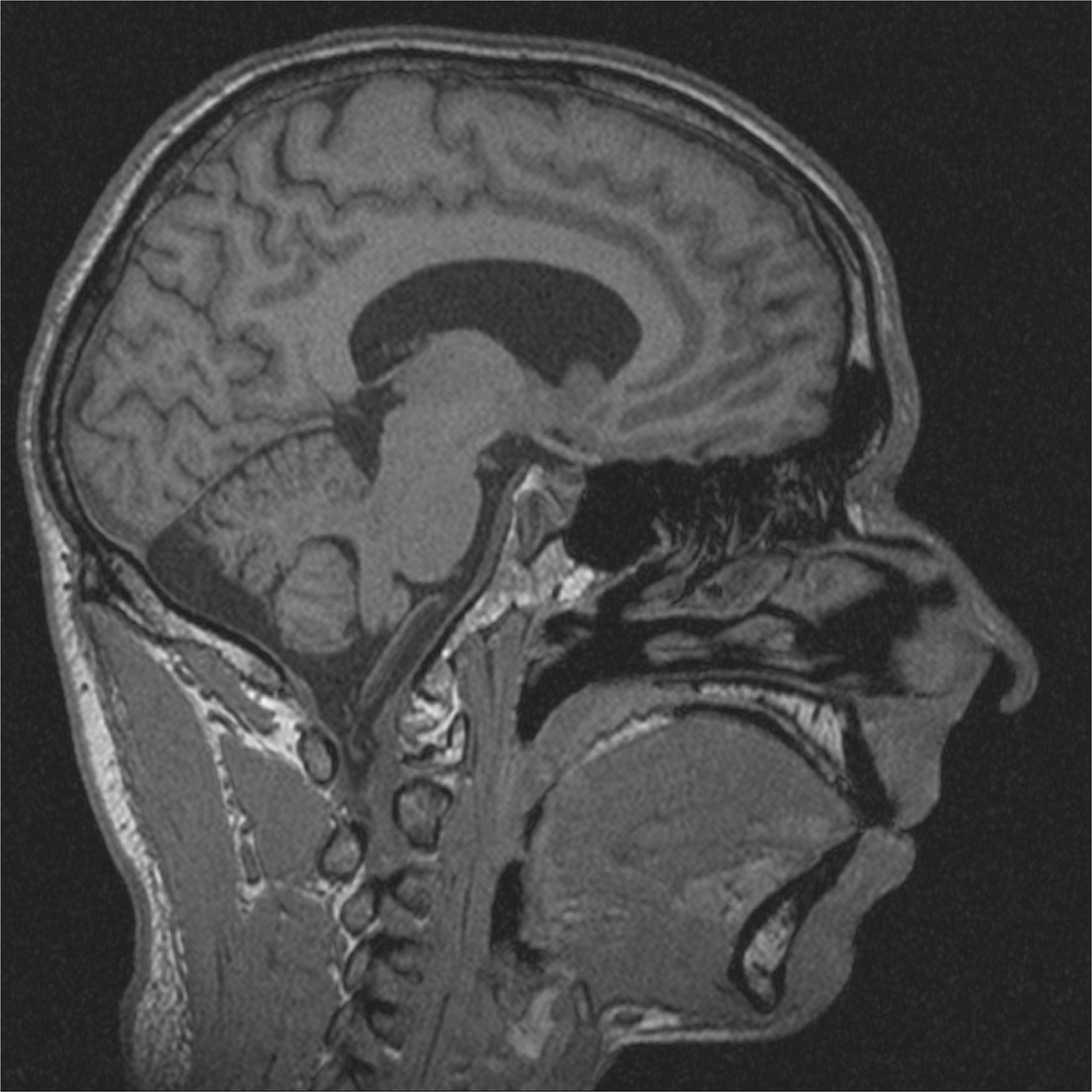}
\caption{MRT-Sagitalschnitt, nahe der Medianebene: Die Abgrenzungen der Siebbeinzellen, Bildmitte halbrechts, die aus dünnen Knochen und Schleimhäuten gebildet wird, wird nicht deutlich.}\label{MRI}
\end{center}
\end{afigure} 

Da das Verfahren ungefährlich, aufgrund des langen regungslosen Verharrens in einer engen Röhre aber wenig angenehm ist, hat der Autor sich dieser Untersuchung selbst unterzogen. Die Untersuchung an einem MRT-System des Universitätsklinikums Frankfurt mit 1,5~Tesla Feldstärke dauert etwa zwei Stunden.\label{MRT-Zeit} Die Aufnahmen geben jedoch die Strukturen des Nasaltrakts nur ungenügend wieder, wie in Bild \ref{MRI} gezeigt, was eine Auswertung nicht mehr sinnvoll erscheinen lässt.
Jedoch sind mittlerweile Geräte mit höherer Feldstärke in Deutschland im Forschungseinsatz, wie bspw. \cite{MPI07} und \cite{Jü09} zeigen, die genauere Untersuchungen ermöglichen. Mit zunehmender Feldstärke sind jedoch auch hier Artefakte zu befürchten, was nach \cite{GrS06} den Vorteil der höheren Feldstärke relativiert. Eine Alternative oder Ergänzung besteht  in der Verwendung von Tieftemperaturspulen nach \cite{Ba08} zur Signalerfassung, sobald diese im humanmedizinischen Bereich verfügbar sind. Dabei wird das thermische Rauschen des ohmschen Spulenwiderstandes und der Verstärkungselektronik durch eine herabgesetzte Betriebstemperatur verringert. Weitere Alternativen sind Multi-Channel-Verfahren, wie in \cite{Bl04} gezeigt. Ebenfalls interessant könnte die in \cite{Eb96} beschriebene Substitution des Stickstoffs in der Atemluft durch hyperpolarisiertes Helium-3 zur Kontraststeigerung der Hohlräume sein. Sobald eines oder eine Kombination dieser Verfahren zur Verfügung stehen, kann die MRT vorteilhaft verwendet werden.

\subsection{Computertomographie}
Eine Computertomographie zeichnet die Röntgenabsorption ortsaufgelöst auf, Bild \ref{CT-Frontal} zeigt ein Beispiel.  
\begin{afigure}[b]
\begin{center}
\epsfig{width=.6\linewidth,file=C\arxiv{T}.eps}
\caption{Computertomographie, frontaler Schnitt: Gut zu erkennen sind die schwarzen Hohlräume der Nasengänge und Nasennebenhöhlen, die sich von dem umliegenden dunkelgrauen Gewebe abheben.  }\label{CT-Frontal}
\end{center}
\end{afigure}%
Die  Messungen der Röntgenabsorption erfolgen entlang verschiedener Geraden, was die räumliche Zuordnung der Absorption an den Schnittpunkten der Geraden ermöglicht. Die zugrundeliegende Berechnung, die \name{Radon-Transformation} \cite{Ra17}, welche zu den Tomographien führt, und die Untersuchungsergebnisse für einige Gewebetypen werden ausführlich beispielsweise in \cite{Hs03} und in \cite{Le04} beschrieben.  Der Absorptionsgrad oder --- im medizinischen Sprachgebrauch --- die Röntgendichte wird dabei in der Einheit \name{Hounsfield} angegeben, abgekürzt HE oder auch HU für \name{Hounsfield Unit}. Die Skala wird durch den Wert von Luft auf -1000~HE und Wasser auf 0~HE definiert.  

\begin{afigure}[t]
\begin{center}
\epsfig{width=.7\linewidth,file=Nasaltrakt3\arxiv{D}.eps}
\caption{CT-Daten: Oberflächendarstellung der nasalen Hohlräume. Die Stirnhöhlen sind oben zu erkennen, die Kieferhöhlen rechts und links. Weiß hervorgehoben ist ein Schnitt durch die vorderen Nasengänge.}\label{CT3D} 
\end{center}
\end{afigure}%
Durch den Einsatz ionisierender Strahlung ist die Methode nicht ungefährlich, ihr Risiko wird in \cite{ICRP00} diskutiert: Die Strahlenbelastung für die untersuchte Person ist prinzipbedingt um ein Vielfaches höher als bei einer Röntgenaufnahme. In Absprache mit dem Universitätsklinikum Frankfurt wurden deshalb die Daten zusammen mit einer medizinisch indizierten Untersuchung und mit dem Einverständnis des Patienten erfasst. In diesem Bild ist der hohe Kontrast zwischen den Hohlräumen, welche schwarz dargestellt sind, und der Schleimhaut, grau, zu sehen; ebenfalls zu erkennen sind die scharfen Ränder zwischen diesen Gebieten.
Von diesen Daten wurde zunächst ein Volumendatensatz mit einer Auflösung von 0,5~mm in allen Raumrichtungen abgeleitet, bei dem Werte der Röntgenabsorption größer 0~HE dem Gewebe zugeordnet wurden, während kleinere Werte als Hohlraum klassifiziert wurden. Die sich so ergebende Oberfläche wurde zur visuellen Kontrolle dreidimensional dargestellt, Bild \ref{CT3D} zeigt eine Ansicht. Eine repräsentative Auswahl der zugrundeliegenden Computertomographien ist im Anhang, Abschnitt \ref{CT-Slices}, abgebildet.

\newpage
\section{Partielle Volumen}\label{partVol}
Trotz des verbesserten Datensatzes können bestimmte, wichtige Strukturen mit dem beschriebenen Verfahren nicht quantitativ erfasst werden. Dazu gehören insbesondere die bereits erwähnten Querschnitte der Verbindungsgänge zwischen den Nasengängen und den Nasennebenhöhlen sowie in einigen Bereichen die Querschnitte der Nasengänge selbst. Die Verbindungsgänge haben einen Durchmesser von rund einem Millimeter, somit in der Größenordnung der Diskretisierung. Würde man eine Klassifizierung anhand eines Schwellwertes durchführen, sollten sich die resultierenden Diskretisierungsfehler zwar im Mittel ausgleichen und beispielsweise die aus dem Volumen der Verbindungsgänge bestimmten Helmholtzresonanzen (im Zusammenwirken mit den Nasennebenhöhlenvolumen) nicht verändern. Es kann jedoch zu diskretisierungsbedingten Eigenresonanzen in den Gängen kommen. Gravierender sind darüber hinaus aus dem gleichen Grund unzutreffende Querschnitte in der Ankopplung der Verbindungsgänge, vgl.{} Abschnitt~\ref{Mehrtor}, da diese die Güte der Resonanz beeinflussen, oder im Extremfall ein Verschluss des Ganges durch eine ungünstige Lage im Diskretisierungsraster. 

In diesem Abschnitt wird eine Methode entwickelt, die diese Artefakte beseitigt, indem weitere Informationen aus den CT-Daten genutzt werden. Hierbei wird ausgenutzt, dass die Computertomographie eine \name{mittlere} Dichte eines Volumenelements liefert.

Der Nasaltrakt ist mit der Nasenschleimhaut ausgekleidet, die die Grenzschicht zwischen den Hohlräumen und dem umliegenden Gewebe bildet. Der Nasenschleimhaut kann in guter Näherung eine Röntgendichte von 0 HE zugeordnet werden, da sie größtenteils aus Wasser besteht. Betrachtet man ein Volumenelement  der Computertomographie an einem Ort $\vec{x}$, welcher im Randbereich des Hohlraums und der Nasenschleimhaut sitzt, so kann nun für dieses Volumenelement anhand der gemessenen Röntgendichte $\mu_\vec{x}$ bestimmt werden, zu welchem Teil $\chi$ es noch mit Luft erfüllt ist beziehungsweise wie viel von dem Volumen durch die Nasenschleimhaut eingenommen wird. Dies geschieht über einen linearisierten Ansatz, wobei $\mu_\text{Luft} = -1000$~HE auf ein leeres Volumen und $\mu_\text{Wasser}=0$~HE ein vollständig gefülltes Volumen abgebildet werden. Werte darüber und darunter werden der Überlegung entsprechend begrenzt:
\[ 
\chi_\vec{x} =\begin{cases} 
1\phantom{\dfrac{R}{R}} &, \text{für } \mu_\vec{x} \le \mu_\text{Luft}\\
 \dfrac{\mu_\vec{x}-\mu_\text{Wasser}}{\mu_\text{Luft}-\mu_\text{Wasser}} & , \text{für } \mu_\text{Luft} <  \mu_\vec{x} < \mu_\text{Wasser} \\
0\phantom{\dfrac{R}{R}} &, \text{für } \mu_\vec{x} \ge \mu_\text{Wasser}\:.
\end{cases}
\]
In Bild \ref{CT-Vol} ist ein Ausschnitt aus dem Bild \ref{CT-Frontal} gezeigt, in dem diese Zuordnung vorgenommen ist.%
\begin{afigure}[t]
\begin{center}
\fboxsep0mm
\fbox{\epsfig{width=.45\linewidth,file=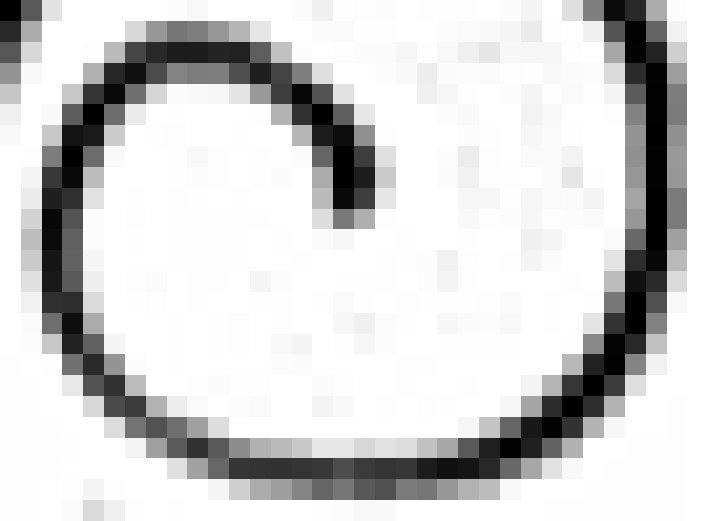}}
\caption{Skalierter Ausschnitt aus den CT-Daten, Bild~\ref{CT-Frontal}, Mitte. Gezeigt ist der Nasengang, wobei -1000~HE schwarz und 0~HE weiß dargestellt sind, die Zwischenwerte sind entsprechenden Grautöne zugeordnet.}\label{CT-Vol}
\end{center}
\end{afigure}%

Die je nach Grad der Füllung der Volumen geänderten akustischen Eigenschaften lassen sich mit Finiten-Differenzen durch die Randbedingungen nach Abschnitt \ref{TheoRand} nicht unmittelbar berücksichtigen. Hierfür  müsste die räumliche Auflösung um die Quantisierung der CT, also um den Faktor 1000, erhöht und der genaue Verlauf des Randes in der erhöhten Auflösung rekonstruiert werden. Dieser Umstand wird durch eine Anleihe aus dem Formalismus des Rohrmodells vermieden; es werden die mittleren Eigenschaften eines Volumens in Form der akustischen Impedanz in die Differenzengleichung übertragen. Der akustischen Impedanz $Z^{ak}$ des Volumens  wird analog zu Abschnitt~\ref{Leitungselement} formuliert: 
\[ 	
Z^{ak}_\vec{x} = \frac{1}{\chi_\vec{x} }Z^0 \:.
\] 
Die teilgefüllten Volumen im Randbereich weisen eine erhöhte Impedanz auf; vollständig zur Nasenschleimhaut gehörende Volumen sind schallhart. 

%

Dieser in \cite{RaL03a} entwickelte und untersuchte Formalismus wird im Folgenden eingehender betrachtet.  Der Impedanzsprung an der Grenzschicht zweier Elemente kann nach Abschnitt \ref{Theo2Adapt} auch durch den Reflexionsfaktor   
\[ r= -\frac{Z^{ak}_x-Z^{ak}_{x+1}}{Z^{ak}_x+Z^{ak}_{x+1}}= \frac{\chi_x-\chi_{x+1}}{\chi_x+\chi_{x+1}}  \]
beschrieben werden, konkretisiert für zwei nebeneinanderliegende Volumen mit $\vec{x}=x$ und mit $\vec{x}=x+1$. In Tabelle \ref{emp} ist die Ausbreitung eines auf einen Impedanzsprung auftreffenden Druckimpulses notiert.
\begin{atable}[t]
\caption{Vier Zeitschritte des Auftreffens eines normierten Druckimpulses auf einen Impedanzsprung, verdeutlicht durch eine dünne vertikale Trennlinie zwischen $x$ und $x+1$. Der Impedanzsprung ist durch den Reflexionfaktor $r$ charakterisiert. \vspace{-5mm}}
\newcommand{\row}{\;\;}
\begin{center}
\setlength{\doublerulesep}{0pt}
\newlength{\rowl}
\setlength{\rowl}{5ex}
\begin{tabular}{c||cc|cc}\label{emp}
& $x-1$ & $x$ & $x+1$ & $x+2$\\ 
\hline\hline &&&&\\[-2ex]
$t_0$&$\row 1\row$& & &\\
$t_1$&&$\row 1\row$& &\\
$t_2$&& $r$ &$\negthinspace 1+r\negthinspace$& \\
$t_3$&$r$ & & &$\negthinspace 1+r\negthinspace$ 
\end{tabular}\vspace{-2mm}
\end{center}
\end{atable}
Ein Vergleich der Zeitschritte $t_1$ und $t_2$ mit den Koeffizienten eines eindimensionalen Differenzen-Operators
\[
p_{t-1,x}-2p_{t,x}+p_{t+1,x} = p_{t,x-1}+\alpha p_{t,x}+\beta p_{t,x+1}
\]
ergibt
\[ \alpha = -2+r\thickspace,\qquad \beta=1-r\thickspace. \]
\begin{afigure}[b]
\begin{center}
\epsfig{width=.3\linewidth,file=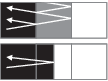}
\caption{Verlauf der Schallwellen an einem Volumenelement mit gemittelter akustischer Impedanz, oben, und an einem scharfen Rand, unten. Die Zeitachse ist vertikal aufgetragen.}\label{BildPartVol}
\end{center}
\end{afigure}%

Aufgrund ihrer Ableitung aus dem Rohrmodell ist diese Erweiterung der Finiten Differenzen für transmittierte Wellen, wie sie in den bereits genannten rohrartigen Verbindungskanälen zwischen Nasengang und Nasennebenhöhlen auftreten, physikalisch zutreffend. Hier korrespondiert $\chi$ direkt mit der Querschnittsflächeninhalt des Verbindungsrohrs,  wie es sich auch in der Äquivalenzbetrachtung in Abschnitt \ref{Äquivalenz} zeigt. Die Reflexion einer Schallwelle an einem schallharten Rand, der willkürlich zwischen das Diskretisierungsraster gelegt ist, wird ebenfalls zutreffend beschrieben: Untersucht werden für eine senkrecht auf den Rand auftreffende Welle zwei  Fälle, $\chi_a = 0,2$ und $\chi_b= 0,5$, wobei die jeweils daneben liegenden Volumen links mit Luft, $\chi=1$, und rechts vollständig mit Wasser, $\chi=0$, gefüllt sind. Die gewonnene Beschreibung des Randbereichs $H^H$ wird mit der exakten Lösungen der Wellengleichung des Randbereichs $H^A$ verglichen:
\[H^H_{a/b}(z) = \frac{1 + r_{a/b}z}{r_{a/b}+z}\; \text{mit } r_a=2/3 \text{, }r_b=1/3\:, \]   
\[H_a^A(z) = z^{-1/5},\quad H_b^A(z)= z^{-1/2}.\]
Eine geometrische Interpretation zeigt Bild~\ref{BildPartVol}. Gleichung $H^H$ ergibt sich aus einer Betrachtung ebener Wellen nach den Abschnitten~\ref{WelleEle}-\ref{Theo2Adapt} mit $H^H= b_1/a_1=t_{12}/t_{22}$, mit der zugrundeliegenden Betriebskettenmatrix $\mat{T}$, gebildet durch eine Abfolge von Impedanzsprung mit $r_{a/b}$, einfacher Laufzeit und Impedanzsprung $r=1$ für den schallharten Abschluss. Alle Übertragungsfunktionen sind Allpässe mit Einheitsverstärkung und stimmen folglich im Betragsgang überein. 
Die Übertragungsfunktionen sind im $\cal{Z}$-Bereich definiert, dessen zugrundeliegende Zeitdiskretisierung gleich der Schallaufzeit für das Durchqueren zweier Volumenelemente ist. In Abbildung \ref{OperaVerg} werden die Gruppenlaufzeiten von $H^H$ und die reinen Laufzeiten $H^A$ verglichen.  Es zeigt sich, dass die jeweiligen Abweichungen  durch die hier vorgestellte Beschreibung des Randbereichs für den Frequenzbereich unter 50 kHz gering und in dem für die Sprachakustik relevanten Frequenzbereich bis 8 kHz vernachlässigbar ist. 
\begin{afigure}[b]%
\begin{center}
\psfrag{1,0}[r][r]{\footnotesize 1,0}
\psfrag{0,9}{}
\psfrag{0,8}{}
\psfrag{0,7}{}
\psfrag{0,6}{}
\psfrag{0,5}[r][r]{\footnotesize 0,5}
\psfrag{0,4}{}
\psfrag{0,3}{}
\psfrag{0,2}[r][r]{\footnotesize 0,2}
\psfrag{0,1}{}
\psfrag{0}[r][r]{\footnotesize 0}
\psfrag{a}{\put(-0.2,0.3){\footnotesize $H^A_a$}}
\psfrag{b}{\put(-0.2,0.3){\footnotesize $H^A_b$}}
\psfrag{c}{\put(-1,0){\footnotesize $H^H_b$}}
\psfrag{d}{\put(-.9,0){\footnotesize $H^H_a$}}
\psfrag{p}{\put(0,-.2){\footnotesize$\pi$}\put(-3.3,-.40){\footnotesize$\omega$}}
\psfrag{t}{\put(0,-1){\footnotesize$\tau_g$}}
\epsfig{width = 0.4\linewidth, file=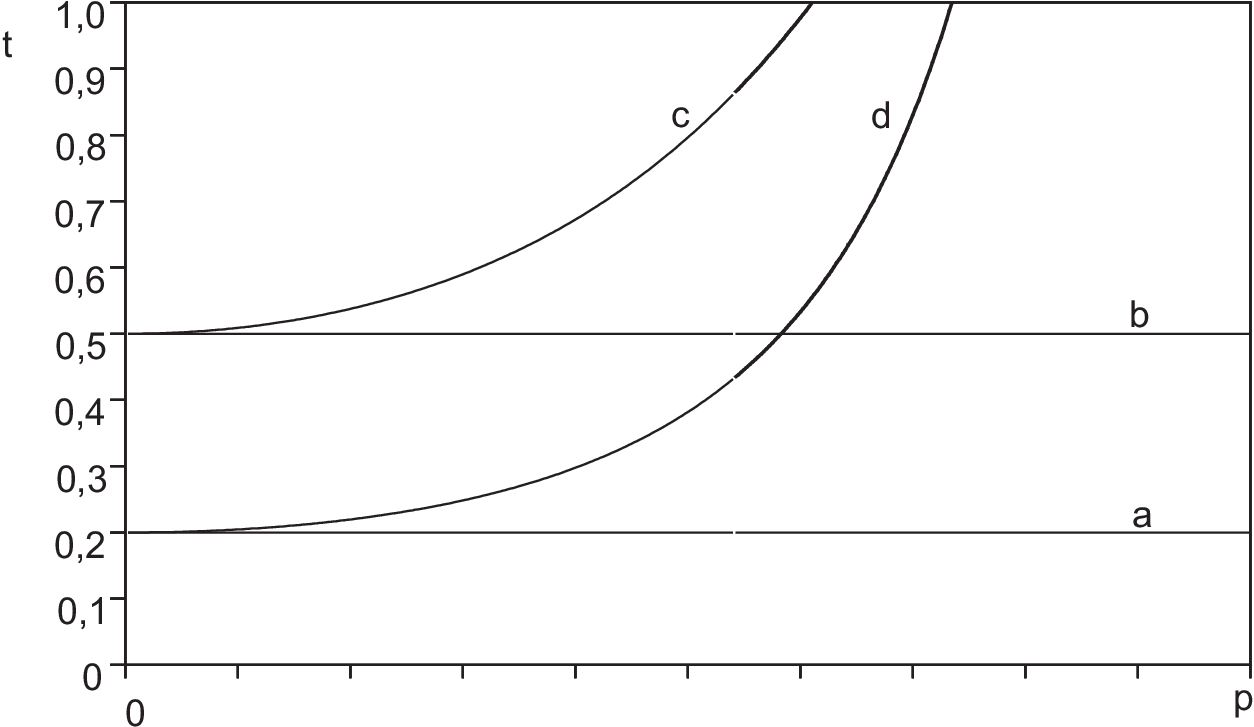}%
\vspace{-1.5ex}%
\caption{Vergleich der Gruppenlaufzeiten $\tau_g$ zweier idealisierter Übertragungsfunktionen $H^A$ mit durch Filter approximierter Übertragungsfunktionen  $H^H$ in Abhängigkeit der Kreisfrequenz $\omega$. Bei der hier verwendeten Diskretisierung von 0,5 mm entspricht $\pi$ einer Frequenz von 340 kHz bei einer Schallgeschwindigkeit von 340 m/s; der Bereich der Gruppenlaufzeit ist einem Schallweg von 1 mm äquivalent.
}\label{OperaVerg}
\end{center}
\end{afigure}%

Der dreidimensionale Differenzen-Operator wird auf dem gleichen Weg erstellt. Die Indizes bei $r$ charakterisieren die Lage des Impedanzsprungs, wobei die tiefgestellten Indizes das Bezugselement spezifizieren und die hochgestellten Indizes das Nachbarelement bezeichnen:
\begin{align}
&p_{t+1,x,y,z}-2p_{t,x,y,z}+p_{t-1,x,y,z}\thickspace\nonumber \\
&=\thickspace K\Bigl[(-6 +r^{x+1,y,z}_{x,y,z}+r^{x-1,y,z}_{x,y,z}+r^{x,y+1,z}_{x,y,z}\nonumber\\
&\makebox[10.4ex]{}+r^{x,y-1,z}_{x,y,z}+r^{x,y,z+1}_{x,y,z-1}+r^{x,y,z-1}_{x,y,z})p_{t,x,y,z}\nonumber\\
&\phantom{=}{}+(1-r^{x+1,y,z}_{x,y,z})p_{t,x+1,y,z}+(1-r^{x-1,y,z}_{x,y,z})p_{t,x-1,y,z}\nonumber\\
&\phantom{=}{}+(1-r^{x,y+1,z}_{x,y,z})p_{t,x,y+1,z} +(1-r^{x,y-1,z}_{x,y,z})p_{t,x,y-1,z}\nonumber\\
&\phantom{=}{}+(1-r^{x,y,z+1}_{x,y,z})p_{t,x,y,z+1} +(1-r^{x,y,z-1}_{x,y,z})p_{t,x,y,z-1}\Bigr]\:.\nonumber
\end{align}
Er stellt eine Erweiterung der in Abschnitt \ref{AnFin} diskutierten Finiten Differenzen dar. Insbesondere bleiben durch die Erweiterung die dort gezeigten Stabilitätsüberlegungen unberührt und sind weiterhin gültig, da sowohl Schallgeschwindigkeit als auch rechts- und linksseitige Koeffizientensumme unverändert bleibt.   

\section{Dämpfung}\label{Daempf}
Die Dämpfung der Schallausbreitung im Sprechtrakt erfolgt zu einem Großteil durch den Schallaustritt am Mund und an den Nasenlöchern. Dieser wird durch eine Reflexion der Wellen mit einem endlichen Verhältnis der akustischen Impedanzen zwischen dem Querschnitt der Schallaustrittöffnung und dem sich daran anschließendem Halbraum Rechnung getragen. In der einfachsten Form erfolgt das durch einen abschließenden Reflexionskoeffizienten, dessen Betrag entsprechend kleiner 1 ist.

Der Sprechtrakt und insbesondere der Nasaltrakt weist zudem eine innere Dämpfungen auf, welche die Resonanzeigenschaften beeinflussen. Diese treten überwiegend an den Wänden auf und werden in den folgenden Abschnitten genauer betrachtet und in die Simulation der Schallausbreitung über ein zweckmäßiges phänomenologisches Modell mit einbezogen. 

\subsection{Dämpfungsursachen}
Für die Dämpfung der Schallausbreitung sind eine Reihe von Ursachen bekannt, vgl.~\cite{Ra03, Morse Ingard, Wolf}. Für kleine Lautstärken, wie sie im Sprechtrakt auftreten, sind dies:
\begin{itemize}
\item{Wärmediffusion}
\item{Wärmekapazität der Wände}
\item{Viskose Reibung.}
\end{itemize}
Der erste Effekt beruht auf der thermischen Diffusion der in den Schallwellen inhärenten Temperaturunterschiede, der Abweichung von der adiabatischen Beziehung zwischen Druck und Temperatur. Er wird bspw.~in \cite{Morse Ingard} diskutiert und trägt wenig zur Dämpfung akustischer Systeme der hier betrachteten Größen und Frequenzen bei: Sie ermitteln bei 1000 Hz eine Dämpfung von 10 dB auf einer Entfernung von 10 km. Auch wenn diese Dämpfung bei mehratomigen Gasen durch die Anregung von Molekülrotation und \mbox{-schwingung} stärker ist, wird sie aufgrund ihres letztlich kleinen Beitrags vernachlässigt. 

Dieser Effekt tritt jedoch bedeutend stärker zutage, wenn die Luft mit einem anderen Medium höherer Wärmekapazität im Kontakt ist, hier den Wänden des Nasaltrakts. Verstärkt wird dieser Effekt durch die viskose Reibung der Schallschnelle an den ruhenden Wänden\footnote{Grundlegene Untersuchungen stammen von Helmholtz, der in \cite{He1863} den Reibungsmechanismus beschreibt, Le Roux, der in \cite{Ro1862,Ro1867} eine abweichende Schallgeschwindigkeit in Röhren beobachtet, Regnault, der in \cite{Re1868} zudem eine Dämpfung erkennt und eine Elastizitätsabnahme der Luft durch Wechselwirkung mit den Rohrwände vermutet, Kundt, der in \cite{Ku1868} die Abhängigkeit der Schallgeschwindigkeit von u.~a.~Durchmesser und Frequenz experimentell untersucht und schließlich Kirchhoff, der in \cite{Ki1868} eine passende mathematische Beschreibung dieser Effekte  unter Berücksichtigung der Theorie reibungsbehafteter Strömungen von Stokes \cite{St1845} bzw.~der Gastheorie von Maxwell \cite {Ma1867} zeigt. So findet beispielsweise Kundt a~a.~O., S.~370, dass die Schallgeschwindigkeit bei einem Rohrdurchmesser von \mbox{3,5 mm} und einer Wellenlänge von \mbox{18 cm} um \mbox{9 \%} reduziert ist.}, welche ihre Ursache ebenfalls in der Diffusion der Gasmoleküle hat --- mit dem Unterschied, dass hier nicht mehr ihre mittlere Bewegungsenergie, sondern der mittlere Impuls betrachtet wird, wie \cite{Ki1868} ausführt. 

Ausgehend von den Navier-Stokes-Gleichungen und der Wärmeleitungsgleichung, kann man eine erweiterte Differentialgleichung für die Schallausbreitung entwickeln, vgl.~\cite{Ki1868, Morse Ingard}:
\[
\nabla^2p=\frac{\kappa}{c^2}\left( \frac{\partial^2}{\partial t^2}-l_vc\frac{\partial}{\partial t}\nabla^2\right)(p-\alpha \tau)
\]
und die Wärmeleitungsgleichung erweitert sich zu
\[
l_hc\nabla^2\tau= \frac{\partial}{\partial t}\left(\tau-\frac{\kappa-1}{\kappa\alpha} p \right)\quad.
\] 
Hierbei ist $\tau$ die Differenz zur mittleren Temperatur, $\kappa$ der Adiabatenkoeffizient, $\alpha$ der Volumenausdehnungskoeffizient, $l_h$ und $l_v$ sind die mittleren freien Weglängen der Gasmoleküle. Man findet hierin eine Reihe von bekannten Gleichungen zur Schallausbreitung, beispielsweise:
\[
\nabla^2p=\frac{\kappa}{c^2}\frac{\partial^2}{\partial t^2}p
\]  
für reibungsfreie isotherme Schallausbreitung ($l_v=0$, $\tau=0$) mit der um $\sqrt{\kappa}$ verringerten Ausbreitungsgeschwindigkeit; für die reibungsfrei adiabatische Schallausbreitung ($l_v= 0$ und  $l_h=0$, letzteres führt zu $\tau=\frac{\kappa-1}{\kappa\alpha} p$)  die akustische Wellengleichung aus Abschnitt \ref{Wellengleichung}:
\[
\nabla^2p=\frac{\kappa}{c^2} \frac{\partial^2}{\partial t^2}\left(p-\alpha \frac{\kappa-1}{\kappa\alpha}p\right)=\frac{1}{c^2}\frac{\partial^2}{\partial t^2}p
\]
Mit einer isobaren Betrachtung ($p=0$) erhält man schließlich
\[
\nabla^2\tau=\frac{1}{l_vc}\frac{\partial}{\partial t}\tau\quad, 
\]
die Wärmeleitungsgleichung und $l_v = l_h$.  Die Randbedingungen für die Differentialgleichung sind näherungsweise
\[
	\vec{u} = 0,\qquad \tau= 0,
\]
welche die ruhenden Wände und die höheren Wärmekapazität und -leitfähig\-keit der Wände erfassen. In \cite{Morse Ingard} ist eine schrittweise Entwicklung der allgemeinen Lösung angegeben.


%
Darüber hinausgehende Dämpfungsursachen, wie die Schalleinkopplung in das das umliegende Gewebe oder aus der turbulenten Reibung der Schallausbreitung, bleiben im Folgenden unberücksichtigt.  Da der Nasaltrakt ein aus Knochen und Knorpeln umgebener Hohlraum ist, was ihm eine hohe Steifigkeit verleiht, ist der Beitrag ersterer entsprechend gering. Auch der nichtlineare Dämpfungsterm der turbulenten Reibung trägt für geringe Schallpegel wenig bei. 
 
\subsection{Modellierung}\label{DämpfModel}
Die lineare Differentialgleichung im Abschnitt zuvor führt zu einer Größenordnung der Konstante des exponentiellen Abfalls unterhalb der Gitterdiskretisierung, wie \cite{Morse Ingard} zeigt. Aus diesem Grund würde die direkte Umsetzung der Differentialgleichungen mittels finiter Methoden eine wesentliche Verfeinerung des Gitters nach sich ziehen und damit eine in der vierten Potenz wachsende Rechenzeit. Darüber hinaus wären wiederum Kenntnisse des genauen Oberflächenverlaufs notwendig.  

Es wird deshalb ein phänomenologisches Modell entwickelt, das diese Erfordernisse nicht hat und sich vergleichsweise einfach in das Diskretisierungsschema einfügen lässt. Im Bereich der Oberfläche erweitert sich die Wellengleichung zu 
\[
\frac{1}{c^2}\left(\frac{\partial ^{2}p}{\partial t^{2}}+R'\frac{\partial p}{\partial t}\right) =\Delta p\quad,\]
wobei $R'$ die Dämpfung charakterisiert. 
Entsprechend erweitert sich die in Abschnitt \ref{AnFin} gefundenen Differenzengleichung mit $R=R'g/2c^2$ zu
\[
	(1+R)p_{x,y,z,t+1}-2p_{x,y,z,t}+(1-R)p_{x,y,z,t-1}= ... \quad .
\]

Die Dämpfung hat verschiedene Ursachen, deren Beiträge unterschiedlichen Gesetzmäßigkeiten gehorchen: Betrachtet man die Schallausbreitung längs eines zylindrischen Rohrs, so ist die aus Wärmeleitung und viskoser Reibung resultierende Dämpfung proportional zu der Wurzel der Frequenz und umgekehrt proportional zu der Wurzel der Querschnittsfläche. Weitere Dämpfungen sind frequenzunabhängig und umgekehrt proportional der Querschnittsfläche oder der Wurzel der Querschnittsfläche. Für den Nasaltrakt ist die Dämpfung im Bereich kleiner Querschnittsflächen maßgeblich. 
Entsprechend wird der Dämpfungskoeffizient
$R=5\cdot10^{-4}$
gewählt, so dass die Differenzengleichungen in diesem Bereich das reale Verhalten approximieren, wie in Tabelle \ref{dtab} dargestellt.
\begin{atable}
\begin{center}
\begin{tabular}{r|rr}
$A/[\mbox{mm}^2]$ & Modell & Literatur\\[-2.1ex]
& & \\\hline
& & \\[-1.8ex]
3,00 & 11,0 & 12,3  \\
9,25 & 7,2 & 6,6 \\
34,25 & 4,2 & 3,3 
\end{tabular}
\end{center}
\caption{Dämpfung eines Rohrs mit der Querschnittsfläche $A$ verglichen mit Literaturwerten aus \cite{MüM04} in [dB/m] bei $f=1\mbox{ kHz}$. \label{dtab}}
\end{atable}

\subsection{Frequenzabhängigkeit}\label{frequenzDaempf}
Die Frequenzabhängigkeit der Dämpfung ist in Bild \ref{D_A_F} gezeigt. Durch eine zwei- oder mehrfache Berechnung der Übertragungsfunktion des Nasaltrakt mit Dämpfungskoeffizienten, die für die jeweilige Frequenz zutreffend sind, kann dieser Abhängigkeit Rechnung getragen werden. Die gesamte Übertragungsfunktion ergibt sich dann durch eine gewichtete Überlagerung der einzelnen Übertragungsfunktionen.  Alternativ oder ergänzend kann die Ordnung des Differenzen-Operators in Zeitrichtung erhöht und dessen Frequenzverhalten angepasst werden.
\begin{afigure}%
\begin{center}
\psfrag{F}{\put(4.3,.3){\footnotesize $f/{\rm [Hz]}$}}
\psfrag{D}{\put(-2,.2){$\frac{D}{\rm [Np/m]}$}}
\psfrag{A3}[r][]{\footnotesize 3~mm${}^2\;$}
\psfrag{A9}[r][]{\footnotesize 9~mm${}^2\;$}
\psfrag{A34}[r][]{\footnotesize 34~mm${}^2\;$}
\psfrag{10000}{{\footnotesize 10000}}
\psfrag{1000}{{\footnotesize 1000}}
\psfrag{100}{{\footnotesize 100}}
\psfrag{8}[r][r]{{\footnotesize 8}}
\psfrag{4}[r][r]{{\footnotesize 4}}
\psfrag{2}[r][r]{{\footnotesize 2}}
\psfrag{1}[r][r]{{\footnotesize 1}}
\psfrag{0.5}[r][r]{{\footnotesize 0,5}}
\psfrag{0.25}[r][r]{{\footnotesize 0,25}}
\psfrag{0.125}[r][r]{{\footnotesize 0,125}}
\epsfig{width = 0.7\linewidth, file=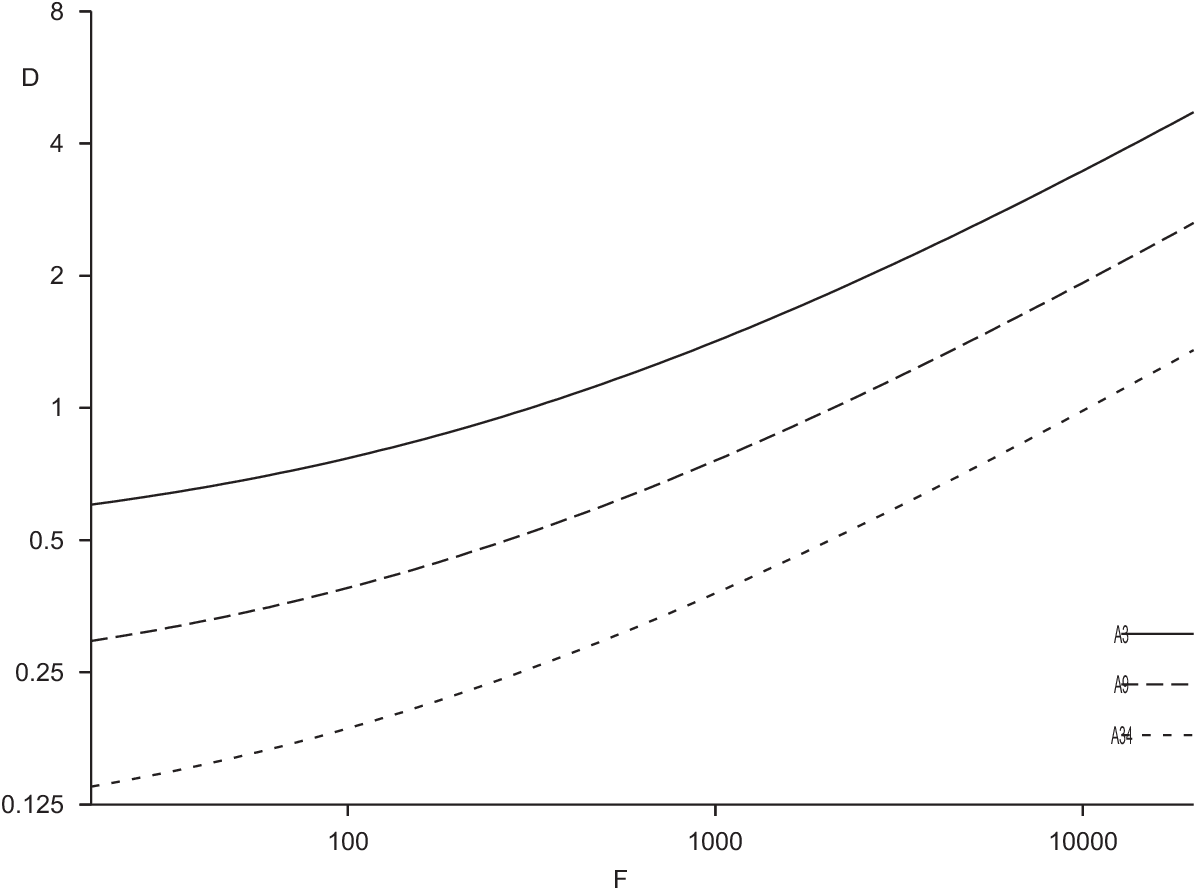}%
\caption{Frequenzabhängige Dämpfung $D$ der Schallausbreitung in Rohren unterschiedlicher Querschnittsfläche, gekennzeichnet durch unterschiedliche Linientypen.}\label{D_A_F}
\end{center}
\end{afigure}%

\subsection{Gedämpfte Wellenausbreitung}
Da die Dämpfungskoeffizienten klein sind, kann man die Approximation der Dämpfung auch in den Termen $1+R$ und $1-R$ durch $e^\gamma$ und durch $e^{-\gamma}$ darstellen, also
\[
	\leadsto\quad e^\gamma p_{x,y,z,t+1}-2p_{x,y,z,t}+e^{-\gamma} p_{x,y,z,t-1}= ...
\]
mit 
\[
	\gamma = \frac{\ln(1+R)-\ln(1-R)}{2}\approx R \quad .
\]
Man gewinnt dadurch eine Separation von Dämpfung und Schallausbreitungsgeschwindigkeit, die sonst miteinander verknüpft sind. Die Separation wird ersichtlich, wenn man die Funktion 
\[
	p_{x,t} = e^{-a t}\sin (\omega t  - k x) 
\] 
als Lösungsansatz für den auf eine Dimension vereinfachten Fall betrachtet. Man erhält
\[\scriptstyle 
	e^\gamma e^{-a(t+1)}\sin (\omega (t+1) - k x)\; - \; 2e^{-at}\sin (\omega t  - k x)\; + \;e^{-\gamma} e^{-a(t-1)}\sin (\omega (t-1) - k x)\; = \;... \quad \textstyle ,
\]
was durch Dividieren mit $e^{-a t}$ und durch $a=\gamma$ in die aus Abschnitt \ref{AnFin} bekannte Form übergeht. Eine Abhängigkeit zwischen $\omega$ und $\gamma$ ist nicht entstanden. 

\subsection{Repräsentation}
Die Dämpfungskoeffizienten werden ähnlich den Volumenkoeffizienten des CT-Datensatzes dem Programm kodiert als Stapel tomographischer Dateien bereitgestellt. Jede Datei korrespondiert dabei einer Datei der CT-Daten, sowohl in Lage als auch in Auflösung. Aus praktischen Erwägungen sollten die Koeffizienten derart kodiert sein, dass die physikalische Beschreibung bei einer Translation oder Rotation der Datensätze invariant ist.

Anhand der hier gefundenen Beziehungen lässt sich die eingangs gestellte Forderung nach Invarianz untersuchen. Betrachtet man ein Volumenelement mit Dämpfung, umgeben von anderen ohne Dämpfung, so werden die Schallwellen in diesem um $e^{-a t}$ gedämpft. Verschiebt man nun das Gitter der Volumenelemente exemplarisch um eine halbe Gitterlänge, so sollen die beiden jetzt beteiligten Volumen die gleiche Dämpfung verursachen, jede einzelne folglich $\sqrt{e^{-a t}}=e^{-a t/2}$. Es ist sinnvoll, die Dämpfung durch den Koeffizienten $a$ zu repräsentieren und diese bei Translation, Rotation und Skalierung linear zu interpolieren, da dann die physikalischen Eigenschaften näherungsweise unverändert bleiben. Diese Parameterform hat zudem den Vorteil, dass sich sowohl Bereiche geringer Dämpfung, wie die hier diskutierten Wände, als auch Bereiche hoher Dämpfung für reflexionsfreie Abschlüsse an den Schall\-austrittsöffnungen einheitlich in Festkommadarstellung abbilden lassen. 

\subsection{Berücksichtigung partieller Volumen}
Die bisherigen Betrachtungen zur Dämpfung sind davon ausgegangen, dass der Rand mit dem Gitter zusammenfällt. Zur Berücksichtigung von partiellen Volumen nach \ref{partVol} muss die Betrachtung erweitert werden. Hierbei steht weniger eine Verbesserung des Dämpfungsmodells im Vordergrund, als vielmehr das Ziel, beide Modelle gleichzeitig nutzen zu können. %

Betrachtet man hierzu die Fälle, dass der Rand auf dem Gitter liegt, und dass der Rand ein Volumenelement mittig durchquert. Iim ersten Fall wird die in Abschnitt \ref{DämpfModel} ermittelte Dämpfung zugewiesen. Eine einfache Verallgemeinerung für den zweiten Fall ist:
\[
R_x = R_0\left((1-\chi_{x-1}) + (1-\chi_{x+1})\right)\quad,
\]
die für die anderen Raumrichtungen durch entsprechende Summanden erweiter wird. $R_0$ ist der in dem genannten Abschnitt halbempirisch bestimmte Dämpfungskoeffizient.

\pagebreak
\section{Implementierung und Optimierung}
Zur Anwendung der in den Abschnitten zuvor entworfenen Finite-Dif\-feren\-zen  wird in diesem Abschnitt ihre programmtechnische Realisierung betrachtet. Ziel ist es, ein Werkzeug zu entwickeln, mit dessen Hilfe auf Personal Computern die erweiterte Wellengleichung für beliebige Randbedingungen aus Tomographien gelöst werden kann. 
Diese Zielsetzung lässt sich in drei Komponenten unterteilen:\begin{itemize}
\item die Daten-Schnittstelle,
\item die Berechnung der Wellendifferentialgleichung,
\item die graphische Benutzeroberfläche.
\end{itemize}
Zur Realisierung des Programms wurde ein objektorientierter Ansatz unter Verwendung der Programmiersprache C++ gewählt. Dies erlaubt zum einen den Zugriff auf verschiedene Bibliotheken für eine graphische Benutzerschnittstelle und Datenschnittstelle, zum anderen eine Optimierung und Parallelisierung der aufwendigen Berechnung der Wellendifferentialgleichung. 

Die Komponente zur Berechnung der Wellendifferentialgleichung wird in Abschnitt \ref{BerechWell}~insbesondere im Hinblick auf die Effizienz der Berechnung betrachtet, da die zur Berechnung herangezogenen Daten einen erheblichen Umfang haben. Die Daten-Schnittstelle bindet digitale Tomographiedaten ein und liefert die zu verarbeitenden Randbedingungen. Das Format der Daten-Schnittstellen wird deshalb in Abschnitt \ref{daten-schnittstelle} entwickelt.  Die Anforderung an die Daten-Schnittstelle ist dabei eine möglichst hohe Integrierbarkeit mit bestehenden Anwendungen, um die anatomischen Daten letztendlich dem Werkzeug zugänglich zu machen. Die analoge Anforderung ergibt sich für die gewonnen Ergebnisse, die für eine über die in der Benutzeroberfläche integrierte  Darstellung hinausgehende Analyse   exportiert werden müssen.

\subsection{Graphische Benutzeroberfläche}
Die graphische Benutzeroberfläche stellt eine interaktive Verbindung zwischen dem Anwender und dem Werkzeug selbst her. Der Anwender kann mit Hilfe der Benutzeroberfläche die Berechnung kontrollieren, indem sie deren Ergebnisse in Form eines zeitlichen Schalldruckverlaufs an einem oder mehreren ausgewählten Punkten visualisiert. Darüber hinaus bietet die Benutzeroberfläche die Möglichkeit, die untersuchte Geometrie wahlweise mit einer Überlagerung  Schallausbreitung in Form eines  zeitschrittweisen Verlaufs  von verschiedenen Perspektiven zu betrachten. Um die dreidimensionale Struktur zu erkunden, wird die Möglichkeit gegeben, diese in Form von \mbox{Sagittal-,} Frontal- oder Transversalschnitten darzustellen, welche senkrecht zu der Schnittebene verschiebbar sind. Da die graphische Benutzeroberfläche lediglich ein Mittel zum Zweck ist und keine darüber hinausgehenden Ergebnisse liefert, wird auf ihre Realisierung  nachfolgend nicht eingegangen.

\subsection{Daten-Schnittstelle 
}\label{daten-schnittstelle}

Die aus anatomischen Untersuchungen gewonnene geometrische Beschreibung oder synthetische Geometrien werden über die Datenschnittstelle dem Werkzeug eingangsseitig  zugänglich gemacht; ausgangseitig müssen die gewonnenen Impulsantworten zur Analyse und Darstellung weiteren Programmen zur Verfügung gestellt werden. Dafür ergibt sich eine Reihe von Anforderungen, die und deren Lösung durch ein intermediäres Datenformat im Folgenden betrachtet werden.

Das Format der Quelldaten ist abhängig von dem datenliefernden System. So verwendet der Computertomograph proprietäre Dateiformate, während die Kryoschnitte in einem Format gespeichert sind, das keine Metainformationen zur Interpretation der Daten enthält und einer Komprimierung nach \cite{We84} unterzogen ist;
Kernspinresonanz-Daten liegen im in \cite{DICOM} beschriebenen und ebenso genannten Datenformat vor. Für die Verifizierung der Modelle, der daraus abgeleiteten Algorithmen und ihrer programmtechnischen Umsetzung sind zudem Tests anhand von synthetischen geometrischen Strukturen mit bekannten akustischen Eigenschaften nützlich. Die Datensätze dieser Teststrukturen sollten möglichst einfach zu generieren sein und müssen in das Programm übernommen werden können. 

Für alle diese Datenformate hätte zur Integration der zugehörigen Schnittstellen in das Werkzeug eine erhebliche Zeit aufgewendet werden müssen und die Anwendung des Werkzeugs bliebe auf eben diese Datenformate beschränkt. Effizienter und flexibler ist das stattdessen verwendete \name{Adapter-Pattern} nach \cite{Gamma}.	
Diesem Entwurfsmuster folgend bilden verschiedenen Adapter jeweils die unterschiedlichen Datenformate auf das einheitliche Schnittstellenformat des Werkzeugs ab. Das hierfür entwickelte Schnittstellenformat orientiert sich dabei an dem Prinzip der Tomographien: eine lineare Abfolge von Bildern in dem weit verbreiteten, zweidimensionalen \name{Bitmap}-Format nach \cite{Microsoft-BMP} wird mit einer die Anordnung in er dritten Dimension beschreibende, klartext-basierten und parsebaren Metainformationsdatei ergänzt, wie in Ausschnitt \ref{excerpt_schnittstelle} gezeigt. Weitere Teile der Metainformationsdatei steuern die Interpretation der zweidimensionalen Bitmap-Dateien.
\begin{excerpt}[t]
\begin{verbatim}
98 107 124
28 79 143
219 231 17
0 255 0
255 255 255
..\VISMAN\VM0043.BMP
..\VISMAN\VM0044.BMP
..\VISMAN\VM0045.BMP
...
..\VISMAN\VM0166.BMP
..\VISMAN\VM0167.BMP
\end{verbatim}\vspace{-1ex}%
\caption{Daten-Schnittstelle,  durch die Datei-Endung {\tt .bnd} gekennzeichnet.  Die erste Zeile liefert die Ausdehnung in $x$-, $y$- und $z$-Richtung. In den folgenden Zeilen werden die Wertetripel bzw.{} RGB-Darstellung der Bereiche Anregung, Aufzeichnungspunkt, schallweicher und schallabsorbierender Rand festgelegt. Diesen schließt sich eine Liste von Pfadangaben für Dateien im Bitmap-Format an. Die Reihenfolge der Datei-Angaben entspricht dem Aufbau der dreidimensionalen Struktur.}\label{excerpt_schnittstelle}  
\end{excerpt}

Bestehende Programme bilden die Adapter. Daten der Kryosektionen können mittels Photoshop, erweitert um eine einfache Skriptsteuerung, gelesen, skaliert und in das Bitmap-Format übertragen werden, wie bereits in Abschnitt \ref{Kryo} angesprochen. Die Daten der Computertomographie wurden mittels EasyVision konvertiert, Daten aus dem MRT durch \name{DicomWorks} aus \cite{PuDICOM}. Das in \cite{Fr98} beschriebenen Programm \name{NMRWIN} liest proprietäre Datenformate von Computer- oder Kernspintomographen und konvertiert sie in das Bitmap-Format, vgl.{} S.{} \pageref{Dank}. Die Bitmap-Dateien sind für die Schnittstelle auf eine Bit-Anzahl von 24 pro Punkt in der zweidimensionalen Bildebene festgelegt, die üblicherweise jeweils dyadisch zu 8~Bit als rote, grüne und blaue Farbkomponente dargestellt werden.\footnote{
Für die Analyse der Kryosektionen in \cite{Ra99,RaSL99,RaL00b} werden 8 Bit verwendet, da die partiellen Volumen aus den in Abschnitt \ref{Kryo} genannten Gründen nicht angewendet werden. Die Kryosektionen werden in vier Bereiche kategorisiert:
\begin{itemize}
\item Hohlraum,
\item absorbierende Rand,
\item Anregungsstelle und
\item schallharte Bereich, dem alle anderen Werte zugewiesen sind.
\end{itemize} 
} 
Durch eine Requantisierung in den Adaptern werden durch die Tomographien bestimmten Volumenparameter $\chi$ auf diese 8 Bit, entsprechend einem Wertebereich von 256 abgebildet,\footnote{Die Requantisierung verursacht keinen signifikanten Fehler.
Die Daten der Computertomographie, deren genutzter Wertebereich sich zwischen -1000 und 0 erstreckt, sind mit geringfügigen Messfehlern behaftet, wie sie auch in Bild \ref{CT-Vol} nach der Requantisierung erkennbar sind. Die Messfehler dominieren offenbar gegenüber der Requantisierung.}
und in allen drei Komponenten abgelegt. Mit dieser Repräsentation ist eine visuelle Überprüfung und die die Erzeugung von Teststrukturen mittels vorhandenen, betriebssystemeigenen  Bildbearbeitungsprogrammen möglich und das Einlesen der Daten in das Werkzeug wird durch vorhandene Programmbibliotheken vereinfacht. In einem zweiten Schritt werden Bereiche der Schallanregung, schallabsorbierende und gegebenenfalls schallweiche Flächen sowie Aufzeichnungspunkte  für das Simulationsergebnis durch bestimmte Wertetripel gekennzeichnet. Dies geschieht wiederum mit den vorhandenen Bildbearbeitungsprogrammen. Durch eine geeignete Wahl der Wertetripel treten darin die besonderen Bereiche mit einem hohen farblichen Kontrast hervor.

%
%


Die Ausgabe der an den Aufzeichnungspunkten gewonnenen Simulationsdaten\footnote{Für die Simulationsdaten wird ein Gleitkommaformat einfacher Genauigkeit verwendet.} erfolgt über standardisierte File-Streams, in denen der dezimalkodiert Wert jedes Zeitschritts zeilenweise abgelegt wird. Diese Folge von Wert kann von anderen Applikationen, wie \name{edit}, \name{Matlab} und \name{Gnuplot} zur Überprüfung, weiteren Analyse und Darstellung unmittelbar gelesen werden.
 Der Dateiname wird durch eine Nummer gebildet, die im Falle mehrerer Aufzeichnungspunkten die Zuordnung erlaubt, und durch die Datei-Endung {\tt .out} gekennzeichnet.

\subsection{Berechnung der Wellengleichung}\label{BerechWell}
Die über die Datenschnittstelle eingelesenen Tomographien werden in einem dreidimensionalen Array	 abgelegt, vgl.{} Ausschnitt \ref{simpleLoop} des Programmtextes.
Anhand der Volumenparameter $\chi$ und der durch Wertetripel ausgezeichneten Bereiche werden die Koeffizienten zur Berechnung der Finiten-Differenzen nach Abschnitt \ref{partVol} und \ref{Daempf} über die Reflexionsfaktoren festgelegt, Speicheradressen der Aufzeichnungs- und Anregungspunkte bestimmt, sowie die zu teilweise oder vollständig mit Luft gefüllten Raumpunkte in dem quaderförmigen Datensatz ermittelt, auf die die rechenintensive Anwendung des Finite-Differenzenoperators begrenzt wird. Im Anschluss erfolgt die Einprägung eines Einheitsimpulses in den Anregungspunkten.  Zur Bestimmung des Übertragungsverhalten werden meist $2^{16}=65536$ Iterationen durchgeführt, was nach einer Fouriertransformation der Impulsantwort zu einer Frequenzauflösung für das Beispiel im Ausschnitt \ref{excerpt_schnittstelle}  von 8 Hz führt.

Einer der wichtigsten und interessantesten Aspekte des Programms ist die Implementierung des Operators zur Berechnung der Wellengleichung, nicht zuletzt deshalb, weil  im Vergleich die Ausführungszeiten aller anderen Programmteile vernachlässigbar sind. Der zu analysierende Datensatz hat in dem genannten Beispiel  eine Größe von $98*107* 124 \approx 1.300.000$ Volumenelementen, deren Speicherbedarf sich von jeweils 8 Byte für ein nicht am Rand liegendes Volumenelement auf andernfalls 44 Byte erstreckt.\footnote{Die Werte ergeben sich im ersten Fall aus zwei Druckwerten zu je 4 Byte, im zweiten Fall kommen die Koeffizienten für alle neun Punkte des Finiten-Differenzen-Operators mit jeweils der gleichen Größe hinzu.}
Für eine kurze  Ausführungszeit der Berechnung ist zum einen die Datenmenge bzw.{} die Speicherzugriffe möglichst gering zu halten, um die Ausführung nicht durch Zugriffslatenzzeiten\footnote{Moderne Betriebssysteme ermöglichen ein automatisches Auslagern von Daten aus dem Hauptspeicher auf Festplatten, falls dessen Kapazität überschritten wird. Ein erneuter Zugriff auf ausgelagerte Daten verursacht jedoch eine erhebliche Wartezeit (typ.~10ms), bis die Daten wieder bereitgestellt sind. Ebenso verhält sich der Speicher des Prozessors selbst, der sogenannten Cache, in Relation zum Hauptspeicher. Auch hier ist die Auslagerung von Daten in den Hauptspeicher um rund 2 Größenordnungen langsamer.} fortwährend zu verlangsamen.  Zum anderen ist Berechnung selbst möglichst optimal zu gestalten.  Erst mit der Erfüllung beider Forderungen ergibt sich eine geringe Laufzeit des Programms.

Die Forderung nach einer geringen Datenmenge hat einen direkten Einfluss auf die Berechnung der Wellengleichung. Eine Klasse speichereffizienter Algorithmen sind \name{In-Place}-Verfahren, bei denen die Ergebnisse einer Iteration auf dann nicht mehr benötigte Variablen zurückgeschrieben  werden. Dies kann --- ein weiterer Vorteil expliziter Zeitschrittverfahren --- genutzt werden, in dem man die Zeitebenen
\[\ws_t  \text{, } \ws_{t+2}\text{, }\ws_{t+4} \dots \]  
sowie
\[\ws_{ t+1} \text{, } \ws_{t+3}\text{, }\ws_{t+5} \dots \]  
jeweils auf die gleiche Stelle im Hauptspeicher abbildet. Desweiteren vermeidet dieses Verfahren die Notwendigkeit, Datensätze zu kopieren beziehungsweise zu verschieben. Die programmtechnische Umsetzung erfolgt durch eine Erweiterung des Array um eine vierte Dimension, die die Zeitrichtung in Form der beiden Zeitebenen umfasst.

\subsection{Programmoptimierung}\label{opt}
Die Ausführungszeit des Programms zur Simulation der Wellenausbreitung mittels Finiter Differenzen liegt mit grundlegenden Optimierungen, vgl.{} Ausschnitt \ref{simpleLoop} und eine ausführlichere Analyse eines Aspekts in \cite{Ra99}, anfangs bei rund fünfzig Stunden. Wenngleich diese Zeit für bestimmte Berechnungen erträglich ist, so ist sie doch störend, wenn man Parameter optimieren möchte oder eine größere Menge von Datensätzen untersuchen will. 
\begin{excerpt}[t]\footnotesize
\begin{verbatim}                                           
for( x = 1; x < Xmax - 1; x++)          for( x = 1; x < Xmax - 1; x++)
for( y = 1; y < Ymax - 1; y++) {        for( y = 1; y < Ymax - 1; y++) 
    float *uL = u[1-t][x][y];           for( z = 1; z < Zmax - 1; z++)               
    float *ul = u[t][x][y];             switch( B.Typ( x, y, z ) ) {                 
    float *ulx = u[t][x-1][y];                                             
    float *ulX = u[t][x+1][y];              case Bound::Inner:                       
    float *uly = u[t][x][y-1];                                             
    float *ulY = u[t][x][y+1];              u[1-t][x][y][z] = kFlaeche * (            
                                                  u[t][x-1][y][z] 
    for( zp = zi[x][y]; z = *zp; zp++)          + u[t][x+1][y][z]        
        uL[z] = kFlaeche * (                    + u[t][x][y-1][z]   
            ulx[z] + ulX[z]                     + u[t][x][y+1][z]
          + uly[z] + ulY[z]                     + u[t][x][y][z-1]
          + ul[z-1] + ul[z+1] )                 + u[t][x][y][z+1] )                         
          + kZentrum * ul[z]                    + kZentrum * u[t][x][y][z]                      
          - uL[z];                              - u[1-t][x][y][z];      
                            
    for( zp = zr[x][y]; z = *zp; zp++)      ...                     
        uL[z]=...;                      }                                         
                                        
    ...
}
\end{verbatim}
\caption{Implementierungen zweier Kerne der zeitlichen Iterationen zur Berechnung der Finiten Differenzen in C++. Der optimierte Kern ist links dargestellt, rechts ist zum Vergleich der nur bezüglich des effizienten Zugriffs auf den Level-1-Cache entworfenen Kern gezeigt. Der Zeitparameter {\tt t} alterniert von  Zeitschritt zu Zeitschritt zwischen 0 und 1.\\
Der Schalldruck ist in dem vierdimensionalen Feld {\tt u} in Gleitkomma-Darstellung hinterlegt, das sich aus einer Zeit- und drei Raumdimensionen ergibt. Der Zugriff erfolgt durch den Dereferenzierungsoperator {\tt []}. In beiden Implementierungen führt die innerste {\tt for}-Schleife in der Raumdimension {\tt z} die letzte Dereferenzierung durch und greift somit auf dicht bei\-einander liegende Daten zu, die deshalb überwiegend in dem latenzarmen Level-1-Cache vorrätig sind. Der links gezeigte Kern nimmt zudem die mehrfache Dereferenzierung aus der innersten Schleife heraus, wodurch die verbleibenden Operationen zum Zugriff auf das Datenfeld parallel zu den Gleitkomma-Berechnungen der Finiten-Differenzen durchgeführt werden und letztere nicht mehr wesentlich verlangsamen.\\ 
Die {\tt switch..case}-Fallunterscheidung, die optimierte Ausführungspfade für die gezeigten inneren, aufwendigeren randnahen und sonstigen Finite-Differenzen bereitstellt, kann zur Vermeidung von \name{Branch-Prediction}-Fehlern ebenfalls aus der inneren Schleife herausgenommen werden. Durch Indexfelder {\tt zi} für innere und {\tt zr} für Randelemente wird sie in die Initialisierungsphase des Programms verschoben, wodurch die Befehls-Abarbeitungskette (\name{Pipeline}) des Prozessors im Kern ungestört ist und die Ausführungsgeschwindigkeit des Programms nochmals deutlich erhöht wird.}\label{simpleLoop}
\end{excerpt}

Demgegenüber ermöglichen Fortschritte in der Halbleiterherstellung die Herstellung von immer schnelleren und eine größere Anzahl von Elementen umfassenden Schaltungen auf einem Chip. Insbesondere bei Prozessoren erlaubt die Integration dieser zusätzlichen Elemente durch eine Erweiterung der Architektur einen Geschwindigkeitszuwachs weit über die Steigerung der Taktrate hinaus. Im Folgenden werden die Ergebnisse zweier Methoden betrachtet, Fortschritte in Rechnerarchitekturen vorteilhaft zu nutzen, um damit eine Reduzierung der Rechenzeit zu erzielen.

\subsection{Parallelisierung für SMP und NUMA}
Während der ursprünglich eingesetzte Prozessor vom Typ \name{Pentium} bereits über eine -- bezogen auf seine Taktfrequenz -- leistungsfähige Recheneinheit für Gleitkommazahlen besitzt, verfügen neuere Prozessoren vom Typ \name{Pentium III} und nachfolgende zudem über ein vielfaches größeren integrierten Zwischenspeicher, der häufig benutzte Daten auf dem Chip vorrätig hält und schnell verfügbar macht. Dieser als Level-2-Cache bezeichnete Zwischenspeicher erlaubt eine sinnvolle und preiswerte Verwendung mehrerer Prozessoren. Hierbei greifen die Prozessoren auf einen gemeinsamen Hauptspeicher zu und können so gleichzeitig einen Algorithmus auf gemeinsamen Daten ausführen; der Hauptspeicher wird dabei von übermäßig vielen Zugriffen der Prozessoren durch ihre Caches entlastet.  Eine derartig aufgebaute Rechnerarchitektur bezeichnet man als \name{Symmetric Multi Processing}, abgekürzt \name{SMP}.

Für die mehrere Megabyte umfassenden Datensätze der Finiten Differenzen  muss diese Architektur durch eine geeignete Partitionierung der Daten unterstützt werden, damit sie ihren Vorteil voll entfalten kann: Die Daten müssen so angeordnet und aufgerufen werden, dass sie möglichst häufig dem Cache entnommen werden, sonst würde die begrenzte Datenübertragungsrate zum Hauptspeicher den Geschwindigkeitsgewinn vereiteln. Hierfür wird in \cite{RaL00a} eine räumliche und zeitliche Unterteilung der Berechnungen in Quader untersucht, vgl.~Bild \ref{Quad}. Dabei wird ausgenutzt, dass sich der Folgewert des berechneten Schallfeldes an einem Raumpunkt aus diesem selbst und nur den benachbarten Werten bestimmen lässt: Wenn man in einem kubischen oder quaderförmigen Volumen sämtliche Werte zu einem Zeitpunkt kennt, kann man daraus für den folgenden Zeitpunkt alle Werte bis auf die Ränder bestimmen; die Ausdehnung der Volumen verringert sich in jede Raumrichtung also jeweils um 2. Sobald in einer Raumrichtung keine Randwerte mehr für eine Berechnung vorliegen, sind die Berechnungen für dieses quaderförmige Volumen abgeschlossen. Die nächsten Quader können durch eine passende Wahl der Reihenfolge auf die vorausberechneten Werte der abgeschlossenen Quader zugreifen und so aufgebaut werden und so fort. Wird die Größe der Quader und Kuben nun so gewählt, dass sie jeweils  komplett im Cache untergebracht werden können, und so sich die  Hauptspeicherzugriffe auf den Auf- und Abbau reduzieren.  Die Anzahl der Speicherzugriffe reduziert sich auf $O(1/\sqrt[3]{n})$, wie in \cite{RaL00a} ausgeführt, wobei $n$ die Anzahl der Raumpunkte der Teilvolumen bei maximaler Ausdehnung ist.
\begin{afigure}[bt]
\begin{center}
\epsfig{file=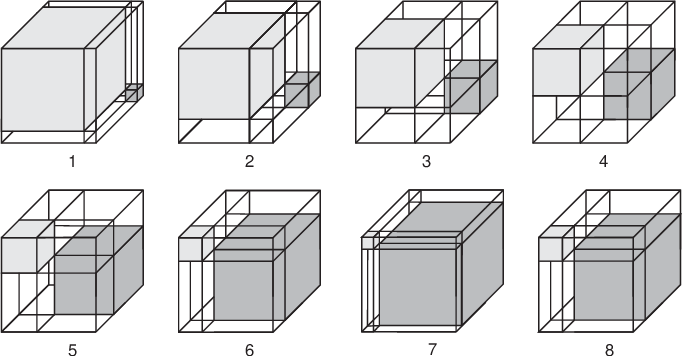}
\caption{Schema der Datenunterteilung in Kuben, hervorgehoben, und Quader in aufeinanderfolgenden Zeitschritten der Simulation. In den Abschnitten 1-7 ist die Abbauphase eines Kubus hellgrau hervorgehoben, dunkelgrau die nachfolgend berechnete Aufbauphase gefolgt von dem anscließenden Abbau in Abschnitt 8.
Die Zentren der Kuben sind ortsfest.}\label{Quad}
\end{center}
\end{afigure}

Dieses Prinzip der Unterteilung nutzt die Caches und die Prozessoren effizient und ist wohl auch als Datenstruktur für weitere Optimierungen wie in Abschnitt \ref{Vektor} geeignet. Für die hier betrachten Geometrien lässt das Verfahren zudem weiter vereinfachen, da die Daten mehrere tomographische Schichten vollständig in dem Level-2-Cache untergebracht werden können. Bei diesen ist der Rand in Schichtebene durch Randbedingungen abgeschlossen, so dass das Prinzip lediglicht senkrecht zu den Schichten angewendet werden braucht.

Die Struktur der Datenzugriffe des Programms sind auch auf modernen, asymmetrischen  \name{NUMA}-Architekturen,  ein Akronym von \name{Non Unified Memory Access},  vorteilhaft. Bei diesen Rechner\-architekturen ist zusätzlich die Schnittstelle zum Hauptspeicher auf dem Prozessorchip integriert, so dass sich die Datenübertragungsrate zum Hauptspeicher mit jedem Prozessorchip erhöht. Die Kommunikation zu anderen Prozessoren und daran angebundene Speicher erfolgt über dedizierte Schnittstellen. So nutzt die Software jetzt ein System mit acht \name{Opteron} Prozessoren, jeweils mit einer Taktrate von zwei Gigahertz jeweils zwei Gleitkommaoperationen durchführen können. Die gesamte Berechnung von Impulsantworten des Vokaltrakts dauert mit den Optimierungen typisch nur wenige Minuten. 

\subsection{Vektorisierung}\label{Vektor}
Ein weiterer Weg, die Rechenleistung zu steigern, liegt in der Verwendung eines Vektorrechners bzw.{} nach \cite{Fl72} einer \name{Single Instruction Multiple Data}{}-Architektur. Diese Architektur war lange die einzige Möglichkeit zu einer hohen Rechenleistung für Gleitkommazahlen \cite{Cr76}. In integrierten Prozessoren wird sie seit dem Pentium III ebenfalls unterstützt, jedoch in einer reduzierten Form auf Vektoren der Länge vier. Diese ist zudem auf eine sehr stringente Platzierung der Daten angewiesen. Eine Untersuchung in \cite{RaL00a} zeigte  eine Geschwindigkeitssteigerung um den Faktor zwei für die Verwendung dieser Erweiterung.

Eine deutlich höhere Steigerung ermöglicht der Einsatz moderner Grafikprozessoren. Zur Unterstützung von dreidimensionalen Darstellungen besitzen diese eine Vielzahl  parallel arbeitender Gleitkommaeinheiten. Diese werden ebenfalls über eine SIMD-Architektur programmiert und leisten über eine Billion Gleitkommaoperationen pro Sekunde. Der an diese Prozessoren angebundene Speicher besitzt eine hohe Bandbreite und hinreichende Größe für diese Applikation \cite{Ma08b, LiNOM08}, so dass sie sich hiermit weiter beschleunigen lässt.

\pagebreak

\section{Validierung}\label{Valid}
Eine Validierung der Rechnungen  hat hier mehrere Ziele. Ein Ziel ist das Aufdecken trivialer Fehler, bspw.~in der programmtechnischen Umsetzung. Ein weiteres Ziel ist die physikalische Überprüfung, hier insbesondere die zutreffende Modellierung der Randbedingungen. Letztlich kann die Validierung auch Fehlerquellen aufzeigen und lässt eine Beurteilung der Aussagekraft der Simulationen zu. 
Zur Validierung wurde ein geschlossenes akustisches System gewählt, in das definiert Schall eingekoppelt wird und in dem an einer bestimmten Stelle der Schalldruckverlauf erfasst wird. Die folgenden Unterabschnitte geben einen Überblick über die verwendeten Methoden und Resultate.
 
\subsection{Schallwandler}\label{Wandler}
Die Umwandlung von Schall in elektrische Signale kann durch elektrodynamische oder elektrostatische Mikrofone erfolgen. Elektrostatische Mikrofone zeichnen sich durch eine einfachere Bauform aus und sind daraus resultierend mit höherer Präzision zu fertigen. Solche Mikrofone wurden auch für die hier vorgenommenen Messungen verwendet, da sie kleinere Abmessungen besitzen, wodurch sie sich besser an den hier untersuchten Geometrien anordnen lassen. Der Hersteller \name{Bruel \& Kj\ae r} hat zudem in \cite{BK95} ihre akustische Rückwirkung untersucht, wodurch sich die Randbedingungen im Bereich der Mikrofone definieren lassen. 

An eine zur Anregung von akustischen Systemen verwendete Schallquelle werden bei einer quantitativen Messung besondere Anforderungen gestellt. Meist werden elektroakustische Wandler eingesetzt, da diese sehr weit entwickelt sind und diesen Anforderungen sehr nahe kommen: Zunächst muss das Übertragungsverhalten von elektrischen Signalen zu akustischen Signalen bekannt sein, ebenso ihre akustische Rückwirkung. Desweiteren sind hohe Linearität und hohe Schallpegel vorteilhaft.

Übliche Lautsprecher wurden aufgrund ihrer Größe nicht in Erwägung gezogen, da ihr Durchmesser zur Erzeugung niedriger Frequenzen meist über 5~cm liegt. Hier lässt sich zwar der Größenunterschied durch einen Adapter kompensieren, dessen akustische Eigenschaften sind aber nur schlecht zu bestimmen. Die Verwendung von miniaturisierten elektrodynamischen Schallwandlern aus Kopfhörern, welche in der Ohrmuschel platziert werden, erbrachten keine befriedigenden Resultate.

Die Zielsetzung konnte unter inversem Betrieb eines weiteren Mikrofons als Schallquelle erreicht werden. Hierzu wird die Elektrode mit einer dem Signal proportionalen Spannung von typisch über 100 Volt betrieben, welche von einer Gleichspannung von 200 Volt überlagert wird. Bauartbedingt gibt diese Schallquelle jedoch insbesondere bei niedrigen Frequenzen nur geringe Schallleistungen ab.
 
\begin{afigure}[ht]
\begin{center}
\epsfig{width=.85\linewidth,file=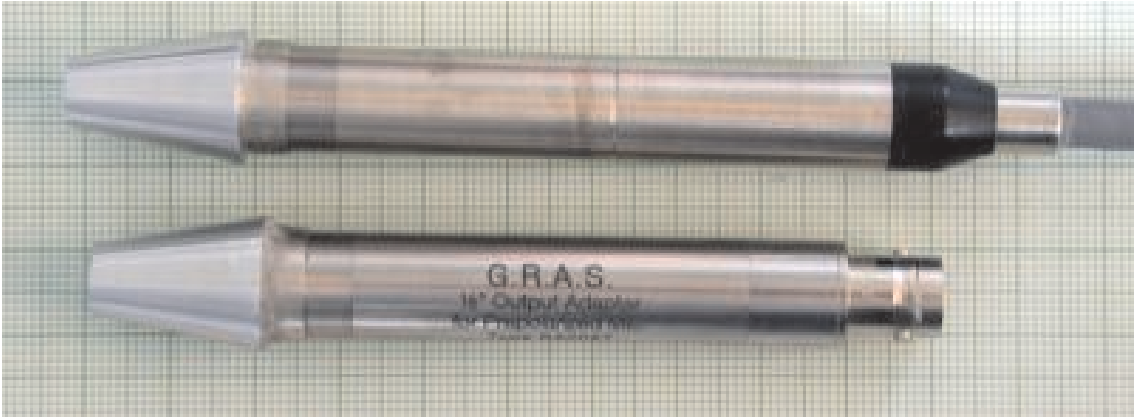}
\caption{Schallwandler: oben Mikrofon mit angeschraubten Vorverstärker, unten Signalgeber. Beiden sind Konen aufgesetzt, um eine schalldichte Schalleinleitung in den Nasaltrakt zu gewährleisten.}\label{Schallwandler}
\end{center}
\end{afigure}%

\subsection{PC-basierte Signalgenerierung und -erfassung}\label{SignalErzeugungErfassung}
Aufgrund der geringen Schallleistungen der Schallquelle ergeben sich niedrige Schall- und Signalpegel an der Mikrofonkapsel,  die das thermische Eigenrauschen des Mikrofonvorverstärkers nicht völlig überdecken. Dieser störende Effekt wird durch Mitteln über eine wiederholt durchgeführte Messung reduziert. Die Rauschleistung reduziert sich hierbei reziprok zur Anzahl der Messungen, während die Signalleistung konstant bleibt. Der Signalrauschabstand, definiert durch den Quotienten der Leistungen, wächst folglich proportional zu der Anzahl der Messungen. Diese Mittelung reduziert zudem auch Störungen durch andere, nicht korrelierte Quellen und liefert so ein im Frequenzbereich von 1 - 10 kHz störarmes Signal.

Zur automatisierten Durchführung der Messung, deren Mittelung und Auswertung mittels FFT wird ein Personal Computer genutzt, bei dem  die Signalausgabe und -erfassung durch ein handelsübliche Audiointerface, \name{Emu10k}, erfolgt. Für die Mittelung  ist eine reproduzierbare Anregung und Messung erforderlich, insbesondere darf kein zufälliger Unterschied, weder in Abtastrate noch Phase, zwischen Ausgabe und Aufzeichnung auftreten. Während die Abtastraten durch das Prinzip der Audiointerfaceschaltung sich von dem gleichen Taktgenerator ableiten und diese Bedingung erfüllt ist, zeigt sich, dass Phasenunterschiede auftreten. Diese resultieren nicht aus einer ungleichmäßigen Reaktionszeit des Betriebssystems \name{Linux}, sondern aus einem nicht an diese Anwendung angepassten Treiber. Die im Betriebssystem enthaltenen Treiber bewirken einen kleinen, aber variablen Zeitversatz zwischen Start der Aufnahme und Wiedergabe. Der Zeitversatz lässt sich durch eine Modifikation des Treibers beheben, wie in \cite{RaL01} gezeigt, bei dem die Startzeitpunkte unter Rückgriff auf die Zeitbasis des Aduiointerfaces synchronisiert werden. Alternativ wird das ausgegebene Signal auf einen Eingang der Soundkarte zurückgeführt, so dass das aufgezeichnete Zweikanalsignal Anregungssignal und Systemantwort enthält. 

\subsection{Versuchsaufbau}
Um die Ergebnisse der Simulation mit denen einer Messung vergleichen zu können, wird zur Untersuchung eine Hohlraumgeometrie gewählt, deren Eigenschaften auch analytisch bestimmbar sind. Am einfachsten lässt sich dies durch einen zylinderförmigen Hohlraum, gebildet durch ein Messingrohr, realisieren. Volumen und Länge werden so gewählt, dass sie dem Nasaltrakt entsprechen. In das Messingrohr werden Schallgeber und Mikrofon eingeschoben.

Als Anregungssignale wurden frequenzmodulierte sinusförmige Signale verwendet, bei denen die Frequenz exponentiell mit der Zeit erhöht wurde, wodurch das Signal  bei den problematischeren tiefen Frequenzen mehr Energie enthält. Von der Soundkarte ausgehend wurden die Signale mittels eines Verstärkers \name{G.R.A.S.~14AA} auf die in Abschnitt \ref{Wandler} genannten Pegel verstärkt. Das Signal des Verstärkers wird mittels des in Bild \ref{Schallwandler} gezeigten Adapters \name{G.R.A.S.~RA0067} an der 1/2-Zoll Mikrofonkapsel \name{Bruel\&Kj\ae r BK4134} angelegt. Die zweite Mikrofonkapsel des gleichen Typs diente zusammen mit dem Vorverstärker \name{BK5678} und dem Pegelverstärker \name{BK2610} zur Schallerfassung; deren Signale werden in den Line-Eingang des Audiointerfaces zurückgeführt. 

\begin{afigure}[b]
\begin{center}
\psfrag{[dB]}[r][r]{\footnotesize [dB]}
\psfrag{A}[t][t]{\footnotesize $f/\text{[kHz]}$}
\psfrag{0}[r][r]{\footnotesize 0}
\psfrag{10}[r][r]{\footnotesize 10}
\psfrag{20}[r][r]{\footnotesize 20}
\psfrag{30}[r][r]{\footnotesize 30}
\psfrag{40}[r][r]{\footnotesize 40}
\psfrag{50}[r][r]{\footnotesize 50}
\psfrag{1}[t][t]{\footnotesize 1}
\psfrag{2}[t][t]{\footnotesize 2}
\psfrag{3}[t][t]{\footnotesize 3}
\psfrag{4}[t][t]{\footnotesize 4}
\psfrag{5}[t][t]{\footnotesize 5}
\psfrag{6}[t][t]{\footnotesize 6}
\psfrag{7}[t][t]{\footnotesize 7}
\psfrag{8}[t][t]{\footnotesize 8}
\epsfig{width=.7\linewidth,file=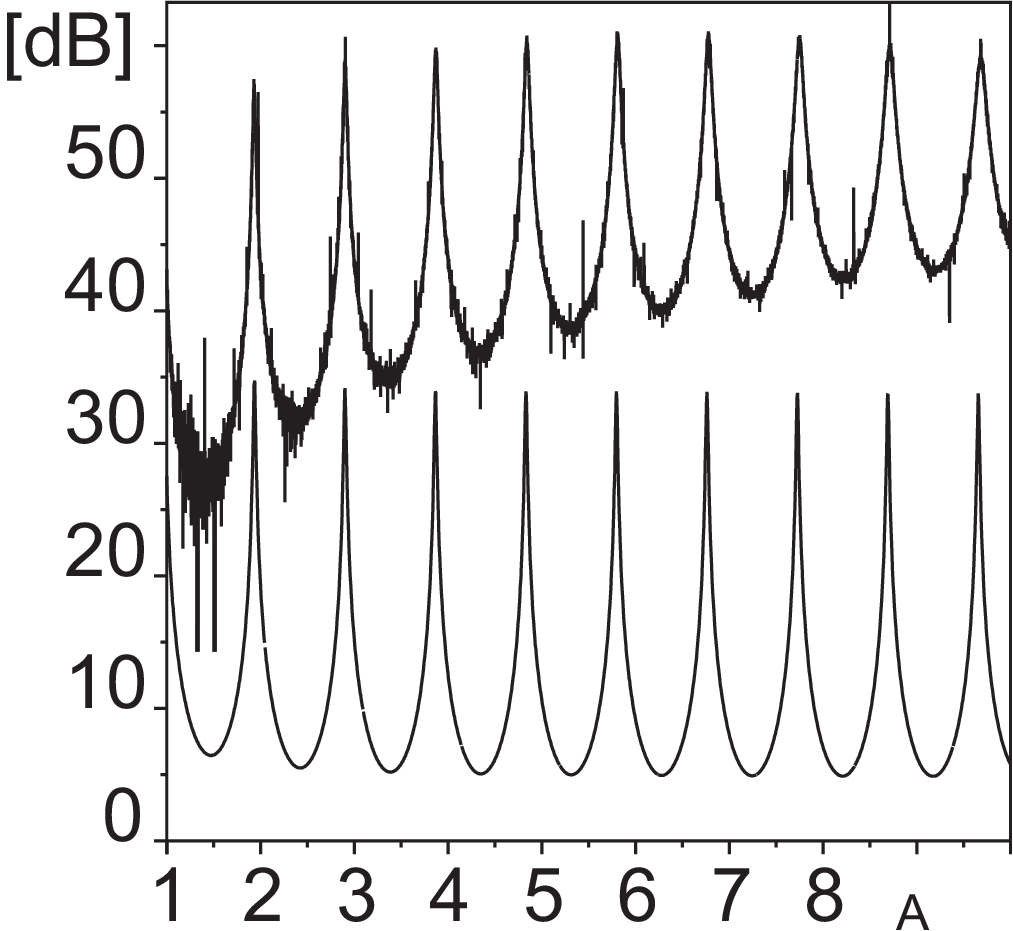 ,height=.4\linewidth}\hspace{1cm}
\caption{Vergleich des Betragsganges einer Messung, oben, mit dem Betragsgang einer Simulation, unten, der Schallausbreitung in einem Rohr mit einer dem Nasaltrakt ähnlichen Proportion. Die übereinstimmenden Resonanzfrequenzen werden deutlich. Durch den Verzicht auf Kompensation verbleiben in der Messung die zu niedrigen Frequenzen hin abfallende Schallleistung der unterhalb ihrer Eigenresonanzfrequenz betriebenen elektrostatischen Schallquelle und deren zu höheren Frequenzen zunehmende Dämpfung. }\label{Simulation_Messung_Rohr} 
\end{center}
\end{afigure}%

Für die Simulation wird der Querschnitt des Rohrs in einem Bildbearbeitungsprogramm gezeichnet und diese Daten in die Simulationsumgebung hineingeladen. Die Anregung erfolgt mit einem Dirac-Impuls. In Bild \ref{Simulation_Messung_Rohr} sind die Ergebnisse von Simulation und Messung gezeigt, wobei der Betragsgang der Simulation durch eine Division der diskreten Fouriertransformation des gemessenen Signals durch diejenige des Anregungssignals und einer anschließenden Betragsbildung ermittelt wird; auf weitere Kalibrierungen wird verzichtet. Gut zu erkennen ist die übereinstimmende Lage der Resonanzfrequenzen. 

Versuchsweise erfolgt eine ähnliche Messung an der Nase eines Probanden durchgeführt, wobei der Schall in ein Nasenloch eingekoppelt und an dem anderen Nasenloch erfasst wird. Hierfür sind die Schallwandler mit Konen ausgestattet, wie in Bild \ref{Wandler} zu sehen ist. Jedoch zeigen die Frequenzgänge nur eine bereichsweise Übereinstimmung mit einer entsprechenden Simulation \cite{RaL02a, RaL02b}, was sicherlich zum Teil an den Unterschieden des Nasaltrakts des Probanden zu dem der  der Probandin aus der CT-Untersuchung liegtg, möglicherweise aber auch an einem nicht vollständig geschlossenen Gaumensegel während der akustischen Messung. Naturgemäß lassen sich auch Unvollkommenheiten in dem Datensatz selbst nicht ausschließen.

\pagebreak
\part{SPEAK}\label{SPEAK}
Wie bereits eingangs aufgezeigt, ist im Bereich der Mundhöhle der Aufbau des Vokaltrakts  zwar anatomisch komplexer; er besitzt jedoch eine einfachere Geometrie. Daher sollten die Untersuchungsmethoden durch finite Approximationen der dreidimensionalen Schallwellenausbreitung auch für diesen Bereich geeignet sein. Da sich die Geometrie des Vokaltrakts beim Sprechen schnell ändert, sind jedoch die vorgenannten tomographischen Verfahren wenig geeignet, um die Geometrie zu erfassen.

Eine gute Alternative besteht darin, die räumliche Konfiguration des Sprechtraktes aus den akustischen Eigenschaften des Sprachsignals zu schätzen. Dabei werden Verfahren wie in Abschnitt \ref{TheoLPC} und \ref{SPEAK_Rohr} beschrieben eingesetzt, die treffende Querschnittsverläufe ergeben, wie zahlreiche Untersuchungen beispielsweise in \cite{MaG72, La05, Sc09} belegen. Bis dato ist eine Reihe von Algorithmen entwickelt und implementiert worden, um Signale mit dieser Zielsetzung zu analysieren. Diese Programme verfügen jedoch über jeweils eigene Schnittstellen, sind in unterschiedlichen Programmiersprachen und für unterschiedliche Betriebssysteme verfasst, und unterscheiden sich zudem in ihrem Bedienkonzept. Um die daraus resultierenden Umstände zu vermeiden, wurden wichtige Verfahren in dem Programm \glqq SPEAK\grqq, ein Akronym von Sprechakustik, verbessert und zusammengefasst. Durch die einheitliche und vereinfachte Bedienung sowie die Möglichkeit, an vielen Stellen interaktiv einzugreifen, eignet sich dieses Programm auch in der Didaktik der Akustik und Phonetik.\footnote{Einige Facetten dieser Anwendung werden anhand des Vorgängers {\sc TubeDesigner} in \cite{RaL03c} betrachtet.}
Die folgenden beiden Abschnitte zeigen die Möglichkeiten dieses Programms auf und geben Beispiele für deren Anwendung.

\section{Analyse und Visualisierung}\label{Speak-Quelle}
Im Folgenden wird ein Überblick über die wichtigsten Analysefunktionen und die korrespondieren Visualisierungsmöglichkeiten gegeben. Hervorzuheben ist hierbei, dass diese Analysefunktionen -- soweit sinnvoll -- sowohl analytisch anhand der Polynome der Übertragungsfunktionen als auch numerisch auf äquidistant zeitdiskretisierten Folgen arbeiten. Auch eine Kombination ist möglich, wie in dem folgenden Abschnitt \ref{SPEAK_Rohr} gezeigt.
Für die numerische Untersuchung sind die Signalquellen 
\begin{itemize}
\item{weißes Rauschen,}
\item{periodische Pulse mit wählbarer Frequenz,}
\item{ein parametrierbares, typisches Glottissignal,}
\item{wahlweise periodische und abtastratenkonvertierte gespeicherte Folge,}
\item{mit einem Mikrofon erfasster Schall} 
\end{itemize}
vorgesehen.

\subsection{Zeit- und Frequenzbereich}
Im Zeitbereich kann der Signalverlauf und für Filter die Impulsantwort angezeigt werden; im Frequenzbereich werden neben des häufig benötigten Betragsgang und der Gruppenlaufzeit auch modellbasierte Analysen verwendet. 

Durch eine Fast-Fourier-Transformation mit vorhergehender Fenstergewichtung können Sprachsignale im Frequenzbereich analysiert werden; zur Verfügung stehen Rechteck-, Dreieck-, Hamming-Fenster. Mit diesen Fensterfunktionen wird dabei zum einen der betrachtete zeitliche Abschnitt des Sprachsignals festgelegt. Zum anderen mildern die beiden letztgenannten Fensterfunktionen Auswirkungen der Abweichung zwischen wirklichen Sprachsignalen, vgl.{} Abschnitt \ref{aktEig}, und der der Fourier-Transformation zugrunde liegende hypothetische Periodizität der Signale, wie es sich durch Anwenden der verschiedenen Fenster in SPEAK unmittelbar zeigt. Ebenso ist die Analyse von Filterstrukturen möglich, hierbei kann zudem als Anregungssignal ein unkorreliertes Rauschen verwendet werden und die Darstellung des durch das Filter hervorgerufenen Betragsgang durch eine Mittelung in der Varianz vermindert werden, wie in Abschnitt \ref{SignalErzeugungErfassung} ausgeführt.

Für Sprachsignale  sind im Besonderen  die in SPEAK integrierten modellbasierten Analyseverfahren geeignet. Bei diesem wird durch die in den Abschnitten \ref{DigitalSynthese} und \ref{Parameter} behandelten  Maximum-Likelihood- oder Maximum-Entropy-Analyse ein Pol-Modell des zugrundeliegenden Prozesses parametrisiert. Die Lage der Pole wir durch eine Nullstellenbestimmung des reziproken System nach \cite{Mu56} iterativ ermittelt und dargestellt. Wie in Bild \ref{Rohr_i_fft_z} zu erkennen ist, zeigt die Poldarstellung  ein deutlich klareres Bild als die überlagerte Projektion der Pole auf den Einheitskreis, dem Betragsgang im Frequenzbereich. In Abschnitt \ref{ZT} ist die zugrundeliegende mathematische Beschreibung skizziert und in  Abschnitt \ref{Z-Ebene} wird auf die in SPEAK hierauf aufbauenden Möglichkeiten eingegangen. 

\subsection{FIR-Filter}\label{FIR-Filter}
Filter endlicher Impulsantwort oder kurz FIR-Filter werden üblicherweise in der Form $y_k=\sum_mb_mx_{k-m}$ durch $b_m$ parametrisiert, wobei $m$ entsprechend auf einen endlichen Bereich beschränkt ist; $x_k$ ist hierbei die Eingangsfolge und $y_k$ die gefilterte Ausgangsfolge.
Das besondere Merkmal dieser Filterform ist, dass jeder Koeffizient einem Wert der Impulsantwort entspricht, wodurch es möglich ist, direkt eine endliche Impulsantwort vorzugeben.

Aufgrund dieser Eigenschaft ist es beispielsweise möglich, die in den Untersuchungen aus Teil II ermittelten Impulsantworten für Transmittanz und Reflektanz des Nasaltraktes nach einer Abtastratenkonvertierung vollständig zu übernehmen  und über einen Dreitoradapter an ein Rohrmodell des übrigen Vokaltrakts anzukoppeln. Die in Abschnitt \ref{TheoBaumRohr} diskutierten Abweichungen aufgrund einer vereinfachten Nachbildung durch Rohrsysteme wird vermieden.

Desweiteren können diese Filter wahlweise im Zeit- und Frequenzbereich manipuliert werden, wobei der jeweils andere Bereich durch eine Fast-Fourier-Transformation aktualisiert wird. Bild \ref{SPEAK-FIR} illustriert diese Darstellung.  

\noindent
\begin{afigure}[ht]
\begin{center}
\epsfig{file=FIR_col\arxiv{2}.eps, width=6cm}
\end{center}
\caption{FIR-Filter in SPEAK. In der linken Fensterhälfte ist die Zeitbereichs-, in der rechten Hälfte die Frequenzbereichsdarstellung angeordnet; jeweils oben Real- und unten Imaginäranteil. In der interaktiven Darstellung kann jeder der vier Quadranten kann mit dem Maus-Zeiger manipuliert werden, wobei sich die Auswirkung in den anderen Quadranten unverzögert zeigt. Eine Beschränkung auf reelwertige Filter, wie sie in diesem Bild gezeigt ist, ist mittels Tastenfeld ($\mathbb{R}$) möglich.}\label{SPEAK-FIR}
\end{afigure}

\subsection{${\cal Z}$-Ebene}\label{Z-Ebene} 
Die Darstellung des Übertragungsverhaltens in der ${\cal Z}$-Ebene ist, wie bereits angedeutet, eine über die Darstellung im Frequenzbereich hinausgehende Ansicht. Hier ist es möglich, die charakteristischen Eigenwerte des Filters zu zeigen, aus welchen Frequenzgang, Phasengang und Gruppenlaufzeit resultieren. Die Eigenwerte werden hierfür aus den Nullstellen des Nenner- und Zählerpolynoms des Filters mittels des bereits genannten  Muller-Verfahrens in \cite{Mu56} bestimmt sofern diese nicht bereits in einer Produktform vorliegen, und durch die üblichen Symbole, Kreuze für Pole bzw.{} Kreise für Nullstellen, in der Ebene dargestellt. Um den Zusammenhang zwischen Pol- und Nullstellenschema in der ${\cal Z}$-Ebene und dem Betragsgang auf der Frequenzachse aufzuzeigen, kann die dreidimensionale Darstellung der Betragsfunktion auf der ${\cal Z}$-Ebene genutzt werden, wie in Bild \ref{Hz3D} gezeigt.

Umgekehrt ist es möglich, Filter durch Positionierung von Polen und Nullstellen in der ${\cal Z}$-Ebene zu definieren. Die Position wird mit der Maus festgelegt oder verändert, wobei die Darstellungen im Zeit- oder Frequenzbereich praktisch unverzögert folgen. Für die interne Repräsentation der Filter wird eine verkette Form elementarer Filter erster Ordnung für reelwertige Extremstellen und Filter 2.{} Ordnung für konjugiert komplexe Extremstellen verwendet. Die ergänzend Erzeugung einer 
dreidimensionale Ansicht der Betragsfunktion in der ${\cal Z}$-Ebene ist dadurch besonders einfach möglich.
\noindent
\begin{afigure}[ht]
\begin{center}
\epsfig{file=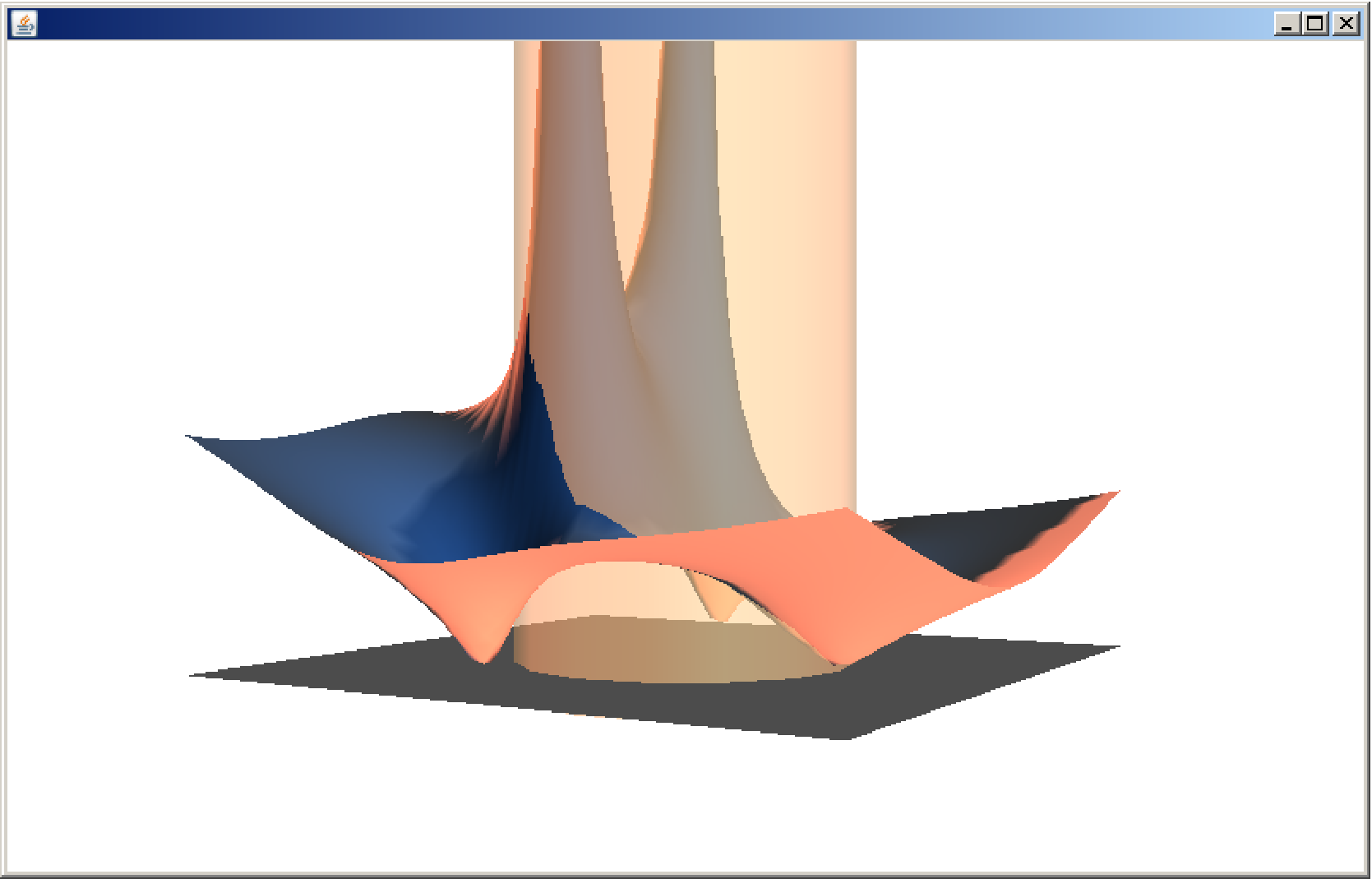, width=9cm}
\end{center}
\caption{Dreidimensionale Darstellung der Betragsfunktion eines Filters mit zwei jeweils konjugiert komplexen Polen und Nullstellen, links, und einer reellen Nullstelle, rechts, auf der ${\cal Z}$-Ebene. Der Einheitskreis ist als transparenter Zylinder dargestellt, so dass der Schnitt der Betragsfunktion mit dem Einheitskreis, der Betragsgang, deutlich wird.}\label{Hz3D}
\end{afigure}

\subsection{Rohrmodell}\label{SPEAK_Rohr}
\begin{afigure}[ht]
\begin{center}
\epsfig{file=Rohr_i_ff\arxiv{t_z}.eps, width=\linewidth}
\end{center}
\caption{Darstellung eines Rohrquerschnittverlaufs ähnlich des Vokaltraktes, bei dem links die Glottis sitzt und rechts die Mundabstrahlung durch einen ausgeprägten Querschnittssprung realisiert ist. Darunter die Impulsantwort des entsprechenden Kreuzgliedkettenfilters, rechts daneben Gruppenlaufzeit, darüber Frequenzgang und Darstellung der Pole in der Übertragungsfunktion auf der ${\cal Z}$-Ebene.}\label{Rohr_i_fft_z}
\end{afigure}
Die Analyse von Rohrmodellen des Sprechtraktes kann anhand von manuell vorgegebenen Querschnittsverläufen erfolgen, wie in Bild \ref{Rohr_i_fft_z} dargestellt. Hierbei kann mit der Maus der Querschnitt jedes einzelnen Segmentes verändert und die Auswirkung auf bspw. Impulsantwort, Frequenzgang und Position der Pole betrachtet werden. Ebenso leicht ist es möglich, den Querschnittsverlauf in einem Bereich abzuändern, indem man diesen mit der Maus vorgibt. Anhand des Querschnittsverlaufs kann wie in Abschnitt \ref{Rohr} das Übertragungsverhalten des Rohrsystems analytisch bestimmt werden. Die so erhaltene Funktion kann zudem in ihre Eigenwerte zerlegt werden, welche Pole in der ${\cal Z}$-Ebene bilden. Man erkennt auf diese Weise, wie sich die Pole und damit das Resonanzverhalten in Abhängigkeit dieser Bewegungen --- teilweise überraschend --- verschieben, aufteilen oder zusammenfallen. Man kann bspw.{} ersehen, welche Schwierigkeiten die frühen Formantensynthesizer (vgl.~Abs.~\ref{Formant}) gehabt haben müssen, dies nachzubilden.

Alternativ besteht die Möglichkeit, den Vokaltraktverlauf anhand von gesprochener Sprache zu ermitteln. Das kann unmittelbar durch ein an den Rechner angeschlossenes Mikrofon erfolgen, oder über digital aufgezeichnete Sprachproben. In beiden Fällen lassen sich Preemphasen schätzen oder vorgeben, um die spektrale Färbung durch das  Abstrahlverhalten des Mundes und durch die Anregung aus der Glottis zu berücksichtigen. Anhand des bereinigten Signals können mittels des Burg-Algorithmus oder eines neueren, hierfür entwickelten Verfahrens die Reflexionskoeffizienten des Rohrmodells geschätzt und daraus der Querschnittsverlauf bestimmt werden. Die Darstellung des Querschnittsverlaufs erfolgt unmittelbar und ohne wahrnehmbare Verzögerung, so dass man beim \glqq Hineinsprechen\grqq~seine eigene Artikulation beobachten und studieren kann. Das in \cite{ScRL04} eingehender gezeigte, neuere Verfahren berechnet dabei den Querschnittsverlauf unter Beachtung der Schallabstrahlung nach Laine und einer uniformen Schalldämpfung.

\newpage
\section{Besonderheiten der Implementierung}
Im Folgendem werden einige Aspekte der Implementierung betrachtet, die sich als nützlich für die umfangreichen Funktionen des Programms herausgestellt haben. Diese Aspekte betreffen verschiedene Ebenen, angefangen bei der Systemtopologie, die den grundlegeneden Datenfluss und darauf einwirkende Verknüpfungselemente umfasst, bis hin zur Auswahl einer Programmiersprache, die die Implementierung praktikabel macht. 

\subsection{Systemtopologie}
Wie in dem vorangegangenen Abschnitt gezeigt, können mit SPEAK eine Vielzahl von Untersuchungen durchgeführt werden. Um dies nicht auf einzelne Systeme zu beschränken, hat SPEAK sechs Elemente, die über zwei Klassen von Verbindungen miteinander kombiniert werden können: 
\begin{itemize}
\item{Signalquellen}
\item{Analysewerkzeuge --- Signalsenken}
\item{Lautsprechersymbol --- Signalsenken}
\item{Filter}
\item{Rohre --- Kreuzgliedkettenfilter}
\item{Knotenpunkt --- Mehrtoradaptoren}
\end{itemize}
Während Filter und die meisten Signalquellen und -senken über einen gerichteten Signalfluss miteinander verbindbar sind, erfordern Rohrsysteme und Mehrtoradaptoren Verbindungen, die hin- und rücklaufende Größen beinhalten, wie in Abschnitt \ref{Rohr} ausgeführt. Damit diesen Wellengrößen eine physikalische Bedeutung zugemessen werden kann, enthält die bidirektionale Verbindung zudem Informationen über den Querschnitt des zugehörigen Rohrsegmentes,  aus denen an Mehrtoradaptoren Reflektionen und Transmissionen berechnet werden. 

An jeden Ausgang eines Elements können zudem ein oder mehrere Signalsenken oder Filter angeschlossen werden. Erstere, insbesondere die Analysewerkzeuge, erlauben die Betrachtung einer oder mehrerer Signaleigenschaften über ein komplexeres System hinweg. Mit letzteren besteht beispielweise die Möglichkeit, komplexere Abschlussbedingungen an einem Rohr oder Rohrsystem nachzubilden, wenn der Ausgang des Filters wieder zu dem Rohrsystem zurückgeführt wird. Ein anderes Beispiel ist ein Preemphasefilter zwischen Signalquelle und Rohrbeginn oder Rohrende und Signalsenke, um realistische Signalverhältnisse zu erzielen.

\noindent
\begin{afigure}[ht]
\begin{center}
\epsfig{file=NasalSimp\arxiv{le}.eps, width=\linewidth}
\end{center}
\caption{Darstellung eines Rohrquerschnittverlaufs ähnlich des Vokaltraktes, bei dem links die Glottis sitzt und rechts die Mundabstrahlung durch einen ausgeprägten Querschnittssprung realisiert ist. Darunter die Impulsantwort des entsprechenden Kreuzgliedkettenfilters, rechts daneben Gruppenlaufzeit, darüber Frequenzgang und Darstellung der Pole in der Übertragungsfunktion auf der $z$-Ebene.}\label{Rohr_Vokal-Nasal-Synth}
\end{afigure}

\subsection{Synthese}
Eine der für die Sprechakustik wichtigsten Funktionen ist die Signalsynthese und Wiedergabe, die dieses Programm fortwährend und mit vernachlässigbarer Verzögerung realisieren kann. Damit kann nicht nur ein visuelles Feedback über die genannte mehr oder weniger abstrakte Darstellung gegeben werden, sondern es wird auch eine direkt auditiv wahrnehmbare Repräsentation gegeben. Das gehörte Synthesesignal beinhaltet eine Vielzahl von Informationen, die von dem menschlichen Gehirn auf unterschiedlichen Ebenen ausgewertet werden. Gerade der letztlich nur subjektiv mögliche Vergleich mit einer Vielzahl von gehörten Sprachäußerungen erlaubt so eine Beurteilung der Natürlichkeit. Des weiteren werden auch Zusammenhänge von Phänomenen in spektraler oder zeitlicher Darstellung, ihre Wirkung und ihre perzeptive Bedeutung nachvollziehbar.

\subsection{Java}
Ein wesentlicher Schritt zu der erfolgreichen und effizienten Realisierung eines Programms ist die Wahl der passenden Werkzeuge. Wie eingangs erwähnt waren einige der Algorithmen bereits anderweitig implementiert; ein gutes Beispiel ist in \cite{Ra99} der \name{TubeDesigner} in \name{C++}. Bei dieser ursprünglichen Implementierung erweisen sich insbesondere Datenstrukturen in der unter C++ üblichen \name{Standard Template Library} in der Implementierung zeitaufwendig. Darüber hinaus sind sie, wie der Name bereits andeutet, als Formvorlage der Sprache hinzugefügt und kein eigentliches Sprachmittel. Dies erschwert das sogenannte \name{Refactoring}, eine semiautomatische Restrukturierung des Programms, was im Laufe der Entwicklung einer umfangreichen Software häufig erforderlich ist. Auch die Programmiersprache \name{C}, in der eine Reihe umfangreicherer Projekte realisiert wurden, wird aus diesem Gründen nicht in Erwägung gezogen, zudem unterstützt sie eine zeitgemäße objektorientierte Programmierung nicht.  Andere Sprachen, wie etwa \name{Delphi}, scheiden aufgrund ihrer geringen Verbreitung aus.

Die gewählte Programmiersprache \name{Java} besitzt die die genannten Einschränkung nicht. Insbesondere im Zusammenspiel mit der Entwicklungsumgebung \name{Eclipse} zeigt sich, dass es problemlos möglich ist, ein Programm mit etwa 30.000 Zeilen Quelltext zu handhaben. Die Einschränkung von Java, keine Definitionen von Operatoren für komplexe Zahlen zu beinhalten oder für neue Datentypen zu erlauben, fiel nicht allzu sehr ins Gewicht, da die mathematischen Ausdrücke der Algorithmen nur einen geringen Anteil am Quelltext haben. Ebenso erweist sich die Befürchtung über ein Java inhärentes Merkmal letztlich als unbegründet, dass die häufige Instanziierung von Daten- bzw.{} Objektstrukturen oder der \name{Garbage-Collector}, der deren Speicherplatz nach Gebrauch wieder freigibt, das Programm verlangsamt. Das  Laufzeitverhalten ist durch passend gewählte und optimierte Algorithmen nicht beeinträchtigt. 

Als besonders leistungsfähig erweisen sich die Grafikfunktionen von Java, welche durch durchgängiges \name{Double Buffering}, aber auch durch passende Schnittstellen die Implementierung der dynamischen Benutzerschnittstellen vereinfachten. Ebenso leicht ist die dreidimensionale Darstellung der $\cal{Z}$-Ebene möglich. Wichtig für die Struktur des Programms ist zudem das \name{Visitor}-Schema\cite{Gamma}, welches eine Separation nach \cite{Re78} von Visualisierung und Berechnung ermöglicht; dies lässt sich mit den \name{Inner Classes} von Java umsetzen. Letztlich erlaubt die gewählte Klassenstruktur durch Vererbung eine schnelle und einfache Erweiterung, falls für andere Anwendungen weitere Funktionen benötigt werden.

\pagebreak
\addtocontents{toc}{\protect\newpage}
\part{Akustik des Vokaltraks}
Man kann mit Finite-Differenzen auch die Akustik des Mundhöhle und des Rachens vorteilhaft simulieren, wie die nachfolgenden Abschnitte zeigen. Es zeigt sich das Rohrmodel und die Finiten-Differenzen bei kleinen Querschnittflächen hier perfekt übereinstimmen, während mit zunehmenden Querschnitt mit den Finiten Differenzen Effekte erfasst werden, die  die laterale Wellenausbreitung beeinflussen und bei einer natürlichen Querschnittskontur dann Quermoden hervorrufen. Abschließend werden vereinfachte Modelle gezeigt, die diese Effekte treffend berücksichtigen.

\section{Äquivalenz zwischen\\ Finite-Differenzen und Kreuzgliedkettenfilter}\label{Äquivalenz}
Eine Identität von Finite-Differenzen und Kreuzgliedketten-Filter in der Beschreibung der Schallausbreitung lässt sich nur bei einer eindimensionalen Betrachtung zeigen, da einerseits Kreuzgliedkettenfilter nur die Schallausbreitung entlang einer Raumrichtung beschreiben, andererseits Finite-Differenzen zur Beschreibung einer mehrdimensionalen Schallausbreitung Dispersion aufweisen würden.

Die Identität für den eindimensionalen Fall lässt sich analytisch gewinnen. Hierfür eignet sich ein übersichtliches, aber nicht triviales akustisches Rohr mit drei äquidistanten Querschnittssprüngen, quantifiziert durch die Reflexionsfaktoren $r_1$, $r_2$, $r_3$, vgl.{} Bild~\ref{Rohr3R}. Mit der in den Abschnitten \ref{WelleEle} -- \ref{TheoLinRohr} gezeigten Betriebsketten\-matrixgleichung kann man die Übertragungsfunktion bestimmten.  In Druckdarstellung ergibt sich die Betriebskettenmatrix $\mat{T}$ zu
\begin{align*}
\mat{T} = {}& \begin{pmatrix} z^{-1} & 0 \\ 0 & z \end{pmatrix}
\frac{1}{1+r_1}\begin{pmatrix}1&r_1\\r_1&1\end{pmatrix} 
\begin{pmatrix} z^{-1} & 0 \\ 0 & z \end{pmatrix}
\frac{1}{1+r_2}\begin{pmatrix}1&r_2\\r_2&1\end{pmatrix}
\begin{pmatrix} z^{-1} & 0 \\ 0 & z \end{pmatrix}\\
{} & \cdot \;
\frac{1}{1+r_3}\begin{pmatrix}1&r_3\\r_3&1\end{pmatrix}
\begin{pmatrix} z^{-1} & 0 \\ 0 & z \end{pmatrix}  \\
= {}& \frac{1}{(1+r_1)(1+r_2)(1+r_3)}\\ 
{} &\phantom{mm}\begin{pmatrix}
r_1r_3+(r_1r_2+r_2r_3)z^{-2}+z^{-4} & r_1z^{2}+(r_2+r_1r_2r_3)+r_3z^{-2}\\ 
r_3z^{2}+(r_1r_2r_3+r_2)+r_1z^{-2} & z^{4}+(r_1r_2+r_2r_3)z^{2}+r_1r_3  \end{pmatrix}\;.
\end{align*}
Die Übertragungsfunktion $H(z)$ ist dabei der Kehrwert des Matrixelements $t_{22}$. Erweitert mit $z^{-4}$ gewinnt sie die übliche Form:
\[H(z)=\frac{(1+r_1)(1+r_2)(1+r_3)\,z^{-4}}{1+(r_1r_2+r_2r_3)\,z^{-2}+r_1r_3\,z^{-4}}\;.
\]

\begin{afigure}[t]
\begin{center}
\begin{picture}(20,9)
\thicklines
\put(1,4){\line(1,0){6}}
\put(1,6){\line(1,0){6}}
\put(7,4){\line(0,-1){2}}
\put(7,6){\line(0,1){2}}
\put(7,2){\line(1,0){3}}
\put(7,8){\line(1,0){3}}
\put(10,2){\line(0,1){1.5}}
\put(10,8){\line(0,-1){1.5}}
\put(10,3.5){\line(1,0){3}}
\put(10,6.5){\line(1,0){3}}
\put(13,3.5){\line(0,-1){.5}}
\put(13,6.5){\line(0,1){.5}}
\put(13,3){\line(1,0){6}}
\put(13,7){\line(1,0){6}}
\thinlines
\put(4,5){\line(0,1){3.5}}
\put(16,5){\line(0,1){3.5}}
\put(3.65,8.7){$X$}
\put(15.75,8.7){$Y$}

\put(2.15,4.8){$p_0$}
\put(5.15,4.8){$p_1$}
\put(8.15,4.8){$p_2$}
\put(11.15,4.8){$p_3$}
\put(14.15,4.8){$p_4$}
\put(17.15,4.8){$p_5$}

\put(7,1.8){\line(0,-1){1}}
\put(10,1.8){\line(0,-1){1}}
\put(13,2.8){\line(0,-1){2}}
\put(6.65,0.1){$r_1$}
\put(9.65,0.1){$r_2$}
\put(12.75,0.1){$r_3$}


\end{picture}
\caption{Schema eines Rohrabschnitts mit drei äquidistant aufeinanderfolgenden Querschnittssprüngen. Oben ist die Lage der Einkopplungsstelle $X$ und Auskopplungsstelle $Y$ des Kreuzgliedkettenfilters gekennzeichnet, in der Mitte die Positionen der Bezugsgrößen $p_0\cdots p_5$ der Finite-Differenzen. Die Reflexionsfaktoren $r_1 \cdots r_3$ charakterisieren die Querschnittssprünge.}\label{Rohr3R}
\end{center}
\end{afigure}

Dieses Rohr lässt sich auch eindimensional mit Finite-Differenzen beschreiben. Für die unterschiedlichen Querschnitte werden die in Abschnitt~\ref{partVol} entwickelten partiellen Volumen verwendet. Es entsteht dadurch eine direkte Korrespondenz zwischen den verwendeten Reflexionsfaktoren. Die Finite-Differenzen-Gleichung für partielle Volumen, das Resultat aus Abschnitt~\ref{partVol}, vereinfacht sich für den eindimensionalen Fall: Die schallharte Begrenzung senkrecht zur Schallausbreitungsrichtung $x$ wird durch Reflexionfaktoren mit dem Wert 1 berücksichtigt, also $r_{x,y,z}^{x, y\pm 1, z} = 1 $, $r_{x,y,z}^{x, y, z\pm 1}=1$ und somit   
\begin{align*}
&p_{t+1,x}-2p_{t,x}+p_{t-1,x}\\
&\phantom{mmm}=\thickspace K[ (1-r^{x+1}_{x})p_{t,x+1}-(2 -r^{x+1}_{x}-r^{x-1}_{x})p_{t,x}+(1-r^{x-1}_{x})p_{t,x-1}]\:.\nonumber
\end{align*}
Mit der Stabilitätsanalyse in Abschnitt \ref{AnFin} ergibt sich durch $a_1=4$ und $a_2=4K$ der Stabilitätsbereich  $K\leqslant 1$, womit $K=1$ gesetzt werden kann, um Dispersion zu vermeiden. Diese Gleichung lässt sich weiter bezüglich des zentralen Summanden und durch die gekürzte Indizierung $r_x^{x+1} = r_x $  und entsprechend $r_{x}^{x-1} = -r_{x-1}^{x} = -r_{x-1} $ vereinfachen, so dass 
\[
p_{t+1,x}+p_{t-1,x}\thickspace=\thickspace (1-r_{x})p_{t,x+1}-( -r_{x}+r_{x-1})p_{t,x} +(1+r_{x-1})p_{t,x-1}
\]
verbleibt. Diese bekommt im $\cal{Z}$-Bereich die Form: 
\[
zP_{x}+z^{-1}P_{x}\thickspace=\thickspace (1-r_{x})P_{x+1}-(- r_{x}+r_{x-1})P_{x} +(1+r_{x-1})P_{x-1} \;.
\]
Löst man die Gleichung nach $P_x$ auf, ergibt sich:
\[
P_{x}\thickspace=\thickspace \frac{(1-r_{x})P_{x+1} +(1+r_{x-1})P_{x-1}}{z+(-r_{x}+r_{x-1})+z^{-1}}\;.
\]

Die  reflexionsfreie Schallleitung rechtsseitig und linksseitig lässt sich durch absorbierende Randbedingungen nach Abschnitt 
\ref{TheoRand} erfassen, die rechtseitige und spiegelbildliche linkseitige Gleichung lauten:
\begin{align*}
p_{t,x+1} = p_{t-1,x} \;,\\
p_{t,x-1} = p_{t-1,x}\;.
\end{align*}
Diese Randbedingungen haben im $\cal{Z}$-Bereich die Form 
\begin{align*}
P_{x}\thickspace&=\thickspace z^{-1}P_{x-1}\;, \\
P_{x}\thickspace&=\thickspace z^{-1}P_{x+1}\;.
\end{align*}

Zur Einkopplung des Anregungssignals $X$ wird zwischen den Elementen $P_0$ und $P_1$ ein passend skalierter Druckunterschied eingeprägt, 
womit sich für das Rohr  aus den gezeigten Beziehungen das Gleichungssystem
\begin{align*}
P_{0}\thickspace&=\thickspace z^{-1}(P_{1} - 2X)\;, \\
P_{1}\thickspace&=\thickspace \frac{(1-r_1 )P_2 + P_0+ 2X}{z-r_1+z^{-1}}\;,\\
P_{2}\thickspace&=\thickspace \frac{(1-r_2 )P_3 +(1+r_1)P_1}{z+(-r_2+r_1 )+z^{-1}}\;,\\
P_{3}\thickspace&=\thickspace \frac{(1-r_3 )P_4 +(1+r_2)P_2}{z+(-r_3+r_2 )+z^{-1}}\;,\\
P_{4}\thickspace&=\thickspace \frac{P_5  +(1+r_2)P_3 }{z+r_3+z^{-1}}\;,\\
P_{5}\thickspace&=\thickspace z^{-1}P_4\;
\end{align*}
ergibt. Das zwischen $P_4$ und $P_5$ liegende Ausgangssignal $Y(z)$ lässt sich entsprechend sich aus dem Mittelwert der beiden Größen bestimmen.
Dieses Gleichungssystem liefert nach sukzessivem Einsetzen und Auflösen nach $Y(z)/X(z)$ die Übertragungsfunktion des Rohrsystems,
\[
H(z)=\frac{Y(z)}{X(z)}= \frac{(1+r_1)(1+r_2)(1+r_3)\,z^{-4}}{1+(r_1r_2+r_2r_3)\,z^{-2}+r_1r_3\,z^{-4}} \;.
\]
Diese ist erwartungsgemäß identisch mit der Beschreibung des Rohrs durch Kreuzgliedkettenfilter.  Die hier auf kurzem Weg erzielten Ergebnisse decken sich mit den Überlegungen in \cite{Mc87}, bei denen hin- und rücklaufende Wellen separat betrachtet werden.

\pagebreak
\section{Reflexionsfreier Abschluss}\label{ReflexionsfreierAbschluss}
Ein reflexionsfreier Abschluss ist in üblichen Modellen der Vokaltraktakustik als Randbedingung zur Beschreibung des glottalen bzw.{} subglottalen Abschlusses notwendig.  Der reflexionsfreie Abschluss in der gerade durchgeführten Äquivalenzbetrachtung eignet sich in dortiger Form jedoch nur zur Beschreibung von eindimensionalen Strukturen. Im Folgenden wird untersucht, wie sich dieser Abschluss auf eine dreidimensionale Formulierung der Finite-Differenzen übertragen kann.

Ein reflexionsfreies Verhalten lässt sich auf verschiedene Weise approximieren. Es ist sicher naheliegend, sich an der Geometrie der Trachea zu orientieren und den nahezu reflexionsfreien Verlauf durch ein hinreichend ausgedehntes Rohr mit geringer Dämpfung wie in Abschnitt~\ref{DämpfModel} nachzubilden. Diese Herangehensweise hat jedoch den Nachteil, dass das Volumen des Abschlusses und damit auch der Rechenaufwand gegenüber dem des Vokaltrakts überwiegen. Vorteilhaft ist die Formulierung einer Randbedingung, die das Verhalten des Abschlusses widerspiegelt. Die zusätzlichen Rechnungen bleiben damit auf den deutlich kleineren Randbereich beschränkt.
\subsection{Anforderungen an den Operator}
Der Operator, der diese Randbedingung approximieren soll, muss näherungsweise, d. h. bei niedrigen Frequenzen, folgende Eigenschaften haben:
\begin{enumerate}
\item Er leitet den Schall mit der Geschwindigkeit c zum Rand hin.
\item Er verändert die Schallamplitude nicht. 
\end{enumerate}
Diese beiden Forderungen ergeben sich aus der ungestörten Schallleitung im Randbereich. Betrachtet man den Finite-Differenzen-Operator direkt neben dem Rand-Operator, so ist es erforderlich, dass auch sein Randelement die senkrecht auf den Rand auftreffende,  durch ihn hindurch propagierende Schallwelle möglichst ungestört erhält.

Der Operator, betrachtet an dem Rand $x$ zum Zeitpunkt $t$, hat drei Freiheitsgrade, da einer der Koeffizienten von $\ws_{x+1,t}$, $\ws_{x,t-1}$, $\ws_{x,t}$ und $\ws_{x,t+1}$ durch Normalisierung entfällt: Der Koeffizient von $\ws_{x+1,t}$ wird gleich 1 gewählt. Die anderen Koeffizienten erhalten in der genannten Reihenfolge die Bezeichnungen $p$, $o$, $q$. Bild \ref{Struk-Rand} illustriert die Struktur des Operators.
\begin{afigure}[t]
\begin{center}
\setlength{\unitlength}{5mm}
\vspace{5mm}
\begin{picture}(8,8)
\put(4,0){\circle{.3}}
\put(4,8){\circle{.3}}
\put(8,4){\circle{.3}}
\put(4,4){\circle{.3}}
\thicklines
\put(4,4.15){\line(0,1){3.7}}
\put(4.15,4){\line(1,0){3.7}}
\put(4,0.15){\line(0,1){3.7}}

\put(4.3,4.4){\scriptsize $o\,\ws_{x,t}$}
\put(3.3,-0.7){\scriptsize $p\,\ws_{x,t-1}$}
\put(6.7,4.4){\scriptsize $\ws_{x+1,t}$}
\put(3.3,8.4){\scriptsize $q\,\ws_{x,t+1}$}
\end{picture}
\caption{Struktur und Koeffizienten des absorbierenden Rand-Operators, die Zeitachse $t$ ist in Blattlängsrichtung projiziert, die mit $x$ bezeichnete Raumrichtung in Blattquerrichtung. Der von rechts kommende Schall hat durch die Orientierung des  Koordinatensystems eine negative Geschwindigkeit.}\label{Struk-Rand}
\end{center}
\end{afigure}

\subsection{Koeffizientenbestimmung}
Die Koeffizienten der allgemeinen Form
\[q \ws_{x,t+1} = o\,\ws_{x,t} + p\,\ws_{x,t-1} + \ws_{x+1,t}\]   
lassen sich anhand einer harmonischen Welle weiter eingrenzen. Wählt man 
\[\ws_{x,t}=a_x(\omega)e^{i(\omega t-kx)}\]
mit der orts- und frequenzabhängigen Amplitude $a$ und setzt dieses ein, ergibt sich mit der Vereinfachung der Indizierung  $x=0$ und der Normierung $a_0(\omega)=1$ 
\[q e^{i(\omega t+\omega)} = oe^{i\omega t} + pe^{i(\omega t-\omega)} + a_1(\omega)e^{i(\omega t-k)}\;.\]   
Eine weitere Vereinfachung ergibt sich durch die beiderseitige Division durch $e^{i\omega t}$, so dass 
\[q e^{i\omega} = o + p e^{-i\omega} + a_1(\omega) e^{-ik}\]   
verbleibt.

Aus der 1.{} Forderung, konstante Schall- bzw. Transportgeschwindigkeit $c$ bei niedrigen Frequenzen, folgt $k=\omega/c$ für $\omega, k \to 0$. Setzt man ersteres ein und entwickelt die Gleichung nach Taylor um $\omega = 0$, ergibt sich:

\[q + qi\omega = o + p - pi\omega +a_1(0) - a_1(0)i\omega/c + a_1'(0)\;. \]

Aus der 2.{} Forderung, der unveränderten Schallamplitude  bei niedrigen Frequenzen, ergibt sich die notwendige Bedingung $a_1(\omega)=a_0(\omega)=1$ und $a_1'(\omega)=0$ für $\omega=0$ und somit
\[q  = o + p  + 1 \quad.\]   

Die Gleichung zuvor vereinfacht sich durch Einsetzen dieser Abhängigkeiten und einer beidseitigen Division durch $i\omega$ zu

\[q = -p - 1/c\quad,  \]
die Koeffizienten sind somit bis auf den freien Parameter $p$ bestimmt.

\subsection{Eigenschaften}
Der Operator hat nun die Form
\[-(p + 1/c) \ws_{x,t+1} = -(2p+1/c+1)\ws_{x,t} + p\ws_{x,t-1} + \ws_{x+1,t}\quad,\]   
die durch die folgende Darstellung besser strukturiert wird:
\begin{eqnarray*}
0=1/c&( \ws_{x,t+1}-\ws_{x,t})\quad + \quad(\ws_{x+1,t} - \ws_{x,t})  \\
 + p &(\ws_{x,t+1} -2\ws_{x,t} + \ws_{x,t-1})\quad. 
\end{eqnarray*}
Die erste Zeile enthält mit ihren beiden Differenzenoperatoren erster Ordnung die Diskretisierung der Transport-Differentialgleichung $cf'+\dot f=0$. Deren Lösungen $f(x+ct)$ erfüllen beide Forderungen. Der Parameter $p$ legt das zeitliche Zentrum der Differenzen-Approximation des Terms $\dot f$ fest. Mit $p=0$ liegt, wie aus der Gleichung ersichtlich, das Zentrum mittig zwischen $t$ und $t+1$, eine rechtseitige Approximation. Mit $p=-1/2c$ ergibt sich $0=1/2c\,( \ws_{x,t+1}-\ws_{x,t-1})\; +\; (\ws_{x+1,t} - \ws_{x,t})$, eine zentrale Approximation, und mit $p=-1/c \Rightarrow 0=1/c\,( \ws_{x,t}-\ws_{x,t-1})\; + \;(\ws_{x+1,t} - \ws_{x,t})$, eine linksseitige Approximation.

Betrachtet man den Operator als zeitdiskretes und -invariantes Filter, wobei $\ws_{x+1}$ den Eingang und $\ws_{x}$ den Ausgang darstellt, ergeben sich die Eigenschaften des Operators  aus der Übertragungsfunktion $H$ des Filters. Wie in Bild \ref{Filt-Rand} zu sehen, handelt es sich um ein rekursives Filter 2.{} Ordnung. Sein Übertragungsverhalten $H$ in der $z$-Ebene ist
\[H(z)=\frac{-z^{-1}}{p+1/c-(2p+1/c+1)z^{-1}+pz^{-2}}\;.\]
Mit $p=0$ hat es beispielsweise eine Polstelle bei $z=1+c$ und ist somit für Transportgeschwindigkeiten $c$ im Intervall $]-2,\,0[$  stabil, da dann die Polstelle im Einheitskreis liegt (Stabilitätskriterium aus Abschnitt \ref{ZT}). 
\begin{afigure}
\begin{center}
\begin{picture}(10,11)
\setlength{\unitlength}{.3cm}
\put(0,8){\thinlines\framebox(2,2){$z^{-1}$}}
\put(0,2){\thinlines\framebox(2,2){$z^{-1}$}}
\put(12,8){\thinlines\framebox(2,2){$z^{-1}$}}
\put(7,6){\thinlines\circle{2}\makebox(0,0){$+$}}
\thicklines
\put(13,14){\vector(0,-1){4}}
\put(13,8){\line(0,-1){2}}
\put(13,6){\vector(-1,0){5}}
\put(1,8){\vector(0,-1){4}}
\put(1,6){\vector(1,0){5}}
\put(7,7){\vector(0,1){7}}
\put(7,12){\line(-1,0){6}}
\put(1,12){\vector(0,-1){2}}
\put(1,2){\line(0,-1){2}}
\put(7,0){\line(-1,0){6}}
\put(7,0){\vector(0,1){5}}
\put(7,12){\circle*{.3}}
\put(1,6){\circle*{.3}}

\put(4,6.2){\makebox(0,0)[cb]{\scriptsize $\frac{2p+1/c+1}{p+1/c}$}}
\put(4,0.2){\makebox(0,0)[cb]{\scriptsize $-\frac{p}{p+1/c}$}}
\put(10,6.2){\makebox(0,0)[cb]{\scriptsize $\;-\frac{1}{p+1/c}$}}
\put(6.3,13){\makebox(0,0)[cb]{\scriptsize $Y$}}
\put(12.3,13){\makebox(0,0)[cb]{\scriptsize $X$}}

\end{picture}
\caption{Struktur und Koeffizienten der Filteranalogie zum absorbierenden Rand-Operator.}\label{Filt-Rand}
\end{center}
\end{afigure}

Die Abweichung von der gewünschten Übertragungsfunktion $H_T$, der Laufzeit $\tau=-1/c$, die durch den Zeitschritt und den Transport mit der Geschwindigkeit $c$ entsteht,
\[H_T(z)=z^{1/c}\;,\]
ergibt sich nun als Betragsquadrat der Differenzen der Übertragungsfunktionen und ist für $c=-\sqrt{K}=-1/\sqrt{3}$ in Bild \ref{Arefl} logarithmiert für zwei Parameter $p$ dargestellt.  
\begin{afigure}[t]
\begin{center}
\psfrag{0.05}{\put(-3.5,9.7){$\frac{D}{\rm [dB]}$}}
\psfrag{0.1}[t][t]{\put(-2.2,0){\footnotesize 0}\footnotesize 0,1}
\psfrag{0.15}{}
\psfrag{0.2}[t][t]{\footnotesize 0,2}
\psfrag{0.25}{}
\psfrag{0.3}[t][t]{\footnotesize 0,3}
\psfrag{0.35}{}
\psfrag{0.4}[t][t]{\footnotesize 0,4}
\psfrag{0.45}{\put(-.0,-.5){\footnotesize $\omega/\pi$}}
\psfrag{0}[r][r]{\footnotesize}
\psfrag{-10}[r][r]{\footnotesize -10}
\psfrag{-20}[r][r]{\footnotesize -20}
\psfrag{-30}[r][r]{\footnotesize -30}
\psfrag{-40}[r][r]{\footnotesize -40}
\psfrag{-50}[r][r]{\footnotesize -50}
\psfrag{-60}[r][r]{\footnotesize -60}
\epsfig{file=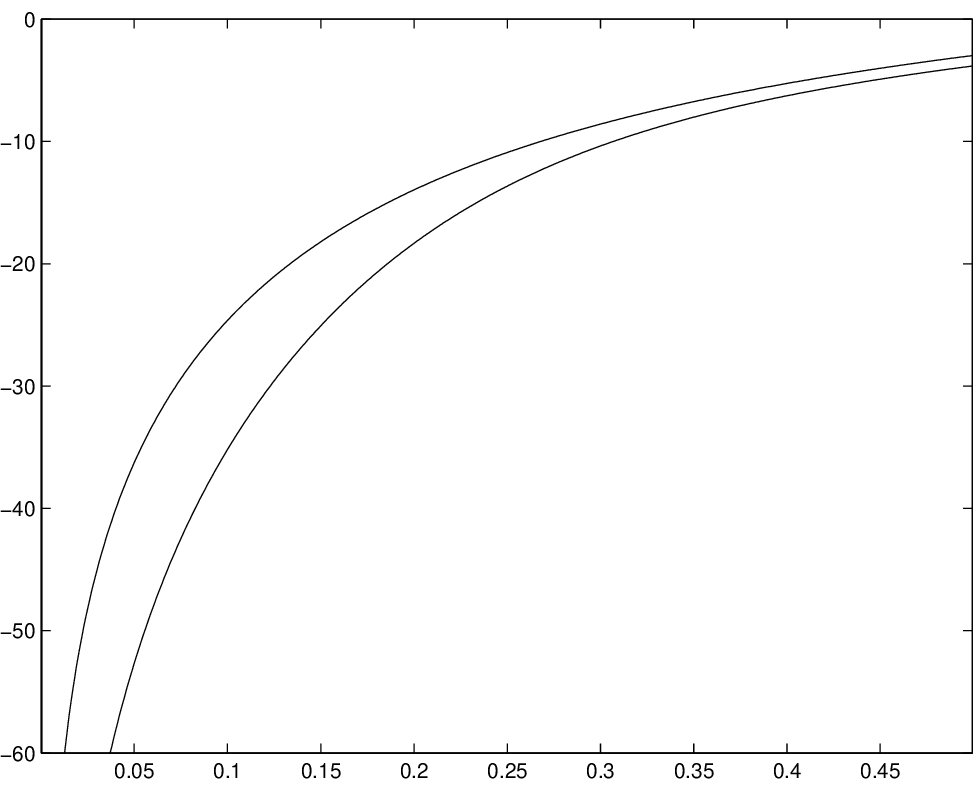, width=.43\linewidth}
\caption{Logarithmische Darstellung der Reflexionsdämpfung $D$ bei $c=1/\sqrt{3}$ für $p=0$ (linke Kurve) und $p=-0,6$ (rechte Kurve), aufgetragen über den Kreisfrequenzbereich $\omega=0..\pi/2$}\label{Arefl}
\end{center}
\end{afigure}
Man erkennt, dass sich der nutzbare Frequenzbereich bei $p=-0,6$ gegenüber $p=0$ je nach gewünschter Reflexionsunterdrückung  verdoppelt oder verdreifacht. Dieser beträgt bei 40~dB etwa $0,1\pi$. Verwendet man eine andere Zielfunktion, wie die dispersionsbehaftete Wellenausbreitung aus Abschnitt \ref{Dispersion}, variiert das Ergebnis geringfügig. In Bild~\ref{ZvZRA} 
\begin{afigure}[b]
\begin{center}
\psfrag{20}[r][r]{\footnotesize 20}
\psfrag{40}[r][r]{\footnotesize 40}
\psfrag{60}[r][r]{\footnotesize 60}
\psfrag{80}[r][r]{\footnotesize 80}
\psfrag{100}[r][r]{\footnotesize 100}
\psfrag{120}[r][r]{\footnotesize 120}
\psfrag{140}[r][r]{\footnotesize }
\psfrag{0}[t][t]{\footnotesize 0}
\psfrag{3}[t][t]{\footnotesize 0,1}
\psfrag{6}[t][t]{\footnotesize 0,2}
\psfrag{9}[t][t]{\footnotesize 0,3}
\epsfig{file=ZeroVsZeroR\arxiv{A}.eps, width=.8\linewidth}
\put(-25.3,7.75){\makebox(0,0)[cb]{$\frac{|Y|}{\rm dB}$}}
\put(-24.5,-0.25){\makebox(0,0)[cb]{\footnotesize 0}}
\put(-2,-0.8){\makebox(0,0)[cb]{\footnotesize $\omega /\pi$}}
\vspace{-1.5ex}
\caption{Vergleich der logarithmierten Betragsgänge zweier Rohre. Die kamm\-artig deutlich sichtbaren Eigenresonanzen bei einem schallharten Abschluss, grau, werden bei einem absorbierenden Abschluss, schwarz dargestellt, fast vollständig unterdrückt. Der Ordinatenbereich umfasst 160~dB; das \glqq Grundrauschen\grqq{} lässt sich auf die Quantisierung der verwendeten 24~Bit-Gleitkommamantisse zurückführen.}\label{ZvZRA}
\end{center}
\end{afigure}
ist eine Simulation im Zeitbereich gezeigt, die die Dämpfungseigenschaft in einem breiten Frequenzbereich bestätigt: Die Betragsgänge zweier einseitig offener ($r=-1$) Rohre mit Einheitsquerschnittsfläche und einer exemplarischen Länge von 72 werden für den reflektierenden und den diskutierten absorbierenden Abschluss auf der Gegenseite verglichen. Als Systemantwort $y$ wird der Signalverlauf direkt vor dem schallweichen Abschluss aufgezeichnet. Der schallweiche Abschluss bewirkt eine entgegengesetzte Schallwelle, die zu einer Interferenz führt, welche durch eine inverse Filterung in Form einer Integration des Zeitsignals kompensiert wird.  Die Simulation umfasst $2^{16}$ Zeitschritte, die durch eine Fouriertransfomation in den Frequenzbereich übertragen werden. Mit einem ausgeprägten Kaiser-Fenster nach \cite{Ka74}, $\beta = 15$, werden dabei Blockgrenzeffekte praktisch gänzlich vermieden. Das Fenster bewirkt jedoch eine schwache Gewichtung der anfänglichen Impulsantwort, weshalb die Übertragungsfunktion des absorbierenden Rohres einen niedrigeren Pegel aufweist.  

\begin{afigure}[t]
\begin{center}
\psfrag{50}{}
\psfrag{150}{}
\psfrag{250}{}
\psfrag{350}{\put(0,-.7){$\omega/\pi$}}
\psfrag{0}[t][t]{\footnotesize 0}
\psfrag{100}[t][t]{\footnotesize 1/8}
\psfrag{200}[t][t]{\footnotesize 1/4}
\psfrag{300}[t][t]{\footnotesize 3/8}
\psfrag{3}[r][r]{\footnotesize 0}
\psfrag{8}[r][r]{\footnotesize 5}
\psfrag{13}[r][r]{\footnotesize 10}
\psfrag{18}[r][r]{\footnotesize 15}
\psfrag{23}{\put(-1.5,-1){$\frac{H}{\rm [dB]}$}}
\epsfig{file=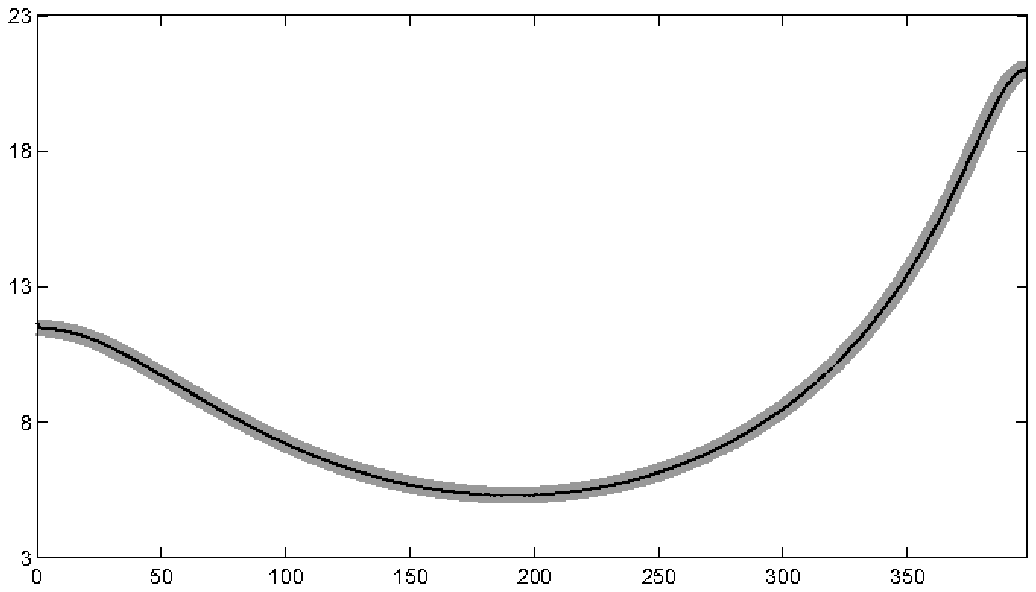, width=.75\linewidth}
\caption{Vergleich der logarithmierten  Betragsgänge eines Rohrmodells mit den durch die Reflexionsfaktoren $r_1= -1$, $r_2=1/2$ und $r_3=-1/2$ charakterisierten äquidistanten Querschnittssprüngen, grau, mit einer Finite-Differenzen-Simulation mit den entsprechenden Querschnittsflächen 1, 1/3 und 1 bei einer exemplarischen Rohrlänge von jeweils 24, schwarz dargestellt. Die Verläufe stimmen nahezu perfekt überein, nur bei genauer Betrachtung erkennt man eine winzige, aus der inversen Filterung rührende Abweichung bei $\omega =0 $. Die hohe Genauigkeit der Übereinstimmung ergibt sich aus Rohrlänge. Erlaubt man rechtsseitig eine Frequenzabweichung von 1\%, kann die Rohrlänge nach Abschnitt~\ref{AnFin} auf jeweils 6 reduziert werden. Die Abszissenskalierung bezieht sich auf die Systemfunktion des Rohrmodells $H=[1-\frac{3}{4}z^{-2}-\frac{1}{2}z^{-4}]^{-1}$ mit $z=e^{i\omega}$.}\label{zzz}
\end{center}
\end{afigure}
Die gerade betrachtete Struktur stimmt mit der üblichen Rohrkonfiguration zur Modellierung des Vokaltrakts überein, die einerseits mit einem schallweichen Abschluss die Schallabstrahlung am Mund beschreibt und andererseits mit einem nicht reflektierenden bzw.{} absorbierenden Abschluss den subglottalen Bereich nachbildet. Das Ergebnis einer nicht minder relevanten Simulation an dieser Struktur zeigt  Bild~\ref{zzz}: Die Genauigkeit der Berechnung von Resonanzen bei variierenden Rohrquerschnittsflächeninhalten. Das zuvor in Abschnitt \ref{Äquivalenz} diskutierte Rohrmodell mit 3~Querschnittssprüngen wird als Referenz  verwendet und mit dem Betragsgang einer Finite-Differenzen-Simulation verglichen. Es zeigt sich, dass auch für eine dreidimensionale, mit Dispersion behaftete Formulierung der Finite-Differenzen eine praktisch perfekte Übereinstimmung erzielt wird. Da die Impulsantwort in den $2^{16}$ Zeitschritten der Simulation schnell abfällt, ist  eine Fensterfunktion zudem nicht erforderlich.

\subsection{Perfectly Matched Layers}
Wenngleich der gefundene Operator bereits gute Resultate liefert, sei abschließend noch eine weitere Verbesserungsmöglichkeit angesprochen. Bei den sogenannten \name{Perfectly Matched Layers}\cite{Be94,KaTT94} wird in einer Schicht nur eine endliche Dämpfung erzielt und der absorbierende Rand durch eine hinreichend große Anzahl von Schichten angenähert. Eine Schallreflexion beim Eintritt und innerhalb dieser Schichten wird durch gleiche Schallkennimpedanz vermieden. 

\pagebreak
\section{Anwendung am Beispiel eines Laterallauts}
In diesem Abschnitt werden mit den vorangegangenen Überlegungen Akustiken des realitätsnahen Vokaltrakts untersucht. Auch hier wird ein Vergleich zu einer eindimensionalen Schallausbreitung in einem Rohrmodell gezogen und die zusätzlichen Erkenntnisse betrachtet, die man durch eine dreidimensionale Berechnung der Schallausbreitung gewinnt.  Interessant sind hierfür insbesondere Vokaltraktkonfigurationen für Laute, bei denen der Vokaltrakt stark von der für Rohrmodelle in Abschnitt~\ref{WelleEle} verwendeten Hypothese eines runden Querschnitts abweicht und sich im Sprachfrequenzbereich Quermoden bilden können. Wenngleich das bereits, wie in Abschnitt~\ref{nasaltrakt} gezeigt, bei dem Vokal /a/ auftritt, ist es naheliegend, dass der Effekt aufgrund des höheren Aspektverhältnisses von Weite zu Höhe des Vokaltraktquerschnitts bei Approximanten und Laterallauten ausgeprägter ist. 

Um auf bestehende MRT-Aufnahmen zurückgreifen zu können, wird eine Kontur des Laterallauts [l] verwendet, die in \cite{ZhEWT03} publiziert ist. \arxiv{
Bild~\ref{latL} zeigt diese Aufnahmen aus dem Mittelbereich der Mundhöhle.} Von der Betrachtung und Nachbildung der für Laterallaute typischen Aufspaltung der Schallausbreitung um den vorderen Zungenbereich wird dabei abgesehen, um die hiervon hervorgerufenen Effekte\footnote{Die Effekte aus der bereichsweisen Aufspaltung der Schallpassage im Vokaltrakt sind u.~a.~in \cite{ZhEWT03} erörtert. Sie äußern sich in Nullstellen im Schallsignal, wie sie auch in Bild~\ref{mnl} als Vertiefung im Frequenzgang zu erkennen sind.} nicht mit denen zu vermischen, die für eine Mehrzahl von Lautklassen erheblich sind.

Es zeigt sich, dass die Auswirkung auf das Spektrum von separaten Effekten herrührt, die im Folgenden in zwei Schritten betrachtet werden.
\arxiv{
\begin{afigure}[b]
\begin{center}
\epsfig{file=Lateral_L1.eps, width=.66\linewidth}\\[.8ex]
\epsfig{file=Lateral_L2.eps, width=.66\linewidth}
\caption{Frontale Magnetresonanztomographien der Mundhöhle während der Artikulation des Lauts [l]. Die erste Aufnahme ist etwa 14,7 cm von der Glottis entfernt, die weiteren Bilder (zeilenweise) sind jeweils 6 mm beabstandet. Der Hohlraum der Mundhöhle, weiß, tritt aufgrund der gröberen Struktur auch mit  Magnetresonanz-Bildgebung klar hervor. Bildquelle: \cite{ZhEWT03}.}\label{latL}
\end{center}
\end{afigure}}

\subsection{Natürlicher Flächeninhalt}
Der Übergang von kleinen Querschnittsflächeninhalten zu natürlichen Flächeninhalten wird schrittweise an einer runden Querschnittskontur vorgenommen, vgl.~Bild~\ref{SpeakL}. Auf diese Weise ergeben sich graduelle Veränderungen der Übertragungsfunktion, die sich bestimmten akustischen Effekten zuordnen lassen. Bild~\ref{latFD-Rohr} zeigt die Ergebnisse. 
\begin{afigure}[t]
\begin{center}
\epsfig{file=Speak-\arxiv{L}4.eps, width=\linewidth}
\caption{Durch ein Rohrmodell  approximierter Querschnittsverlauf des Vokaltrakts (Glottis links, hervorgehoben) während der Artikulation des Lauts~[l]. Der Querschnittsverlauf  in {\sc Speak} ist anhand des Sprachsignals aus Abschnitt \ref{aktEig} unter Anwendung einer zweifachen, adaptiven Preemphase und der Burg-Methode berechnet. Das Rohrmodell hat eine Länge von 19 cm und einen Durchmesser von 3,6 cm bei einer hypothetischen runden Querschnittskontur.  Das Verhältnis von  Querschnittsflächeninhalt zu Rohrlänge entstammt den MRT-Aufnahmen aus \cite{ZhEWT03}, wobei auch der gesamte Querschnittsverlauf mit diesen Aufnahmen  gut übereinstimmt. }\label{SpeakL}
\end{center}
\end{afigure}

Für kleine laterale Ausdehnungen ergibt sich eine sehr gute Übereinstimmung mit der Übertragungsfunktion der Kreuzgliedkettenfilter des Rohrmodells. Die immer bessere Übereinstimmung bei abnehmender lateraler Ausdehnung hat sich bei einer Reihe weiterer Simulationen bestätigt. Dieser erwartungsgemäße Befund resultiert aus der für diesen Fall guten Approximation der Schallwellenausbreitung durch das Rohrmodell, das als eindimensionaler Wellenleiter für geringe laterale Ausdehnung adäquat ist. 

Eine deutliche Abweichung tritt mit zunehmender lateraler Ausdehnung  auf, insbesondere bei der ausgeprägtesten Resonanz um 2~kHz. Eine weitere Untersuchung mit sowohl verdoppelter lateraler Ausdehnung als auch verdoppelter Rohrsegmentlänge weist praktisch keinen Unterschied zu dem in Bild~\ref{latFD-Rohr} oben gezeigten Betragsgang, grenzt andere Abhängigkeiten aus und bestätigt den Zusammenhang zwischen Resonanzverschiebung und Aus\-dehnungs-Längen-Verhältnis. Diese Resonanzverschiebung lässt sich mit einer Mündungskorrektur nach \cite{Co1860, Mo02, MüM04} erklären, wie sie an Impedanz- bzw.{} Querschnittsflächensprüngen auftritt, die hier innerhalb des Rohres eine Resonanz  bewirken. Eine beispielhafte Berechnung verdeutlicht dies: Ausgehend von einer engen, sich rasch erweiternden Querschnittsfläche, wie sie in Bild~\ref{SpeakL} an siebter Stelle von rechts zu finden ist, die einen Durchmesser $d$ von rund 1~cm aufweist und einer effektiven Resonanzrohrlänge $l$ von 3,5~cm für einen $\lambda/4$-Resonator mit einer Resonanzfrequenz von 2,4~kHz, lassen sich Größenordnung und Relevanz der Resonanzverschiebung  mit der Formel $\Delta l=\frac{5}{3}d$ von Cavaillé-Coll abschätzen: 
der korrigierte Resonator hat eine Länge von 5,2~cm und eine Resonanz bei 1,7~kHz. Selbige Abschätzung gilt auch für einen $\lambda/2$-Resonator, der die doppelte effektive Länge aufweist, aber auch eine zweiseitige Mündungskorrektur bedingt. Experimentell wird dieser Effekte für  Rohrsysteme    in \cite{Sc90} bestätigt.

Die Pol-Nullstellenkombination bei etwas über 6~kHz korrespondiert offensichtlich mit der in Abschnitt~\ref{WelleEle} notierten Radialmode. Eine exakte Übereinstimmung der Frequenz mit der Zylinder- oder Kugelmode ist nicht zu erwarten, da die Mundhöhle von beiden geometrischen Idealen  abweicht.
\begin{afigure}[t]
\begin{center}
\psfrag{-10}[r][r]{\footnotesize 0}
\psfrag{A}[r][r]{\footnotesize 10\hspace{.1em}}
\psfrag{10}[r][r]{\footnotesize 20}
\psfrag{20}[r][r]{\footnotesize 30}
\psfrag{30}[r][r]{\footnotesize 40}
\psfrag{40}[r][r]{\footnotesize 50}
\psfrag{0}[t][t]{\footnotesize 0}
\psfrag{1}[t][t]{\footnotesize 1}
\psfrag{2}[t][t]{\footnotesize 2}
\psfrag{3}[t][t]{\footnotesize 3}
\psfrag{4}[t][t]{\footnotesize 4}
\psfrag{5}[t][t]{\footnotesize 5}
\psfrag{6}[t][t]{\footnotesize 6}
\psfrag{7}[t][t]{\footnotesize 7}
\psfrag{8}[t][t]{\footnotesize 8}
\epsfig{file=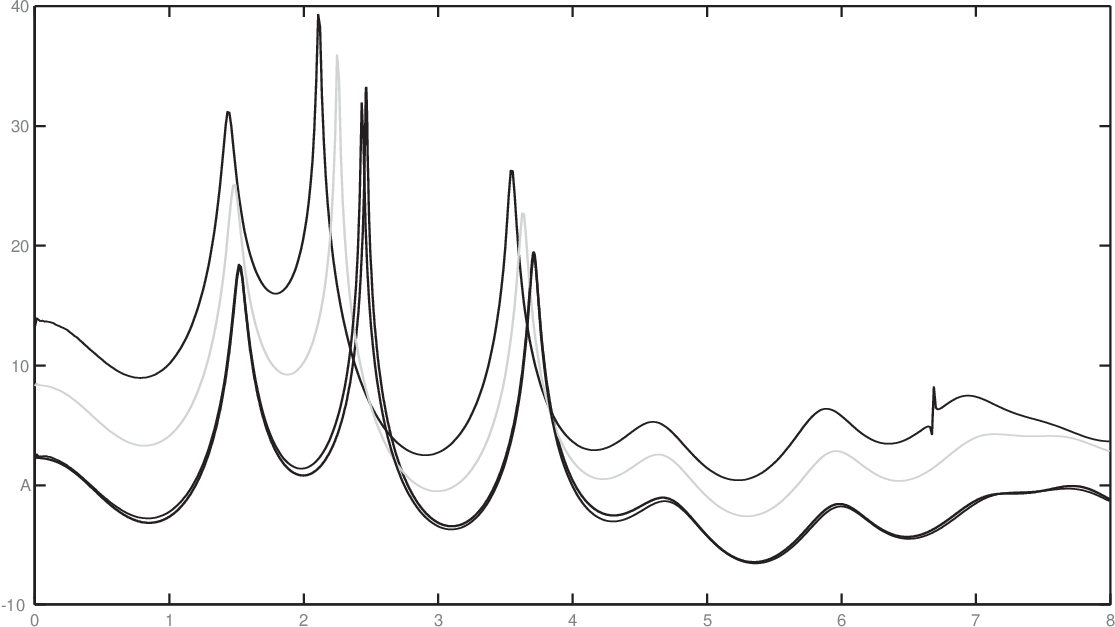, width=.92\linewidth}
\put(-29.4,13.9){\makebox(0,0)[cb]{$\frac{{\rm log}|\hspace{-.06em}H\hspace{-.04em}|}{\rm [dB]}$}}
\put(-2,-0.8){\makebox(0,0)[cb]{\footnotesize $f/\rm [kHz]$}}
\caption{Vergleich von Betragsgängen des Vokaltrakts, bestimmt durch Finite-Differenzen und durch ein Rohrmodell-Kreuzgliedkettenfilter. Die obere Kurve zeigt das Ergebnis der FD-Simulation für zylindrische Rohrsegmente gemäß Bild~\ref{SpeakL}, deren (maximale) Querschnittsfläche $A_m$ und Länge $l$ mit den MRI-Untersuchungen in Bild~\ref{latL} übereinstimmt, $\sqrt{A_m}/l=$~0,17; die Kurven darunter zeigen FD-Simulationen mit verringertem Verhältnis der lateralen Ausdehnung zur Länge, $\sqrt{A_m}/l=$~0,068 und  0,007. Deutlich tritt die sich mit zunehmenden Querschnittsverhältnis von 2,4~kHz auf 2,1~kHz verschiebende Resonanz hervor, sowie die bei 6,7~kHz hinzukommende Pol-Nullstellenkombination. Zum Vergleich ist der Betragsgang des Rohrmodells aus Bild~\ref{L-BURG-PRE} gezeigt, der sich merklich nur in der etwas höheren Frequenz der ausgeprägtesten Resonanz bei 2,4 kHz und in Folge davon im niederfrequentere Betragsgang geringfügig unterhalb und im höherfrequenteren geringfügig oberhalb liegenden Kurvenverlauf von der Finite-Differenzen-Simulation mit kleinem Querschnittsverhältnis unterscheidet.   }\label{latFD-Rohr}
\end{center}
\end{afigure}

\subsection{Querschnittsform}\label{Querschnittsform}
\begin{afigure}[t]
\begin{center}
\psfrag{0}[r][r]{\footnotesize 0\hspace{.1em}}
\psfrag{10}[r][r]{\footnotesize 10}
\psfrag{20}[r][r]{\footnotesize 20}
\psfrag{30}[r][r]{\footnotesize 30}
\psfrag{40}[r][r]{\footnotesize 40}
\psfrag{50}[r][r]{\footnotesize 50}
\psfrag{A}[t][t]{\footnotesize 0}
\psfrag{B}[t][t]{\footnotesize 1}
\psfrag{C}[t][t]{\footnotesize 2}
\psfrag{D}[t][t]{\footnotesize 3}
\psfrag{E}[t][t]{\footnotesize 4}
\psfrag{F}[t][t]{\footnotesize 5}
\psfrag{G}[t][t]{\footnotesize 6}
\psfrag{H}[t][t]{\footnotesize 7}
\psfrag{I}[t][t]{\footnotesize 8}
\epsfig{file=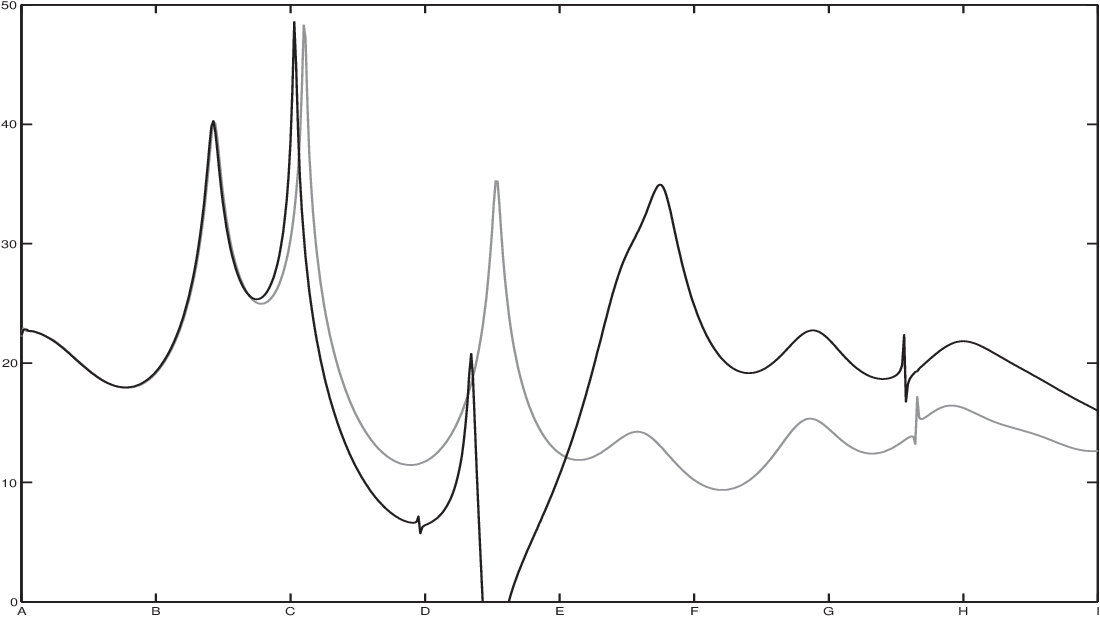, width=.91\linewidth}
\put(-29.4,13.9){\makebox(0,0)[cb]{$\frac{{\rm log}|\hspace{-.06em}H\hspace{-.04em}|}{\rm [dB]}$}}
\put(-2,-0.8){\makebox(0,0)[cb]{\footnotesize $f/\rm [kHz]$}}
\caption{Vergleich von Betragsgängen der Vokaltraktkonfiguration mit einer runden Querschnittskontur, grau, und einer aus MRT gewonnenen Querschnittskontur, schwarz. Beide weisen bereichsweise Übereinstimmung auf, so ist der Verlauf bis über die Resonanz bei 1,3 kHz hinweg praktisch identisch und die Resonanzen bei 2,1 kHz, 5,8 kHz und 6,9 kHz sind in Frequenz und Güte nur wenig verändert. Im Frequenzbereich zwischen 3 und 5 kHz zeigen sich hingegen deutlich Unterschiede.}\label{FD-lat-mri}
\end{center}
\end{afigure}
Die beim Übergang von einer runden zu einer natürlichen Querschnittsform auftretende Veränderung des Resonanzschemas ist im Bild~\ref{FD-lat-mri} zu sehen. Die vom Verlauf des Querschnittsflächeninhalts hervorgerufenen Längsmoden bleiben in weiten Bereichen nahezu unverändert. Die geringen Verschiebungen der Resonanzen bei 5,8 kHz und 6,9 kHz zu höheren Frequenzen lässt sich durch eine aufgrund der Querschnittsform abgeschwächte Mündungskorrektur erklären, wie in \cite{In53} untersucht. 

Auffällig sind die starken Abweichungen im Bereich zwischen 3 und 5 kHz, die aus der Veränderung der Querschnittsform herrühren. Auch hierbei handelt es sich augenscheinlich um Pol-Nullstellenkombinationen aus Quermoden, deren niedrige Frequenzen sich aus der großen lateralen Ausdehnung des Querschnitts ergeben: Geht man schrittweise von der runden Querschnittsform zu einer zunehmend langgezogenen Ellipse über, bis diese die durch MRI bestimmte Querschnittsform approximiert, kann man den Verlauf der Resonanzen und Nullstellen verfolgen und die Extrempunkte zuordnen.  So spaltet sich die Pol- und Nullstellenkombination bei 6,7 kHz in ein Paar auf, dass offensichtlich mit einer vertikalen und horizontalen Mode korrespondiert: Mit zunehmender Exzentrizität der Ellipse und entsprechend zunehmender lateraler Breite verringert sich die Resonanzfrequenz der einen Kombination bis hinab zu 3 kHz, die andere  verschiebt sich zu größeren Frequenzen mit abnehmender Resonatorhöhe. In gleicher Weise gelangt eine ausgeprägte Nullstelle aus einer höherfrequenten Mode des runden Rohrs von 12 kHz hinunter zu 3,5 kHz. Diese niedrigen Resonanzen haben auch eine Auswirkung auf darunterliegende Längsresonanzen. Sie verursachen durch eine zusätzliche Laufzeit eine effektive Verlängerung des akustischen Wegs und damit eine Reduzierung der Resonanzfrequenz der Längsmoden, wie sie bei  2 kHz zu beobachten ist. 

Diese Effekte treten auch bei Vokalen auf, eine Mehrzahl von ihnen kann auch mittels Finite-Elemente-Modells nachgewiesen werden. So hat bereits Bild \ref{vokaltrak_FEM} die komplexe Modenstruktur gezeigt. Die zugehörige Publikation geht insbesondere auf die Pol-Nullstellen-Komplexe ein, die in sehr ähnlicher Form --- wenn auch bei etwas höheren Frequenzen --- in der dort untersuchten Vokalkonfiguration des Sprechtrakts auftreten. Speziell für den hier betrachteten Laterallaut [l] ist die gute Übereinstimmung der ausgeprägten Nullstelle bei 3,5 kHz mit dem Spektrum des Sprachsignals bemerkenswert. Die Nullstelle, die auch von der Resonanzschätzung nicht erfasst wird, vgl.~Bild \ref{L-BURG-PRE}, deckt sich wesentlich besser mit dem Sprachsignalspektrum, als es die sonst herangezogenen, von der Laterallaut-typischen ringförmigen Aufspaltung hervorgerufenen Nullstellen tun. Für diese Ring-Antiresonanzen, die stark von einer willkürlichen Rechts-Links-Asymmetrie beeinflusst sind, werden in \cite{ZhEWT03} beispielsweise Frequenzen von 2,4 kHz und 4,2 kHz genannt. Auch in dem dort gezeigten Sprachspektrum treten die Frequenzen nicht deutlich als Antiresonanzen hervor. 

\subsection{Dimensionalität}
Bei der Lösung des inversen Problems, aus einem Sprachsignal die erzeugende Vokaltraktkonfiguration zu schätzen, sind Modelle mit einer möglichst kleinen Parameteranzahl gewünscht, wie in Abschnitt \ref{Parameter}  diskutiert. Motiviert durch die erfolgreiche Verwendung des elliptischen Querschnitts zur Klärung der Resonanzeigenschaften wird abschließend betrachtet, wie gut sich das aus der Querschnittsform resultierende Betragsspektrum durch ein vereinfachtes Modell approximieren lässt, dessen longitudinaler Verlauf des Querschnittsflächeninhalts vorgegeben ist und dessen laterale Querschnittskontur nur durch einen Parameter, die Exzentrizität, beschrieben wird. Das Ergebnis ist in Bild \ref{FD-el-mri} gezeigt. Mit Ausnahme von nur einer Resonanzfrequenz, die um 5\% Prozent verschoben ist, ergibt sich eine sehr gute Übereinstimmung.

\begin{afigure}[t]
\begin{center}
\psfrag{A}[r][r]{\footnotesize 0\hspace{.1em}}
\psfrag{B}[r][r]{\footnotesize 10}
\psfrag{C}[r][r]{\footnotesize 20}
\psfrag{D}[r][r]{\footnotesize 30}
\psfrag{E}[r][r]{\footnotesize 40}
\psfrag{F}[r][r]{\footnotesize 50}
\psfrag{0}[t][t]{\footnotesize 0}
\psfrag{1}[t][t]{\footnotesize 1}
\psfrag{2}[t][t]{\footnotesize 2}
\psfrag{3}[t][t]{\footnotesize 3}
\psfrag{4}[t][t]{\footnotesize 4}
\psfrag{5}[t][t]{\footnotesize 5}
\psfrag{6}[t][t]{\footnotesize 6}
\psfrag{7}[t][t]{\footnotesize 7}
\epsfig{file=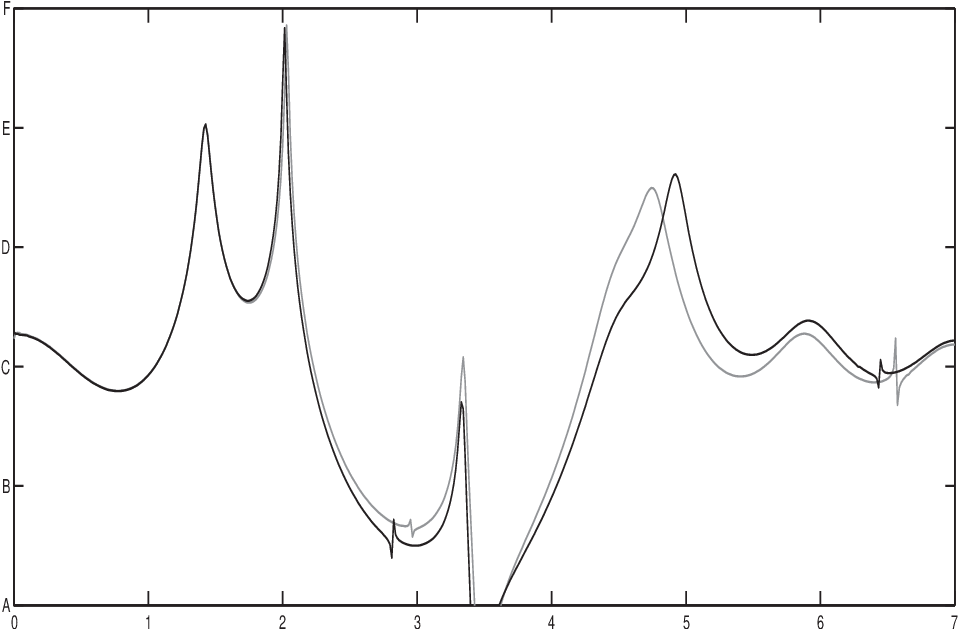, width=.8\linewidth}
\put(-26.,13.9){\makebox(0,0)[cb]{$\frac{{\rm log}|\hspace{-.06em}H\hspace{-.04em}|}{\rm [dB]}$}}
\put(-2,-0.8){\makebox(0,0)[cb]{\footnotesize $f/\rm [kHz]$}}
\caption{Vergleich von Betragsgängen der Vokaltraktkonfiguration mit einer elliptischen Querschnittskontur, schwarz, und einer aus MRT gewonnenen Querschnittskontur, grau. Die Verläufe stimmen gut überein. Eine nennenswerte Abweichung tritt nur bei der 4,8 kHz-Resonanz auf und beträgt weniger als 5\%. Die Proportion der Ellipse $\sqrt{A}/b$ = 0,34 unterscheidet sich leicht von der des MR-Schnittes, 0,38.}\label{FD-el-mri}
\end{center}
\end{afigure}
Insbesondere für iterativ arbeitende Algorithmen für inverse Probleme und die dafür erforderliche vielmalige Bestimmung des Betragsspektrums ist eine möglichst einfache, mit geringem Aufwand verbundene Berechnung vorteilhaft. Es sei deshalb abschließend noch auf eine diese Vorteile realisierende Beschreibung hingewiesen, die sich aus dem Formalismus der partiellen Volumen ergibt. Die Approximation durch Ellipsen mit deutlicher Exzentrität und die in dem Abschnitt zuvor gefundene Aufspaltung der Resonanzen, von denen offenbar nur die Horizontalmoden in dem hier betrachteten Frequenzbereich relevant sind, legt eine zweidimensionale Beschreibung des Vokaltrakts mit Finiten Differenzen nahe. Die Fläche liegt dabei senkrecht zur Sagittalebene und folgt der Trajektorie des Vokaltrakts, ist also im vorderen Mundbereich näherungsweise horizontal. Die senkrechten Luftsäulen über jedem Flächenelement werden als partielle Volumen kodiert, wobei die \glqq Füllung\grqq{} $\chi$ auf eine größte oder größtmögliche Luftsäule bezogen wird. Durch den Verbleib nur einer Schicht, dem für zwei Dimensionen größeren Wert für $K$ sowie den vereinfachten Differenzenoperator ergibt sich eine Laufzeitverbesserung der numerischen Simulation von ein bis zwei Größenordnungen.

\pagebreak
\section*{Zusammenfassung}
\addcontentsline{toc}{part}{
Zusammenfassung}
\markboth {Zusammenfassung}{Zusammenfassung}%
In dieser Arbeit werden verschiedene Methoden zur Untersuchung der Akustik des Sprechtrakts gezeigt und exemplarisch auf bestimmte Laute geeigneter Lautklassen angewendet. Dabei wird das bekannte Rohrmodell mit der Finite-Differenzen-Methode im Zeitbereich zu einem eigenen neuen Modell für den Sprechtrakt zusammengeführt.

Die Entwicklungsgeschichte der akustischen Modelle des Sprechtrakts wird in dieser Arbeit eingehend betrachtet; dabei werden wichtige bekannte und auch weniger bekannte, relevante  Untersuchungen auf diesem Gebiet erörtert. Diese  Betrachtung gibt einen Überblick über die akustischen Prozesse des Sprechens und zeigt im Besonderen einen ungebrochener Trend, dass jede substantielle Verfeinerung des akustischen Sprechtraktmodells wesentliche neue Erkenntnisse erbringt. Die essentiellsten Fortschritte ergeben sich beim Übergang von dem Resonatormodell einzelner Resonanzen \cite{Kra1781, Vi1783} zu den Rohranalogien des Sprechtrakts, zuerst bei der Reproduktion von Lauten anhand von Verläufen der  Querschnittsflächeninhalte aus Röntgenaufnahmen \cite{Du50, KeL62}, kurz darauf bei der Analyse des gesamten Resonanzspektrums aus Sprachsignalen.  
Diesen Trend fortführend wird durch eine Reihe von Modellen  die Schallausbreitung dreidimensional im Sprechtrakt oder in Bereichen des Sprechtrakts erfasst. Überwiegend sind das Finite-Elemente-Methoden, seltener erweiterte Rohrmodelle und Wave-Guide-Meshes. Die Diskussion in Abschnitt \ref{DetailMod} zeigt Vorteile der ersten beiden Herangehensweisen, die akkurate Nachbildung der Akustik, aber auch die spezifischen Nachteile aller drei Modelle auf. 

Die im Rahmen dieser Arbeit entwickelte Modellierung des Sprechtrakts vereint die Vorteile von Finiten-Differenzen mit den Vorzügen des Rohrmodells: 
die Charakterisierung von Sektionen durch deren akustische Impedanz, bei der die darüber hinausgehende Gestalt einzelner Sektionen unberücksichtigt bleibt. Diese hier als partielle Volumen bezeichneten Sektionen\footnote{in der Rohranalogie sind es Rohrabschnitte konstanten Querschnittsflächeninhalts} korrespondieren direkt mit den Voxeln tomographischer Datensätze. Schallabsorption durch die prominente Wechselwirkung mit den Hohlraumrändern des Sprechtrakts wird durch eine Nachbildung der linearen Grenzschichtprozesse an den Rändern Rechnung getragen. Hierbei zeigt sich, dass diese einfache phänomenologische Nachbildung ohne bedeutenden Mehraufwand in die Formulierung der FDTD integriert werden kann, und dass sie für typische Querschnittsflächen gut mit Literaturwerten übereinstimmt.  

Die in dieser Arbeit geschaffene unmittelbare Korrespondenz zwischen der Datenstruktur der Finiten Differenzen und den Voxeln erlaubt die direkte Übernahme verschiedener tomographischer Untersuchungsmethoden. Zur Untersuchung der Akustik des Nasaltrakts werden in dieser Arbeit drei verschiedene tomographische Verfahren miteinander verglichen: Kryosektionen, Kernspinresonanztomographie und die röntgenabsorptionsbasierte Computertomographie. Die Computertomographie zeichnet sich dabei als das mit Abstand vorteilhafteste Verfahren hinsichtlich Auflösung und Artefaktfreiheit aus, um die filigranen geometrischen Strukturen des Nasaltrakts zu erfassen. Die akustische Impedanz jedes Voxels wird aus der Röntgendichte durch eine lineare Gleichung bestimmt, womit der gesamte Informationsgehalt der Computertomographie übernommen und genutzt wird. 

Die gefundenen akustischen Eigenschaften des Nasaltrakts lassen sich mit einem Vergleich zu bereits veröffentlichten akustischen Untersuchungen verifizieren, beispielsweise anhand von Resonanzfrequenzen von Nasennebenhöhlen. Für eine über die bereits im Vorfeld dieser Arbeit durchgeführtw punktuelle Betrachtung hinausgehende akustische Untersuchung wird in dieser Arbeit ein eigenes Messverfahren entwickelt: Mit der Verwendung von kapazitiven Schallwandlern gelingt die Erzeugung präziser Schallpegel mit bekannter Quellimpedanz. Die inhärent geringen Schallpegel dieser Wandler werden mittels eines PC-basierten Messsystems ausgeglichen, das mit einem hierfür modifizierten Betriebssystem Einzelmessungen phasenstarr akkumuliert. Dieses  sehr kompakte  Messsystem (verglichen mit zuvor bekannten, Bild \ref{Lip-Imped}) liefert eine bereichsweise Übereinstimmung des Übertragungsverhaltens mit der Simulation; ein erwartungsgemäßes Ergebnis, da akustische Messung und tomographische Untersuchung nicht an den gleichen Probanden durchgeführt werden konnte.   

Zur Untersuchung und Demonstration von akustischen Sachverhalten  auch über den Nasaltrakt hinaus wurde im Rahmen dieser Arbeit das Programm {\sc Speak} geschaffen. {\sc Speak} bietet zahlreiche Signalquellen, Filtertypen und Analysemöglichkeiten, die frei kombinierbar sind. So wird gezeigt, 
wie man damit einfache und verzweigte Rohrmodelle des Sprechtrakts bilden kann, 
wie man die mittels Finiten Differenzen berechneten, Transmittanz und Impedanz des Nasaltrakts beschreibenden Impulsantworten mit einem Rohrmodell des Vokaltrakts kombinieren kann, 
und wie man ein Rohrmodell des Sprechtrakts anhand von Sprachsignalen parametrisiert. Letzteres ist insbesondere nützlich für eine Modellierung des zeitvariablen Sprechtraktbereichs, der Mundhöhle. Durch die implementierten Entropiemaximierungs- und Partial-Correlation-Methode sowie mit einem erweiterten Verfahren, welches Dämpfung und Abstrahlung des Vokaltrakts mit einbezieht, wird der relative Querschnittsverlauf eines Rohrmodells bestimmt. Aus diesem wird der absolute Querschnittsverlauf durch Skalierung anhand einer Querschnittsfläche aus einer MRI-Untersuchung gewonnen. Auf diese Weise wird zum einen ein Referenz-Rohrmodell mit bekannter Akustik, zum anderen ein geometrisches Modell für die weitere Untersuchung durch Finite-Differenzen bestimmt.

Durch den Bezug auf die Akustik einer einfachen Rohranalogie gelingt  in dieser Arbeit sowohl die wechselseitige Verifikation beider Modelle bei kleinen Querschnittsflächen als auch  eine Quantifizierung der Vorteile der dreidimensionalen Betrachtung der Schallausbreitung in der Mundhöhle bei natürlichen Querschnittsflächen. Einzelne Formanten verschieben sich durch den Übergang auf natürliche Flächeninhalte um bis zu 20 \% in der Resonanzfrequenz.
Diese Untersuchungen der Mundhöhle werden in perfekter Weise durch Ergebnisse akustischer Untersuchungen in \cite{TaMK} bestätigt: Die dort durchgeführten aufwendigen  Messungen zeigen eine ausgezeichnete Übereinstimmung bereits zu einer  einfacheren für Vokale nutzbare FDTD-Simulation.

Das in dieser Arbeit entwickelte Modell zur Beschreibung des Sprechtrakts eignet sich für die meisten Lautklassen, wie exemplarische Untersuchungen zeigen. Die  partiellen Volumen erfassen physikalisch treffend die Akustik von Strukturen des Nasaltrakts, die für die Entstehung der Nasale ([m], [n], [\ng]) und der nasalierten Vokale wesentlich sind. Die Artikulationsgeometrie und Akustik des Lateralapproximanten [l] wird eingehend betrachtet. Die gewonnenen Erkenntnisse lassen sich auf die einfachere Geometrie der Approximanten ([j] etc.) und, wie auch im Vergleich mit \cite{Mo02} deutlich wird, auf die Artikulation der Vokale übertragen. Unbenommen ist auch die Übertragbarkeit auf Diphthonge und Plosive oder -- genereller -- auf zeitvariable Vokaltraktkonfigurationen durch Verzicht auf eine zeitunabhängige Darstellung. Gerade die kleinen Zeitschritte der Zeitbereichsbetrachtung und die daraus folgenden kleinen Schritte in der Geometrieänderung lassen für diesen hier nicht näher untersuchten Fall  eine  Übereinstimmung zwischen natürlicher Akustik und Modell erwarten.

Es wird in dieser Arbeit zudem eine Reihe von effektiven Vereinfachungen gezeigt, die den Rechenaufwand und die Ausführungsgeschwindigkeit der erweiterten FDTD-Modelle verbessern. Grundlegend und essentiell sind in der Anfangsphase dieser Arbeit Optimierungen hinsichtlich einer parallelisierten Ausführbarkeit über mehrere Prozessoren hinweg und zur Reduzierung der Rechenzeit jedes einzelnen \name{Thread}. Die damit möglich gewordene Berechnung \glqq über Nacht\grqq{} beschleunigt auch die Weiterentwicklung des Modells erheblich --- mittlerweile sind mit diesen und  weiteren Fortschritten Berechnungen der Impulsantworten in wenigen Minuten möglich. 

Bedeutsamer als diese eher praktische Sicht ist für die Untersuchung der Sprechtraktakustik eine Betrachtung von Möglichkeiten, das Modell an sich zu vereinfachen. Wie in dieser Arbeit ausgeführt wird, heben diese Vereinfachungen wesentliche Merkmale der Sprechtraktakustik hervor, solange weiterhin eine gute Übereinstimmung des akustischen Verhaltens bestehen bleibt. Basierend auf diesem Kriterium zeigt sich, dass für den pharyngal-oralen Bereich Skalierung des Querschnittsflächeninhalt und Exzentrität einer elliptischen Kontur zusätzlich zu dem relativer Querschnittsverlauf wichtige Merkmale sind. Mit diesen beiden weiter Merkmalen vergrößert sich der Parametersatz, der bereits bei einem einfachen Rohrmodell typischerweise mehr als 10 Parameter für den relativen Querschnittsverlauf umfasst, nicht erheblich und das Modell bleibt gängigen Schätzverfahren zur Bestimmung dieser Parameter aus dem Sprachsignal zugänglich.
Insbesondere für eine Anwendung von iterativen Verfahren ist zudem ein weiter vereinfachtes, zweidimensionales Modell gezeigt; die damit erreichbare nochmalige Reduzierung des Berechnungsaufwandes ebnet den Weg für deren Anwendung trotz der für diese Verfahren erforderlichen mehrfachen Berechnung des Übertragungsverhaltens --- und somit für eine akustisch und geometrisch treffende Bestimmung der Vokaltraktkonfiguration aus Sprachsignalen.

Im Ergebnis erweist sich das in dieser Arbeit entwickelte Verfahren der erweiterten Finiten Differenzen im Zeitbereich für die Untersuchung der Akustik des Sprechtrakts sowohl für den Nasal-Bereich als auch für Mundhöhle und Pharynx als gut geeignet. Es ist einfach und flexibel zu handhaben, bildet mit hoher Genauigkeit die akustischen Prozesse nach und kann zur Beantwortung einer Reihe von Fragestellungen verwendet werden, wie sie in dieser Arbeit exemplarisch gezeigt sind und durch hierfür entwickelte Messmethoden untermauert werden.

\pagebreak
\part{Anhang}

\pagebreak
\selectlanguage{french}

\section[têtes parlantes de l’abbé Mical]{M M.~de Milli, le Roy, Lavoisier, Laplace, Ferrier 
et Vicq d’azir ont rendu Compte des deux têtes parlantes de 
M.~l’abbé Mical}\label{Tetes}

\lnr L’examen des machines de ce genre est curieux parceque, faites pour imiter la nature dans la prononciation des sons,
\lnr elles peuvent aussi jetter quelques yeux sur le mecanisme de la voix. Nous avons donc consideré dans le plus grand detail toutes
\lnr les pieces de la machine dont-il s’agit. M.~l’abbé Mical a eu la complaisance de les demonter devant nous. Il a même permis
\lnr que nous infissions une courte description : Il desire seulement que le compte que nous allons curendre ne soit point publié, au moins
\lnr Sans Son aveu  ; et nous avons cru pouvoir prendre cet engagement avec lui au nom de l’academie. Tous les mouvents de la
\lnr machine sont disposés de manière à faire prononcer par deux têtes, comme en dialogant les deux phrases suivantes.
\lnr Le Roi a donné la paix à L’Europe. La paix fait le bonheur des peuples. Avant d’aller plus loin, nous croyons devoir
\lnr dire que ces deux phrases ne sont pas prononcées distinctement dans toutes leurs parties ; Surtout la derniere : ce qui tient sans doute(ant)
\lnr que le sonds de la voix produite par cette machine est très different de la voix humaine ; à ce que certaines Syllabes resultant de la
\lnr combinaison de plusieurs Sons, leuer réunion ne se fait pas avec toute la précision possible ; Et aussi à ce que la prononciation de plusieures
\lnr consonnes a besoin encore d’être perfectionnée. Malgré ces defauts que M.~l’abbé Mical lui même ne se dissesseule pas, le mechanisme de
\lnr cette machine nous a paru interessant. On peut y considerer deux parties très differentes. 1\ieme{}. Une chambre à vent, dans laquelle un soufflet
\lnr porte l’air et de laquelle ce fluide s’echappe lorsque differentes soupapes s’elevent. L’air est alors dirigé par des conduits vers les cavités, 
\lnr ou il est modifié, et ou il devient fondre. 2\ieme{}. Un cylindre qui ucent des leviers, et qui leuer donne l’impulsion nécessaire, soit pour lever a
\lnr propos les soupapes de la chambre à vent, soit pour donner aux differentes cavités où le son se modifie les formes necessaires a ses diverses
\lnr changemens.  Nous decrirons sommairement chacune des parties de la machine. Le Soufflet, la chambre à vent, les soupapes qui ferment les conduits, 
\lnr et les condiuits eux-mêmes n’ont rien de particulier. Leur Structure est la même que celle que l’on observe dans les
\lnr orgues. Les cavités on [en?] boîtes dans lesquelles le son est modifié meritent une attention plus particuliere. Toutes ces boîtes sont formees
\lnr dans leuer partie inferieure par une cloison au Diaphragme très tendue, formé d’une peau très fine, située horizontalement au milieu
\lnr de laquelle est un trou  ellyptique, qui repond au conduit à vent, et se trouve place immediatement au dessus. Cette ouverture est recouvert
\lnr par une languette, dont une des extremités est attachée à un des points de la circonference de la Boîte, tandi que l’a.t.. qui depasse une
\lnr peuletruce, peut vibrer lorsque l’air du tuyau à vent est dirigé vers cette ouverture. Ce sont les vibrations de cette languette qui
\lnr produisent le son. M.~l’abbé Mical à observé que la plus ou moins grande tension de la membrane au milieu de laquelle est le trou
\lnr influe peu sur le son. Mais il n’en est pas de même de la languette vibrante, une petite plaque de metal est placé sur celles de
\lnr ses extremités qui tient á la circonference de la boîte, et peut parle moyen d’une verge être plus ou moins avancé sur cette
\lnr languette, vers le trou Ellyptique, qu’elle recouvre. Plus cette plaque de metal s’avance sur la languette qui devient alors
\lnr plus courte, plus le son qu’elle produit est aigu ; et au contraire il est d’autant plus grave, que la longeuur de la languette est plus
\lnr grande. C’est par ce moyen que que l’abbé Mical rend uniformes les differens sons de chaque boîte qui sans cela seraient
\lnr dissonans. Cette circonstance nous a paru remarquable, parce qu’elle est la seule qui puisse, dans la machine dont nous 
\lnr avons examiné le mécanisme produire des tons [sons?] differens ; tous les autres details qui nous exposerons n’a[...] etes destiné
\lnr qu’a modifier le même ton de maniere à prononcer des sillabes mais sans changer l’inflixion de la voix. M.~l’abbé Mical
\lnr a essayé determiner ses conduits à vent, par une ouverture plus ou moins etroite, qui produisait des sons du même genre
\lnr que ceux des flûtes ou des jeux d’orges à Bizeau. Mais ces tentatives ne lui ont point réussi. Il n’a pu obtenir des sons
\lnr analogues a ceux de la voix humaine et susceptibles des modifications dont il a fait usage, que par le moyen d’une languette qui
\lnr ressemble évidemment à la plaquette qui vibre dans le jeu à Au[...] de l’orgue et dont le ton est changé comme celui de la langue
\lnr par une tension plus ou moins grande ; d’ou il resulte que si on trouve le moyen d’avancer plus ou moins et dans des proportions
\lnr determinées la plaque de métal sur la languette, on pourrer changuer de ton et faire chanter la machine, il est au moins
\lnr probable qu’elle est susceptible dace dagré de perfection. Les boîtes dans lesquelles les sons se modifient et dont nous avons decrit 
\lnr Diaphragme ou la cloison, sont construites de differente manière. 1\ieme{}. Les unes sont formées de deux moitiés a peu près égales tou..
\lnr deux concaves, arrondies et ajoustées l’une sur l’autre de manière à s’ouvrir par le moyen d’une charniere  et formant un angle
\lnr plus ou moins grand. 2\ieme{}. Les deux moities des autres boîtes sont disposées de façon que la superieure peut s’enlever tout à fait et ensuite
\lnr en contact avec l’inferieure dans laquelle est toujours le Diaphragme percé d’un trou et recouvert d’une languette.
\lnr 3\ieme{}. D’autres boîtes sont toutes d’une piece, alles ont la même forme que les précedentes qui sont ovoiides. Leur partie superieure est percée dans
\lnr quelques unes d’un seul trou rond, dans d’autres de plusieurs qui sont tous recouverts par des Soupapes. 4\ieme{}. Il y a quelques Boîtes qui
\lnr different de celles-cy, en ce qu’etant d’une seule piece, elles sont beaucoup moins elevées. Il semble que pour les former on ait tronqué
\lnr les premieres ; elles sont percées d’un trou qu’une Soupape recouvre et le diaphragme disposé comme il a été dit plusieurs fois, est tendu
\lnr le bas de especes de gadets. 5\ieme{}. Une de ces boîtes réunit le mécanisme des autres c’est à dire que la moitie supérieure peut se mou[...]
\lnr dessus de l’inferieure en formernt un angle plus ou moins ouvert et étant retenue par une charuiore tandis que la region la plus elevée de
\lnr cette moitie est percée d’un trou qu’une Soupape recouvre. 6\ieme{}. Enfin l’interieure de ces boîtes vû au-dessous du Diaphragme, n’est pas b[...]
\lnr dans chacune et les variations contribuent encore à modifier le son. Cette premiere partie de la machine composée de la chambre à vent,
\lnr des conduits et des gadets on Boîtes sonores, etait la plus importante à considerent Leur donner une idée de la seconde partie qui est
\lnr composée d’un cylindre et de leviers, il suffira de dire que les leviers [...] par le cylindre paraissent être divisés en trois ordres. Les uns
\lnr levent et baissent les Soupapes de la chambre à vent ; les autres recouvent les moities superieures des boîtes sonores ou les Soupapes qui
\lnr recouvrent leurs ouvertures ; les troisiemes enfin repondent aux têtes et ne contribuent en rien aux sons. Nous [...],  l’academie
\lnr a portée de juger de ces pieces en lui exposant leur [...] dans la prononciation de quelques lettres ou Syllabes.  Nous avons
\lnr choisi celles que l’on entend de la maniere la plus distincté. 1\ieme{}. La.. A se pronence dans une des grandes boîtes (1) [...]
\lnr de deux moities mobiles l’une sur l’autre pour que l’on entende cette lettre il faut que la boîte restent immobile, sa moitie[? ...]
\lnr supérieure etant demeurant ouverte a 40 dégrès à peu près.

\noindent\setcounter{linenumbercounter}{0}
\lnr Le son de la lettra a dans la prononciation naturelle résulte d’une disposition analogue, pendant que la ..ngue est
\lnr fixée dans le fonds de la bouche, son dos se relevant un peu, les deux machoires sont et demeurent ouvertes tant que l’on entend
\lnr le même son. 2\ieme{}. La lettre o se modifie dans une boîte de la même grandeur et de la même forme que la lettre a, avec cette
\lnr difference que la moitie superieure n’est point mobile, mais seulement percée d’une ouverture ronde (2). en effet lorsqu’on
\lnr prononce la lettre a, si on retrecit l’ouverture de la bouche, sans changer la situation de la langue le son o se fait
\lnr entendre au lieu du premier. 3\ieme{}. L’ouverture de la bouche lorsque l’on prononce la lettre e tient le milieu entre celles requises par
\lnr la lettre a et pour la lettre o ; aussi le vase dans lequel (I) la lettre e se fait entendre, at-il une ouverture plus grande que
\lnr celui de la’article précedent et plus petite que celui de la lettre a dont-il diffère encore en ce qu’il n’y a point de partie superieure
\lnr detailée et mobile, et en ce qu’il est en total plus court que les deux premiers. La proportion de ces ouvertures est d’accord avec
\lnr celles observées et determinées par M.~Kratzenstein qui a remporté le prix de l’academie de Petersbourg en 1781. Sur un
\lnr sujet analogue (page 15 de ce Memoire). 4\ieme{}. Il est facile de prononcer la avec la Boîte destinée à la voyelle  a. Il suffit pour
\lnr cela que l’air partoant parle conduit et soulevant la languette mobile un moment avant qu’aucun autre mouvement s’execut
\lnr ce que M.~l’abbé Mical appelle préparation ; la moitie superieure s’eleve (d) et passe un angle avec l’inferieure. L’angle
\lnr etant de 25 degrès le son est distinct et il devient plus net encore ; si l’ouverture est plus grande ; pour prononcer la avec la
\lnr bouche, on ouvre de même cette cavité et l’on écarte les machoires. 5\ieme{}. La Syllabe pe se forme dans un vase court qui n’a
\lnr qu’une ouverture couverte d’une Soupape. (I). Il faut aussi une préparation. L’air est poussé vers la cavité du vase ; la
\lnr Soupape s’eleve prusquement (2) et le cours de l’air est interrompu dans le conduit à vent. On fait absolument la même chose avec
\lnr la bouche lorsque l’ouvert prononce cette Syllabe. 6\ieme{} La Syllabe fai s’entend aussi très bien. M.~l’abbé Mical à employè un moyen
\lnr particulier pour sa formation. Il se sert d’une boîte divisée en deux moities articulées par une charniere (3) mais l’extremité du levier  très
\lnr mince souleve la languette vibrante. L’air passe et produit un sifflement ; alors la languette cessant d’être soulevé retombe et vibre. La
\lnr partie supérieure du vase s’ouvre et l’on entend fa, fè ou fai. Suivant les degrès d’ouverture. sans le sifflement produit par le mécanisme
\lnr dont nous avons parlé, on aurait a ou e ou la ou le suivant que les deux parties du vase aurai enteté ouvertes ou qu’elles se 
\lnr écartées dans le moment de la prononciation. En reflechissant sur la maniere dont on prononce pai, il est facile de se convaincre que
\lnr cette syllabe est en effet composée de ai et d’une sifflement. C’est ce que M.~l’abbé Mical à executé. 8\ieme{}. Oa se prononce en deux tems
\lnr dans un vase (i) dont les deux moities sont articulées, la moitie supérieure etant de plus percée d’un trou rond, recouvert d’une soupape.
\lnr cette soupape se leve et on entend la lettre O ; alors la motie supérieure du vase se leve. sur l’inférieure et on entend la lettre A. ces deux
\lnr neanmoins se sucéedent avec rapidité ; l’oreille entend oa. la bouche dans cette prononciation fait apeuprès la même chose. 9\ieme{}. Nous
\lnr terminerons en détails en exposant comment M.~l’abbé Mical fait prononcer la lettre R par la machine que nous avons éxaminés
\lnr au dessus de la chambre à vent est une cavité prticuliere ou une languette plus forte que celle des Boîtes sonores est mise en resonance[?]
\lnr par une Colonne d’air. Elle vibre et produit des frémissemens ou battemens qui expriment le son de la lettre R et suivant que telle
\lnr et telle Boîte sonore joue en même tems on en obtient le son RA, RO ye la langue produit dans la bouche le même effet que le
\lnr mécanisme emploié par M.~l’abbé Mical. Nous en avons dit assez pour faire voier qu’il a toujours cherché à imiter la nature et 
\lnr c’est sans ce rapport que son travail nous a paru si interessant. La chambre à vent fait l’office des poumons ; le conduit a
\lnr veut fait celui de la trachée artère ; le trou de la cloison repond à la glotte, la cloison et la [...]me vibrante aux lèvres de la glotte
\lnr et aux ligamens  du Larianx ; la cavité de la boîte sonore doit être comparée aux fosses nazales,
\lnr palatines et Buccales et les differentes ouvertures de la Boîte à celles de la bouche elle-même, comme nous l’avons
\lnr en parlant de chaque son en particulier. Nous pensons que l’academie doit applaudier aux efforts de M.~l’abbé M.~que 
\lnr sa machine est ingénieuse, que ses travaux meritent d’être encouragés et que son essai quoi qu’imparfait encor
\lnr est très digne de l’approbation de l’Academie.

\vfill
\noindent
{\tiny Aus \cite{Vi1783, Lü10}, ergänzt um Zeilennummern}
\pagebreak

\selectlanguage{german}

\section{${\cal Z}$-Transformation}\label{ZT}
In diesem Abschnitt wird die ${\cal Z}$-Trans\-formation vorgestellt, die verwendeten Eigenschaften gezeigt und kurz erörtert. Abschließend wird ihre Anwendung auf lineare Differenzengleichungen mit konstanten Koeffizienten skizziert, die häufigen zeitdiskreten Systemen\footnote{Unter einem  zeitdiskreten  System wird hier ein Tupel aus dem Zustand $s\in \mathbb K^n$ mit dem Anfangswert $i\in \mathbb K^n$, dem Eingangswert $x\in \mathbb K^m$, dem Ausgangswert $y\in \mathbb K^l$, der Abbildung $A\in \mathbb K^{n+m}\negthickspace\to\negthickspace \mathbb K^{n+l}$ verstanden, wobei  initial $s=i$ und die folgenden Zeitschritte mittels $(s,y)=A(s,x)$ gebildet werden; $n,m,l \in \mathbb N$.  } entsprechen.

Die ${\cal Z}$-Transformation\footnote{
Die ${\cal Z}$-Transformation wurde 1952 zur Analyse abtastender Systeme als eine Spezialisierung der Laplace-Transformation $\cal{L}\{\cdot\}$ vorgestellt, $\trans{R}(z)={\cal L}\{ r(t)\sum_{n=-\infty}^\infty\delta(t-nT)\}$ mit dem Zeitsignal $r(t)$, vgl.~\cite{RaZ52}. Die Autoren weisen darauf hin, dass bereits in \cite{Hu47} die gleiche Transformation, wenngleich namenlos und  über \name{erzeugende Funktionen} hergeleitet, zur Stabilitätsanalyse zeitdiskreter Systeme gezeigt ist; sie irren aber darin, die Einführung dieser Analyse von Differenzengleichungen Laplace zuzuschreiben, vgl.~\cite{Mo1730}. 
}  
ist eine lineare Abbildung einer reell- oder komplexwertigen Folge\footnote{Die Beschränkung auf Folgen ${\mathbb N} \rightarrow {\mathbb R} $ ist zulässig, da die in dieser Arbeit durchgeführten physikalischen Betrachtungen einen Anfang haben und kausal sind. Eine weitere Beschränkung auf endliche Folgen, wie sie sich durch definitionsgemäß endliche Messung begründen liese, würde jedoch einige Aspekte verbergen. } $x$ auf eine Funktion über der komplexen Ebene. Eine übliche Definition\cite{RVí64} ist
\[
\trans{X}={\cal Z}\{x\}\quad \Leftrightarrow \quad\trans{X}(z) = \sum_{k=0}^{\infty} x_kz^{-k}\;,\;\;  z\in{\mathbb C}\;,
\]
wobei die Bildfunktion durch eine Majuskel gekennzeichnet wird. Die ${\cal Z}$-Transformation enthält die zeitdiskrete Fouriertransformation der Folge, \linebreak $\trans{X}(\omega)=\sum_{k=0}^{\infty} x_ke^{-i\omega k}$, $\omega\in{\mathbb R}$, als Spezialfall: Die Reihe ergibt sich für $z=e^{i\omega}$, also als Funktionswerte von $\trans{X}(z)$ auf dem Rand des Einheitskreises der komplexen Bildebene.
Mit der  ${\cal Z}$-Transformation lässt sich eine größere Menge von Folgen betrachten. Damit die Fouriertransformation auf einem Gebiet konvergiert, muss die Folge $x$ den Grenzwert null haben. Die ${\cal Z}$-Transformation konvergiert bereits, wenn $x$ durch eine Exponentialfolge $a^k$ majorisiert wird --- die Konvergenz erfolgt im Gebiet $|z| > a$. 

Beiden Transformationen gemein ist der \name{Faltungssatz}. Wird eine Folge $y_k$ durch die Faltung der Folgen $w_k$ und $x_k$ bestimmt, so ergibt sich die \mbox{${\cal Z}$-Transformierte} aus dem Produkt im Bildbereich:
\[
\trans{Y}(z)=\trans{W}(z)\trans{X}(z)\;.
\]  

Eine in dieser Arbeit nützliche Eigenschaft ist der \name{Verschiebungssatz}. Eine Folge $y_k=x_{k+n}$ die um $n$ Glieder gegenüber der Folge $x$ verschoben ist, hat die Bildfunktion  
\[
\trans{Y}(z)=\sum_{k=0}^{\infty} y_{k}z^{-k}= \sum_{k=0}^{\infty} x_{k+n}z^{-k} = \sum_{k=0}^{\infty} x_{k}z^{-k+n} = z^n\trans{X}(z)\;,
\]
vereinfacht unter der Annahme, dass nicht in beiden Folgen enthaltene Glieder gleich null sind. Eine Verschiebung der Folge entspricht also im Bildbereich der Multiplikation mit einer entsprechenden Potenz von $z$.\footnote{ Der Verschiebungsatz ist nicht auf $n\in \mathbb N$ beschränkt: eine Erweiterung auf $\mathbb Q$ ergibt sich über ein kürzeres Abtastintervall gefolgt von einer ganzzahligen Unterabtastung.  }

Damit können lineare Differenzengleichungen mit konstanten Koeffizienten $a_i$  untersucht werden. Diese Gleichungen lassen sich in der Form\footnote{Der Name \glqq Differenzengleichung\grqq{} stammt von der Darstellung mittels Differenzenoperators $\triangle y = y_{k+1}-y_k$ und dessen Potenzen in Form wiederholter Anwendung, die zu der hier gezeigten Form äquivalent ist, vgl.{} englische Ausgabe von \cite{RVí64}. } 
\[
\sum_{n=0}^N a_n y_{k-n} = w_k
\]
darstellen. Die Folge $y$ gibt die Entwicklung der Differenzengleichung wieder, und die Folge $w$ bestimmt, ob es sich aufgrund $w_k=0\; \forall\; k\in{\mathbb N}$ um eine homogene oder andernfalls inhomogene Differenzengleichung handelt. Durch die ${\cal Z}$-Transformation gewinnt man
\[
\sum_{n=0}^N a_n z^{-n} \trans{Y} = \trans{W}\;.
\]
Einsichten über die Eigenschaften der Differenzengleichung gewinnt man anhand der Eigenwerte $\lambda_i$, die Nullstellen des \name{charakteristischen Polynoms}\linebreak $\sum_{n=0}^N a_n z^{N-n}$ sind, und den dazugehörigen Eigenfolgen $\lambda_i^k$.
\footnote{Der Zusammenhang wird durch Darstellung der homogenen Differenzengleichung \mbox{$N$-ter} Ordnung (mit $a_0 =1$) als System von Differenzengleichungen 1.~Ordnung offensichtlich:
\[
\underline{y_k}= \underline{\underline{A}}\,\underline{y_{k-1}}\;\; \mathrm{mit}\;\;
\underline{y_k}=
\begin{pmatrix}
y_k \\
y_{k-1} \\
y_{k-2} \\
\vdots  \\
y_{k-N+1}
\end{pmatrix}\;,\;\;
\underline{\underline{A}}=
\begin{pmatrix}
-a_1 & -a_2 & \cdots  & -a_{N-1} & -a_{N}\\
1 & 0 & \cdots  & 0 & 0\\
0 & 1 & \cdots  & 0 & 0\\
\vdots & \vdots & \ddots & \vdots & \vdots \\
0 & 0 & \cdots  & 1 & 0
\end{pmatrix}\;.
\]
Dieses System führt mit dem Ansatz $\underline{y_k}=\lambda\, \underline{y_{k-1}}$, $\lambda \in{\mathbb C}$  zu der Eigenwertgleichung $\lambda\,\underline{y_{k-1}}= \underline{\underline{A}}\,\underline{y_{k-1}}$. Die Lösungen der Eigenwertgleichung,  die Eigenwerte $\lambda_i$,  ergeben sich aus der charakteristischen Funktion  $\det(\underline{\underline{A}}-\lambda_i\underline{\underline{E}}) = \sum_{n=0}^N a_n \lambda_i^{N-n} = 0 $. Die dazugehörigen Eigenfolgen haben aufgrund des Ansatzes die Gestalt $y_k=\lambda_i^k$. Ausgenommen hiervon sind Nullstellen der charakteristischen Funktion im Ursprung; um diese Spektralwerte zu erfassen, muss der Eigenwertbegriff erweitert werden.  
} 
Die Eigenwerte lassen sich in Polarkoordinaten $re^{-i\omega}$ darstellen;  hierbei ist $\omega$ die Eigen- oder Resonanzfrequenz und $r=|\lambda|$ ein Maß der Resonanzgüte und der Stabilität. 
Für komplexwertige Eigenwerte  ergeben sich reelle Eigenfolgen aus der Überlagerung mit dem konjugierten Eigenwert $y_k = (re^{i\omega})^k \pm (re^{-i\omega})^k = 2r^k\sin(\omega k)$ und $2r^k\cos(\omega k)$. Ist $r>1$ nimmt die Amplitude der Schwingung exponentiell zu, das System ist instabil. Für $r=1$ erhält man eine ungedämpfte, für  $r<1$ eine gedämpfte Schwingung.  

Für die Betrachtung inhomogener Differenzengleichungen in Form von linearen zeitinvarianten zeitdiskreten Systemen wird die Differenzengleichung häufig dahingehend umformuliert, dass auch die Folge $w$ durch eine Differenzengleichung aus einer Eingangsfolge $x$ gebildet wird. 
Die Differenzengleichung und deren ${\cal Z}$-Transformation haben dann die Form
\[
\sum_{n=0}^N a_n y_{k-n} = \sum_{m=0}^M b_m x_{k-m} \;\;\overset{\cal Z}\longrightarrow\;\;\sum_{n=0}^N a_n \trans{Y}(z)z^{-n} = \sum_{m=0}^M b_m \trans{X}(z)z^{-m}\;\;,
\]
wobei die Koeffizienten $b_m$ der Eingangsfolge zugeordnet sind.
Die Übertragungseigenschaften H dieses Systems lassen sich anhand des Verhältnisses der Bildfunktionen von  Ausgangs- zur Eingangsfolge bestimmen:
\[
\trans{H}(z)= \frac{\trans{Y}(z)}{\vphantom{\sum^N}\trans{X}(z)}=\frac{\sum_{m=0}^M b_m z^{-m}}{\sum_{n=0}^N a_n z^{-n}}
\]  
Insbesondere ist die Impulsantwort -- wenn also zur Anregung als Eingangsfolge die Einheitsimpulsfolge  $\{1, 0, 0, ... \}$  verwendet wird, deren Bildfunktion $\trans{X}(z)=1$ ist --  des Systems gleich der Übertragungsfunktion.

Damit ist die Analyse dieser Systeme auf die Analyse einer rationalen Funktion zurückgeführt. Diese rationale Funktion lässt sich, abgesehen von einem Faktor, durch Produkte ihrer Pole und Nullstellen darstellen. Dies sind die Nullstellen des Nenner- und des Zählerpolynoms, erstere folglich die bereits diskutierten Eigenwerte. Die Faktoren des Zählerpolynoms, $(1-n_iz^{-1})$, sind im Zeitbereich gewichtete gleitende Mittlungen, $y_k-n_iy_{k-1}$, die Signale oder Signalkomponenten des Typs $y_k=n_i^k$ auslöschen.

\clearpage

\pagebreak

\vfill
\noindent
Anmerkung:
Die Referenzen sind bei selteneren Werken mit einer Angabe zu einer Bezugmöglichkeit versehen.

\pagebreak
\section*{Danksagung}
\addcontentsline{toc}{part}{
Danksagung}\label{Dank}
\markboth {}{}%

Die Vielseitigkeit in der Thematik dieser Arbeit, die selbst bei der Betrachtung nur einer Schicht der Sprachentstehung verblieben ist, wäre in dieser  Breite und Tiefe nicht ohne ein Mitwirken vieler bewerkstelligbar gewesen, für das sich der Autor an dieser Stelle bedanken möchte:

Der Autor dankt Herrn Prof.{} Lacroix für die Aufnahme in die Arbeitsgruppe und für die Betreuung der Dissertation. Für die vielzähligen konstruktiven Gespräche und das angenehme Umfeld in der Arbeitsgruppe bedankt sich der Autor bei den Herren Karl Schnell, Ralf Thomas Pietsch und Martin Eichler; ein besonderer Dank gilt darüber hinaus Herrn Christian Lüke, der für diese Arbeit den Bericht über die \name{Têtes Parlantes} entziffert und übersetzt hat und Frau Hermine Reichau, die mit Hinweisen zur sprachübergreifenden Grammatik zu dieser Arbeit beigetragen haben. Ebenso dankt der Autor den Herrn Alexander Weber, Hals-Nasen-Ohrenklinik, Johann Wolfgang Goethe-Universität, für die Unterstützung bei der Interpretation der Kryosektionen; Herrn Bernd Turowski für die Anfertigung des CT-Datensatzes und  die Einweisung und Bereitstellung des \name{Easy Vision} zur Analyse und Aufarbeitung des Datensatzes im Institut für Neuroradiologie, Klinikum der Johann Wolfgang Goethe-Universität; Herrn Nasredin Abolmaali für die Durchführung der MRT-Untersuchung im Institut für Diagnostische und Interventionelle Radiologie, Johann Wolfgang Goethe-Universität. NMRWin wurde freundlicherweise von Herrn Wang, Mitarbeiter des Instituts für Computer Graphik der Fraunhofer Gesellschaft, zur Verfügung gestellt.
Der Autor dankt Frau Karolina Ostapkowicz, Frau Isolde Asbeck und Herrn Martin Eichler für das Korrekturlesen der Arbeit.

Für die außerhalb des wissenschaftlichen Umfeldes erhaltene Unterstützung dankt der Autor Herrn Reinald Pasedag und Herrn Roland Pasedag, die ihm in ihrer Firma den Freiraum   zur Vollendung diese Arbeit gegeben haben. Ein umfassender Dank des Autors gilt seiner Familie und Sandra Wegener, die ihn -- nicht nur während dieser Arbeit -- fortwährend unterstützt haben und den Familien von der Heyden, Wojdyno und Heine für das motivierende Interesse an dieser Arbeit.

\arxiv{
\pagebreak
\section*{Lebenslauf}
\addcontentsline{toc}{part}{
Lebenslauf}
\markboth {}{}%

\noindent
Frank Ranostaj\\
geboren am 23.~Oktober 1970 in Frankfurt am  Main\\ \\
\noindent
\begin{tabular}{ll}
1977-1981& Grundschule Kalbach, Frankfurt am Main\\
1981-1990& Ziehen-Gymnasium, Frankfurt am Main\\
         & Abschluss: Abitur\\
1990-1992& Zivildienst\\
1992-1999& Studium der Physik an der \\
         & Johann Wolfgang Goethe-Universität, Frankfurt am Main\\
         & Abschluss: Diplom Physiker\\
2000-2005& Wiss.~Mitarbeiter im Institut für Angewandte Physik,\\ 
         & Johann Wolfgang Goethe-Universität, Frankfurt am Main\\
2005-    & Entwicklungsleiter, Firma RP-Technik, Rodgau
\end{tabular}\vspace{5ex}

\noindent
Akademische Lehrer:\\[2ex]
\noindent
\begin{tabular}{ll}
Prof.~Dr.~K.~Bethge&
Prof.~Dr.~R.~Bieri\\
Prof.~Dr.~F.~Constantinescu&
Prof.~Dr.~Th.~Elze\\
Prof.~Dr.~W.~Greiner&
Prof.~Dr.~R.~Kemp\\
Prof.~Dr.~J.~Kummer&
Prof.~Dr.-Ing.~A.~Lacroix\\
Prof.~Dr.~W.~Martienssen&	
Prof.~Dr.-Ing.~R.~Mester\\
Prof.~Dr.~E.~Mohler&
Prof.~Dr.~M.~Reichert-Hahn\\
Prof.~Dr.~A.~Schaarschmidt &
Prof.~Dr.~G.~Schnitger\\
Prof.~Dr.~K.~Waldschmidt&
Prof.~Dr.~D.~Wolf
\end{tabular}
}

\pagebreak
\addcontentsline{toc}{part}{Nachtrag}
\part*{Nachtrag}
\section*{Beiträge von Galen und Mersenne.}

Nach dem Verfassen dieser Arbeit ist der Autor auf weitere Dokumente gestoßen, mit denen sich ein zusammenhängenderes Bild\footnote{Eine weitere Untersuchung, etwa zeitgleich zu Kempelen, Mical und Kratzenstein, vgl. Abschnitt~\ref{mech}~-- hier verdeutlichend, dass die Suche nach Sprechmaschinen seinerzeit halb Europa überspannt hat -- stammt von Erasmus Darwin. Den daraus resultierenden Aufbau, der Silben mit Plosiv, Vokal und Nasal hervorbringt,  beschreibt er in \cite{Da1803}, Note XV kurz:

[...] {\it I contrived a wooden mouth with lips of soft leather, and with a valve over the back part of it for nostrils, both which could be quickly opened or closed by the pressure of the fingers, the vocality was given by a silk ribbon about an inch long and a quarter of an inch wide stretched between two bits of smooth wood a little hollowed; so that when a gentle current of air from bellows was blown on the edge of the ribbon, it gave an agreeable tone, as it vibrated between the wooden sides, much like a human voice. This head pronounced the p, b, m, and the vowel a, with so great nicety as to deceive all who heard it unseen, when it pronounced the words mama, papa, map, and pam; and had a most plaintive tone, when the lips were gradually closed.} } 
der Entwicklungsgeschichte ergibt, und die deshalb nicht unerwähnt bleiben sollen. 

Bereits in der Antike hat Galen gemäß \cite{Ba62, Ma68} den Larynx detailiert beschrieben und dabei  ausgeführt, dass die Stimme im Kehlkopf entstehe und dafür die Glottis  notwendig sei. Die Bezeichnung Glottis überträgt Galen  dabei  von dem Rohrblatt-Mundstück des Blasinstruments {\it Aulos} auf das menschliche Organ aufgrund der Ähnlichkeit in der Gestalt und der Unabdingbarkeit beiderorts zur Tonentstehung. Er erkennt, dass die Stimme durch Nerven und Muskeln gesteuert werde, die auch die Öffnung der Glottis formen und so die Tonhöhe bestimmen. Der Ton werde erst nachfolgend, vor allem durch die Zunge zur Sprache geformt. Zur Akustik indes ergeben Galens überlieferte Ausführungen keine zusammenhängende Darstellung, die sich mit der modernen Vorstellung deckt: Er beschreibt zwar zutreffend, dass bei der Stimmentstehung die Membranen des Kehlkopfes durch einen Luftstoß geöffnet werden können und sich dann sofort wieder schließen, aber an anderer Stelle beispielsweise, dass der knorpelige Teil der Trachea die Stimme vorbereite/hervorbrächte und zum hörbaren Anschlagen der Luft durch die Knorpel ihre genaue Härte ganz wesentlich wäre.

Als im 17.{} Jahrhundert Marin Mersenne zur Stimmentstehung recherchiert, erhält er von Pierre Trichet in \cite{Tr1631} Hinweise auf diese Untersuchungen von Galen und auf funktionelle Ähnlichkeiten zwischen den Elementen des menschlichen Sprechtraktes und denen einer Orgelpfeife. Zur Lokalisierung der Tonentstehung zieht Trichet einen Vergleich zwischen der Larynx und und der Stimmzunge ({\it languette}) einer Lingualpfeife, während er deren Resonator dann dem Mundraum zuordnet. Ein weiterer Korrespondent, Christophe Villiers, teilt in \cite{Vi1633} diese Ansicht und schreibt Mersenne, welche Parallelen er zwischen dem Sprechtrakt des Menschen und dem Mechanismus einer Orgel sieht, insbesondere des bereits in dieser Zeit verbreiteten Registers {\it vox humana}:
\begin{quote} 
{[...] \it\selectlanguage{french} de tous les instruments nul n'aproche de si prez les organes  des la voix de l'homme que l'orgue qui a, ce semble, les souflets pour poulmon, le porte-vent pour trachea artere, et pour le larinx, glotte, epiglotte et cavité depuis iceux jusqu'au palais, le tuyau de l'orgue et ses partyes, en sorte mesme que de cette analogie, je conclurois l'orgue plus antienne que pas un autre instrument, n'ayant esté fait sur d'autre prototype que celuy des partyes dediees à la} voix humaine. \selectlanguage{german}
\end{quote}

Marin Mersenne greift Galens Untersuchungen in Teilen auf und stellt sie im Jahr 1636 in \cite{Me1636} anderen Überlegungen\footnote{Von Hippokrates, Aristoteles und weitern, s.~a.~\cite{Li1846}, wie dem Vergleich des Sprechtrakts mit einer Flöte, bei dem die Trachea dem Resonator und der Larynx dem Mundstück entsprechen sollen, vielleicht motiviert durch die Beobachtung, dass ein großer Mann (im Mittel) tiefer spricht als ein kleines Kind. Mersenne entzog dabei dem Vergleich jegliche Plausibilität, da für die Tonhöhenunterschiede eines Sprechers von mehreren Oktaven keine entsprechende Längenveränderung der Trachea denkbar ist.} gegenüber. Dabei favorisiert Mersenne die Rohrblattpfeifen-Analogie zur Sprachschallentstehung, da bei einer Rohrblattpfeife nur durch Veränderung der Stimmung der große Grundfrequenzbereich der menschlichen Stimme realisiert werden kann. Ein weiteres Argument für die Analogie sieht er in der klanglichen Ähnlichkeit der Stimme zu dem auf  Rohrblattpfeifen basierenden Orgelregister {\it vox humana}.
Er schließt in Proposition~XVI dann, dass die Luft von den Schwingungen des Rohrblattes bzw.~der Glottis bewegt werde, und folglich die Glottisschwingung[sfrequenz] der Tonhöhe entspreche:\footnote{Die Glottisschwingung wurde knapp 100 Jahre später erneut von Denis Dodart und kurz darauf von Antoine Ferrein untersucht, und häufig wird ihnen die Entdeckung zugeschrieben. Einzig Ferrein zieht wie Mersenne einen Vergleich zwischen Stimmlippen und Saiteninstrumenten und hebt dann darüber hinaus hervor, dass die Spannung der Stimmlippen maßgeblich für die Tonhöhe ist, vgl. \cite{Fe1741, Ke1791, Ge94}. }
\begin{quote} 
{\it\selectlanguage{french} 
Il faut donc conclurre que l'air ou le vent doit trembler, ou se mouuoir autant de fois que la chorde d'vn Luth, ou la languette du larynx ou des flustes, pour faire vn bruit Vnisson à ladite chorde, \& consequemment que la petit tambour, c'est à dire la membrane de l'oreille, doit estre frappé autant de fois par sortes de bruits Vnissions.
\selectlanguage{german}}
\end{quote}

Im Geist des Experimentalismus und der beginnenden Aufklärung untersucht Mersenne in \cite{Me1635}, Pro\-positio~XXXVI, wie sich aus dem Ton der Rohrblattpfeifen [die  Klänge der verschiedenen] Vokale formen lassen. Basierend auf Analysen der Zungen- und Lippenstellung beim Sprechen von Vokalen bildet er an einer Pfeife die Artikulatoren  mit den Händen nach. Damit gelingt es ihm, den Laut e der {\it vox humana~}in die Laute u, indem er die Hände konusförmig vor die Öffnung hält, o, die Öffnung trompetenförmig ausbildet und die Pfeife mit drei- oder vier Löchern versieht, und a, indem er die Pfeife an der Öffnung weitet und gleichermaßen locht, zu verwandeln. Er überlegt abschließend, dass man für Silben eine Vielzahl von Pfeifen bräuchte, wobei jede mit unterschiedlichen zusätzlichen Mechanismen  die verschiedenen an- und auslautende Konsonanten der zugehörigen Silbe reproduziert.\footnote{Dies zu realisieren gelang im gewissen Umfang Ende des 18.~Jahrhunderts Abbé Mical mit seinen {\it tetes parlantes}, s.~S.~\pageref{mical}.}

Ein Jahrhundert später fasst Leonhard Euler in seinen {\it  Lettres à une princesse d'Allemagne\footnote{Friederike Charlotte von Brandenburg-Schwedt} sur divers sujets de physique et de philosophie} den damaligen Kenntnisstand auf diesen Gebieten zusammen. Der 137.~Brief widmet sich dem Schall und der Sprachentstehung. Euler führt aus, dass die Vokale \glqq a, e, i, o, u\grqq~nur durch unterschiedliche Gestalt der Mundhöhlung artikuliert werden und erklärt die Bedeutung der Nasaltrakts für die Nasale m und n. Der Brief schließt mit der wiederum auf das Orgelregister {\it vox humana} gestützten Überlegung, dass die Konstruktion einer des Sprechens mächtigen Maschine möglich und bewundernswert wäre. 

Nach Eulers Rückkehr zur Sankt Petersburger Akademie wurde dort die Bestätigung der Vokalartikulation mit einem  Preisgeld ausgelobt, und wie in Abschnitt \ref{mech} ausgeführt, von Kratzenstein erneut nachgewiesen. 

\pagebreak

\end{document}